\documentclass[useAMS,usenatbib]{mn2e}
\usepackage{times,mathptm}
\usepackage{lscape}
\usepackage{psfig}
\usepackage{subfigure}
\usepackage{multirow}

\usepackage{psfig}
\usepackage{times}
\usepackage{amsmath}
\usepackage{amssymb}
\usepackage{graphicx}
\usepackage{graphics}
\usepackage{rotating}
\usepackage{lscape}
\usepackage{morefloats}
\usepackage{hyperref}

\def\apj{ApJ}
\def\apjl{ApJL}
\def\mnras{MNRAS}
\def\pasp{PASP}

\def\araa{ARAA}
\def\aap{A\&A}
\def\aj{AJ}
\def\apjs{APJS}

\def\nat{Nature}
\def\pasj{PASJ}
\def\nar{New Astron. Rev.}

\def\gs{\mathrel{\raise0.35ex\hbox{$\scriptstyle >$}\kern-0.6em
\lower0.40ex\hbox{{$\scriptstyle \sim$}}}}
\def\ls{\mathrel{\raise0.35ex\hbox{$\scriptstyle <$}\kern-0.6em
\lower0.40ex\hbox{{$\scriptstyle \sim$}}}}

\def\kms{\,\hbox{km\,s$^{-1}$}}
\def\Msol{\,\hbox{M$_{\odot}$}}
\def\Lsol{\,\hbox{L$_{\odot}$}}
\def\Msolyr{\,\hbox{M$_{\odot}$\,yr$^{-1}$}}
\def\Wm2{\,\hbox{W}\,\hbox{m}^{-2}}
\def\gsim{\mathrel{\raise0.35ex\hbox{$\scriptstyle >$}\kern-0.6em\lower0.40ex\hbox{{$\scriptstyle \sim$}}}}
\def\lsim{\mathrel{\raise0.35ex\hbox{$\scriptstyle <$}\kern-0.6em\lower0.40ex\hbox{{$\scriptstyle \sim$}}}}
\def\ltsima{$\; \buildrel < \over \sim \;$}
\def\simlt{\lower.5ex\hbox{\ltsima}}
\def\gtsima{$\; \buildrel > \over \sim \;$}
\def\simgt{\lower.5ex\hbox{\gtsima}}

\def\mum{$\mu$m}

\begin{document}

\title[The prevalence of kpc-scale outflows among AGN]{
Kiloparsec-scale outflows are prevalent among luminous AGN: outflows
and feedback in the context of the overall AGN population}

\author[C.\ M.\ Harrison et al.]
{ \parbox[h]{\textwidth}{ 
C.\ M.\ Harrison,$^{\! 1\, *}$
D.\ M.\ Alexander,$^{\! 1}$
J.\ R.\ Mullaney$^{\! 1,2}$
and A.\ M.\ Swinbank$^{\! 3}$
}
\vspace*{6pt} \\
$^1$Department of Physics, Durham University, South Road, Durham, DH1
3LE, UK \\
$^2$Department of Physics \& Astronomy, University of Sheffield,
Sheffield, S3 7RH, UK\\
$^3$Institute for Computational Cosmology, Durham University, South
Road, Durham, DH1 3LE, UK \\
$^*$Email: c.m.harrison@durham.ac.uk \\
}
\maketitle
\begin{abstract}
We present integral field unit (IFU) observations covering the
  [O~{\sc iii}]$\lambda\lambda4959,5007$ and H$\beta$ emission lines of
sixteen $z<0.2$ type~2 active galactic nuclei (AGN). Our targets are selected from a well-constrained
parent sample of $\approx24,000$ AGN so that we can place our observations into the context of the overall
AGN population. Our targets are radio-quiet with star formation rates
($\lesssim$[10--100]\,\Msolyr) that are consistent with normal
star-forming galaxies. We decouple the kinematics of galaxy dynamics
and mergers from outflows. We find high-velocity ionised gas (velocity widths
$\approx600$--1500\kms; maximum velocities $\le1700$\,km\,s$^{-1}$)
with observed spatial extents of $\gtrsim$\,(6--16)\,kpc in all targets and observe signatures of spherical outflows and
bi-polar superbubbles. We show that our targets are representative of $z<0.2$, luminous (i.e.,
$L_{{\rm [O~III]}}>10^{41.7}$\,erg\,s$^{-1}$) type~2 AGN and that ionised outflows are not only common but also in $\ge$70\% (3$\sigma$ confidence)
of cases, they are extended over kiloparsec scales. Our
study demonstrates that galaxy-wide energetic outflows are not
confined to the most extreme star-forming galaxies or radio-luminous AGN; however, there may be a higher incidence of the most extreme outflow
velocities in quasars hosted in ultra-luminous infrared galaxies. Both star formation and AGN activity appear to be
energetically viable to drive the outflows and we find no definitive evidence
that favours one process over the other. Although highly uncertain, we derive
mass outflow rates (typically $\approx$10$\times$ the SFRs), kinetic energies ($\approx0.5$--10\% of $L_{{\rm AGN}}$) and momentum rates (typically $\gtrsim10$--$20\times L_{{\rm AGN}}/c$) consistent with theoretical models that predict AGN-driven outflows play a significant role in shaping the evolution of galaxies.
\end{abstract}

\begin{keywords}
  galaxies: active; --- galaxies: kinematics and dynamics; ---
  quasars: emission lines; --- galaxies: evolution
\end{keywords}

\section{Introduction}
One of the most remarkable discoveries in modern astronomy is that all massive galaxies host supermassive black holes (BHs) with
masses that are proportional to that of their host galaxy spheroid (e.g.,
\citealt{Kormendy95}; \citealt{Magorrian98}; \citealt{Tremaine02};
\citealt{Gultekin09}; see \citealt{Kormendy13} for a review). These
BHs primarily grow through mass accretion that is visible as active galactic nuclei (AGN) in the
centre of galaxies. Theoretical models of galaxy
formation have found it necessary to implement AGN ``feedback''
processes, during which AGN activity injects energy into the gas in the larger-scale
environment, in order to reproduce the properties of local massive
galaxies, intracluster gas and the intergalactic medium (e.g.,
BH-mass--spheroid mass relationship; the sharp
cut-off in the galaxy luminosity function; colour bi-modality; metal enchrichment; X-ray
temperature--luminosity relationship; e.g., \citealt{Silk98};
\citealt{Churazov05}; \citealt{Bower06}; \citealt{Hopkins06};
\citealt{McCarthy10}; \citealt{Gaspari11}). Placing observational
constraints on how AGN activity couples to the gas in galaxies and
halos, and where these processes are most prevalent, is an
important area of ongoing research (for reviews see: \citealt{Cattaneo09}; \citealt{Alexander12}; \citealt{Fabian12}; \citealt{McNamara12}). 

Several of the most successful galaxy formation models invoke a
dramatic form of energy injection (sometimes called the ``quasar
mode'' or ``starburst mode'') where AGN drive galaxy-wide (i.e., $\gtrsim0.1$--10\,kpc) energetic
outflows that expel gas from their host galaxies and consequently this
shuts down future BH growth and star formation and/or enriches the larger-scale
environment with metals (e.g., \citealt{Silk98}; \citealt{Fabian99};
\citealt{Benson03}; \citealt{King03}; \citealt{Granato04}; \citealt{DiMatteo05}; \citealt{Springel05}; \citealt{Hopkins06,Hopkins08a};
\citealt{Booth10}; \citealt{DeBuhr12}). This is in contrast to the
``maintenance mode'' (or ``hot-halo'') feedback where radio jets,
launched by AGN, control the
level of cooling of the hot gas in the most massive halos (see
\citealt{Bower12} and Harrison~2014\nocite{Harrison14a} for a discussion on the two modes). While there is
little doubt that star formation processes (e.g., stellar winds and
supernovae) drive galaxy-wide outflows (e.g., \citealt{Heckman90};
\citealt{Lehnert96}; \citealt{Swinbank09}; \citealt{Genzel11};
\citealt{Newman12b}; \citealt{Bradshaw13}; see review in
\citealt{Veilleux05}) and are an integral part of galaxy evolution
(e.g., \citealt{DallaVecchia08}; Hopkins
et~al. 2013a\nocite{Hopkins13a}), it is believed that AGN activity is
required to drive the highest velocity outflows and are particularly
important for the evolution of the most massive galaxies (e.g., \citealt{Benson03};
\citealt{McCarthy11}; Hopkins et~al. 2013b\nocite{Hopkins13b}; \citealt{Zubovas14}).

X-ray and ultraviolet spectroscopy has shown that a high-fraction, and
potentially all, of high-accretion rate AGN drive high-velocity outflows
($v\approx0.1c$) close to their accretion disks (i.e., on sub-parsec
scales; e.g., \citealt{Blustin03}; \citealt{Reeves03}; \citealt{Ganguly08}; \citealt{Tombesi10}; \citealt{Gofford11}). However, are AGN
capable of driving outflows over galaxy scales as is required by
galaxy formation models? A diagnostic that is commonly used to identify outflowing gas over
large scales is broad (i.e., exceeding that expected from galaxy
dynamics), asymmetric and high-velocity [O~{\sc iii}]$\lambda$5007
emission-line profiles. This emission line traces the kinematics of
the ionised gas; however, we briefly note that outflowing gas has been
observed in multiple gas phases in some AGN
 (e.g., \citealt{Rupke05b}; \citealt{Martin05};
\citealt{Fischer10}; \citealt{Feruglio10}; \citealt{Alatalo11};
\citealt{Veilleux13}; \citealt{Cimatti13}; \citealt{Rupke13}). As a forbidden transition
the [O~{\sc iii}]$\lambda$5007 emission line
cannot be produced in the high-density sub-parsec scales of the AGN broad-line region (BLR) making it a good tracer of the
kinematics in the narrow-line region (NLR) and can be observed over parsecs to tens of kiloparsecs (e.g.,
\citealt{Wampler75}; \citealt{Wilson85}; \citealt{Boroson85}; \citealt{Stockton87};
\citealt{Osterbrock89}). The [O~{\sc iii}]$\lambda$5007 emission line
has long been used to identify outflowing ionised gas in small samples of
local and low redshift AGN (e.g., \citealt{Weedman70};
\citealt{Stockton76}; \citealt{Veron81}; \citealt{Heckman81,Heckman84}; \citealt{Feldman82};
\citealt{Vrtilek85b}; \citealt{Whittle85,Whittle88}; 
\citealt{Veilleux91,Veilleux95}; \citealt{Boroson92};
\citealt{Nelson96}); however, the small sample sizes makes it
difficult to know how representative these observations are. More recently, large systemic spectroscopic surveys (e.g., the Sloan
Digital Sky Survey [SDSS]; \citealt{York00}) have enabled the study of
NLR kinematics in hundreds to tens of thousands of AGN (e.g.,
\citealt{Boroson05}; \citealt{Greene05a}; \citealt{Komossa08};
\citealt{Zhang11}; \citealt{Wang11};
\citealt{Mullaney13}) that can constrain both the ubiquity of these
outflow features and study them as a function of key AGN
properties. 

\cite{Mullaney13} used the SDSS spectroscopic database to study
the one-dimensional kinematic properties
of [O~{\sc iii}]$\lambda$5007 by performing multi-component
fitting to the optical emission-line profiles of $\approx24,000$, $z<0.4$ optically selected
AGN. They showed that $\approx17$\% of the AGN have emission-line profiles that indicate their ionised gas kinematics are
{\em dominated} by outflows and a considerably larger fraction are likely to host ionised outflows at lower levels. The fraction of AGN
with ionised gas kinematics dominated by outflows increases to
$\gtrsim40$\% for the more radio-luminous AGN (i.e., those with $L_{\rm{1.4
    GHz}}>10^{23}$\,W\,Hz$^{-1}$), in contrast, when taking into account
intrinsic correlations, this fraction shows little dependence on
[O~{\sc iii}] luminosity or Eddington ratio (\citealt{Mullaney13}). In agreement with smaller studies
(e.g., \citealt{Heckman81}; \citealt{Whittle85,Whittle92};
\citealt{Gelderman94}; \citealt{Nelson96}; \citealt{Nesvadba11};
\citealt{Kim13}; see also \citealt{Greene05a}), this
result shows that ionised outflows are most common in AGN that have
moderate-to-high radio luminosities. However, while insightful, the
origin of the radio emission is often unknown, particularly at the
moderate radio luminosities (i.e., $L_{\rm{1.4
    GHz}}\approx10^{23}$--10$^{24}$\,W\,Hz$^{-1}$) where AGN cores,
radio jets, shocks and high-levels of star formation could all
contribute (e.g., \citealt{DelMoro13}; \citealt{Condon13}; \citealt{Zakamska14}). It
is therefore vital to measure star-formation rates (SFRs) and properly
investigate the origin of the radio emission in the sources that host
these outflows to properly interpret these results.

The one-dimensional spectra discussed above provide no insight
into the spatial extent or structure of the outflows, for this, we must
appeal to spatially resolved spectroscopy. Both longslit and integral-field unit (IFU) observations of AGN, over
a large range of redshifts, have identified disturbed and
high-velocity ionised gas over kiloparsec scales (e.g., \citealt{McCarthy96};
\citealt{VillarMartin99}; \citealt{Colina99}; \citealt{Holt08};
\citealt{Nesvadba07a,Nesvadba08}; Lipari
et~al. 2009a,b\nocite{Lipari09a,Lipari09b}; \citealt{Fu09}; \citealt{Alexander10};
\citealt{Humphrey10}; \citealt{Greene11}; \citealt{Rupke11,Rupke13}; \citealt{Harrison12a};
\citealt{Westmoquette12}; \citealt{CanoDiaz12}; \citealt{Husemann13};
\citealt{Liu13a,Liu13b}; \citealt{ForsterSchreiber13}). Several of
these studies have revealed considerable masses of outflowing
ionised gas with velocities higher than the galaxy escape velocities,
in apparent agreement with basic predictions from galaxy formation
models. However, a key limitation of these studies is that it is often
difficult to place these observations into the context of the overall
AGN and galaxy populations as the samples are small,
inhomogeneous and/or only represent the most extreme AGN or
star-forming systems in the Universe. 

In this paper we are interested in measuring the prevalence, properties, and
the potential impact of galaxy-wide energetic outflows. We present Gemini (South) Multi-Object Spectrograph
(GMOS; \citealt{AllingtonSmith02}) IFU observations of sixteen $0.08<z<0.2$ type~2 AGN drawn from the parent
sample of \cite{Mullaney13}. Importantly, this means that we can place our IFU observations into the context of the overall
AGN population. So that we can properly interpret our results we
perform SED fitting and analyse the available radio data to measure
SFRs, AGN luminosities and to search for evidence of radio jets in all of our
targets. In Section~\ref{Sec:obs}, we give details of the IFU observations, data reduction and SED
fitting. In Section~\ref{Sec:Analysis} we provide details of our
analysis of the ionised gas kinematics and in
Section~\ref{Sec:Results} we present our results. In
Section~\ref{Sec:Discussion} we discuss our results and their
implication for understanding galaxy evolution and in
Section~\ref{Sec:conclusions} we give our conclusions. We provide background information and a discussion of the
results on individual sources in Appendix~\ref{Sec:appA}.  We have
adopted $H_0 = 71$\kms\,Mpc$^{-1}$, $\Omega_{\rm{M}} = 0.27$ and $\Omega_{\Lambda}
= 0.73$ throughout; in this cosmology, 1 arcsecond corresponds to
$\approx$\,1.5--3.3\,kpc for the redshift range of our sample ($z=0.08$--0.2).


\section{Targets, observations and data reduction}
\label{Sec:obs}

\begin{figure}
\centerline{\psfig{figure=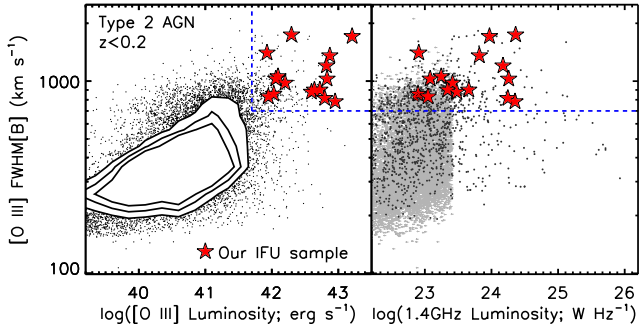,width=3.4in,angle=90}}
\caption{{\em Left:} The FWHM of the broadest, luminous (contributes at least 30\% of
  the total flux; FWHM[B])  [O~{\sc iii}]  emission-line
  component versus total [O~{\sc iii}] luminosity for our IFU targets (red
  stars) compared to the overall population of $z<0.2$ type~2 AGN (black data points and contours;
  \protect\citealt{Mullaney13}). The dashed lines show the selection
  criteria used to select the luminous AGN ($L_{\rm{[O~III]}}>5\times10^{41}$\,erg\,s$^{-1}$)
  with the spectral signatures of ionised outflows
  ([O~{\sc iii}] FWHM[B] $>$ 700\kms\,). {\em Right:} FWHM[B] versus 1.4\,GHz luminosity; the symbols are the same as in
  the left panel (with the addition of upper limits plotted in
  grey).}
\label{fig:selection}
\end{figure}

\begin{figure}
\centerline{\psfig{figure=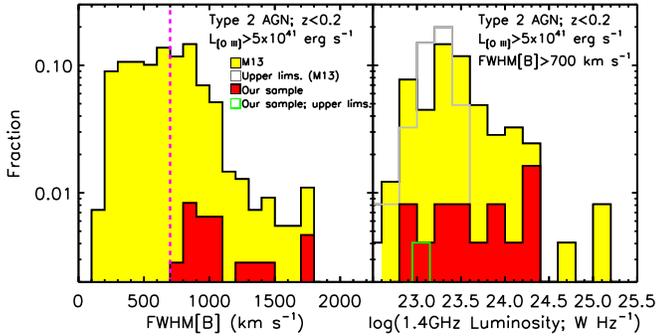,width=3.5in,angle=90}}
\caption{Histograms demonstrating how representative our IFU targets are of the
  overall luminous ($L_{{\rm [O~III]}}>5\times10^{41}$\,erg\,s$^{-1}$)
  type~2 AGN population in the redshift range $z<0.2$. {\em Left:} Histogram
  of the FWHM of the broadest, luminous [O~{\sc iii}] component
  (FWHM[B]; see Fig.~\ref{fig:selection}) from \protect\nocite{Mullaney13}Mullaney et~al. (2013; M13). The
  filled red histogram represents the sources selected for this work
  and yellow represents the parent sample. The
  vertical dashed line shows our selection criteria
  of FWHM[B]\,$>$\,700\,km\,s$^{-1}$. {\em Right:} Histogram of radio luminosity for the sources
with FWHM[B]\,$>$\,700\,km\,s$^{-1}$ from
the parent sample (yellow; \protect\citealt{Mullaney13}) and our IFU targets (red;
Table~\ref{Tab:observations}). The grey and green empty histograms
represents the sources with upper-limits from the parent sample and
our IFU targets, respectively. Our targets are selected from the 45$\pm$3\% of luminous type~2 AGN in this redshift range have luminous
broad [O~{\sc iii}] emission-line components and their radio
luminosities are representative of the parent population (see Section~\ref{Sec:Selection}).
}
\label{Fig:histograms}
\end{figure}

\begin{table}
\begin{center}
{\footnotesize
{\centerline{\sc Targets}}
\begin{tabular}{cccccc}
\noalign{\smallskip}
\hline
\noalign{\smallskip}
Name & RA (J2000)          & Dec (J2000) &Run&$z_{\rm{sys}}$& FWHM[B]\\
 (1)          & (2)       &(3)                 &(4) &(5) & (6)\\
\noalign{\smallskip}
\hline
J0945+1737 & 09:45:21.33 & +17:37:53.2 & 10 & 0.1283 & 1027\\
J0958+1439 & 09:58:16.88 & +14:39:23.7 & 10 & 0.1092 & 878\\
J1000+1242 & 10:00:13.14 & +12:42:26.2 & 10 & 0.1480 & 815\\
J1010+1413 & 10:10:22.95 & +14:13:00.9 & 10 & 0.1992 & 1711\\
J1010+0612 & 10:10:43.36 & +06:12:01.4 & 10 & 0.0984 & 1743\\
J1100+0846 & 11:00:12.38 & +08:46:16.3 & 10 & 0.1005 & 1203\\
J1125+1239 & 11:25:46.35 & +12:39:22.6 & 12 & 0.1669 & 1028\\
J1130+1301 & 11:30:28.86 & +13:01:19.6 & 12 & 0.1353 & 857\\
J1216+1417 & 12:16:22.73 & +14:17:53.0 & 12 & 0.0818 & 1404\\
J1316+1753 & 13:16:42.90 & +17:53:32.5 & 10 & 0.1505 & 1357\\
J1338+1503 & 13:38:06.53 & +15:03:56.1 & 12 & 0.1857 & 901\\
J1339+1425 & 13:39:56.48 & +14:25:29.6 & 12 & 0.1390 & 830\\
J1355+1300 & 13:55:45.66 & +13:00:51.0 & 12 & 0.1519 & 1058\\
J1356+1026 & 13:56:46.10 & +10:26:09.0 & 10 & 0.1238 & 783\\
J1430+1339 & 14:30:29.88 & +13:39:12.0 & 10 & 0.0855 & 901\\
J1504+0151 & 15:04:20.90 & +01:51:59.4 & 12 & 0.1827 & 978\\
\hline
\hline
\end{tabular}
}
\caption[Positions and redshifts for the type~2 AGN in our IFU sample]{\label{Tab:QSOs:tab1}
{\protect\sc Notes:}\protect\\
Details of the targets that we observed with the Gemini-GMOS IFU. (1) Object name; (2)--(3)
optical RA and DEC positions from SDSS (DR7); (4) the Gemini-GMOS run when we
observed this object (see Section~\protect\ref{Sec:QSOs:Observations}); (5)
systemic redshifts derived from this work (see Section~\protect\ref{Sec:QSOs:FittingProcedure}); (6) FWHM (in
km\,s$^{-1}$) of the broadest, luminous (contributes at least 30\% of
the total flux) [O~{\sc iii}]$\lambda$5007 emission-line component from \protect\cite{Mullaney13}. 
}

\end{center}
\end{table}

\subsection{Target selection}
\label{Sec:Selection}
We are interested in studying the spatially resolved
ionised gas kinematics in luminous AGN, with the aim of understanding the
prevalence and properties of galaxy-wide ionised outflows. For this
study, we selected sixteen $z<0.2$ type~2 AGN from the parent catalogue of
\cite{Mullaney13} to observe with the Gemini-GMOS (South) IFU. We
describe our selection criteria below (also illustrated in Figure~\ref{fig:selection}) and we provide a list
of the targets, along with their positions and redshifts, in
Table~\ref{Tab:tab1}.

Our parent catalogue (\citealt{Mullaney13}) contains fits to the emission-line profiles of 24\,264, $z<0.4$, optical AGN (identified using
a combination of ``BPT'' diagnostics, \citealt{Baldwin81}, and emission-line widths) from the SDSS data
release Data Release 7 (DR7; \citealt{Abazajian09}). \cite{Mullaney13} decompose the
[O~{\sc iii}]$\lambda$5007 emission-line profiles into two
components (``narrow'' and ``broad'') to identify
emission-line profiles that are broad and asymmetric, indicative of outflow kinematics. A primary aim of this work is to establish if the broad
and asymmetric emission-line features observed in one-dimensional
spectra are spatially extended and we therefore selected sources that
have a luminous broad [O~{\sc iii}]$\lambda$5007
emission-line component. Our definition of a luminous broad
component is one that contributes at least
30\% of the total flux and has FWHM\,$>$\,700$\kms$ (see Figure~\ref{fig:selection}). Additionally, we only selected
sources that are classified as type~2 AGN by \cite{Mullaney13}, i.e., the permitted and
forbidden lines are of similar width, implying that we cannot
directly see the broad-line region in these objects (e.g.,
\citealt{Antonucci93}). From these, we selected the sources that have
total observed [O~{\sc iii}] luminosities of
$L_{\rm{[O~III]}}>5\times10^{41}$\,erg\,s$^{-1}$ and have a redshift $z<0.2$ to ensure sufficient
emission-line flux for measuring the spatially resolved kinematics
with reasonable exposure times for our IFU observations. Finally, we chose
sixteen targets with celestial co-ordinates that are accessible with
Gemini-South. Our final target list is given in
Table~\ref{Tab:tab1} and we give notes in Appendix~A on the relevant
multi-wavelength observations of these targets that can be found in the literature. The
majority of our targets have received little or no attention previously in the literature, with the
exceptions of J0945+1737, J1316+1753 and J1356+1026 that have been
studied in some detail (see Appendix~A).

It is important to address where our targets fit into the overall AGN population if we are to draw global conclusions from
our observations. Firstly we note that our targets are all type~2 AGN, which are thought to
constitute at least half of the overall AGN population (e.g.,
\citealt{Lawrence10}), and that they have {\em observed} [O~{\sc iii}]
luminosities in the range $L_{{\rm [O~III]}}=10^{41.9}$--$10^{43.2}$\,erg\,s$^{-1}$
(see Table~\ref{Tab:observations}) that fit the luminosity criteria
for type~2 (``obscured'') quasars (i.e., $L_{{\rm
    [O~III]}}\gtrsim10^{41.9}$\,erg\,s$^{-1}$; \citealt{Reyes08}). The absorption corrected [O~{\sc iii}]
luminosities of our targets have the range $L_{{\rm
    [O~III]}}=10^{42.4}$--$10^{43.2}$\,erg\,s$^{-1}$  (calculated
using the H$\alpha$/H$\beta$ emission-line flux ratios from
\citealt{Mullaney13} and following
\citealt{Calzetti00}). Additionally, we emphasise that all of our targets are classified as
``radio-quiet'' in the $\nu L_{\nu}$(1.4\,GHz)-$L_{{\rm [O~III]}}$ plane
  (\citealt{Xu99}; \citealt{Zakamska04}) which make up the majority of
  the luminous type~2 AGN population (i.e., $\approx$90\% based on
  the sample in \citealt{Zakamska04}). Furthermore, considering all
  galaxies at low redshift (i.e., $0.0<z<0.5$), the radio
  luminosities of our targets [i.e., $L_{{\rm
      1.4\,GHz}}=(<0.8-25)\times10^{23}$\,W\,Hz$^{-1}$]
are known to be common with a space density of
$\approx$10$^{-4}$\,Mpc$^{-3}$ and $\approx$10$^{-5}$\,Mpc$^{-3}$ for sources with $L_{{\rm 1.4\,GHz}}=10^{23}$\,W\,Hz$^{-1}$ and $L_{{\rm
    1.4\,GHz}}=10^{24}$\,W\,Hz$^{-1}$ respectively
(\citealt{Simpson12}). 

The key advantage of selecting our sources from a large parent sample of
AGN is that we can quantitatively define how representative our targets are. In Figure~\ref{Fig:histograms} we
show histograms for the FWHM of the broad [O~{\sc iii}] emission-line components for both the
parent sample and our targets. Considering all of the type~2 AGN from
\cite{Mullaney13} with $z<0.2$ and with an [O~{\sc iii}] luminosity, $L_{{\rm
    [O~III]}}>5\times10^{41}$\,erg\,s$^{-1}$ (i.e., to match our
selection criteria), we find that 45$\pm$3\% (246 out of 546 sources
and assuming $\sqrt{N}$ errors) have a luminous [O~{\sc iii}]
component with FWHM[B]\,$>$\,700\,km\,s$^{-1}$ (Figure~\ref{Fig:histograms}). This indicates
that roughly half of luminous type~2 AGN have a luminous broad
component to their [O~{\sc iii}] emission-line profiles and demonstrates that our targets do not represent a
rare sub-population. Additionally, this fraction can be considered a lower limit, as the fraction of sources that host weaker broad
components will be considerably higher. In
Figure~\ref{Fig:histograms} we also show histograms of the radio luminosities of our targets and the type~2 AGN in the parent
sample that have luminous and broad [O~{\sc iii}] emission-line components. Our targets cover a similar range in radio
luminosities to those in the parent sample further demonstrating that our targets are
representative of the parent population.

\subsection{Gemini-South GMOS observations and data reduction}
\label{Sec:Observations}

The observations of our targets were performed using Gemini-South GMOS (\citealt{AllingtonSmith02})
in IFU mode. The GMOS IFU uses a lenslet array to slice the focal plane
into several small components which are each coupled to a
fibre. We made use of the one-slit mode with the B1200 grating to obtain the required
wavelength coverage to observe the [O~{\sc iii}]$\lambda\lambda$4959,4007
and H$\beta$ emission-lines for all our targets. In this mode the IFU consists of 25$\times$20 lenslets sampling a $5^{\prime\prime}\times3.^{\prime\prime}5$ field-of-view. The GMOS IFU also has a
dedicated set of 250 lenslets to simultaneously observe
the sky $\approx$1$^{\prime}$ away from the target field-of-view. We
determined the instrumental dispersion for the observations of each
source ($\Delta v\approx50$--60$\kms$, with uncertainties $\lesssim$5\,km\,s$^{-1}$) by measuring the widths of several skylines close to the observed wavelengths
of the emission lines. The emission-line profiles we measure are typically
non-Gaussian and the non-parametric line-widths we quote (see
Section~\ref{Sec:VelocityDefinitions}) are very large
($\gtrsim$500\,km\,s$^{-1}$); therefore, we do not correct for
this instrumental broadening. This will result in the quoted line-width
measurements being broadened by $\lesssim$2\%. 

The observations were undertaken during two observing programmes:
GS-2010A-Q-8 (PI: Mullaney) taken between 2010 February 09 and 2010 March 19, and
GS-2012A-Q-21 (PI: Harrison) taken between 2012 February 16 and 2012 June 26. For the
observations in the 2012A run, that covered the fainter targets, the
on-source exposures consisted of 6$\times$1500s while for the 2010A
run, that covered the brighter targets, the exposure times consisted of
4$\times$1300s. Table~\ref{Tab:tab1} indicates during which
run each target was observed. The observations were performed with $V$-band
seeing of $\lesssim0.^{\prime\prime}8$ (typically
$\approx0.^{\prime\prime}7$, from measurements of the acquisition images taken before each IFU observation). 

We reduced the data using the standard Gemini {\sc iraf gmos} pipeline to perform wavelength calibration and flat
fielding and we used a dedicated set of IDL routines to perform sky
subtraction, cosmic ray removal and cube construction. We did not
resample the pixels and therefore the pixel scale of the final cubes
is 0.$^{\prime\prime}$17$\times$0.$^{\prime\prime}$2. To find spatial
centroids for the individual data cubes, emission-line free continuum
images from each exposure were produced that were then spatially
aligned and co-added to create the final mosaics, using a median stack with a 3$\sigma$ clipping threshold to remove the remaining cosmetic
defects and cosmic rays. Flux calibration for each target was performed in two stages. Stage one
involved reducing standard star observations in an
identical manner to the targets and consequently obtaining the
instrumental response using {\sc iraf} routines from the {\sc onedspec} package. Stage two used the SDSS spectra to find the
absolute flux calibration for each of the objects. This was achieved by fitting a low order
polynomial to the emission-line free SDSS spectrum and also an
emission-line free spectrum extracted from our IFU data cubes over the same
spatial region as covered by the SDSS fibers (i.e., $\approx3^{\prime\prime}$
diameter).  We verified that the equivalent widths of the emission lines were consistent
between these two spectra and then used the ratio between these fits
to apply the flux calibration to each of the data cubes. We estimate a
conservative uncertainty of 15\% on the absolute flux calibration using this process (see
\citealt{Abazajian09} and Liu et~al. 2013a\nocite{Liu13a})

\subsection{Aligning the IFU data to SDSS images}
As part of our analyses we compare the location of the kinematic
features we observe in our IFU data, to the host galaxy morphologies based upon
three-colour ($g$, $r$, $i$) SDSS images (Section~\ref{Sec:Kinematics}). We
aligned the SDSS three-colour images to our IFU data by matching the
surface brightness peak in the $g$-band SDSS images to the surface
brightness peak of images constructed from our IFU data cubes, collapsed around
$\approx$5000\,\AA. We then rotated
the SDSS images to match the orientation of our IFU data cubes. We
note that we only use these images for qualitative descriptions of the
galaxy morphology with respect to our IFU data and therefore we do not
require a more accurate alignment process.

\subsection{SFRs, AGN luminosities and the origin of the radio emission}
\label{Sec:SEDs}

\begin{table*}
\begin{center}
{\footnotesize
{\centerline{\sc Target Properties}}
\begin{tabular}{cccrcccrrrc}
\noalign{\smallskip}
\hline
\noalign{\smallskip}
Name & $\log[L_{\rm{[O~III]}}]$ & $\log[L_{\rm{H}\beta}]$ & $S_{1.4}$ &$\log[L_{1.4}]$           &$\theta_{{\rm FIRST}}$&PA$_{{\rm 1.4}}$/$R_{{\rm 1.4}}$&$\log[L_{\rm IR,SF}]$&SFR&$\log[L_{\rm{AGN}}]$&$q_{\rm{IR}}$\\
    & (erg\,s$^{-1}$) &(erg\,s$^{-1}$) & (mJy) &(W\,Hz$^{-1}$) &&$^{\circ}$/$^{\prime\prime}$&(erg s$^{-1}$)&(M$_{\sun}$\,yr$^{-1}$)&(erg s$^{-1}$)&\\
(1)                &(2) & (3)                 &(4)&(5)&(6)&(7)&(8)&(9)&(10)&(11)\\
\noalign{\smallskip}
\hline
J0945+1737 &  42.83 &41.80& 44.5[4] & 24.3 &1.072[3]&112$^{\circ}$/2.9& 45.5$^{+0.1}_{-0.1}$ &81$^{+7}_{-25}$ & 45.5$^{+0.2}_{-0.1}$ & 1.8$^{+0.1}_{-0.1}$\\
J0958+1439 &  42.60 &41.49& 10.4[4] & 23.5 &1.006[9]&-& 45.1$^{+0.1}_{-0.1}$ & 36$^{+12}_{-11}$ & 45.0$^{+0.3}_{-0.3}$ & 2.2$^{+0.1}_{-0.1}$\\
J1000+1242 &  42.80 &41.84& 31.8[4] & 24.2 &1.111[4]&158$^{\circ}$/3.3& 45.0$^{+0.3}_{-0.2}$ & 29$^{+18}_{-16}$ & 45.7$^{+0.1}_{-0.1}$ & 1.3$^{+0.2}_{-0.2}$\\
J1010+1413 & 43.21 &42.15& 8.8[5] & 24.0 &1.044[13]&-& 45.7$^{+0.2}_{-0.1}$ & 122$^{+60}_{-32}$ & 46.0$^{+0.2}_{-0.1}$ & 2.3$^{+0.2}_{-0.1}$\\
J1010+0612 & 42.30 &41.48& 99.3[1.3] & 24.4 &1.038[3]&-& 44.9$^{+0.1}_{-0.1}$ & 23$^{+4}_{-7}$ & 45.6$^{+0.1}_{-0.1}$ & 1.1$^{+0.1}_{-0.1}$\\
J1100+0846 & 42.82 &41.77& 61.3[3] & 24.2 &1.0230[12]&-& $<$46.1 & $<$365& 46.0$^{+0.1}_{-0.1}$ & $<$2.5\\
J1125+1239 &  42.06 &40.94& 1.7[5] & 23.1 &0.96[6]&-& $<$45.4 & $<$66 & 45.2$^{+0.1}_{-0.2}$ & $<$2.9\\
J1130+1301 & 42.03 &40.88& 1.7[4] & 22.9 &1.07[7]&122$^{\circ}$/5.1$^{\dagger}$& 44.8$^{+0.2}_{-0.2}$ & 18$^{+7}_{-7}$ & 45.1$^{+0.1}_{-0.1}$ & 2.5$^{+0.2}_{-0.2}$\\
J1216+1417 & 41.93 &-& 5.1[4] & 22.9 &1.07[2]&97$^{\circ}$/2.7$^{\dagger}$& $<$44.6 & $<$10 & 45.3$^{+0.1}_{-0.1}$ & $<$2.1\\
J1316+1753 & 42.87 &41.81& 11.4[4] & 23.8 &1.04[1]&-& $<$45.5& $<$77& 45.4$^{+0.1}_{-0.1}$ & $<$2.2\\
 J1338+1503 & 42.64 &-& 2.4[4] & 23.3 &1.06[5]&-& 45.5$^{+0.1}_{-0.1}$ & 81$^{+13}_{-13}$ & 45.2$^{+0.3}_{-0.2}$ & 2.7$^{+0.2}_{-0.1}$\\
J1339+1425 & 41.95 &40.84& $<$3.0 & $<$23.2 &-&-& $<$44.9 & $<$23& 44.3$^{+0.1}_{-0.1}$ & -\\
J1355+1300 & 42.09 &41.13& 3.0[5]$^{\star}$ & 23.2 &-&-& $<$45.4& $<$62 & 45.7$^{+0.1}_{-0.1}$ & $<$2.7\\
J1356+1026 & 42.95 &41.99& 59.6[4] & 24.4 &1.014[2]&-& 45.4$^{+0.1}_{-0.1}$ & 63$^{+7}_{-17}$ & 45.1$^{+0.3}_{-0.2}$ & 1.6$^{+0.1}_{-0.1}$\\
J1430+1339 &  42.72 &41.88& 26.4[4] & 23.7 &1.400[9]&77$^{\circ}$/7.8& 44.4$^{+0.2}_{-0.2}$ & 7$^{+3}_{-3}$ & 45.3$^{+0.1}_{-0.1}$ & 1.3$^{+0.2}_{-0.2}$\\
J1504+0151 & 42.20 &-& 3.0[4] & 23.4 &1.06[4]&-& 45.2$^{+0.1}_{-0.1}$ & 44$^{+10}_{-10}$ & 45.2$^{+0.1}_{-0.2}$ & 2.4$^{+0.2}_{-0.1}$\\
\hline
\hline
\end{tabular}
}
\caption{\label{Tab:observations}
{\protect\sc Notes:}\protect\\
Details of our IFU targets that we observed with Gemini-GMOS. (1) Object name; (2)-(3) Total
[O~{\sc iii}]$\lambda$5007 and H$\beta$ 
luminosities derived in this work, uncertainties are typically 15\%;
(4) 1.4\,GHz flux
densities obtained from the FIRST survey (\protect\citealt{Becker95})
and uncertainties that are defined as 3$\times$ the RMS noise of the radio map at
the source position; (5)
Rest-frame radio luminosities using a
spectral index of $\alpha=0.7$ and assuming $S_{\nu}\propto \nu^{-\alpha}$
(we note that a range of $\alpha=0.2$--1.5
introduces an spread of $\lesssim\pm$0.1\,dex on the radio
luminosity); (6) Radio morphology paramater, where sources with $\theta>1.06$ are classified as extended
in the 1.4\,GHz FIRST data (see Section~\ref{Sec:SEDs}); (7) The
deconvolved PA and major-axis radius from the
FIRST survey for the radio-extended sources; (8)-(10) Infrared
luminosties from star-formation, SFRs and bolometric AGN luminosities derived from our SED analyses
(see Section~\protect\ref{Sec:SEDs}); (11) The $q_{\rm{IR}}$
(``radio excess'') parameter
for each source, where sources with
$q_{\rm{IR}}\le1.8$ are classified as ``radio excess'' (see Section~\protect\ref{Sec:SEDs}).
\newline $^{\star}$The 1.4\,GHz flux density for SDSS\,1355+1300 is taken from the
NVSS survey (\protect\citealt{Condon98}).\newline$^{\dagger}$The uncertainty
on $\theta$ is high and therefore these sources may not be truly
extended in their FIRST images.
}

\end{center}
\end{table*}

\begin{figure*}
\centerline{\psfig{figure=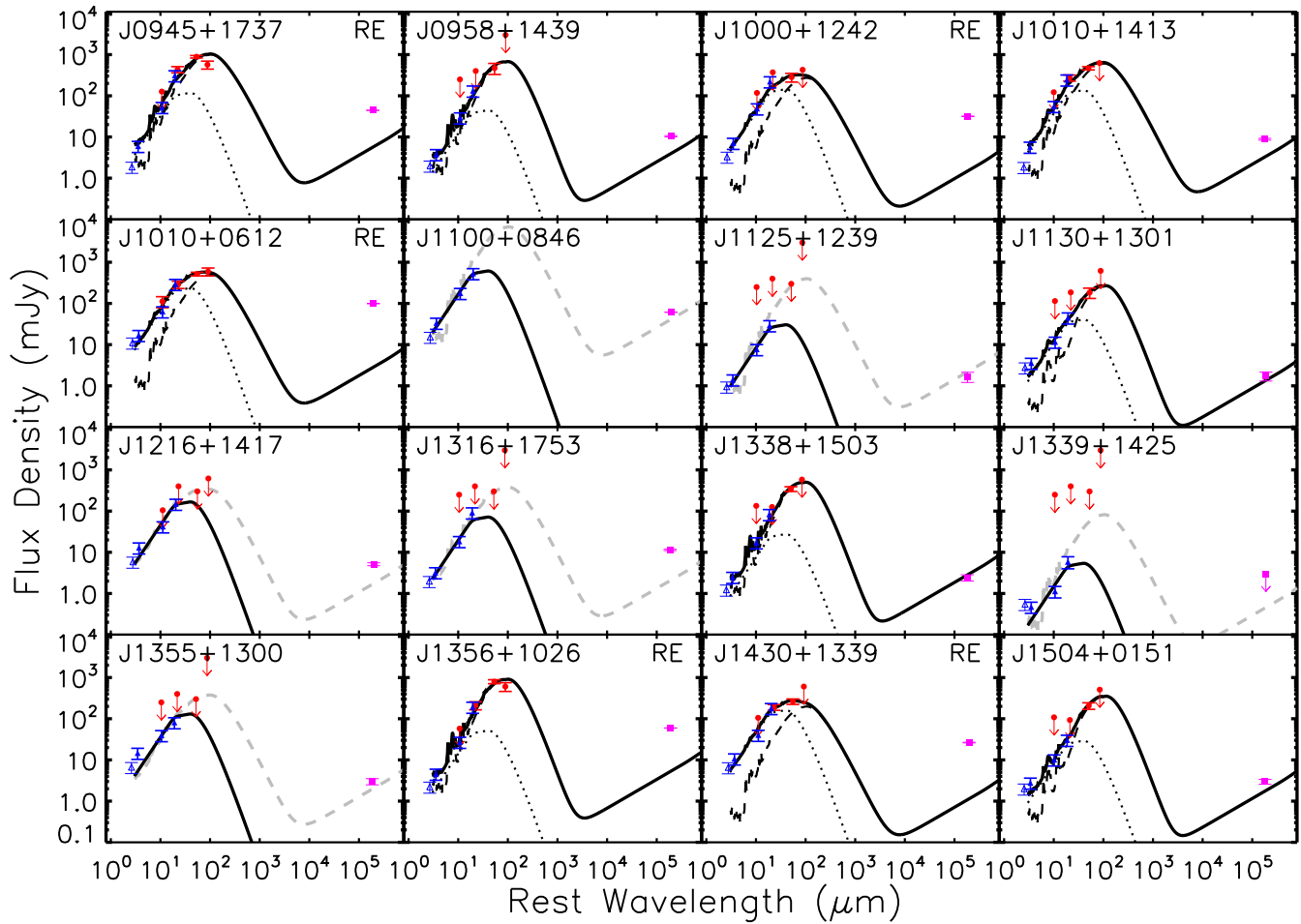,width=7in,angle=90}}
\caption{The infrared SEDs for each target in our sample. The flux
  densities used in the fitting procedure are shown as blue filled
  triangles (WISE data) and red filled circles (IRAS data; photometric data in Table~\ref{Tab:fluxes}). The first WISE band (at 3.4$\mu$m)
 falls outside of the wavelength range of our
  templates and is shown with an empty symbol. Also shown are the overall
  best-fit SEDs: the total SEDs are shown as solid curves, the AGN
  templates are shown as a dotted curves and the starburst templates as
  dashed curves. The grey dashed curves show the upper-limits on the
  starburst templates for the sources that lack far-infrared
  data. Using these SED fits we obtain SFRs and AGN
  luminosities (see Section~\ref{Sec:SEDs} and
  Table~\ref{Tab:observations}). We also plot the 1.4\,GHz flux
  densities (square symbols) that are not included in the fitting but
  do illustrate that five of the sources have
  significant excess radio emission above that expected from star formation
  (indicated with ``RE'' in the top-right of their panels; see
  Section~\ref{Sec:SEDs}; Table~\ref{Tab:observations}).
}
\label{Fig:SEDs}
\end{figure*}

\begin{table*}
\begin{center}
{\footnotesize
{\centerline{\sc Infrared Photometric Data}}
\begin{tabular}{lrrrrrrrr}
\noalign{\smallskip}
\hline
\noalign{\smallskip}
Name & W1 (3.4$\mu$m) & W2 (4.6$\mu$m) & W3 (12$\mu$m) & W4
(22$\mu$m) & IRAS 12$\mu$m & IRAS 25$\mu$m & IRAS 60$\mu$m & IRAS
100$\mu$m \\
\noalign{\smallskip}
\hline
J0945+1737 & 1.9$\pm$0.6 & 6.0$\pm$1.8 & 53$\pm$16 & 310$\pm$90 & $<$127& 440$\pm$70 & 890$\pm$60 & 570$\pm$130\\
J0958+1439$^{\dagger}$ & 2.0$\pm$0.6 & 3.8$\pm$1.1 & 30$\pm$9 & 130$\pm$40 & $<$250& $<$400 & 470$\pm$140 & $<$3000\\
J1000+1242 & 3.3$\pm$1.0 & 7$\pm$2 & 49$\pm$15 & 220$\pm$70 & $<$118 & $<$372& 280$\pm$70 & $<$428\\
J1010+1413 & 1.8$\pm$0.6 & 5.7$\pm$1.7 & 56$\pm$17 & 250$\pm$70 & $<$122& 270$\pm$50 & 470$\pm$40 &$<$625\\
J1010+0612 & 11$\pm$3 & 17$\pm$5 & 64$\pm$19 & 290$\pm$90 & 110$\pm$30 & 290$\pm$60 & 520$\pm$50 & 590$\pm$130\\
J1100+0846$^{\dagger\dagger}$ & 15$\pm$5 & 34$\pm$10 & 180$\pm$50 & 540$\pm$160 & - & - & - & -\\
J1125+1239 & 1.0$\pm$0.3 & 1.4$\pm$0.4 & 8$\pm$2 & 29$\pm$9 & $<$250 & $<$400 & $<$300 & $<$3000\\
J1130+1301$^{\star}$ & 2.8$\pm$0.8 & 3.6$\pm$1.1 & 12$\pm$4 & 45$\pm$14 & $<$114 & $<$186 & 180$\pm$50 & $<$619\\
J1216+1417 & 5.9$\pm$1.8 & 13$\pm$4 & 42$\pm$13 & 150$\pm$50 & $<$105 & $<$400& $<$300 & $<$620\\
J1316+1753 & 2.0$\pm$0.6 & 3.2$\pm$1.0 & 19$\pm$6 & 90$\pm$30 & $<$250 & $<$400 & $<$300 & $<$3000\\
J1338+1503 & 1.2$\pm$0.4 & 2.5$\pm$0.8 & 17$\pm$5 & 80$\pm$30 & $<$134 & $<$126 & 340$\pm$50 & $<$583\\
J1339+1425 & 0.5$\pm$0.2 & 0.5$\pm$0.1 & 1.1$\pm$0.4 & 6$\pm$2 & $<$250 & $<$400 & $<$300 & $<$3000\\
J1355+1300 & 7$\pm$2 & 15$\pm$4 & 40$\pm$13 & 80$\pm$20 & $<$250 & $<$400 & $<$300 & $<$3000\\
J1356+1026 & 2.2$\pm$0.7 & 4.6$\pm$1.4 & 27$\pm$8 & 190$\pm$60 & $<$58 & 210$\pm$50 & 800$\pm$90 & 600$\pm$150\\
J1430+1339 & 7$\pm$2 & 11$\pm$3 & 40$\pm$12 & 180$\pm$50 & $<$105 & 190$\pm$30 & 260$\pm$40 & $<$605\\
J1504+0151 & 2.0$\pm$0.6 & 2.8$\pm$0.8 & 10$\pm$3 & 30$\pm$9 & $<$108 & $<$93 & 210$\pm$40 & $<$508\\
\hline
\hline
\end{tabular}
}
\caption{\label{Tab:fluxes}
{\protect\sc Notes:}\protect\\
The mid-infrared--far-infrared photometric flux densities,
uncertainties and upper limits for the targets in our sample (in units of
mJy) which are used in our SED analysis 
(these photometric data are shown in Fig.~\ref{Fig:SEDs}). These were obtained from WISE and IRAS all-sky catalogues
(\protect\citealt{Moshir92}; \protect\citealt{Wright10}; see
Section~\ref{Sec:SEDs} for details). The first WISE band
(W1) falls outside of the wavelength range of our templates, therefore these flux densities are not included in the
 SED fitting procedure.\protect\newline$^{\dagger}$The IRAS fluxes for
 J0958+1439 are from {\protect\sc Scanpi} (\protect\url{http://irsa.ipac.caltech.edu/applications/Scanpi});
  $^{\dagger\dagger}$J1100+0846 was not covered by IRAS observations; $^{\star}$J1130+1301 did not have
  an entry in the IRAS faint source catalogue; however, inspection of
  the images indicates that it is detected and therefore we obtained the fluxes from the IRAS reject catalogue. 
}

\end{center}
\end{table*}

To thoroughly interpret our results we investigate the
properties of the AGN and host galaxies of our targets in more detail (i.e.,
we measure SFRs and AGN luminosities and explore the origin of the radio emission). To obtain the bolometric
AGN luminosities and SFRs of our targets (see Table~\ref{Tab:observations}) we performed
mid-infrared to far-infrared SED fitting to archival photometric data
that is available for our targets (described below). Our procedure decomposes the emission from AGN activity and
star-formation, so that we do not need to make any assumptions about the relative
contribution from these two processes. At the end of this sub-section,
  we also briefly investigate the possible presence of radio AGN activity
  (i.e., radio cores or radio jets) in our targets by combining the results of our SED fitting with archival radio data.

For our SED fitting, we obtained mid-infrared to far-infrared photometric flux densities,
uncertainties and upper limits (over 3\mum--100\mum; see Table~\ref{Tab:fluxes}) from (1) the
Wide-field Infrared Survey Explorer all-sky survey (WISE;
\citealt{Wright10}) using the closest source to the SDSS position (all
are $<0.^{\prime\prime}6$) and (2) the Infrared Astronomical Satellite (IRAS; \citealt{Neugebaur84})
faint source catalogue (\citealt{Moshir92}) with a matching radius of
$\le$30$^{\prime\prime}$\footnote{We verified these matches by
    following the log likelihood method of matching IRAS to WISE
    counterparts outlined in \cite{Wang14}.}. We added in quadrature 30\% of the measured
flux densities to the quoted WISE uncertainties to account for
calibration uncertainties in the WISE data (\citealt{Wright10}) and the discrete nature of the SED templates used
in our analysis (see below; also see \citealt{DelMoro13}). For
the sources that we were not able to obtain IRAS flux density
measurements (see Table~\ref{Tab:fluxes}) we estimated conservative maximum upper limits of 0.25\,mJy, 0.4\,mJy,
0.3\,mJy and 3\,mJy for the 12\mum\,, 25\mum\,, 60\mum\,, and 100\mum\
bands respectively. One source (SDSS\,J1100+0846) falls into an area
of the sky that was not observed by IRAS observations and we therefore cannot
place photometric constraints on the IRAS bands for this source. The
photometric data we used are shown in Figure~\ref{Fig:SEDs}.

To perform the SED fitting we followed the procedure described in
\cite{Mullaney11} that simultaneously fits an empirically derived AGN
model (paramaterised in an analytical form) and
empirically derived host-galaxy templates to the infrared photometric
data (Table~\ref{Tab:fluxes}). All of the sources in our sample are known to host an AGN (from
optical spectroscopy) and additionally all but one of the sources are
classified as an infrared AGN based upon the WISE
colours (W2$-$W1 and W3$-$W2) by falling inside the ``AGN-wedge'' defined in
\cite{Mateos12}.\footnote{SDSS\,J1339+1425 does not have WISE colours that place
  it inside the ``AGN-wedge''; however, we still found that including an AGN template
  provided a better fit than any of the star-forming templates alone.}
We therefore simultaneously fit these data with the five star-forming
galaxy templates (``SB1''--``SB5'') and the mean AGN model, originally defined in \cite{Mullaney11} and extended
to cover the wavelength range 3--10$^{5}$\mum, by
\cite{DelMoro13}.\footnote{We tested how much of an affect our limited
  range of templates has on our quoted results (Table~\ref{Tab:observations}) by: (1) re-fitting
  including an Arp~220 model SED and (2) allowing the second power-law
  slope of the AGN model to be a free parameter (i.e., $\alpha_{2}$ in \citealt{Mullaney11}). The first of these changes allows for a more extreme
  ULIRG-like SED that is not covered by our main SB templates. The
  second change allows for a very large range in
  instrinsic AGN SEDs, to account for the variations seen in AGN of
  different luminosities (see Fig.~7 in \citealt{Mullaney11}). When
  performing the SED fitting with these two additions we found that
  all the results were consistent within the errors of our quoted
  results (Table~\ref{Tab:observations}) with
  only two minor exceptions: (1) The SFR for J1430+1339 had an upper limit of
  $<2$\Msolyr and (2) the AGN luminosity of J1504+0151 was
  $\log[L_{{\rm AGN}}]=45.4_{-0.3}^{+0.1}$\,erg\,s$^{-1}$. We
  therefore conclude our primary choice of templates are sufficient
  for this work.} To derive the AGN luminosities and SFRs (Table~\ref{Tab:observations}) we use the star-forming template plus AGN template combination that gives
the lowest overall $\chi^2$ value, rejecting all
fits that lie above the photometric upper
limits (given in Table~\ref{Tab:fluxes}; see Fig.~\ref{Fig:SEDs}). For the sources where we have no
detections in the far-infrared bands (i.e., the IRAS bands), we are
unable to measure reliable SFRs and we therefore calculate
conservative upper limits by increasing the normalisation of the star-forming templates until
either the photometric upper limits were reached or the solution was
3$\sigma$ above one of the photometric data points. 

We calculated the SFRs for each source (Table~\ref{Tab:observations})
by measuring the infrared luminosities of the
star-formation components from the best-fitting SED solutions ($L_{\rm{IR,SF}}$; integrated over
8--1000\mum) and using the relationship of \cite{Kennicutt98}
(corrected to a Chabrier IMF by dividing by a factor of 1.7; \citealt{Chabrier03}). To determine the bolometric AGN luminosities for
each source (Table~\ref{Tab:observations}), we first calculated the AGN continuum luminosity at 6\mum\
from the best fitting SED solution and converted this to an AGN bolometric
luminosity using a correction factor of $\times$8
(\citealt{Richards06}; see Table~\ref{Tab:observations}).  The
conservative upper and lower bounds on both the SFRs and AGN
luminosities are derived from summing in quadrature the range in solutions from
the different star-formation templates (only those which did not exceed the upper
limits) and the formal uncertainties from the best-fit solution. We
emphasise that our method decomposes the emission due to an AGN and star
formation, therefore removing the need to assume the relative contributions from
these two processes. Furthermore, this procedure produces consistent SFRs and AGN luminosities with those derived from
mid-infrared spectra where available (see Appendix A and
Footnote~\ref{Foot:RZSEDs}; also see \citealt{DelMoro13}).

The bolometric AGN luminosities for our targets span $L_{{\rm
    AGN}}=(0.2-10)\times10^{45}$\,erg\,s$^{-1}$ (with a median
$L_{{\rm AGN}}=2\times10^{45}$\,erg\,s$^{-1}$) and therefore cover the
classical ``quasar'' threshold of $L_{{\rm
    AGN}}=10^{45}$\,erg\,s$^{-1}$. We note that these mid-infrared
derived AGN luminosities are a factor of $\approx$1--20 lower than
those that would be predicted using the [O~{\sc iii}] luminosity and
the relationship of \cite{Heckman04}  (i.e., $L_{{\rm AGN}}=3500L_{{\rm [O~III]}}$). However, this is consistent
with other studies of luminous ($L_{{\rm [O~III]}}\gtrsim10^{42}$\,erg\,s$^{-1}$) type~2
AGN, at similar redshifts, that show that $L_{{\rm [O~III]}}$ is not
linearly correlated with the nuclear luminosity and can over predict
the AGN luminosity by an order of magnitude or more (\citealt{Schirmer13}; \citealt{Hainline13}).

Of the sixteen sources in our sample, eight have far-infrared
luminosities that classify them as luminous infrared galaxies
(LIRGs; $L_{{\rm IR,SF}}=10^{11}$--10$^{12}\Lsol$\,) and one as an
ultraluminous infrared galaxy (ULIRG: $L_{{\rm IR,SF}}>10^{12}\Lsol$). Of the
remaining seven sources, six of them have upper-limits on their
far-infrared luminosities which are less than or consistent with the
luminosity of LIRGs (see Table~\ref{Tab:observations}).\footnote{The remaining source, SDSS\,J1100+0846,
  which is not covered by IRAS and therefore has poor FIR
  constraints, has an upper limit consistent with a ULIRG.} These infrared luminosities correspond
to SFRs of $\lesssim$\,[7--120]\Msolyr and are typical for AGN of these
luminosities at these redshifts (e.g., \citealt{Zakamska08};
\citealt{Mullaney10}; \citealt{Rosario12}).

Finally, we investigate the origin of the radio emission from
our targets. We plot the 1.4\,GHz radio flux densities (given in
Table~\ref{Tab:observations}) on the SEDs shown in
Figure~\ref{Fig:SEDs}. We identify if there is radio emission significantly above that expected from their ongoing star-formation (e.g.,
\citealt{Helou85}; \citealt{Condon95}), that may indicate radio emission due to an AGN either in form of a radio core or
radio jets (e.g., see \citealt{DelMoro13} and references there-in) or due to shocks (e.g., \citealt{Zakamska14}). To do this we use the definition given
by \cite{Ivison10} (see also \citealt{Helou85}), calculating
the ratio between the far-infrared and radio emission ($q_{{\rm IR}}$) as
\begin{equation}
q_{IR} = \log\left[ \frac{ S_{{\rm IR}} / 3.75\times10^{12} {\rm\,W
    m^{-2}}}{S_{1.4}/{\rm W\,m}^{-2}{\rm\,Hz}^{-1}}\right]
\end{equation}
where $S_{{\rm IR}}$ is the rest-frame far-infrared (8--1000\,$\mu$m)
flux and $S_{1.4}$ is the rest-frame 1.4\,GHz flux density assuming, $S_{\nu}\propto\nu^{-\alpha}$, with
a radio spectral index of $\alpha=0.7$. We give the value of $q_{{\rm IR}}$ for each source in
our sample in Table~\ref{Tab:observations}. The quoted uncertainties
combine the uncertainties on $S_{1.4}$, $S_{{\rm IR}}$ and a range in
the unknown radio spectral indices (we use $\alpha=0.2$--1.5). Star-forming
galaxies have $q_{{\rm IR}}\approx2.4$ and we define ``radio excess'' sources as those with
$q_{{\rm IR}}\le1.8$, i.e., $\gtrsim$\,2.5$\sigma$ away from the peak
seen in the star-forming galaxies (\citealt{Ivison10}; see also
\citealt{DelMoro13} who use a similar definition). Using
this definition we identify five of our targets as being radio excess
and five sources that are classified as radio normal (see
Table~\ref{Tab:observations} and Figure~\ref{Fig:SEDs}); however, we cannot rule out low-level radio
emission above that expected from star formation in these targets as it can be difficult to identify using this
method, especially in systems with high SFRs (e.g., \citealt{DelMoro13}). The remaining six sources have upper limits on
their $S_{{\rm IR}}$ values which results in upper limits on their
$q_{{\rm IR}}$ values that are consistent with them being
either radio normal or radio excess sources. 

Based on FIRST and NVSS images only one of our sources (J1430+1339)
shows clear evidence for luminous extended radio
structures on scales $>$5$^{\prime\prime}$ (see Table~\ref{Tab:observations} and Appendix~A). Therefore, we follow the
method of \cite{Kimball08} to search for evidence of extended radio
structures (on $\approx2$--5$^{\prime\prime}$ scales) that compares
the peak and integrated 1.4\,GHz FIRST flux densities ($F_{{\rm
    peak}}$ and $F_{{\rm int}}$ respectively). They define the
parameter $\theta=(F_{{\rm int}}/F_{{\rm peak}})^{0.5}$ and any source with
$\theta>1.06$ is classed as spatially resolved. We provide $\theta$ for all of
our targets with FIRST detections in
Table~\ref{Tab:observations}. For the five sources with
$\theta>1.06$, i.e., those that pass the criteria to be ``resolved'', we also provide the position angle and radius of the de-convolved
major axis provided by FIRST. Of these five sources, two have very
low 1.4\,GHz flux densities and therefore the uncertainty on $\theta$
is high. The other three have stronger evidence for being
spatially resolved and are also ``radio excess'' sources; therefore,
they are the strongest candidates for having radio jets on
$\approx3$--$8^{\prime\prime}$ scales (i.e., $\approx 7$--12\,kpc;
Table~\ref{Tab:observations}), although shocks due to outflows are
another possible explanation for this excess radio emission (\citealt{Zakamska14}). High-resolution radio imaging
will be instrumental in determining the origin and morphology of the
radio emission in the whole sample. 


\section{Velocity definitions and spatially resolved kinematics}
\label{Sec:Analysis}

\subsection{Non-parametric velocity definitions}
\label{Sec:VelocityDefinitions}

\begin{figure}
\centerline{\psfig{figure=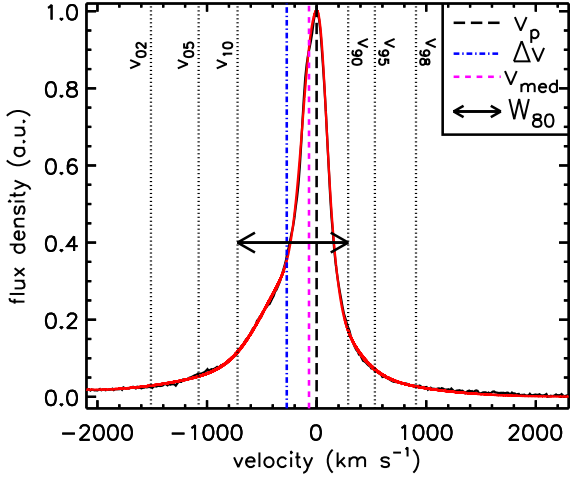,width=3in}}
\caption{Illustration of the different non-parametric velocity definitions
  used in this work which are described in Section~\ref{Sec:VelocityDefinitions}. We show an example [O~{\sc iii}]$\lambda$5007
  emission-line profile from our sample (black curve) and our fit to this profile (red
  curve). The vertical dotted lines show different percentiles to the flux
  contained in the overall emission-line profile (from left to right:
  2nd; 5th; 10th; 90th; 95th and 98th)  The long-dashed line shows the
  velocity of the peak flux density ($v_{p}$). The vertical short-dashed line shows the
  median velocity ($v_{\rm med}$) and the dot-dashed line shows the
  velocity offset of the underlying broad wings ($\Delta v$; see
  Section~\ref{Sec:VelocityDefinitions}). The arrow indicates the line width that
  contains 80\% of the flux ($W_{80}$). 
}
\label{Fig:ExampleSpec}
\end{figure}

We are interested in the spatially-resolved kinematics
of the [O~{\sc iii}]$\lambda\lambda$4959,5007 and H$\beta$ emission
lines for the sixteen type~2 AGN we observed with the Gemini-GMOS
IFU. The combination of these emission-line species allows us to
measure emission-line region sizes (Section~\ref{Sec:EELRs}), 
spatially resolve the ionised gas kinematics (Section~\ref{Sec:Kinematics}) and calculate ionised gas
masses (Section~\ref{Sec:properties}) for our targets. The emission-line profiles in
type~2 AGN are often found to be complex, consisting of multiple narrow and broad components (e.g.,
\citealt{Veilleux91}; \citealt{Greene11}; Villar-Mart\'in et~al. 2011b\nocite{VillarMartin11b}), which is also the case
for our sample (Fig.~\ref{Fig:velmaps}; Fig.~A1--A15). Due the
complexity of the emission-line profiles we opted to use
non-parametric definitions to characterise their widths and velocities
within each pixel of the IFU data. We used an approach that: (1) is consistent for all of the targets in the sample; (2) allows
us to de-couple galaxy dynamics, merger remnants and outflow
features and (3) provides results that are directly comparable to observations in the
literature. The definitions that we have used throughout, for describing
both the galaxy-integrated emission-line profiles and the spatially resolved kinematics, are described
below and are illustrated in Figure~\ref{Fig:ExampleSpec}:
\begin{itemize}
\item The peak velocity ($v_{p}$) is the velocity of the peak
  flux density in the overall [O~{\sc iii}]$\lambda$5007 emission-line profile. For our targets,
  this traces the brightest narrow emission-line component
  (i.e., those with FWHM\,$\lesssim$\,500\,km\,s$^{-1}$). 
\item The line width, $W_{80}$, is the velocity width of the line
  that contains 80\% of the emission-line flux such that $W_{80} = v_{90} -
  v_{10}$, where $v_{10}$ and $v_{90}$ are the
  velocities at the 10th and 90th percentiles, respectively (Fig.~\ref{Fig:ExampleSpec}). For a single Gaussian profile this is approximately
  the FWHM. 
\item To measure the velocity offset of the broad underlying
  wings we define the velocity offset, $\Delta v$, that is measured
  as $\Delta v=(v_{05}+v_{95})/2$, where $v_{05}$ and $v_{95}$ are the
  velocities at the 5th and 95th percentiles of the overall emission-line
  profiles respectively (Fig.~\ref{Fig:ExampleSpec}). For a profile
  that is well characterised by a single narrow Gaussian component and
  a luminous broad Gaussian component, $\Delta v$ is the velocity offset of the broad
  component. For a single Gaussian component $\Delta v=0$. 
\item To enable us to compare to previous work we also measured
$v_{02}$ and $v_{98}$, which are the 2nd and 98th percentiles of the
flux contained in the overall
emission-line profiles (as used in \citealt{Rupke13})\footnote{We note that
  \cite{Rupke13} fit a broad and narrow component to the emission lines and then define $v_{98}$ for the broad component
  only. As a result their values are likely to be higher
  than our values that we defined from the {\em overall}
  emission-line profile so as not to make any assumptions on the kinematic
  structure of the emission lines (i.e., the number of distinct
  velocity components).} and can be considered the
``maximum'' projected velocity of the gas.
\item Finally, we measured the median velocity
($v_{\rm{med}}$; Fig.~\ref{Fig:ExampleSpec}) of the overall
emission-line profile so that we can compare to the type~2 AGN
observed in Liu et~al. (2013b)\nocite{Liu13b}. We find that $v_{\rm{med}}$
is likely to be dominated by galaxy kinematics in many cases (i.e., galaxy rotation or mergers; see Section~\ref{Sec:outflows}) and should
be used with care if interpreting outflow kinematics. 
\end{itemize}
We emphasise that these definitions are used to enable us to
consistently compare all of the targets in our sample as well as compare to
observations in the literature; however, for a few individual cases some care should be taken when
interpreting these measurements due to the uniqueness and complexity of the
emission-line profiles. For this reason we discuss the details of
kinematics of individual sources in Appendix~A. 

\subsection{Emission-line profile fitting procedure}
\label{Sec:FittingProcedure}

As several of the velocity definitions outlined above and illustrated
in Fig.~\ref{Fig:ExampleSpec} are measured
from the broad wings of the emission-line profiles they are subject to the
influence of noise in the spectra, especially in the lower surface
brightness regions of our IFU observations. To circumvent this issue,
we characterise each emission-line profile using a combination of multiple Gaussian
components. We note that we do not draw any physical
meaning from the individual Gaussian components, they are
only used to characterise the emission-line profiles. Below, we describe the
emission-line profile fitting procedure that we use throughout for
fitting the galaxy-integrated spectra and also the
spectra used to produce the spatially resolved maps and velocity
profiles described in Section~\ref{Sec:VelMaps}.  

Initially, we isolated the [O~{\sc iii}]$\lambda\lambda$4959,5007
emission-line doublet from the underlying continuum by linearly interpolating
each spectrum between line-free regions close to the emission lines, a commonly used method for type~2 AGN (e.g.,
\citealt{Husemann13}; Liu et~al. 2013b\nocite{Liu13b}). The emission-line doublet was then
fit with a minimising-$\chi^2$ method, using the IDL routine {\sc mpfit}
(\citealt{Markwardt09}) with Gaussian components. For every Gaussian
component we simultaneously fit the [O~{\sc iii}]$\lambda$5007 and
[O~{\sc iii}]$\lambda$4959 emission lines using the same line-width, a
fixed wavelength separation and the intensity ratio was fixed at
[O~{\sc iii}]$\lambda4959$/[O~{\sc iii}]$\lambda5007=0.33$ (e.g., \citealt{Dimitrijevic07}). Initially we fit a single Gaussian component to
the emission-line profile and consequently additional Gaussian components
were added to the overall fit (up to a maximum of six components) until the
important values for our study (i.e., $\Delta v$, $W_{80}$ and $v_{\rm{p}}$;
see Fig.~\ref{Fig:ExampleSpec}) became stable within $\le$20\%, i.e.,
the addition or removal of a Gaussian component does not change these values significantly. We only fitted more than
one Gaussian component if the emission line was detected
above a S/N threshold of $\approx$30, with this threshold being
motivated by visually inspecting the fits and the reduced $\chi^2$
values (see below). For spectra below this S/N threshold, $W_{80}$
cannot be well characterised and we are also unable to define $\Delta
v$. We included appropriate constraints to the line widths to avoid
fitting excessively broad components to the continuum or narrow
components to noise spikes (i.e., limited by the
spectral resolution). We verified that our fitting procedure
was effective across the whole field-of-view by visually inspecting the spectra and their
best fitting solutions extracted from 25 spatial regions for each object
individually (Fig.~\ref{Fig:velmaps}; Fig.~A1-A15). An additional verification of the success of our fitting procedure was that, in all but
one of the targets,\footnote{The source with a higher
  median reduced $\chi^{2}$=3.5 is SDSS\,J1356+1026 that we attribute
to the highly complex emission-line profile and exceptionally low
noise (Fig.~\ref{fig:1356+1026}).} the median reduced $\chi^{2}$
values of the fits to the individual pixels in the data cubes are between
1.0--1.8 with $\ge$60\% of the pixels having reduced $\chi^{2}$ values between
0.8--2.5. 

Using the overall fits to the emission-line profiles, we measure the
non-parametric velocity definitions defined in
Section~\ref{Sec:VelocityDefinitions} (see Fig.~\ref{Fig:ExampleSpec}). Random errors on these
quantities were calculated by taking the 1$\sigma$ spread of values
calculated from a 1000 random spectra that were generated using the final
models plus Gaussian random noise (based on the standard deviation of
the continuum). To obtain the final quoted uncertainties, we added these
random errors in quadrature with an estimated systematic error that we
calculated from the range in the velocity values derived from adding or removing a Gaussian component to the final
emission-line models. We note that these uncertainties are extremely
small ($\lesssim$10\%) in the high signal-to-noise ratio spectra and therefore the
uncertainties will be dominated by the physical interpretation of these values
(see Section~\ref{Sec:Results}). 

To enable us to measuring ionised gas masses
(Section~\ref{Sec:properties}) we also measured the H$\beta$
emission-line profiles and fluxes from our IFU data. To do this, we extract the H$\beta$ emission-line
profiles from the IFU data cubes in the same manner to that for [O~{\sc iii}]$\lambda$4959,5007 and show them in
Figure~\ref{Fig:velmaps} and Figure~A1--A15. We find that the overall
shapes of the H$\beta$ emission-line
profiles are in excellent agreement with the [O~{\sc iii}] emission-line profiles for all but
three of our sources (the exceptions are SDSS\,J1216+1417; SDSS\,J1338+1503;
and SDSS\,1504+0151). This indicates that for the majority of the
sources, the H$\beta$ emission lines are not significantly affected by
stellar absorption features (which is true for many AGN with luminous emission lines; \citealt{Soto10};
\citealt{Harrison12a}; Liu et~al. 2013b\nocite{Liu13b}) and this emission line traces the same kinematic
components as the [O~{\sc iii}] emission line (see also \citealt{RodriguezZaurin13}). For these thirteen sources we measure
the H$\beta$ emission-line fluxes by fitting the [O~{\sc iii}] emission-line models
(described above) to the data at the observed wavelength of the
H$\beta$ emission lines, allowing only the overall normalisation to
vary. The remaining three sources, which have different profiles for these two
emission-line species, may have varying ionization states for different
kinematic components and/or have higher levels of H$\beta$
stellar absorption features that is certainly plausible given the stellar
continua and absorption features observed in their optical spectra (see Appendix~A). For this work, we wish to avoid using a variety of approaches for our
targets; therefore, we choose to not use the H$\beta$ emission lines in the
analyses of these sources. 

Due to the lack of clearly defined stellar absorption lines in the
spectra across our targets, we define the systemic redshift from the velocity of the peak
flux density in the [O~{\sc iii}]$\lambda$5007 emission line (see
Fig.~\ref{Fig:velmaps}). This choice of
systemic redshift effectively traces the mean velocity of the narrow
components in the emission-line profiles, which are often attributed
to galaxy kinematics (e.g., \citealt{Greene05a}; \citealt{Harrison12a};
\citealt{Rupke13}; also see Section~\ref{Sec:GalaxyKinematics});
however, this might not always be the case and therefore places an
uncertainty on our systemic redshift. For SDSS\,J0958+1439 and SDSS\,J1316+1753 where the
overall emission line profiles lack a
clear single peak (see Fig.~\ref{fig:0958+1439} and
Fig.~\ref{fig:1316+1753}) we take the flux weighted
mean velocity of the narrow components (i.e., those with FWHM\,$<$\,500\,km\,s$^{-1}$) used in the
fit to the galaxy-integrated spectrum (i.e., similar to the procedure used
in \citealt{RodriguezZaurin13}). Across the whole sample, the mean velocity
offset between our systemic redshift and the SDSS redshift is
$-$18\,km\,s$^{-1}$ with a scatter of 54\,km\,s$^{-1}$. We note that any velocity {\em differences}
that are measured (such as $W_{80}$) are unaffected by the choice in systemic
redshift. 

\subsection{Velocity maps and velocity-distance profiles}
\label{Sec:VelMaps}

\begin{figure*}
\centerline{\psfig{figure=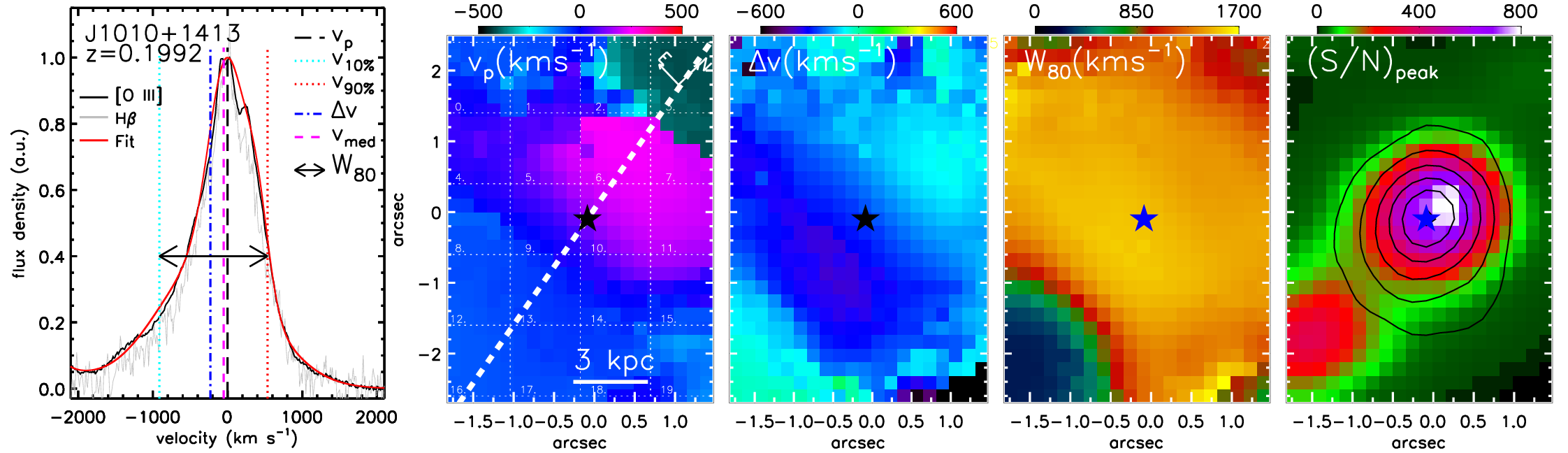,width=6.in,angle=90}}
\centerline{\psfig{figure=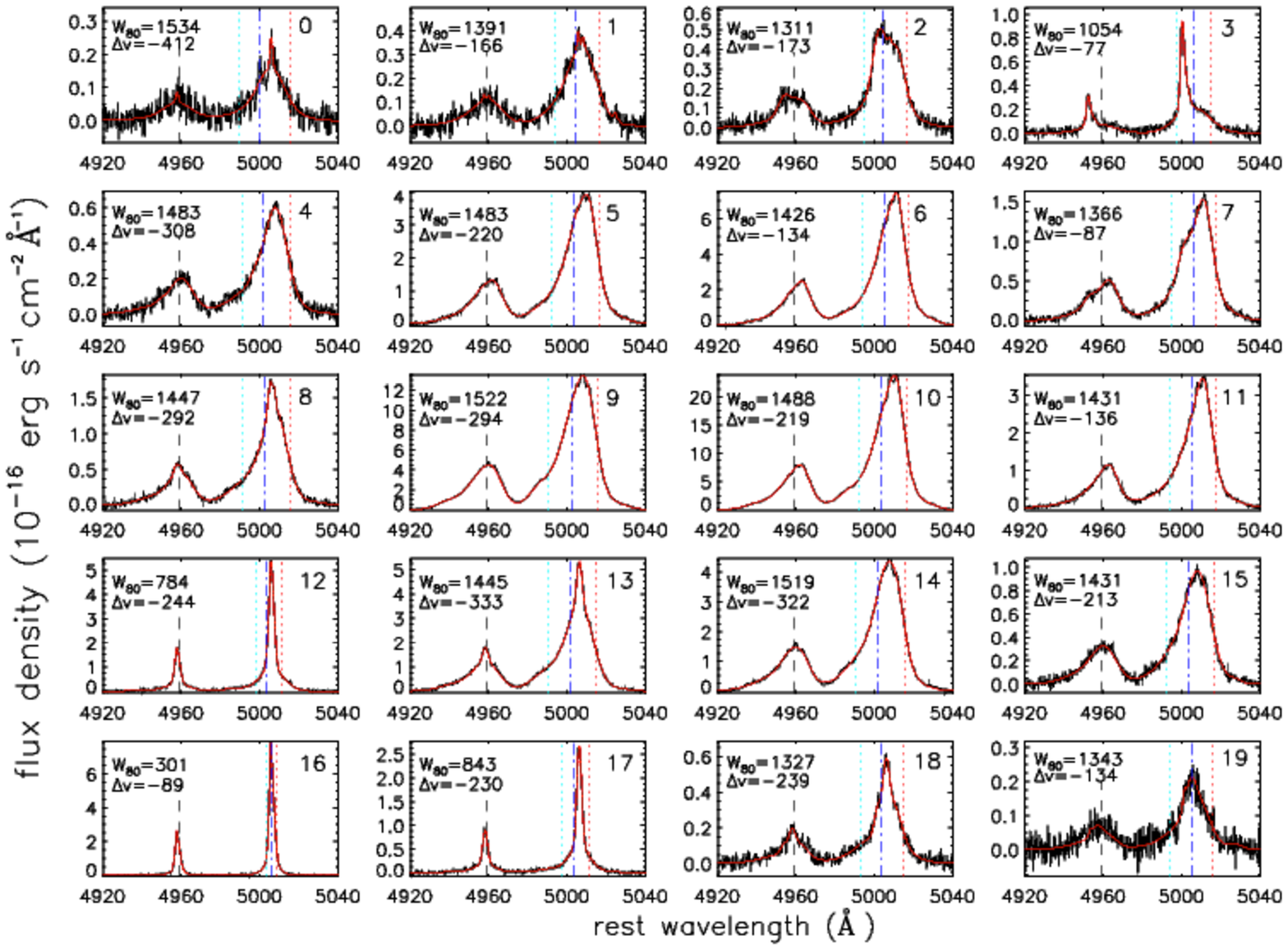,width=6.in,angle=0}}
\caption{Our IFU data for an example object in our sample
  (SDSS\,J1010+1413). The equivalent plots for the rest of the sample can be
  found in Appendix~A. Upper panels from left to right: {\it Panel 1:}
  Galaxy-integrated [O\,{\sc iii}]$\lambda$5007 emission-line profile
  (black curve) extracted from the full field-of-view of the IFU. Zero
  velocity corresponds to the redshift given in Table~\protect\ref{Tab:observations} and the
  spectrum has been normalised to the peak flux density. The solid red
  curves indicate the best fit to the emission-line profile. The
  vertical dashed and dotted lines correspond to the non-parametric velocity
  definitions described in Figure~\protect\ref{Fig:ExampleSpec} and
  Section~\protect\ref{Sec:VelocityDefinitions}. The H$\beta$
  emission-line profile is
  shown in grey; {\it Panel 2:} The velocity of the
  [O~{\sc iii}] emission-line peak ($v_{\rm{p}}$) at each pixel. The solid bars indicate 3\,kpc in spatial extent and the dashed line
  indicates the kinematic ``major axis'' as defined in
  Section~\protect\ref{Sec:VelMaps}. The dotted lines and numbers indicate the spatial regions from which we extracted
  the spectra shown in the lower panels; {\it Panel
    3:}  The value of the velocity offset of the broad emission-line wings ($\Delta v$) at each
  pixel; {\it Panel 4:} The value of the emission-line width ($W_{80}$) at each pixel. {\it
    Panel 5:} Signal-to-noise ratio of the peak flux density of the
  [O~{\sc iii}] emission-line profile at each pixel. The contours show the
  morphology of line-free continuum emission (collapsed around a $\approx$200\,\AA\
  wavelength region centered on [O~{\sc iii}]$\lambda$5007) indicated by lines of
  constant signal-to-noise ratio, starting at 15$\sigma$ with increments of
  10$\sigma$. The stars in each panel
  show the position of the peak of this continuum emission. {\it Lower
    panels:} Continuum-subtracted spectra extracted from the
  individual spatial regions indicated in Panel~2. The definitions of the
  curves and lines are the same as in Panel~1 with the addition
  that the black vertical dashed line fixed at $\lambda=4959$\,\AA\ that indicates the systemic wavelength for
  the [O~{\sc iii}]$\lambda$4959 emission line.
}
\label{Fig:velmaps}
\end{figure*}

\begin{figure*}
\centerline{\psfig{figure=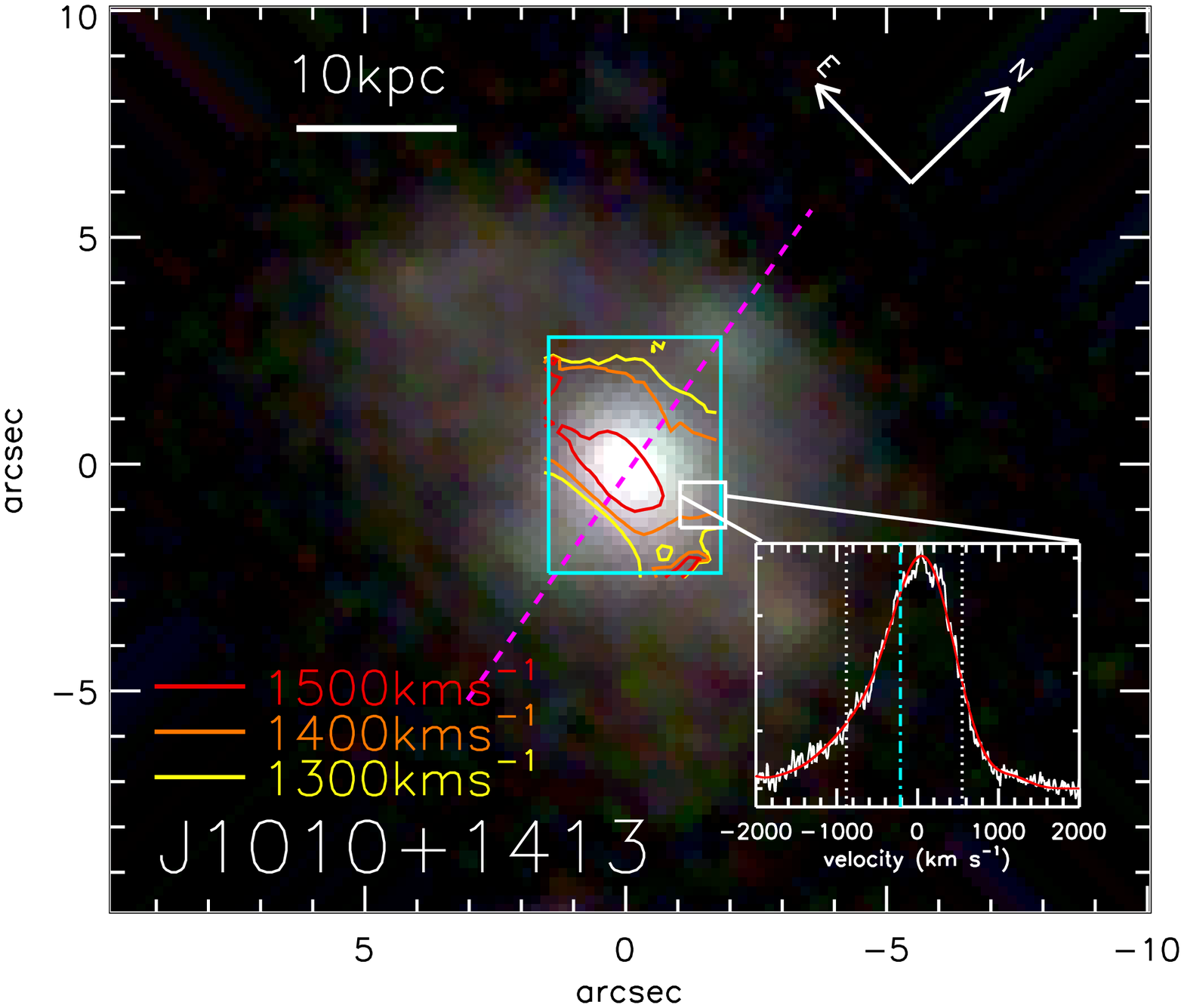,width=3.2in,angle=0}\psfig{figure=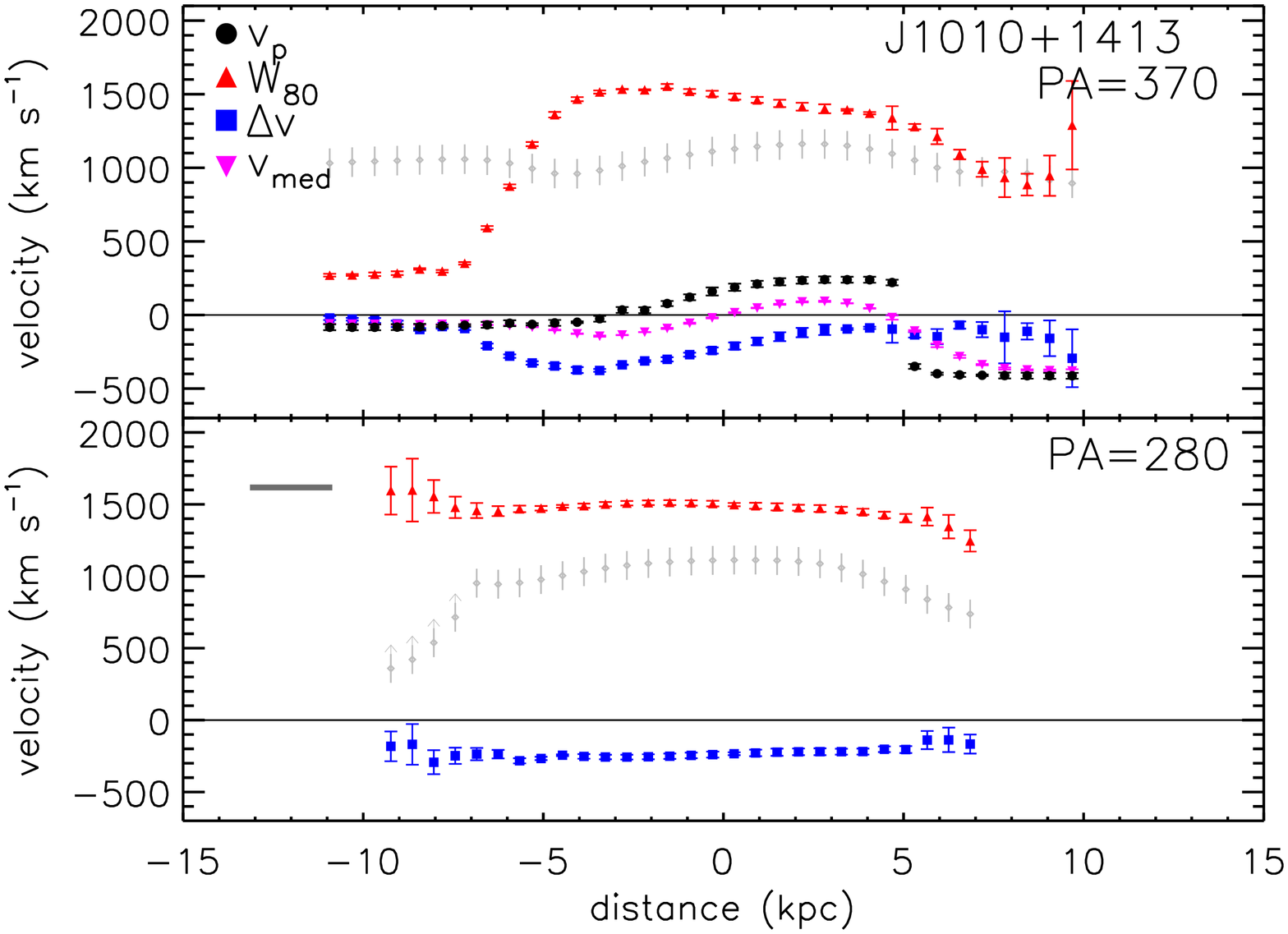,width=3.7in,angle=90}}
\caption{Our IFU data and SDSS image from an example object
  (J1010+1413) in our
  sample. The equivalent plots for the rest of the sample can be
  found in Appendix A. {\em Left:} Three colour ($g$, $r$, $i$) SDSS image. The cyan box shows the field-of-view of our GMOS-IFU
  observations. The contours show values of $W_{80}$ and highlights
  the spatial distribution where the [O~{\sc iii}] emission-line
  profiles are broadest (see Fig.~\ref{Fig:velmaps}). The dashed
  line shows the kinematic ``major axis'' defined in Section~\ref{Sec:VelMaps}. We also
  show an example [O~{\sc iii}] emission-line profile extracted from
  the highlighted spatial
  region (white box; for curve and line definitions see Fig.~\ref{Fig:velmaps}). {\em Right upper:}
  [O~{\sc iii}] velocity profiles (using the velocity definitions
  described in Section~\ref{Sec:VelocityDefinitions}) along the kinematic ``major axis'' indicated in
  the left panel. Empty red triangles in Figures~A1--A15 are where the S/N of the spectra are low
  and the emission-line widths cannot be well characterised (see
  Section~\ref{Sec:FittingProcedure}). The grey diamonds indicate the
  flux ratio: $1000 \times \log\left( {\rm [O~III]} / {\rm H}\beta \right)$. {\em
    Right lower:} Velocity profiles taken perpendicular to the
  kinematic ``major axis''
  indicated in the left panel. The solid horizontal bar indicates the
  typical seeing of the observations (i.e., 0.$^{\prime\prime}$7). For
  this source very broad emission-line profiles
  ($W_{80}\gtrsim800$\,km\,s$^{-1}$) with a velocity
  offset of $\Delta v = -300$\kms\, are found over the full extent of
  the field-of-view ($\approx$\,16\,kpc). Extended broad emission is found in all of the objects
  in our sample but with a variety of morphologies. 
}
\label{Fig:sdss}
\end{figure*}

We significantly detect [O~{\sc iii}]
emission to the edges of the IFU field-of-view in all sixteen
sources (i.e., over $\ge$10--20\,kpc; see discussion in Section~\ref{Sec:EELRs}). These
extended emission-line regions enable us to
trace the ionised gas kinematics over these scales. In
Figure~\ref{Fig:velmaps} we show example velocity maps of $v_{\rm{p}}$,
$W_{80}$ and $\Delta v$ (for the target J1010+1413) that were
created by measuring these quantities across the field-of-view following the procedure outlined in Section~\ref{Sec:FittingProcedure}
(see Fig.~A1--A15 for these maps for the other targets). The spectra
used to calculate these values are from averaging over 3$\times$3
pixels (i.e., $\approx0.^{\prime\prime}6$;
comparable to the seeing of the observations) at the position of every pixel. At each
position we measured the velocities from the fits to the [O~{\sc iii}] emission-line
profiles but only if it was detected in the spectra with a signal-to-noise
ratio $\ge$\,7. In Figure~\ref{Fig:velmaps} we also show the peak
signal-to-noise ratio [(S/N)$_{\rm{peak}}$] map which is the ratio of the peak flux
density of the fitted [O~{\sc iii}] emission-line profile to the noise in the
spectrum at each pixel position. These maps are discussed in Section~\ref{Sec:Results}. 

To further aid our analyses we also produced velocity profiles along two
axes through the IFU data cubes for each source (shown in
Fig.~\ref{Fig:sdss} and Fig.~A1--A15). We firstly defined a
``major axis'' for each source based upon the velocity peak maps
($v_{\rm{p}}$; e.g., Fig.~\ref{Fig:velmaps}). The positional angle (PA) of
the ``major axis'' for each source was chosen
such that the velocity shear of the $v_{\rm{p}}$ map was the maximum. For
a source where $v_{\rm{p}}$ traces galactic rotation this ``major axis'' corresponds to the kinematic major axis of the galaxy (for
a discussion see Section~\ref{Sec:GalaxyKinematics}). The second axis was
taken perpendicular to the ``major axis'' (see
e.g., Fig.~\ref{Fig:sdss}). Using the PAs defined above we produced
pseudo-long slit spectra with a slit-width of five pixels
($\approx0.8^{\prime\prime}$--1$^{\prime\prime}$). We obtained the
spectra along each slit (from averaging over a length of 3 pixels) and
calculated the velocity measurements outlined in Section~\ref{Sec:VelocityDefinitions}. In these velocity
profiles we also plot the ratio of the [O~{\sc iii}] to
H$\beta$ emission line fluxes, plotting the 3$\sigma$ lower-limit on this ratio when H$\beta$ is not detected
above 3$\sigma$.

\section{Results}
\label{Sec:Results}

Using Gemini-GMOS IFU data we study the spatially resolved properties
of the [O~{\sc iii}]$\lambda\lambda$4959,5007 and H$\beta$ emission
lines of sixteen luminous type~2 AGN (Table~\ref{Tab:observations}) chosen from our well constrained parent
sample (described in Section~\ref{Sec:Selection}; see
Fig.~\ref{fig:selection} and Fig.~\ref{Fig:histograms}). In
Figure~\ref{Fig:velmaps} and Figure~A1--A15 we show the
spatially-integrated [O~{\sc iii}]$\lambda$5007 and H$\beta$ emission-line profiles, the velocity
maps and the peak [O~{\sc iii}] signal-to-noise ratio maps for all of the sources in our sample. In
Figure~\ref{Fig:sdss} and Figure~A1--A15 we show SDSS images and
velocity-distance profiles and in Table~\ref{Tab:velmaps} we tabulate the
values that we derive from the galaxy-integrated spectra, the
emission-line images and the velocity maps. In the
following sub-sections we present our results of the sample as a whole,
firstly presenting the sizes and morphologies of the emission-line
regions (Section~\ref{Sec:EELRs}) before presenting the spatially
resolved kinematics (Section~\ref{Sec:Kinematics}). We give notes on
the results from our IFU data for the individual sources in Appendix~A.

\subsection{Extended emission-line regions: sizes and morphologies}
\label{Sec:EELRs}

\begin{figure*}
\centerline{\psfig{figure=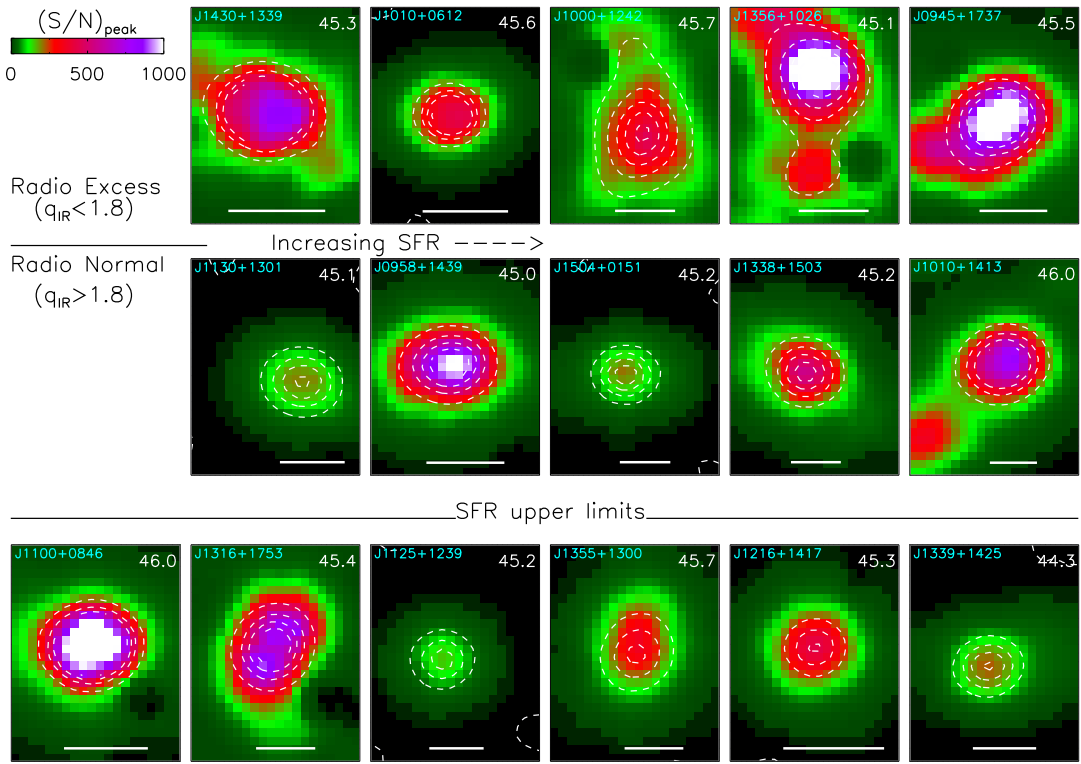,width=5.7in,angle=90}}
\caption{The [O~{\sc iii}] peak signal-to-noise ratio maps (see
  Section~\ref{Sec:VelMaps}) for all sixteen type~2 AGN in our
  sample. We have split the maps for the targets by radio excess (top row), radio
  normal (middle row) and those with only upper limits on their SFRs, for which we cannot determine their radio excess parameter
  (bottom row; see Table~\ref{Tab:observations}). The top two rows are
  arranged by SFR from the lowest (left) to the highest (right). The dashed
  contours show the morphology of the wavelength collapsed [O~{\sc
    iii}]$\lambda$5007 images. The horizontal bars represent 3\,kpc in
  extent. The numbers in the top-right of each map is the AGN
  luminosity for that source ($\log[L_{{\rm AGN}}; {\rm
    erg\,s^{-1}}]$; Table~\ref{Tab:observations}). The [O~{\sc iii}]
  images are predominantly round or moderately elliptical and there is tentative evidence that the
  radio excess sources show a higher incidence of irregular morphologies.
}
\label{Fig:SNmontage}
\end{figure*}

Although the ionised gas kinematics are the focus of this study, it is
useful to make some initial comments on the emission-line region sizes
and morphologies to compare to previous studies and to aid our later
discussion. We detect [O~{\sc iii}] emission over physical spatial extents of $\ge$10--20\,kpc (i.e., in some cases the
[O~{\sc iii}] emission is extended beyond the IFU field-of-view). This is consistent with other
studies that have shown that luminous AGN can have emission-line regions up to tens of kiloparsecs in size (e.g.,
\citealt{Bennert02}; \citealt{Humphrey10}; \citealt{Greene11}; \citealt{Schirmer13}; \citealt{Harrison12a};
\citealt{Husemann13}; Liu et~al. 2013a\nocite{Liu13a};
\citealt{Hainline13}). In Figure~\ref{Fig:SNmontage} we show the peak
[O~{\sc iii}] signal-to-noise maps, for the whole sample, overlaid with contours showing the continuum-free
[O~{\sc iii}]$\lambda$5007 images (i.e., wavelength collapsed images around the emission
line). We fit single ellipses to the [O~{\sc iii}] emission-line images to calculate the
semi-major axis ($R_{{\rm [O~III]}}$; Table~\ref{Tab:velmaps}). The range of sizes we get from this method
are $R_{{\rm [O~III]}}=1.5$--4.3\,kpc, similar to, but generally
higher than, the ULIRG-AGN composites
observed in \cite{RodriguezZaurin13}. We note that if we
  subtract the typical seeing of the observations (see Table~\ref{Tab:velmaps}), in
quadrature, from these size measurements they are reduced
by $\approx$20\% with a {\em worst-case} scenario of $\approx$50\% if the
seeing reached $0.^{\prime\prime}9$.\footnote{The one source that has
  a $R_{{\rm [O III]}}$ value smaller than $0.^{\prime\prime}9$ (i.e., our
  most pessimistic value of the seeing across the whole sample) was
  taken under good conditions of $\approx0.^{\prime\prime}6$ and is therefore
  spatially resolved.} Based on a comparison to
high-resolution imaging \cite{RodriguezZaurin13} suggest
that $R_{{\rm [O~III]}}$ is an over estimate of the emission-line
region size for their sample; however, using the kinematics of
our IFU data we commonly find spatially resolved emission over much larger extents (see
Section~\ref{Sec:Kinematics} and Table~\ref{Tab:velmaps}) and
therefore find that $R_{{\rm [O~III]}}$ is probably an underestimate of the
size of the emission-line regions in our sources. Although we do not use identical methods, our observed sizes are comparable
to, but towards the lower end of, those seen in the more luminous
type~2 quasars in \cite{Liu13a} as is expected by the
[O~{\sc iii}] size-luminosity relationship (e.g., \citealt{Bennert02}; \citealt{Hainline13}). 

In agreement with previous studies of radio-quiet AGN (e.g.,
\citealt{Husemann13}; \citealt{Liu13a,Liu14}) we find that the emission-line
regions in our sample are predominantly round or moderately elliptical (Fig.~\ref{Fig:SNmontage}). However, we find
tentative evidence that radio excess sources (i.e., those with
the strongest evidence of radio emission above that expected from star
formation) have a higher incidence
of extended and irregular morphologies. It has previously been shown that
radio-loud quasars have a higher incidence of irregular morphologies
than their radio-quiet counterparts (\citealt{Liu13a} and references
there-in). Although the radio data presented here are
  of insufficient spatial resolution to establish a direct (or lack there-of) connection, our
results may imply that the presence of radio-AGN activity, even at
modest luminosities, is connected to the morphologies of their
emission-line regions (see also \citealt{Husemann13} and Appendix~A). Alternatively, the ``radio excess'' in these sources could be due to shocks
that are also responsible for the irregular emission line regions that
we observe in Figure~\ref{Fig:SNmontage} (see \citealt{Zakamska14}). Multi-frequency high-resolution
radio imaging is required to determine the origin and morphology of
the radio emission in these objects.

\subsection{Spatially resolved ionised gas kinematics}
\label{Sec:Kinematics}

Multiple kinematic components (i.e., multiple narrow and broad
components) are seen in the galaxy-integrated emission-line profiles
of the sources in our sample (Fig.~\ref{Fig:velmaps}; Fig.~A1--A15). There are several possible
origins for these kinematic components such as merging nuclei, merger
remnants or merger debris, halo gas, rotating gas discs and outflowing or inflowing
gas. Indeed, previous spatially-resolved spectroscopy of type~2 AGN has shown
that multiple kinematic processes are likely to contribute to
the overall emission-line profiles of ionised gas (e.g., \citealt{Holt08};
\citealt{Fischer11}; Villar-Mart\'in et~al. 2011a,b\nocite{VillarMartin11a,VillarMartin11b}; \citealt{Rupke13}). By
studying the spatially resolved velocity maps, velocity profiles and peak signal-to-noise maps from our IFU data (Fig.~\ref{Fig:velmaps};
Fig.~\ref{Fig:sdss}; Fig.~A1--A15) we can distinguish between these scenarios and
characterise individual kinematic components. Although the focus of our study
is to identify and characterise galaxy-wide ionised outflows, in order to isolate the kinematics due to outflows we must also investigate
these other kinematic features. In Section~\ref{Sec:GalaxyKinematics}
we give a qualitative description of the observed kinematics due to galaxy
rotation and merger features, including candidate dual AGN (summarised in
Table~\ref{Tab:velmaps}), and in Section~\ref{Sec:outflows} we describe
the outflow kinematics. The ionised gas kinematics for
individual sources are described in Appendix~A. We note here that most of the analysis
is described on the basis of inspecting our velocity maps and velocity-distance
profiles; however, we checked that these maps are a reasonable
representation of the data by inspecting the spectra extracted from several regions across the IFUs (these
spectra are shown in Fig.~\ref{Fig:velmaps} and Fig.~A1--A15).

\subsubsection{Ionised gas kinematics: galaxy rotation, mergers and dual AGN}
\label{Sec:GalaxyKinematics}

\begin{table*}
\begin{center}
{\footnotesize
{\centerline{\sc Kinematic measurements and emission-line region sizes}}
\begin{tabular}{crrrrrrccc}
\noalign{\smallskip}
\hline
\noalign{\smallskip}
Name & $\langle$$W_{80}$$\rangle$          & $W_{\rm{80,max}}$   &
$\langle$$\Delta v$$\rangle$ &$|\Delta v_{\rm{max}}|$       &
$\langle$$v_{02}$$\rangle$        &$\Delta v_{\rm{med,max}}$ &
Gal. Kinematics& $R_{{\rm [O~III]}}$ & $D_{600}$\\
&(km\,s$^{-1}$) &(km\,s$^{-1}$) &(km\,s$^{-1}$)&(km\,s$^{-1}$)&(km\,s$^{-1}$)&(km\,s$^{-1}$)&&(kpc)&(kpc)\\
(1)       &(2)                 &(3)                &(4)                   &(5)&(6)&(7)&(8)&(9)&(10)\\
\noalign{\smallskip}
\hline
 J0945+1737 & 1009 & 1284 & $-$273 & 284 & $-$1511 & 138 & I&2.7$\pm$1.6 & $\ge$12\\
J0958+1439 & 815 & 904 & $-$46 & 107 & $-$866 & 190 & R &2.6$\pm$1.4 & $\ge$10\\
J1000+1242 & 795 & 873 & $-$58 & 171 & $-$761 & 311& R/I &4.3$\pm$1.8 & 14\\
J1010+1413 & 1449 & 1525 & $-$229 & 350 & $-$1523 & 299 & R/I& 3.9$\pm$2.3 & $\ge$16\\
J1010+0612 & 1280 & 1468 & $-$95 & 216 & $-$1267 & 138 & R& 1.6$\pm$1.3 & $\ge$6\\
J1100+0846 & 1066 & 1367 & $-$30 & 148 & $-$1192 & 55 & R&1.9$\pm$1.3 & $\ge$10\\
J1125+1239 & 1285 & 1574 & $-$265 & 424 & $-$1547 & 73 & F&2.9$\pm$2.0 & $\ge$9\\
 J1130+1301 & 778 & 849 & 149 & 173 & $-$616 & 140 & R& 2.8$\pm$1.7 & 8\\
J1216+1417 & 1228 & 1456 & 124 & 230 & $-$1115 & 40 & F &1.5$\pm$1.1 & $\ge$6\\
J1316+1753 & 1127 & 1169 & $-$191 & 326 & $-$1216 & 500 & R/I&3.1$\pm$1.8 & $\ge$14\\
J1338+1503 & 890 & 1085 & 124 & 182 & $-$813 & 145 & R/I &3.5$\pm$2.2 & $\ge$13\\
J1339+1425 & 672 & 724 & 147 & 178 & $-$505 & 84 & R &2.5$\pm$1.7 & 6\\
J1355+1300 & 667 & 953 & $-$184 & 277 & $-$797 & 71 & R &3.5$\pm$1.8 & $\ge$6\\
J1356+1026 & 900 & 964 & $-$215 & 544 & $-$1049 & 523 & I &3.1$\pm$1.5 & $\ge$11\\
J1430+1339 & 822 & 1042 & $-$152 & 268 & $-$999 & 439 &R/I & 1.8$\pm$1.1 & $\ge$9\\
J1504+0151 & 1184 & 1181 & $-$495 & 520 & $-$1739 & 149 & R&2.9$\pm$2.1 & $\ge$7\\

\hline
\hline
\end{tabular}
}
\caption{\label{Tab:velmaps}
{\protect\sc Notes:}\protect\\
Velocity values and sizes derived from the galaxy-integrated spectra and velocity maps
shown in Fig.~\ref{Fig:velmaps} and Fig.~A1-A15 (see
Section~\ref{Sec:VelocityDefinitions} and Fig.~\ref{Fig:ExampleSpec} for
velocity definitions). (1) Object name; (2) value of the [O~{\sc iii}]
emission-line width, $W_{80}$, derived from the
galaxy-integrated spectra; (3) maximum value of $W_{80}$ from the spatially resolved maps
(with a 95\% clipping threshold to remove spurious pixels); (4) velocity offset, $\Delta v$, derived from
the galaxy-integrated spectrum; (5) absolute value of the maximum $\Delta v$ in the
spatially resolved maps (with 95\% clipping); (6) the value of maximum
[O~{\sc iii}] velocity, $v_{02}$, derived from the integrated spectrum; (7) the maximum
velocity shear of the median velocity, $v_{\rm{med}}$, from the velocity maps (with 95\%
clipping); (8) qualitative description of the peak velocity maps:
R$=$rotation-like velocity fields; R/I$=$red-blue velocity gradient
but irregular velocity field; I$=$
irregular velocity field; F$=$no velocity gradients (see Section~\ref{Sec:GalaxyKinematics}); (9)
observed semi-major axis of the [O~{\sc iii}] emission-line region from fitting an
ellipse with uncertainties that are the typical seeing (i.e., 0.$^{\prime\prime}$7); (10) total projected extent of observed broad
emission-line profiles (i.e., with $W_{80}>600$\,km\,s$^{-1}$) with
conservative uncertainties of $2\times$ the seeing. Measurement
uncertainties on the galaxy-integrated velocities are typically $<10$\% and therefore the dominant source of uncertainty is the physical
interpretation of these values (see Section~\ref{Sec:Kinematics}). 
}

\end{center}
\end{table*}

Rotating gas discs will produce regular velocity fields, with a
gradient of blue to red and relatively narrow emission-line profiles (i.e.,
$W_{80}<$\,600\,km\,s$^{-1}$). As discussed in Section~\ref{Sec:VelocityDefinitions} the emission-line peak
velocity ($v_{\rm{p}}$; see Fig.~\ref{Fig:ExampleSpec}) predominantly
traces the luminous narrow emission-line components in our data. We find
regular $v_{p}$ velocity fields and velocity profiles, indicative of
tracing galaxy rotation (e.g., \citealt{Courteau97}), for seven of the sixteen
objects in the sample. We give these a classification of ``R'' in
Table~\ref{Tab:velmaps} and give a justification for this, for
individual sources, in Appendix~A. This
interpretation of rotating discs is often supported by the kinematic ``major axis'' (defined in
Section~\ref{Sec:VelMaps}) being aligned with the morphological major
axis in the disc-like galaxies revealed in the SDSS images and/or the optical continuum seen in our
IFU data. Indeed, it has previously been observed
that the narrow-components (or ``cores''; i.e., with the broad-wings
removed) of ionised emission-line profiles, can be good tracers of
stellar dynamics (i.e., traces galactic rotation and/or stellar velocity
dispersion) or molecular gas disks (e.g., \citealt{Greene05a};
\citealt{Barth08}; \citealt{Barbosa09}; \citealt{Rupke13}; but see
\citealt{Zakamska14}). Further
  confirmation that the $v_{p}$ velocity fields are tracing rotation would require detailed
kinematic analysis (e.g., \citealt{Courteau97}; \citealt{Krajnovic06})
that is beyond the scope of this paper. In a further five
sources we see a velocity gradient from blue to red in the $v_{p}$ velocity maps and profiles; however, the
velocity fields are more irregular and therefore they may not be
dominated by galaxy rotation but instead may trace kinematics due other
processes such as merging components or outflows (given a
classification ``R/I'' in Table~\ref{Tab:velmaps}; see Appendix~A for
discussion of individual cases). 

We next consider kinematics due to mergers. Mergers are known to create extended emission-line regions
due to companion galaxies or merger debris and have been previously been observed in spatially
resolved spectroscopy of AGN (e.g., Villar-Mart\'in et~al. 2011a,b\nocite{VillarMartin11a};
\citealt{Harrison12a}). Furthermore, they may be expected to be common
in our sample as $\gtrsim$60\% of type~2 AGN at moderate
redshifts $z\approx0.3$--0.4 show signs of morphological disturbance
at some level in $HST$ imaging (\citealt{VillarMartin12}).\footnote{We
note that \citealt{VillarMartin12} refer to low-level morphological disturbances
of several types based on the classifications of
\citealt{RamosAlmeida11}. The number of {\em major mergers} observed in AGN
is much lower (e.g., \citealt{Cisternas11}).} The [O~{\sc iii}]
emission-line peak signal-to-noise ratio maps (Fig.~\ref{Fig:SNmontage}) allows us to search for faint spatially-distinct
emission-line regions which could indicate companion/merging
galaxies, merger debris, tidal features or halo gas. Based on these
maps, five sources show spatially distinct emission-line regions with
distinct velocities to the rest of the host galaxy. One of these has
clear signs of major-merger activity based on the SDSS image and our
IFU data (J1356+1026; Fig.~\ref{fig:1356+1026}), three are likely
to be distinct emission-line regions (i.e., merger
debris or halo gas; J0945+1737; Fig.~\ref{fig:0945+1737}; J1010+1413; Fig.~\ref{Fig:velmaps} and J1000+1242; Fig.~\ref{fig:1000+1242}) and
one could have several interpretations (J1316+1753;
Fig.~\ref{fig:1316+1753}; these objects are discussed in detail in Appendix~A).

Major mergers of galaxies naturally leads to the possibility of
``dual'' AGN;  i.e., two nuclei which are gravitationally bound and
are both actively accreting material. Systematic spectroscopic searches for AGN with
luminous double peaked [O~{\sc iii}] emission-line profiles has become
a popular method to identify candidate dual AGN (e.g., \citealt{Wang09}; \citealt{Smith10};
\citealt{Liu10}; \citealt{Ge12}; \citealt{Comerford13}). However, these double peaks could also be due
to other processes such as rotating gas disks, merging components or outflows,
all illuminated by a single AGN (e.g., \citealt{Fu11,Fu12}; see \citealt{Comerford12} for a discussion). Two of our
sources show luminous double peaks (J1316+1753 and J1356+1026)
in their galaxy-integrated [O~{\sc iii}] emission-line profiles. Based
on our data, the strongest case for a dual AGN is J1356+1026 (see also \citealt{Greene12}),
where we observe two distinct emission-line and continuum regions (see
Appendix~A for further discussion of these two sources). 

\subsubsection{Ionised gas kinematics: galaxy-wide outflows}
\label{Sec:outflows}

\begin{figure*}
\centerline{\psfig{figure=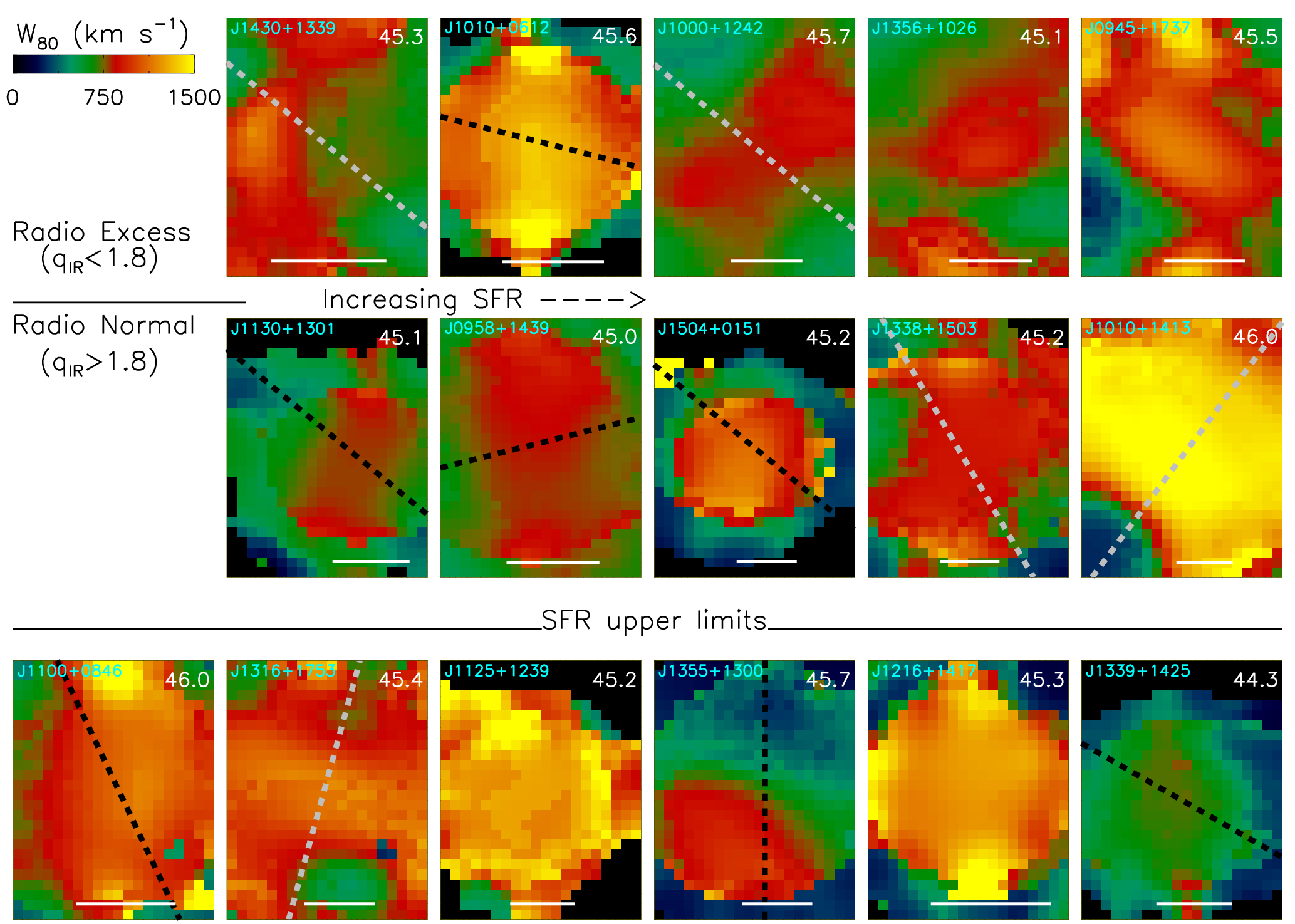,width=5.7in,angle=90}}
\caption{Maps of $W_{80}$ for all sixteen type~2 AGN in our
  sample that illustrate the spatial extent and morphology of the
  outflows (Section~\ref{Sec:outflows}). The maps are ordered in an identical manner to
  Fig.~\ref{Fig:SNmontage}. The dashed
  lines correspond to the ``major axes'' based on the kinematics of
  the narrow emission-line components (Section~\ref{Sec:VelMaps}). Black dashed lines are those
  with blue to red velocity gradients indicative of rotation (i.e., ``R'' in
  Table~\ref{Tab:velmaps}) and grey are those with irregular blue to red
  velocity gradients (i.e., ``R/I'' in
  Table~\ref{Tab:velmaps}). Sources without a clear blue to red velocity
  gradient do not have these ``major axes'' shown. The solid horizontal bars
  represent 3\,kpc in extent and the number in the top-right of each
  map is $\log[L_{{\rm AGN}}; {\rm erg\,s^{-1}}]$ for that source. We observe a wide variety in
  the morphology of the kinematically disturbed ionised gas and there
  is no obvious connection between outflow morphology and SFR, radio excess or AGN luminosity.
}
\label{Fig:W80montage}
\end{figure*}

In the IFU data of all sixteen targets we find broad [O~{\sc
  iii}]$\lambda\lambda$4959,5007 emission-line profiles across the field-of-view
(i.e., $W_{{\rm 80,max}}\approx720$--1600\kms; Table~\ref{Tab:velmaps}) with a range of maximum projected
velocity offsets ($|\Delta v_{\rm{max}}|=110$--540\,km\,s$^{-1}$;
Table~\ref{Tab:velmaps}). Such broad (i.e., $W_{80}\gtrsim600$\kms) and asymmetric [O~{\sc iii}] emission-line profiles, are
difficult to explain with mergers or galaxy rotation, even for the most massive
galaxies, and are attributed to high-velocity outflowing gas
(e.g. \citealt{Heckman81}; \citealt{Veilleux95}; \citealt{Greene05a}; \citealt{Barth08};
\citealt{Nesvadba08}; \citealt{Greene11}; \citealt{Westmoquette12};
\citealt{Harrison12a}; Liu et~al. 2013b\nocite{Liu13b};
\citealt{Rupke13}). As with all of these studies we favour outflow over inflow
due to the exceptionally high velocities and line-widths that we
observe.  In the previous sub-section we isolated kinematic components
due to other features (e.g., mergers and galaxy rotation) and we
find that these have little effect on the observed kinematic structure of the line-width, $W_{80}$ and also in most
cases that of the velocity offset, $\Delta v$ (the exceptions are
J1356+1026 and J1316+1753 for which $\Delta v$ is affected, see
Appendix~A for details); i.e., these maps are dominated by outflow kinematics. We therefore use these maps to describe the
outflow kinematic structures and morphologies. We note that Liu et~al. (2013b)\nocite{Liu13b} use
the velocity structure of the median velocity ($v_{\rm{med}}$;
Fig~\ref{Fig:ExampleSpec}), in which they predominantly see a blue to
red velocity gradient, as part of their outflow analysis of their IFU
data of $z=0.3$--0.6 radio-quiet luminous type~2 AGN. We observe a very
similar range in maximum line-widths and $v_{\rm{med}}$ maximum velocity gradients (i.e., $\Delta
v_{{\rm med,max}}\approx55$--520\kms; Table~\ref{Tab:velmaps}) to the objects in Liu
et~al. (2013b)\nocite{Liu13b}. However, by spatially resolving $v_{\rm{med}}$ we find that this quantity is predominantly dominated by the
kinematics of the $v_{p}$ velocity field (Fig.~\ref{Fig:sdss}; Fig~A1--A15) and consequently $v_{\rm{med}}$
can often be dominated by kinematics due to galaxy rotation and mergers
(see Section~\ref{Sec:GalaxyKinematics}). Therefore, we favour $\Delta
v$ over $v_{\rm{med}}$ to describe the projected outflow velocity
offset for the rest of our analysis.

For an outflow consisting of ionised clouds (observable
in [O~{\sc iii}] emission) embedded in a bulk outflow (e.g., \citealt{Heckman90};
\citealt{Crenshaw00a}) the projected velocity offsets will be sensitive to inclination
effects while the emission-line widths (i.e., $W_{80}$) are more
likely to reflect typical bulk motions and be less sensitive to
inclination (also see discussion in \citealt{Harrison12a}). We
therefore concentrate on the $W_{80}$ maps to investigate the sizes and
morphologies of the outflows (Fig.~\ref{Fig:W80montage}), comparing them
with the $\Delta v$ maps when appropriate (discussed in Appendix~A for
individual objects). The {\em observed} outflow velocities and
morphologies will also depend on other factors. For example, an
outflow with the far side obscured by dust can explain the excess of blue wings in
[O~{\sc iii}] emission-line profiles (e.g., \citealt{Heckman81};
\citealt{Vrtilek85b}; \citealt{Veilleux91}) while, outflowing
redshifted components can be explained if the outflow
is highly inclined and/or is extended beyond the likely obscuring
material (e.g., \citealt{Barth08}; \citealt{Crenshaw10};
\citealt{Greene12}; \citealt{Rupke13}). We observe both red and blue
velocity offsets in our sample (Table~\ref{Tab:velmaps}). A range of structures of
outflows have been observed in galaxies both with and without powerful
AGN, including spherically-symmetric outflows and bi-polar superbubbles (e.g., \citealt{Veilleux94}; \citealt{Greene12}; Liu
et~al. 2013b\nocite{Liu13b}; \citealt{Rupke13}). Detailed modelling of
the outflow kinematics of individual objects is beyond the scope of
this paper; however, we briefly describe the morphology and kinematics
of the outflows that we observe in our sample below (we describe individual objects in more detail in Appendix~A).

We define the quantity $D_{600}$ (values in Table~\ref{Tab:velmaps})
which is the maximum projected spatial extent of the broad [O~{\sc iii}]
emission-line profiles (with $W_{80}>600$\,km\,s$^{-1}$; i.e., where the kinematics are
likely to be dominated by outflows) based upon the velocity profiles (Fig.~\ref{Fig:sdss} and Fig.~A1--A15). The outflow features are located over large extents
in all sources ($D_{600}\gtrsim$\,[6--16]\,kpc; Table~\ref{Tab:velmaps}) with the spatial extents being
mostly lower limits due to the limited field-of-view of our
observations. We note that in most cases the broad [O~{\sc iii}] emission is
  clearly spatially extended, due to the morphological or kinematic
  structure that we observe (see below) and in all cases it is extended beyond the
  seeing; however, a conservative uncertainty on these $D_{600}$
  measurements would be $\approx2\times$ seeing (see Table~\ref{Tab:velmaps}). Although our sample was initially selected
to have luminous and broad [O~{\sc iii}] emission-line components (see Section~\ref{Sec:Selection}), these
  results demonstrate that when these components are seen in
  the one-dimensional spectra of luminous type~2 AGN (as is the case for
  at least 45$\pm$3\% of objects; see Section~\ref{Sec:Selection} and
  Fig.~\ref{Fig:histograms}), they are always found over galaxy-wide
  scales when observed with spatially resolved spectra (i.e., over
  at least a few kiloparsecs). To assess how confident we can be of the fraction of
  extended ionised outflows within our selection criteria we ran Monte Carlo
  simulations. To do this, we took 246 mock galaxies (i.e., the number from the
  parent catalogue that met our selection criteria;
  Section~\ref{Sec:Selection}), assigned them as
  either ``extended'' or ``not-extended'' (in various ratios)
  and randomly drew 16 objects from this sample 10$^{4}$ times. We
  found that, at most, 30\% of the sources can be ``not-extended''
  in the parent population for the probability of selecting 16
  ``extended'' sources by random chance to be $\le$0.3\% (i.e., the
  3$\sigma$ confidence level). In other words,
  when broad and luminous emission-line components are observed in one-dimensional spectra (i.e.,
  when $\gtrsim$30\% of the ionised gas appears to be outflowing;
  Section~\ref{Sec:Selection}) of luminous ($L_{{\rm
      [O~III]}}>5\times10^{41}$\,erg\,s$^{-1}$) $z<0.2$ type~2
  AGN, we can be confident, with 3$\sigma$ significance, that in
  {\em at least} 70\% of instances these features are extended on a few
  kiloparsec scales or greater.

In Figure~\ref{Fig:W80montage} we show the $W_{80}$ maps for the
whole sample to illustrate the morphology of the outflows. From this figure we can see a large diversity in the
morphology. Several of the sources show a preferential axis, which is
orientated away from the galaxy ``major axis'' (albeit not always
perpendicular); however, in some cases  we require a stronger constraint on the
major axis of the host galaxies to confirm
the relative orientations. This is
consistent with the collimated winds emerging along the minor axis
seen in star-forming galaxies (e.g., \citealt{Heckman90};
\citealt{Rupke13}). The most impressive outflows that we observe are
those which have the kinematic features of  $\approx$10--15\,kpc scale
super-bubbles; i.e., high-velocity bi-polar emission-line
regions each side of the nucleus. Combining our IFU data with
archival imaging and spectroscopic data (see Appendix~A) we identify three super-bubble candidates (one was previously
known: J1356+1026; \citealt{Greene12} and, as far as we are aware, two
are new: J1000+1242 and
J1430+1339; see Appendix~A). In two cases (J1356+1026 and J1430+1339)
we only observe the base of these bubbles and therefore require IFU observations with a larger field-of-view to fully characterise the
kinematics of the outflow structure. In the third case (J1000+1242) we see a clear blue and red sided
outflow. Similar outflows have been seen in galaxies both with and
without luminous AGN (e.g., \citealt{Veilleux94}; \citealt{Greene12}; \citealt{Rupke13}). 

In four of our targets the morphology of the outflows are circular
(Fig.~\ref{Fig:W80montage}) with
reasonably constant $W_{80}$ and $\Delta v$ values seen across the
extent of the emission-line regions. This kinematic structure is seen in the majority
of the more luminous type~2 AGN in Liu et~al. (2013b)\nocite{Liu13b}
who argue that these structures are characteristic of spherically symmetric, or
wide-angle bi-conal outflows, and a power-law [O~{\sc iii}] surface-brightness
profile (see \citealt{Liu14} for similar observations of type 1 AGN). We note that circular morphologies could (at least partially) be an
artifact of the fitting routine since in areas of low surface brightness only single
Gaussian components are fit and we are not able to fully characterise
the emission-line profile (see Section~\ref{Sec:FittingProcedure}); however,
this not likely to be the full explanation for the observed
morphologies (see Appendix~A; also see Fig.~5 of Liu
et~al. 2013b\nocite{Liu13b}). Finally, in one source (J1355+1300) we observe a spatially distinct high-velocity
region, offset from the nucleus, that has a few possible
interpretations including a small-scale outflowing bubble (see
Appendix~A for a discussion). 

\subsection{Outflow properties: estimates of mass, energy and momentum}
\label{Sec:properties}
\begin{figure}
\centerline{\psfig{figure=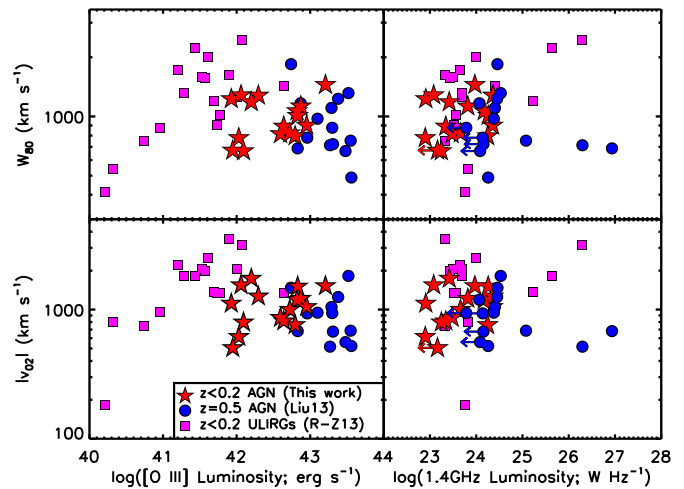,width=3.5in,angle=90}}
\caption{Line-width ($W_{80}$) and maximum velocity ($v_{02}$)
  versus [O~{\sc iii}] luminosity and 1.4\,GHz radio luminosity for our
  targets (red stars; see Table~\ref{Tab:observations} and
  Table~\ref{Tab:velmaps}). Also shown are $z\approx$\,0.3--0.6
  type~2 quasars (blue circles; Liu et~al. 2013b\protect\nocite{Liu13b}) and  $z<0.1$ ULIRG--AGN
(magenta squares; \protect\citealt{RodriguezZaurin13}). High velocity and disturbed
ionised gas is found in AGN over four orders of magnitude in both [O~{\sc
  iii}] and radio luminosity; however, no clear trends are observed
between these quantities from these samples.}
\label{Fig:compare}
\end{figure}

The mass, energy and momentum being carried by galaxy-wide outflows
are important quantities to constrain in order to understand the role that outflows play in
galaxy formation (discussed in Section~\ref{Sec:Impact}) and to also
help determine the physical mechanisms that are
driving them (discussed in Section~\ref{Sec:Drivers}). Outflows are
likely to be entraining gas in multiple phases (i.e., ionised,
molecular and neutral) and multiple gas phase observations are required to
fully characterise the outflow properties (e.g., \citealt{Shih10};
\citealt{Mahony13}; \citealt{Rupke13}). However, the warm ionised gas,
which we observe in our observations,
provides initial information on the outflows in this sample and could represent a large fraction of the overall
mass and energy of the total outflows, as has been shown for some AGN (e.g., \citealt{Rupke13}). The kinematic
structures of the outflows across our sample are diverse and complex
(see Section~\ref{Sec:outflows}) which makes accurate characterisation
of the outflows challenging without doing detailed kinematic modelling
of individual systems. However, as is often invoked
throughout the literature, we can apply some simple outflow models
to the whole sample to provide first order constraints on the mass,
energy and momentum involved in the outflows and to enable a direct comparison to
other studies. As most of the calculations outlined below require measurements of the H$\beta$ luminosity we do not include the three
sources where we did not make this measurement (see Section~\ref{Sec:FittingProcedure}). 

An important quantity for measuring ionised gas masses, and hence mass
outflow rates, is the electron density ($n_{e}$). This quantity is often
measured from the emission-line ratio [S~{\sc
  ii}]$\lambda$6716/$\lambda$6731 which is sensitive to
electron density (e.g., \citealt{Osterbrock06}). The [S~{\sc ii}] doublet is not
covered in our IFU observations; however, using single-component
Gaussian fitting to the SDSS spectra we find that the emission-line ratios for our sources are in the range  [S~{\sc
  ii}]$\lambda$6716/$\lambda$6731\,$=$\,0.8--1.2, corresponding to a
range in $n_{e}\approx200$--1000\,cm$^{-3}$ with a median of
$n_{e}=500$\,cm$^{-3}$ (\citealt{Osterbrock06}; assuming $T_{e}=10^{4}$\,K). This range of $n_{e}$
is in agreement with that previously found for type~2 AGN (e.g.,
\citealt{Greene11}). We note that these are average electron densities
within the SDSS fibre and the values for the {\em outflowing} kinematic components
are unknown and have the potential to be outside of the electron
density range that the [S~{\sc ii}]$\lambda$$\lambda$6716,6731 doublet is sensitive to (i.e.,
$n_{e}\gtrsim10^{4}$\,cm$^{-3}$; \citealt{Osterbrock06}; see discussion in
\citealt{Holt11} and \citealt{RodriguezZaurin13}), providing a source of uncertainty in
the mass and energy outflow rates in {\em all} studies that use this
method. However, we note that \cite{Greene11} provide convincing arguments that the
emission line regions in their $z<0.5$ radio-quiet type~2 AGN are
likely to be clumpy and that standard mass estimates (such as those we
derive below) are likely to be under-estimated.

The observed ionised gas mass (i.e., the gas which is emitting in
H${\beta}$) can be estimated assuming purely photoionized gas
with ``Case B'' recombination (with an intrinsic line ratio of
H$\alpha$/H$\beta=$\,2.9) and an electron temperature of
$T=10^{4}$\,K, following \cite{Nesvadba11} and \cite{Osterbrock06}, using the relation:\footnote{To be consistent with \cite{Liu13b}, Equation (2) is given in the exact form that is provided in \cite{Nesvadba11}. However, we note that if you follow section 4.2 in \cite{Osterbrock06} with the appropriate value for the recombination coefficient of H$\beta$ for the conditions described (i.e., $\alpha_{{\rm H}\beta}^{{\rm eff}} = 3.03\times10^{-14}$ cm$^3$ s$^{-1}$) the normalisation constant in Equation~\ref{Eq:MASSS} is $6.78\times10^8\Msol$. This factor of four difference to the quoted normalisation constant has no significant bearing on our order of magnitude estimates of the gas mass.}
\begin{equation}
\label{Eq:MASSS}
\frac{M_{{\rm gas}}}{2.82\times10^{9}{\rm \Msol}} = \left(\frac{L_{{\rm
      H}\beta}}{10^{43}\,{\rm erg\,s}^{-1}}\right) \left(\frac{n_{e}}{100\,{\rm cm}^{-3}}\right)^{-1}
\end{equation}
This relationship between the observed line-emitting gas mass, $n_{e}$ and $L_{{\rm H}\beta}$ (or
an equivalent using $L_{{\rm H}\alpha}$) is commonly adopted in studies
of outflows (e.g., \citealt{Holt06}; \citealt{Genzel11};
\citealt{RodriguezZaurin13}; Liu et~al. 2013b\nocite{Liu13b};
\citealt{SchnorrMuller14}) with normalisation factors that vary within a
factor of a few, depending on the exact assumptions. Using our
extinction un-corrected $L_{{\rm H}\beta}$ values
(Table~\ref{Tab:observations}), we obtain total observed ionised gas
masses of $M_{{\rm
    gas}}=$\,(2--40)\,$\times10^{7}$\Msol, assuming the commonly
adopted $n_{e}=100$\,cm$^{-3}$ (e.g., Liu et~al. 2013b\nocite{Liu13b}), or $M_{{\rm
    gas}}=$\,(0.4--8)\,$\times10^{7}$\Msol\ assuming our median value
of $n_{e}=500$\,cm$^{-3}$. We note that the ionised gas phase will make up only a
  fraction of the {\em total} gas content. For example, taking a typical
SFR for our targets (40\Msolyr) and assuming a star forming region of $\approx1$\,kpc would imply
molecular gas masses of $\approx10^{10}$\Msol\ (\citealt{Genzel11})

The simplest method that we can use to estimate the total kinetic energy
in the outflows is the following,
\begin{equation}
\label{Eq:Edot}
E_{{\rm  kin}} = \frac{1}{2}M_{{\rm gas}}v^{2}_{{\rm gas}},
\end{equation}
Following Liu et~al. (2013b)\nocite{Liu13b} we assume
$n_{e}=100$\,cm$^{-3}$ and that all of the observed ionised gas is involved in
the outflow\footnote{As discussed in
  Section~\ref{Sec:GalaxyKinematics} not all of the ionised gas
  appears to be involved in an outflow. The broad
  kinematic components (FWHM$>$700\,km\,s$^{-1}$) in the [O~{\sc iii}]
  and H$\beta$ emission lines typically contribute 30--80\% of the total flux (based on
\citealt{Mullaney13}) which may give an estimate of the fraction of
gas involved in the outflows. However, the emission-line
ratios of up to H$\alpha$/H$\beta=8$ (from \citealt{Mullaney13})
implying that the intrinsic $L_{{\rm H}\beta}$ values (and hence
total calculated masses) may be up to a factor of
10 times higher than the observed values (following
\citealt{Calzetti00}). Therefore, these two effects will cancel each other out
to some level and this assumption is sufficient for our
order-of-magnitude estimates} with a bulk-outflow velocity $v_{{\rm
  gas}}=W_{80}/1.3$ (suitable for the spherically symmetric or
wide-angle bi-cone outflow models outlined Liu
et~al. 2013b\nocite{Liu13b}; see also \citealt{Harrison12a} for
different arguments on using line-widths to estimate the bulk velocities). Across
the sample this leads to total kinetic energies of $E_{{\rm kin}}=(0.5$--$50)\times10^{56}$\,erg. Assuming the
typical maximum radial extent of the outflows is 6\,kpc (see
Table~\ref{Tab:velmaps}) and a continuous outflow, this implies outflow
lifetimes of $t_{{\rm out}}\approx6$\,kpc$/v_{{\rm
    gas}}=$\,(5--11)\,Myrs and consequently outflow kinetic energy rates of $\dot{E}_{{\rm kin}}=(0.1$--$30)\times$10$^{42}$\,erg\,s$^{-1}$. 

An alternative approach is to assume a spherical volume of outflowing
ionised gas (following e.g., \citealt{RodriguezZaurin13}; see also \citealt{Holt06}), which gives mass outflow rates of
\begin{equation}
\label{Eq:RZMdot}
\dot{M}_{{\rm out,M1}} = \frac{3 M_{{\rm gas}}v_{{\rm out}}}{r}
\end{equation}
and outflow kinetic energy rates of
\begin{equation}
\label{Eq:RZEdot}
\dot{E}_{{\rm out,M1}} = \frac{\dot{M}}{2}\left(v_{{\rm out}}^2 + 3\sigma^2\right)
\end{equation}
where $\sigma$ is the velocity dispersion and $v_{{\rm out}}$ is the
outflow velocity. To match
\cite{RodriguezZaurin13} as closely as possible\footnote{We note that
  \cite{RodriguezZaurin13} use a different definition of $v_{{\rm out}}$
  to $\Delta v$; however, the two methods agree within a factor of
  $\approx$(1--3) with a median of 1.1. Likewise, their definition for $\sigma$ agrees
  with  $\sigma=W_{80}/2.355$ within a factor of $\approx$(1--2) and
  median of 1.2.} we take the outflow velocity to be $v_{{\rm
    out}}=\Delta v$, the velocity dispersion to be
$\sigma=W_{80}/2.355$, the outflow radius to be $r=R_{{\rm
    [O~III]}}$ (Table~\ref{Tab:velmaps}) and again use
$n_{e}=100$\,cm$^{-3}$. Using this approach (we will call
  ``Method 1'') we obtain $\dot{M}_{{\rm out,M1}}=3$--$70$\Msolyr and
  $\dot{E}_{{\rm out,M1}}=$\,(0.3--30)\,$\times$10$^{42}$\,erg\,s$^{-1}$. This
  range in outflow kinetic energy rate is in excellent agreement with the range
of values using the previous method, which we will now ignore, and the two methods
agree within a factor $\le3$ for all sources. Both of these methods
provide strict lower limits, as we are only observing the
line-emitting gas and hence the total mass involved in the outflow
could be much higher (see also e.g., \citealt{Greene11}; Liu
et~al. 2013b\nocite{Liu13b}).

We also consider the mass and energy injection rates assuming an energy conserving bubble in a uniform medium (e.g.,
\citealt{Heckman90}; \citealt{Veilleux05}; \citealt{Nesvadba06} and
references there-in), which gives the relation:
\begin{equation}
\dot{E}_{{\rm out,M2}} \approx1.5\times10^{46}r^{2}_{10}v^{3}_{1000}n_{0.5}{\rm\,erg\,s}^{-1},
\label{Eq:energy}
\end{equation}
where $r_{10}$ is taken to be half the extent of the observed broad
[O~{\sc iii}] emission (in units of 10\,kpc; we assume $r_{10}=0.6$
for all sources; Table~\ref{Tab:velmaps}), $v_{1000}$ are the velocity offsets in units
of 1000\,km\,s$^{-1}$ (we assume $v_{1000}=W_{80}/1300$;
Table~\ref{Tab:velmaps}) and $n_{0.5}\approx1$, is the ambient density (ahead of the expanding bubble in units of
0.5\,cm$^{-3}$). Using this method  (we will call ``Method 2'') we obtain values of $\dot{E}_{{\rm
    out,M2}}\approx(0.7$--$7)\times10^{45}$\,erg\,s$^{-1}$. The mass outflow rates are then given by $\dot{M}_{{\rm
    out,M2}}=2\dot{E}_{{\rm out,M2}}/v_{c}^{2}\approx(9$--$20)\times10^{3}$\Msolyr.

We consider ``Method 1'' and ``Method 2'' to be lower and upper bounds
of the mass outflow rates and kinetic energy rates (see also discussion in
\citealt{Greene12}) and the fiducial values we will use throughout
are the mean (in log space) of these two values which are $\dot{E}_{{\rm
    out}}=(0.15$--$45)\times10^{44}$\,erg\,s$^{-1}$ and $\dot{M}_{{\rm
    out}}=170$--$1200$\Msolyr . Comparing these values to those found using the approach
followed in Liu et~al. (2013a,b)\nocite{Liu13b} provides confirmation that using these fiducial values
are reasonable. They argue that the
NLR in luminous type~2 AGN can be well described by relatively dense
clouds embedded in a hot, low-density, volume-filling wind. A spatial profile with a constant [O~{\sc iii}]/H$\beta$
emission-line ratio followed by a sharp decline in this value indicates
the break where the clouds become optically thin to ionizing
radiation (see full description in Liu et~al. 2013b\nocite{Liu13b}). In four of our
sources (these are J0958+1439; J1100+0846; J1316+1753; J1339+1425; see
  Appendix~A) we see the same sharp decline in the [O~{\sc
  iii}]/H$\beta$ ratio (in most of the other
sources this transition may happen beyond our field-of-view). For these four
sources, the kinetic energies and mass outflow rates using their
method (see Appendix~A) are all within a
factor of $\approx$1--3 of the fiducial values we chose to adopt
here.\footnote{Although here we use a different method to Liu et~al. 2013b\nocite{Liu13b} we
  verified our results using a similar method to these authors using {\sc iraf ellipse}.} 

Finally, in preparation for our discussion in Section~\ref{Sec:Drivers}, we estimate the mean and upper and lower bounds of outflow momentum rates by taking the mass
  outflow rates calculated above and assuming ${\dot P}_{{\rm out}} = {\dot M}_{{\rm out}} v_{{\rm gas}}$.

\section{Discussion}
\label{Sec:Discussion}
We have presented our IFU data for a sample of sixteen luminous type~2
AGN which traces the ionised gas kinematics (via [O~{\sc iii}] and
H$\beta$ emission lines) over kiloparsec scales. We have decoupled the kinematics of outflows from other
kinematic components, such as galaxy dynamics and mergers, and show that energetic, galaxy-wide ionised outflows are
ubiquitous in our sample. Our sample
is selected from a well constrained parent sample so that we can place our
observations into the context of the overall AGN population. In this section we use our observations, combined with previous
observations from the literature, to discuss the likely driving mechanisms
of the outflows (Section~\ref{Sec:Drivers}) and the role that these outflow may play
in galaxy evolution (Section~\ref{Sec:Impact}). 

\subsection{What drives the outflows?}
\label{Sec:Drivers}

The dominant processes that drive galaxy-wide outflows in massive
galaxies and the efficiency to which they are able couple to the gas are currently
sources of uncertainty in galaxy formation models. Several
possible mechanisms have been suggested to drive galaxy-wide outflows, for example: the mechanical input from stellar winds and
supernovae (e.g., \citealt{Leitherer99}; \citealt{Heckman90}); direct radiation pressure from the luminosity of an AGN or star formation
(e.g., \citealt{Murray05}); the interaction of radio jets (launched by
the AGN) with a clumpy and multiphase ISM (e.g., Sutherland \&
Bicknell 2007; \citealt{Wagner12}); and AGN winds, initially launched
at the accretion disk, that propagate through the galaxy and sweep
up the ISM (e.g., \citealt{King11}; \citealt{FaucherGiguere12b}). In this subsection
we will investigate which of these processes could be responsible for
driving the outflows observed in our targets. 

\subsubsection{Morphology and structure}
\label{Sec:morphology}
A possible method to distinguish between an AGN-driven and
star-formation driven outflow could be the morphology and structure of
the outflow. Galactic-scale star-formation driven outflows are known
to propagate along low density regions, perpendicular to galaxy discs (e.g.,
\citealt{Heckman87}; \citealt{Ohyama02}; \citealt{Veilleux05}). In
contrast, outflows driven by an AGN on {\em small scales} have no
relation between the orientation of the outflow and the disk in the
host galaxy (e.g., \citealt{Fischer13}). However, if AGN-driven
outflows propagate to galaxy-wide scales they may also be forced to propagate away from the
galaxy disk along a path of least resistance (e.g.,
\citealt{FaucherGiguere12b}; \citealt{Gabor14}) which could make galactic-scale outflows
driven from a nuclear star forming region or an AGN
indistinguishable (see also \citealt{Hopkins10}; \citealt{Heckman11}
and \citealt{DiamondStanic12} for other reasons why it could be difficult
to distinguish between AGN and star formation driven outflows). In some observations compact radio
jets ($\lesssim$ few kiloparsec in size) are inferred to play a role, where the combination of high
resolution radio data and spatially resolved spectroscopy has shown
that high velocity gas (in multiple phases) is spatially co-incident with radio emission
(\citealt{Baldwin80}; \citealt{Whittle88}; \citealt{Capetti99};
\citealt{Whittle04}; \citealt{Morganti05b,Morganti13}; \citealt{Emonts05}; \citealt{Leipski06};
\citealt{Barbosa09}; \citealt{Fu09}; \citealt{StorchiBergmann10}; \citealt{Shih13}). However, even
when radio jets {\em are} present it appears that they may not always
be fully responsible for the observed outflows and other processes are
also required (e.g., \citealt{VillarMartin99}; \citealt{Rupke11}; \citealt{Riffel14}). 

The morphologies of the outflows in our targets are illustrated in
Figure~\ref{Fig:W80montage}. In
our sample we observe no obvious relationship
between the morphology of the observed outflows and the SFRs, AGN
luminosities or radio excess parameters of the host
galaxies; i.e., irrespective of the host galaxy or AGN properties we see a
range in outflow morphology including those which are escaping away
from the galaxy discs. This is in broad agreement
with \cite{Rupke13} who find no structural difference between the
outflows in the ULIRGs with and without luminous AGN. Interestingly, all of
superbubble candidates are radio excess sources (see Section~\ref{Sec:outflows} and
Appendix~A). However, deeper and higher-resolution radio data are
required to search for small and low luminosity radio jets, which we cannot
identify in the FIRST data, and to ultimately determine the origin of the radio
emission (see Section~\ref{Sec:SEDs}). 

\subsubsection{Outflow velocities compared to AGN and galaxy properties}

\begin{figure*}
\centerline{\psfig{figure=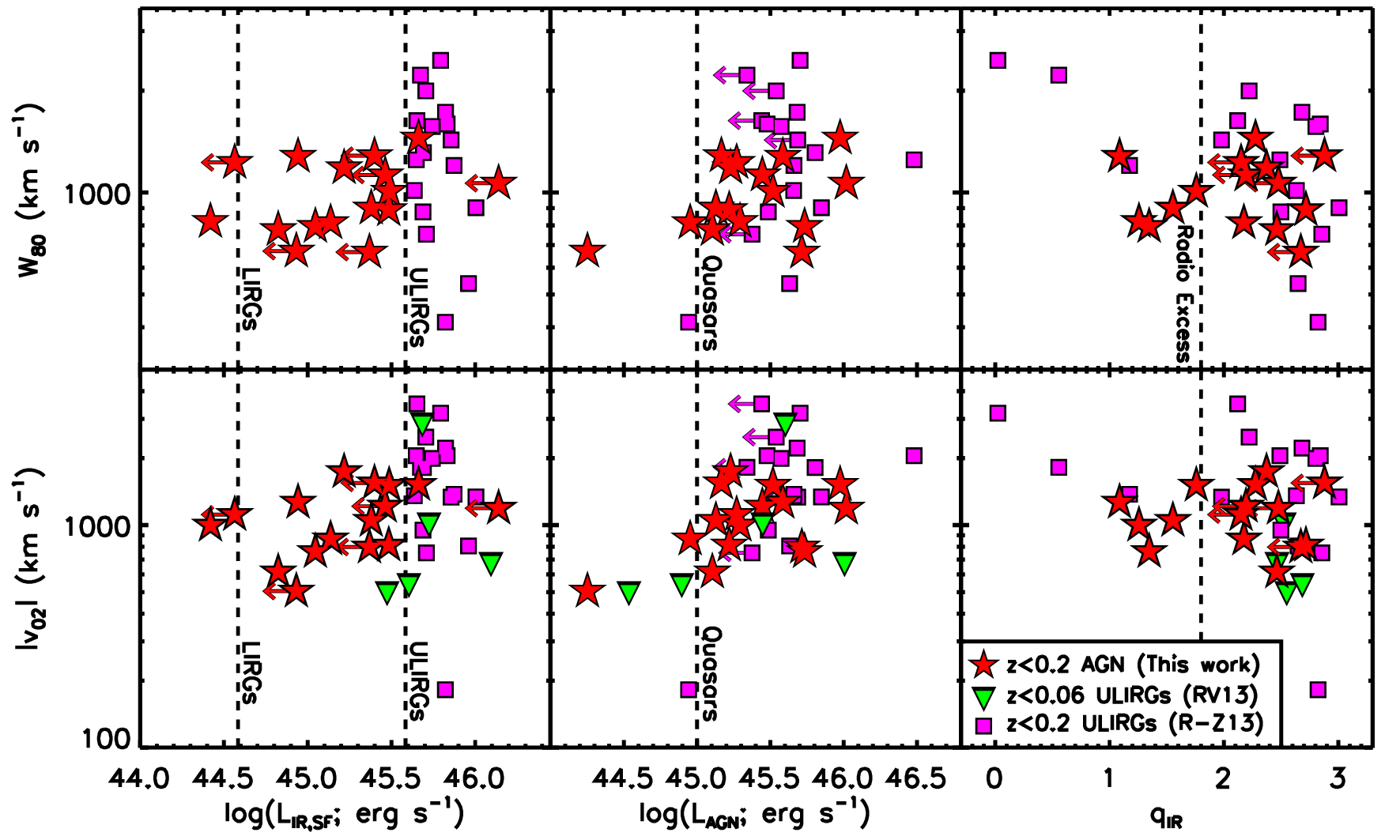,width=6in,angle=90}}
\caption{Line-width ($W_{80}$) and maximum velocity ($v_{02}$)
  versus infrared luminosity from star formation, AGN luminosity and $q_{IR}$ for our
  targets (red stars; see Table~\ref{Tab:observations} and
  Table~\ref{Tab:velmaps}). Also shown are $z<0.2$ ULIRG--AGN
  composite galaxies (magenta squares; \protect\citealt{RodriguezZaurin13}) and $z<0.06$ ULIRGs (green triangles; \protect\citealt{Rupke13}). There is tentative evidence that most
extreme velocities and line-widths are preferentially found in systems
with high star-formation rates (i.e., ULIRGs) that host the most luminous AGN
(i.e., `quasars'). There is no such effect seen when
comparing systems where there is excess radio emission AGN; however,
low-level excess radio emission is
difficult to identify using this method especially in high SFR systems (see Section~\ref{Sec:SEDs}).}
\label{Fig:compareIRagn}
\end{figure*}

Another possible method to determine the driving mechanism of
outflows is to search for trends between outflow properties and properties of the host galaxy or AGN. Positive correlations have been claimed to exist
between outflow properties, albeit using different observations, for
each of SFR (e.g., \citealt{Rupke05b}; \citealt{Martin05}; \citealt{Bradshaw13}), AGN
luminosity (e.g., \citealt{Cicone14}) and radio luminosity (e.g.,
\citealt{Nesvadba11}). Of course these analyses are complicated by the
fact that AGN luminosity, SFR and radio luminosity are closely related
(e.g., \citealt{Tadhunter98}; \citealt{Ivison10};
\citealt{Mullaney12b}; \citealt{Chen13}) and that small duty cycles
may mean that an AGN-driven outflow may be observed when the AGN
itself is no longer observable (e.g., \citealt{King11}; \citealt{Gabor14}). 

In Fig.~\ref{Fig:compare} we show that the line widths ($W_{80}$)
and maximum velocities ($v_{02}$) of our sample are comparable to
those found in more luminous type~2 AGN (Liu
et~al. 2013b\nocite{Liu13b}) and ULIRG-AGN composite galaxies
(\citealt{RodriguezZaurin13}).\footnote{Although
  \cite{RodriguezZaurin13} do not quote the values of $W_{80}$ and
  $v_{02}$ directly, we reconstructed the emission-line profiles (from
  the Gaussian components given in their paper) and then measure these
  values directly.} We observe no clear trends between $v_{02}$ or $W_{80}$
and $L_{\rm{[O~III]}}$, despite four orders of magnitude in [O~{\sc iii}] luminosity. Additionally, although limited in
dynamic range, we do not see any obvious correlation in $W_{80}$ or
$v_{02}$ with radio luminosity (Fig.~\ref{Fig:compare}). These
conclusions are consistent with \cite{Mullaney13} who demonstrate that
the broadest emission lines are prevalent for AGN with radio luminosities
of $L_{1.4\rm{GHz}}\gtrsim10^{23}$\,W\,Hz$^{-1}$ (i.e., the range
that is covered in Fig.~\ref{Fig:compare}), but found no clear
trends with [O~{\sc iii}] luminosity when taking radio-luminosity matched
samples of AGN. We note that Figure~\ref{fig:selection} is not
radio-luminosity matched so gives an apparent positive trend between
[O~{\sc iii}] luminosity and line width; see \cite{Mullaney13} for details.

In Figure~\ref{Fig:compareIRagn} we compare the galaxy-integrated line-widths ($W_{80}$) and
maximum velocities ($v_{02}$) to the infrared luminosities from star
formation ($L_{{\rm IR,SF}}$; a proxy for SFR), bolometric AGN
luminosities ($L_{{\rm AGN}}$) and the radio excess parameter ($q_{IR}$) derived in Section
~\ref{Sec:SEDs}. We also compare to the ULIRG-AGN composite galaxies from \cite{RodriguezZaurin13}\footnote{\label{Foot:RZSEDs}We fit the SEDs
  to the IRAS and WISE data for these sources in a consistent manner
  to that described in Section~\ref{Sec:SEDs}; however, we only accept AGN
  components if the AGN template contributes $\ge$50\% of the flux at
  19$\mu$m due to the difficulty in reliably identifying
  low-luminosity AGN in ULIRGs using this method (see
  \protect\citealt{DelMoro13}). Eight of these ULIRGs also have AGN luminosities
  derived from Spitzer infrared spectra in \protect\cite{Veilleux09} that we use in
  favour of our AGN luminosities. We found that the AGN luminosities
  and IR luminosities are well matched (within a factor of $\lesssim$2) between our method and that in
  \protect\cite{Veilleux09}.} and the $z<0.06$
ULIRGs from \cite{Rupke13} (who make these velocity measurements from
various ionised emission-line species). In
Figure~\ref{Fig:compareIRagn}, the only possible trend we
observe is between the maximum velocity ($v_{02}$) and AGN luminosity, although we are limited in measurements below $L_{{\rm AGN}}=10^{45}$\,erg\,s$^{-1}$. Based on these samples, although
high-velocity outflow features are observed across the samples, the most extreme velocities (i.e.,
$v_{02}\gtrsim2000$\,km\,s$^{-1}$) and line-widths
(i.e., $W_{80}\gtrsim$1500\,km\,s$^{-1}$) appear to be predominantly found in
the quasar-ULIRG composite galaxies
(Fig.~\ref{Fig:compareIRagn}). This is in broad agreement with other
studies of the outflows in ULIRGs and quasars, in multiple different
gas phases, that find the most extreme outflows in the quasar-ULIRG
composites (e.g., \citealt{Harrison12a};
\citealt{Westmoquette12}; \citealt{Veilleux13};
\citealt{Cicone14}; \citealt{Hill14}). Further evidence that the incidence of extreme ionised outflows in quasar-ULIRG composite
galaxies is high, compared to the overall population, is that 16/17 of the
objects in \cite{RodriguezZaurin13} have extreme ionised gas kinematics
(i.e., significant broad [O~{\sc iii}] emission-line components) which is much larger than we
find for the overall optical AGN population (see Section~\ref{Sec:Selection}), and considerably higher than expected
given their moderate [O~{\sc iii}] luminosities (i.e., $L_{{\rm
    [O~III]}}\lesssim10^{42}$\,erg\,s$^{-1}$; Fig.~\ref{fig:selection}). Furthermore, the flux-weighted average
line-widths (FWHM$_{{\rm avg}}\gtrsim800$\,km\,s$^{-1}$) of these
quasar-ULIRG composites are representative
of the top $\approx2$--3\% most extreme objects in
\cite{Mullaney13}. This result is not sufficient to establish which
process is driving the observed outflows, but instead may demonstrate
that the most extreme outflows are most prevalent during an active
star formation and AGN phase as is predicted by some evolutionary
models (e.g., \citealt{Sanders88}; \citealt{Hopkins08a}). 

Finally, we note that \cite{Mullaney13} show that the fraction of sources with their
ionised gas dominated by outflows, was considerably higher for
sources with moderate-to-high radio luminosities ($L_{{\rm
    1.4\,GHz}}\gtrsim10^{23}$\,W\,Hz$^{-1}$). Based on the analysis of the radio emission
in our sources and the comparison samples (Section~\ref{Sec:SEDs};
Figure~\ref{Fig:compareIRagn}), star formation could contribute
considerably to the radio emission. Additionally, as noted by
\cite{Zakamska14}, radio emission could be produced by shocks due to
outflows that were driven by e.g., quasar winds. Further analysis of
the parent population is required to establish the physical process
that is driving the trend between radio luminosity and ionised outflows.
 
\subsubsection{Coupling efficiencies and momentum fluxes}
\begin{figure*}
\centerline{\psfig{figure=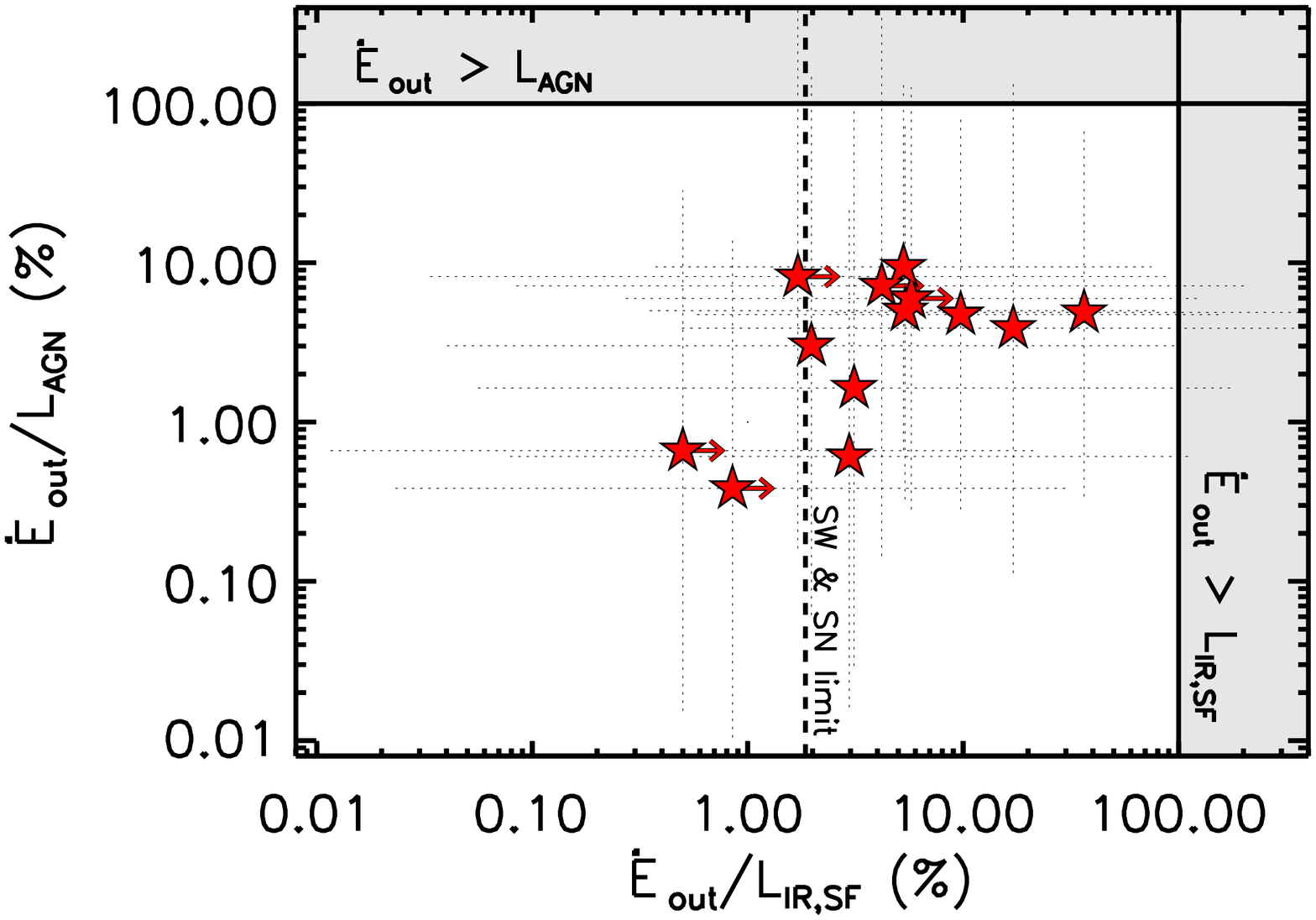,width=3.7in,angle=90}\psfig{figure=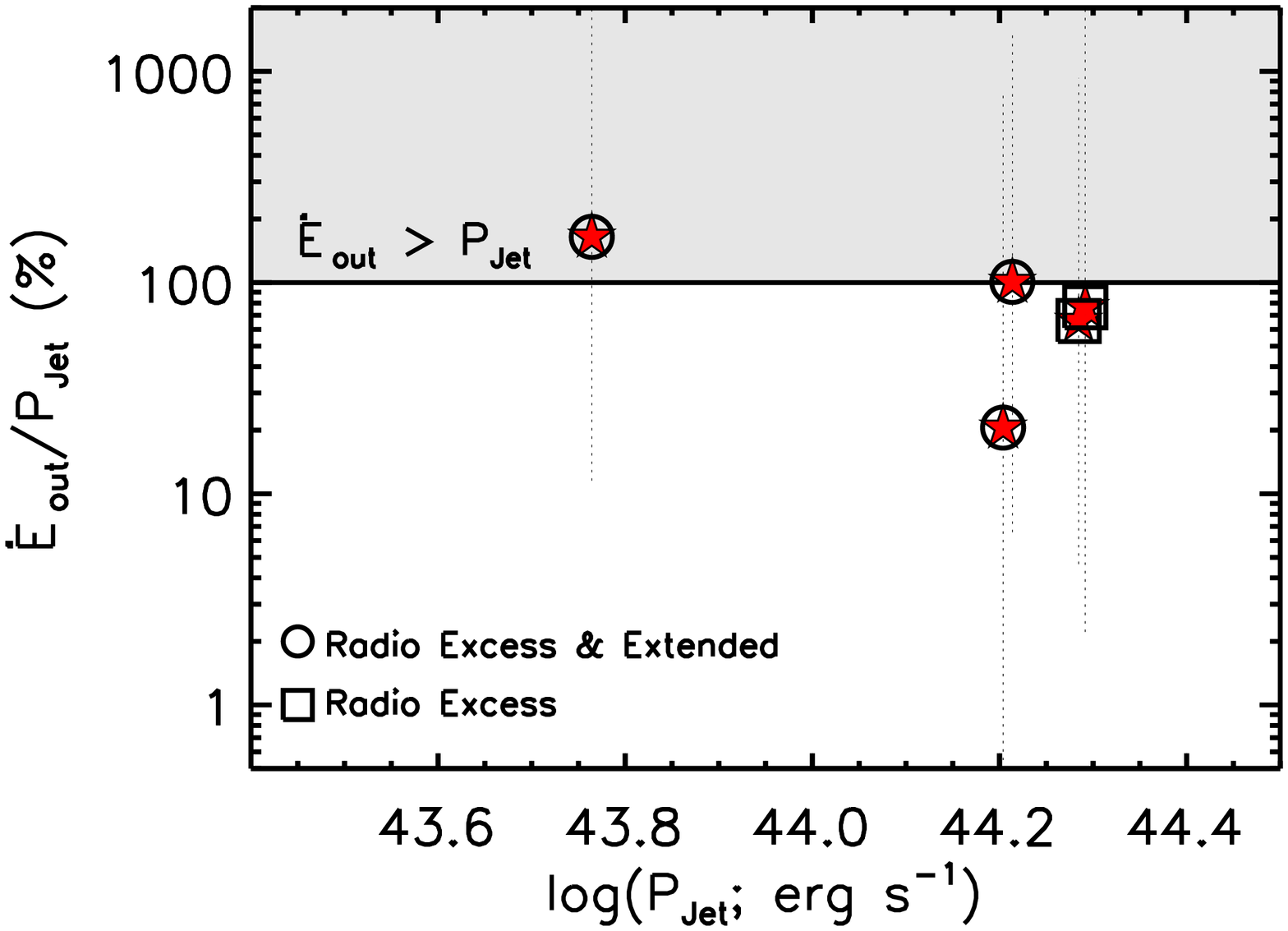,width=3.7in,angle=90}}
\caption{{\em Left panel:} The ratio of our estimated outflow kinetic
  energy rates ($\dot{E}_{{\rm out}}$; Section~\ref{Sec:properties}) to the AGN luminosity
  (ordinate) and to the star-formation luminosity (abscissa) for the
  thirteen sources in our sample for which we estimated $\dot{E}_{{\rm
      out}}$. The dotted lines illustrate the ratios from using our upper and lower bounds on
  the $\dot{E}_{{\rm out}}$ values (Section~\ref{Sec:properties}). The shaded regions
indicate where $>$100\% coupling efficiency is required between the
input energy and the gas to power the outflows. The dashed vertical line is the
estimated maximum mechanical input expected from supernovae and
stellar winds (see Section~\ref{Sec:Drivers}). {\em Right panel:} Similar to the
left panel, but showing the ratio of $\dot{E}_{{\rm out}}$ to the estimated jet power, as a function of jet
power, for the five sources with the strongest evidence of hosting
radio jets (see Section~\ref{Sec:SEDs}). Based on our
simplifying assumptions a similar outflow coupling efficiency is
required for AGN (i.e., $\approx0.5$--10\%) and star formation
(i.e., $\gtrsim0.5$--40\%); however, stellar winds and supernovae are less
likely to be fully responsible for the observed outflows. Radio jets would require uncomfortably high coupling
efficiencies to power the outflows in some cases (although see Section~\ref{Sec:Drivers}).
}
\label{Fig:Energetics}
\end{figure*}

\begin{figure*}
\centerline{\psfig{figure=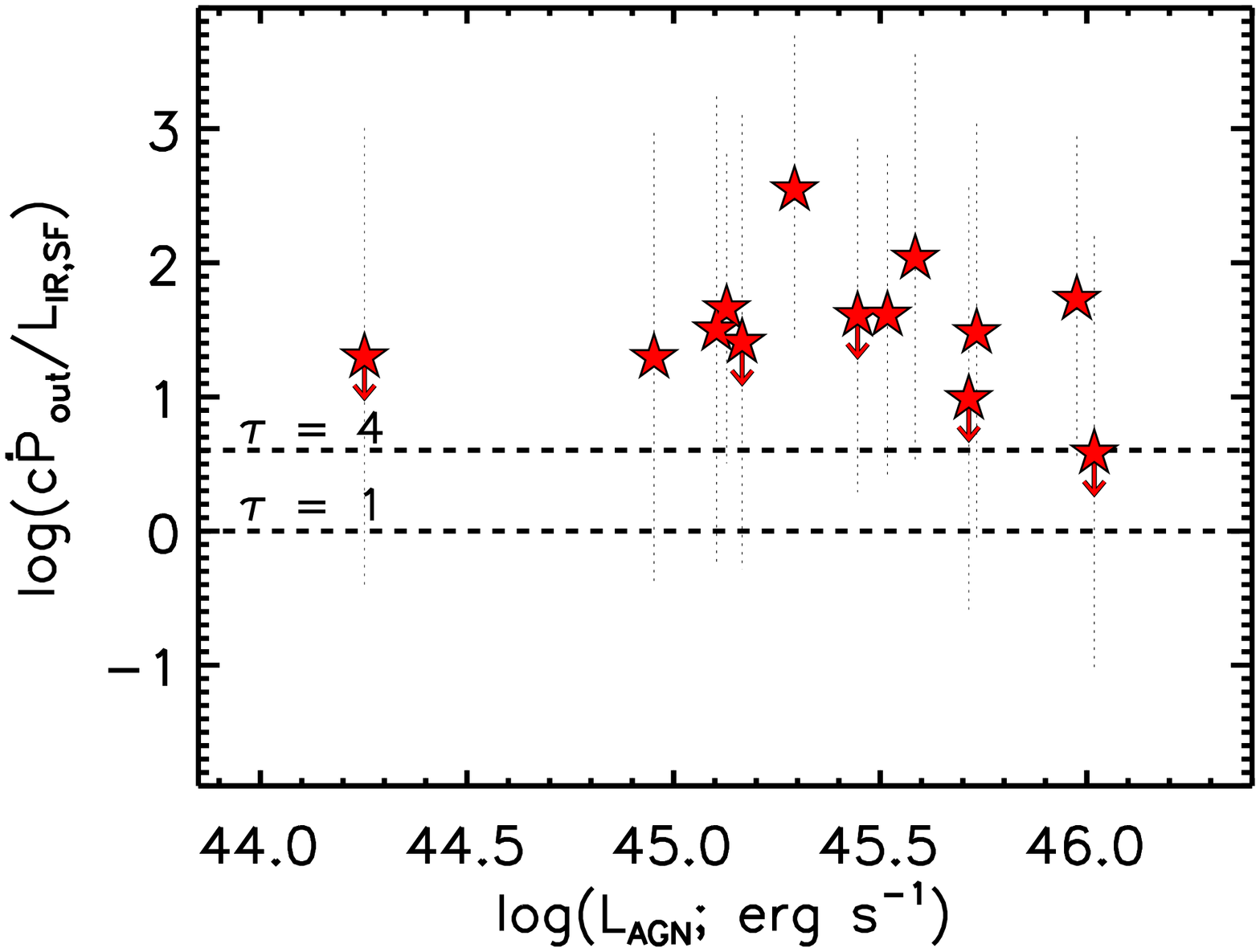,width=3.8in,angle=90}\psfig{figure=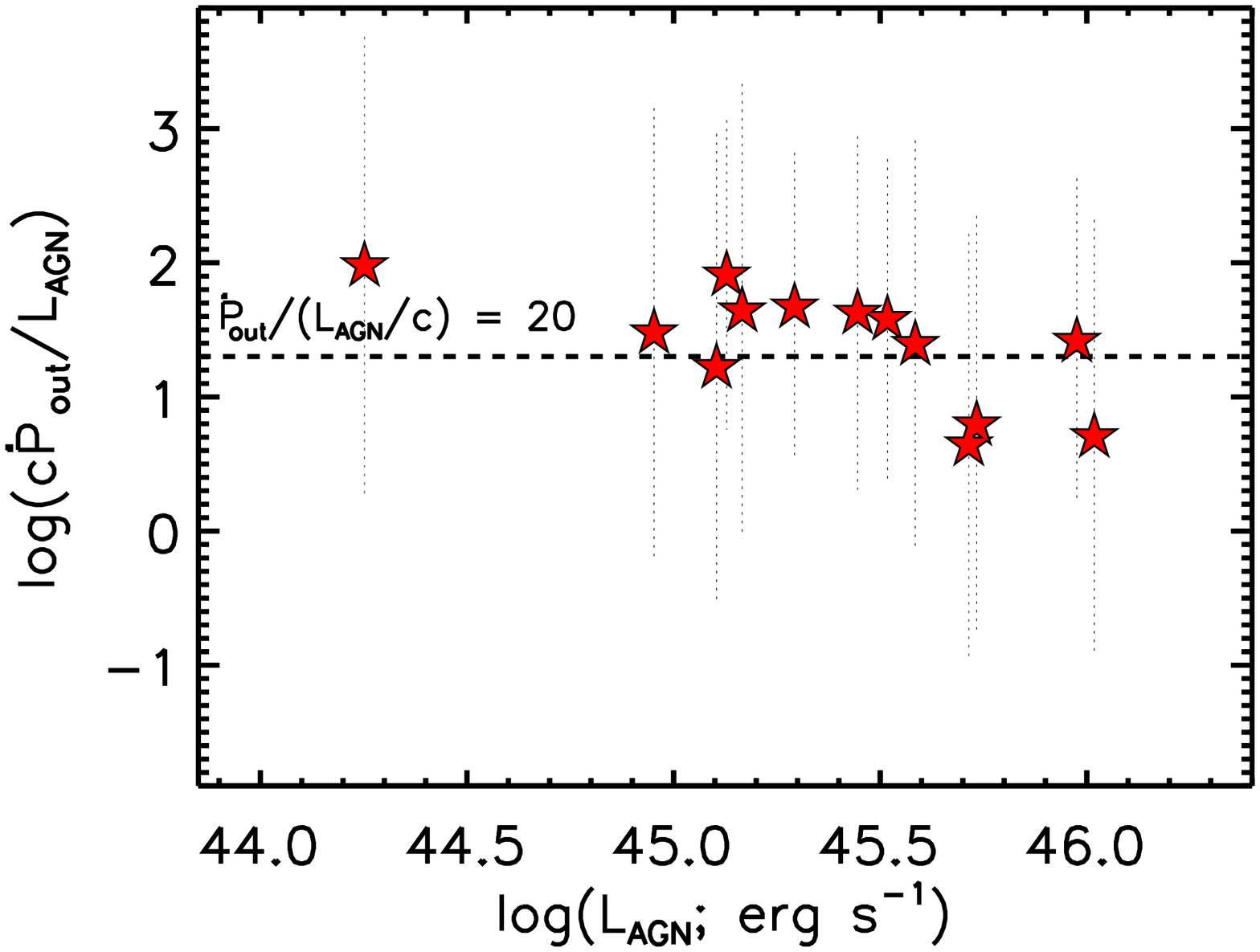,width=3.8in,angle=90}}
\caption{{\em Left Panel:} Momentum rates of the outflows (${\dot P}_{{\rm out}}$) normalised to the
  star formation luminosity ($L_{{\rm
      IR,SF}}/c$) versus AGN luminosity. The dashed lines
  represent the required optical depths if the outflows are driven by
  radiation pressure from star formation (see
  Section~\ref{Sec:Drivers}). {\em Right Panel:} Momentum rate of the
  outflows normalised to the AGN luminosities ($L_{{\rm
      AGN}}/c$) versus AGN luminosity. In both panels the vertical dotted lines represent the upper and
  lower bounds in $\dot{P}_{{\rm out}}$ (see
  Section~\ref{Sec:properties}). Based on our assumptions, the
  outflows are unlikely to be purely radiatively driven but are broadly consistent
  with theoretical predictions of galactic-scale energy-driven
  outflows that are launched by an AGN accretion-disc wind (i.e.,
  $\dot{P}_{{\rm out}}\gtrsim10$--$20L_{{\rm AGN}}/c$; e.g., \citealt{Zubovas12}; \citealt{FaucherGiguere12b};
\citealt{DeBuhr12}).}
\label{Fig:Momentum}
\end{figure*}

A popular method to investigate the likely drivers of galactic-scale outflows is to
compare the energy and momentum available from each potential driver (i.e.,
supernovae and stellar winds, radiation pressure [from AGN or star
formation], AGN winds, or radio jets) to the kinetic energy and
momentum in the outflows. Although we are limited by a
large uncertainty in the outflow kinetic energies (see
Section~\ref{Sec:properties}), we have measured
SFRs and AGN luminosities and have identified the sources that are most likely to
host radio jets (see Section~\ref{Sec:SEDs} and
Table~\ref{Tab:observations}). In Figure~\ref{Fig:Energetics} we compare the ratio of our outflow kinetic
energy rates (${\dot E}_{{\rm out}}$) with (1) the bolometric AGN luminosities; (2) the
infrared star-formation luminosities\footnote{We note that for infrared
luminous galaxies (i.e., $L_{{\rm IR}}\ge10^{11}$L$_{\odot}$), the infrared
luminosity makes up the bulk of the bolometric luminosity; e.g.,
\citealt{Sanders96}; \citealt{Veilleux05}.} and (3) the mechanical radio jet power (estimated using the 1.4\,GHz luminosity
and the relationship in \citealt{Cavagnolo10}). In
Figure~\ref{Fig:Momentum} we compare the momentum rates of the outflows (${\dot
  P}_{{\rm out}}$) with the momentum flux output radiatively by (1) star formation (i.e.,
$L_{{\rm IR,SF}}/c$) and (2) the AGN (i.e., $L_{{\rm AGN}}/c$). Using
these results we now explore the plausibility of different potential
driving mechanisms and compare to theoretical predictions. 

Figure~\ref{Fig:Energetics} shows that a similar outflow coupling
efficiency is required to the radiative output of both the AGN (i.e., $\approx0.5$--10\%) and star
formation (i.e., $\gtrsim0.5$--40\%), indicating that either of these
processes could power the outflows if they are able to couple their
energy to the gas. One way for star formation to drive an outflow is from stellar
winds and supernovae. An estimate of the maximum energy injection from stellar
winds and supernovae, which is used throughout the literature, is
$\approx7\times10^{41}$(SFR/\Msolyr)\,erg\,s$^{-1}$ (applicable for
stellar ages $\gtrsim$40\,Myrs and following \citealt{Leitherer99}; \citealt{Veilleux05}).\footnote{If we were to follow instead
  \citealt{DallaVecchia08} to estimate this energy input, the values
  would be a factor of $\approx$2 lower.} This corresponds to a
maximum ${\dot E}_{{\rm out}}/L_{{\rm IR,SF}}\approx2$\% (following
\citealt{Kennicutt98} but correcting to a Chabrier IMF) that we
indicate Figure~\ref{Fig:Energetics}. Based on these assumptions, stellar
winds and supernovae are unlikely to be fully responsible for the
observed outflows. If we instead consider a momentum driven wind with momentum deposition from the radiation pressure of stars, we
expect ${\dot P}_{{\rm out }}=\tau L_{{\rm SF,IR}}/c$ (following
\citealt{Murray05}) that based on our calculations would require unrealistically high optical depths
(i.e., $\tau \gg 1$; see Figure~\ref{Fig:Momentum}). Further investigation of the possibility
that star formation drives the outflows could be achieved using high
resolution optical-continuum imaging to measure the SFR surface densities in our targets
(e.g., \citealt{Murray11}; \citealt{Heckman11}; \citealt{DiamondStanic12}; \citealt{ForsterSchreiber13}). 

The estimated coupling efficiencies between the bolometric AGN
luminosity and the outflows (i.e., $\approx0.5$--10\%) are consistent with those predicted by various
models (e.g., \citealt{DiMatteo05}; \citealt{Hopkins10};
\citealt{Zubovas12}; \citealt{DeBuhr12}). However, we note that these coupling efficiencies do not tell us {\em how}
the AGN couples to the gas. It is thought that direct radiation
pressure from an AGN is unlikely to drive momentum-driven galaxy-wide outflows (e.g.,
\citealt{King11}; \citealt{DeBuhr12}) and this is supported by the high outflow momentum rates we observe
(Figure~\ref{Fig:Momentum}). However, it has been predicted that for
high-accretion rate AGN, an accretion-disc wind could sweep up
material and become a galaxy-wide energy-driven outflow (e.g.,
\citealt{King11}; \citealt{FaucherGiguere12b}; \citealt{DeBuhr12}; \citealt{Zubovas12}).
These models predict momentum rates of ${\dot P}_{{\rm
    out}}\approx10$--$40L_{{\rm AGN}}/c$ on $\approx$1--10\,kpc scales that is in good agreement
with our observations (Figure~\ref{Fig:Momentum}) as well as multi-gas-phase outflow observations of other AGN (e.g., \citealt{Cicone14};
\citealt{Zubovas12} and references there-in). An AGN could also couple
to the gas in their host galaxies though the interaction of a radio
jet with the ISM. In Figure~\ref{Fig:Energetics} we show that the coupling
efficiencies between the radio jet power and the outflows is estimated to be $\approx$20--160\%
(calculated for the sources with the strongest evidence for radio AGN
activity only; see Section~\ref{Sec:SEDs}). These basic calculations may rule out radio jets being solely responsible for
all of the outflows observed (Fig.~\ref{Fig:Energetics}; but see
Appendix~A for a discussion on the radio jets in individual objects); however these numbers are uncertain
and reasonably high coupling efficiencies are predicted for outflows
driven by jet-ISM interactions (e.g., \cite{Wagner12} predict
$\approx10$--40\%).

In summary, based on our analyses we find no definitive evidence that the outflows
we observe are universally driven by the AGN (either launched by accretion-disk winds or radio
jets) or by star formation. However, using the arguments that we have
presented we find that supernovae, stellar winds, radio jets or radiation pressure are
unlikely to be solely responsible for all of the outflows that we
observe. Although uncertain, we have shown that the outflows we observe have
properties that are broadly consistent with predictions of energy-conserving outflows that are
initially driven by accretion-disc winds. It is certainly viable that multiple
processes contribute to driving the observed outflows as has been seen in local systems, where
star-formation winds, nuclear winds (potentially AGN-driven) and radio
jets all are playing a role (e.g., \citealt{Cecil01}; \citealt{StorchiBergmann10};
\citealt{Rupke11}; also see discussion on our individual targets in
Appendix~A). As has been highlighted throughout this section, multi-phase (i.e.,
ionised, atomic and molecular) observations of the outflows, combined with high resolution
multi-wavelength imaging will provide the required information to
determine the relative role of these different outflow driving mechanisms in
our targets.

\subsection{What role do these outflows play in galaxy evolution?}
\label{Sec:Impact}
Galaxy-wide outflows are required in galaxy formation models, driven
by both star formation and AGN activity, to reproduce the observable
properties of galaxies and the intergalactic medium (e.g., \citealt{Silk98};
\citealt{DiMatteo05}; \citealt{Hopkins08a}; \citealt{DallaVecchia08}
\citealt{Booth10}; \citealt{DeBuhr12}; Hopkins
et~al. 2013a\nocite{Hopkins13a}). In these models some fraction of the
gas escapes from the potential of the host galaxy, regulating future star
formation and black hole growth and enriching the larger-scale environment with
metals. In this section we will assess what impact the outflows that
we observe may have on the evolution of their host galaxies.

If we consider the estimated mass outflow rates (derived in
Section~\ref{Sec:properties}) we find that they are $\approx$6--20
times greater than the SFRs of our targets
(with one exception at 100 times [J1430+1339] and excluding upper limits). These ``mass loading''
values are typically higher than those observed in star-forming galaxies but similar
to those seen in luminous AGN, derived from a variety of observations
of outflows in different gas phases (e.g., \citealt{Martin99};
\citealt{Heckman00}; \citealt{Newman12b}; \citealt{Cicone14}). We do
not know the multi-phase outflowing gas properties of our targets; however, these observations imply that the star forming material may
be rapidly depleted, as has also been suggested from
molecular gas observations of a few nearby ULIRG-AGN composite galaxies (\citealt{Cicone14}). 

We can assess if the outflows will be able to escape the
potential of their host galaxies. The escape velocities of typical type~2 AGN within the luminosity range
of our sample are $\approx500$--1000\,km\,s$^{-1}$ (assuming an
isothermal sphere of mass out to 100\,kpc; \citealt{Greene11}). These escape velocities correspond to circular
velocities of $\approx200$--400\,km\,s$^{-1}$, which are likely to be
representative of the intrinsic values for our galaxies, based on
those in which we identify rotation (see Section~\ref{Sec:GalaxyKinematics}). If we assume the intrinsic bulk
velocities of the outflows are $v_{{\rm out}}=W_{80}/1.3$ (see Section~\ref{Sec:properties})
then we have a range of outflow velocities $v_{{\rm
    out}}=510$--$1100$\,km\,s$^{-1}$ (with a median of
780\,km\,s$^{-1}$) which are comparable to the escape velocities. The
maximum projected velocities are even higher (up to
$v_{02}\approx-1700$\,km\,s$^{-1}$; Table~\ref{Tab:velmaps}). If we
instead consider a $\approx$10$^{11}$\,M$_{\odot}$ galaxy inside a
$\approx$10$^{13}$\,M$_{\odot}$ dark matter halo with Navarro-Frenk-White
(\citealt{Navarro96}) density profile, the gas is unlikely
to escape unless it is travelling at $\gtrsim$1000\,km\,s$^{-1}$ (see
the calculation in Section 5.2 of \citealt{Harrison12a}). Indeed, some models have shown that
even massive outflows may stall in the galaxy halo, re-collapse and cool at later times (along with new fuel supplies), resulting in re-ignition
of star formation and black hole growth (e.g., \citealt{Lagos08};
\citealt{McCarthy11}; \citealt{Gabor11}; \citealt{RosasGuevara13}; Hopkins
et~al. 2013b\nocite{Hopkins13b}; see also \citealt{Gabor14}).

In summary, while it is not possible to constrain the ultimate fate of
outflowing gas from observations, we have observed outflows with
extremely high velocities and the estimated outflow kinetic energy rates (i.e., $\approx0.5$--10\% of $L_{{\rm AGN}}$) and momentum rates
($\gtrsim10$--$20\times L_{{\rm AGN}}/c$) are in broad agreement with the
requirements of models that invoke AGN-driven outflows to regulate star
formation and black hole growth in luminous AGN (e.g.,
\citealt{DiMatteo05}; \citealt{Hopkins10}; \citealt{DeBuhr12}). Even if the
outflows do not escape the galaxy halos, they may be the required
pre-cursor to the postulated ``maintenance mode'' of feedback that appears to be necessary
to control the level of cooling in massive halos at later times (e.g., \citealt{Churazov05}; \citealt{McCarthy11}; \citealt{Gabor11}; \citealt{Bower12}).


\section{Conclusions}
\label{Sec:conclusions}

We have presented optical IFU observations covering the [O~{\sc iii}]$\lambda\lambda$4959,5007 and H$\beta$ emission lines of sixteen
$z=0.08$--0.2 type~2 AGN. Our targets were selected from a parent sample of
$\approx24,000$ $z<0.4$ AGN (\citealt{Mullaney13}) and we demonstrate
that they are representative of the $45\pm$3\% of luminous
($L_{\rm [O~III]}\ge5\times10^{41}$\,erg\,s$^{-1}$) $z<0.2$ type~2 AGN that have a significant fraction ($\gtrsim$30\%) of their ionised
gas outflowing (Section~\ref{Sec:Selection}). The fraction of
AGN with ionised outflows at lower levels will be much higher. We use infrared SED decomposition on our
targets to derive SFRs ($\lesssim[10$--100]\,\Msolyr) and AGN
luminosities ($L_{{\rm
    AGN}}=10^{44.3}$--10$^{46.0}$\,erg\,s$^{-1}$) for our targets. We also show that the
radio emission in our targets (luminosity range: $L_{{\rm
    1.4\,GHz}}\le10^{24.4}$\,W\,Hz$^{-1}$) is due to a combination of
star formation and other processes, with five targets showing a ``radio
excess'' above that expected from star formation and at least three of these showing spatially resolved radio
emission (on scales $R_{1.4}\gtrsim7$\,kpc) that
could be due to radio jets or shocks (Section~\ref{Sec:SEDs}). In summary, our targets are taken
from a well constrained parent sample and are not extreme star forming
systems or radio loud AGN that have been the focus of
many similar studies. 

The main results from our analysis are:
\begin{itemize}
\item We find extended [O~{\sc iii}] emission-line regions over the
  full-field-of-view of our IFU observations (i.e., total extents of
  $\ge$10--20\,kpc). In most of the sources these emission-line
  regions are circular or moderately elliptical; however, we
  observe some irregular morphologies, particularly in the
  radio excess sources (see Section~\ref{Sec:EELRs}). 
\item By tracing narrow [O~{\sc iii}] emission-line components across
  the field-of-view, we identify a range of kinematic features
  associated with galaxy dynamics including galaxy rotation,
  merger debris and double nuclei. We isolate these kinematic
  components from any kinematics due to outflows
  (see Section~\ref{Sec:Kinematics}). 
\item We find high-velocity and disturbed ionised gas (velocity widths of
$W_{80}\approx600$--1500\,km\,s$^{-1}$) extended over $\gtrsim(6$--16)\,kpc in
all of our targets. With our knowledge of the parent sample we
conclude that, $\ge$70\% (3$\sigma$ confidence level) of the outflow
features observed in the [O~{\sc iii}] emission-line profiles of
$z<0.2$ luminous type~2 AGN are extended on kiloparsec
scales (see Section~\ref{Sec:Kinematics}). 
\item The bulk outflow velocities across the sample are $v_{{\rm
      out}}\approx510$--$1100$\,km\,s$^{-1}$, which are comparable to the
galaxy escape velocities. The maximum projected gas velocities reach
up to $v_{02}\approx$\,1700\,km\,s$^{-1}$ (see Section~\ref{Sec:Kinematics}). These velocities indicate that
  ionised gas is currently being unbound from their host
  galaxies but it is not clear if it will permanently escape their
  host galaxy halos (see Section~\ref{Sec:Impact}).
\item We observe a range of kinematic structure in the
  outflows including signatures of spherical outflows and bi-polar superbubbles. In several cases the outflows are preferentially oriented
away from the plane of the host galaxy. We observe no obvious
relationship between the outflow kinematic structures as a function of
AGN luminosity, SFR or the presence or absence of radio AGN
activity (see Section~\ref{Sec:Kinematics} and
Section~\ref{Sec:Drivers}). However, we do find that all of our
  three superbubble candidates are radio excess sources.
\item Based on our analyses we find that both star formation and AGN
  activity may contribute to driving the outflows that we observe, with no
  definitive evidence that favours one over the other in the sample as
  a whole; however, we discuss individual objects in
more detail (see Section~\ref{Sec:Drivers}). By combining our observations with those from the literature,
  we show that kiloparsec-scale ionised outflows are not confined to the
  most luminous AGN, extreme star-forming galaxies or radio-loud AGN; however,
  we find that the most extreme ionised gas velocities (with maximum velocities:
  $v_{02}\gtrsim$\,2000\,kms$^{-1}$ and line-widths $W_{80}\gtrsim$\,1500\,km\,s$^{-1}$) are
  preferentially found in quasar-ULIRG composite galaxies (see Section~\ref{Sec:Drivers}). 
\item Although based on simplifying assumptions, we estimate kinetic energy outflow rates
  (${\dot E}_{{\rm out}}\approx(0.15$--$45)\times10^{44}$\,erg\,s$^{-1}$) and mass
  outflow rates (typical $\approx10\times$ the SFRs) that imply that
  considerable mass and energy are being carried in the observed outflows
  (see Section~\ref{Sec:properties}). 
\item It is not possible to provide {\em direct} evidence of the long-term
impact of individual outflows; however, although uncertain, we find
that the mass outflow rates, the outflow kinetic energy rates
($\approx0.5$--10\% of $L_{{\rm AGN}}$) and outflow momentum rates
(typically $\gtrsim$10--20$\times L_{{\rm AGN}}/c$) are
in broad agreement with theoretical predictions of AGN-driven outflows that
are postulated to play significant role in shaping the evolution of galaxies (see Section~\ref{Sec:Impact}). 
\end{itemize}

In this paper we have investigated the prevalence, properties and
potential impact of galaxy-wide ionised outflows. By selecting targets for detailed observations from
our well constrained parent sample we have been able to place our
observations into the context of the overall AGN population. We
have established that galaxy-wide ionised outflows are prevalent
in AGN. Our investigation was based upon outflows of ionised gas since it is currently is the
only suitable gas phase for performing large statistical studies of
outflows; however, it is imperative that we now obtain multi gas-phase
observations of outflows from representative samples of objects, such
as our targets, to fully characterise the properties and impact of galaxy-wide
outflows in the global population. 


\subsection*{Acknowledgments}

We thank Michael Hogan for useful discussions on interpreting the
FIRST radio data. We thank the referee, Jenny Greene, for her prompt
and constructive comments. We gratefully acknowledge support from the Science and Technology
Facilities Council (CMH through grant ST/I505656/1; DMA through grant
ST/I001573/1 and AMS through grant ST/H005234/1) and the Leverhulme Trust (DMA;
JRM). JRM also acknowledges support from the University of Sheffield via
its Vice-Chancellor Fellowship scheme. This publication makes use of data products from the Wide-field
Infrared Survey Explorer, which is a joint project of the University
of California, Los Angeles, and the Jet Propulsion
Laboratory/California Institute of Technology, funded by the National
Aeronautics and Space Administration. The data used for this work
are available through the Gemini Science Archive under the programme IDs:
GS-2010A-Q-8 and GS-2012A-Q-21.



\appendix

\section{Notes on individual sources}
\label{Sec:appA}

In this appendix we provide some notes on individual sources by
highlighting relevant previous multi-wavelength observations from the
literature and describing some of the
key features in our IFU observations. The majority of these sources have
received little or no attention previously in the literature, with the
exceptions of J0945+1737, J1316+1753 and J1356+1026 that have been
studied in some detail. The IFU data for all of our
sources are displayed in this Appendix (Fig.~A1-A15), except for J1010+1413 that can
be found in Figure~\ref{Fig:velmaps} and Figure~\ref{Fig:sdss} as it
is used as an example in the main text.

\subsubsection*{J0945+1737}
In agreement with our classifications (see
Table~\ref{Tab:observations}) the source J0945+1737 has previously been identified as a LIRG with a
type~2 AGN nucleus (\citealt{Kim95}; \citealt{Veilleux95}) and as a
type~2 (``obscured'') quasar by \cite{Reyes08}. It has been shown to have a single nucleus
based upon {\em HST} observations (\citealt{Cui01}). The infrared
luminosity that we measure (Table~\ref{Tab:observations}) is consistent
 with previous measurements (\citealt{Cui01}) and our AGN luminosity
 measurement is consistent with that derived from {\em Spitzer} mid-infrared spectroscopy (\citealt{Sargsyan08}). This
 source has evidence for a galaxy-scale radio jet due to its
 radio excess, above that expected from star formation, and the extended radio emission observed in FIRST
 ($R_{1.4}\approx$\,6.6\,kpc; see Section~\ref{Sec:SEDs} and
 Table~\ref{Tab:observations}).

\begin{figure*}
\centerline{\psfig{figure=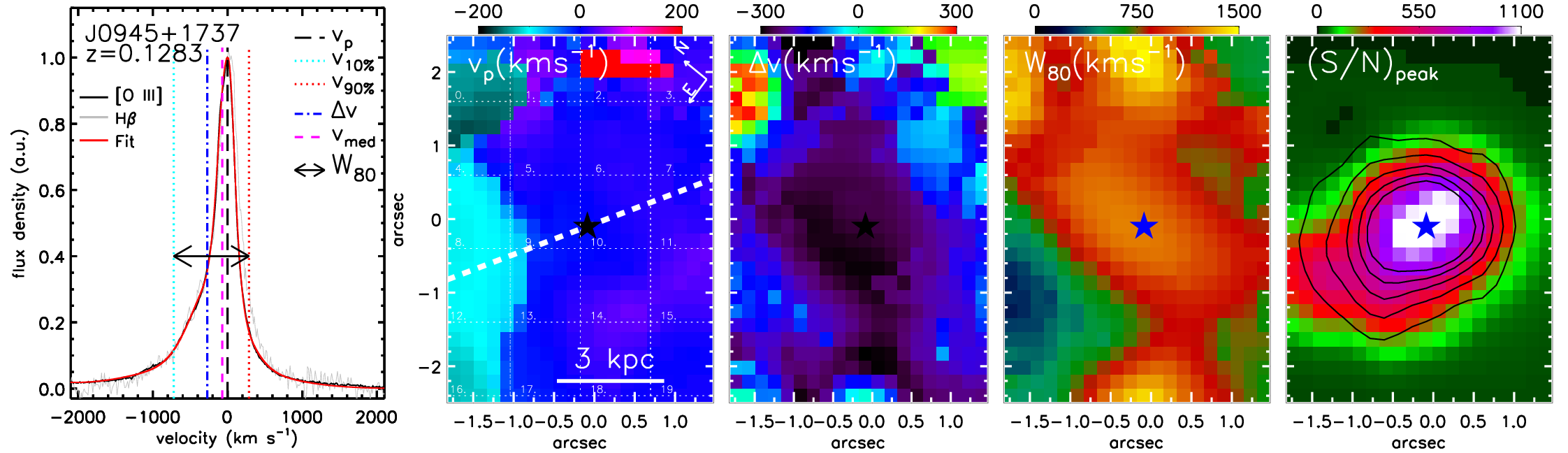,width=6.5in,angle=90}}
\centerline{\psfig{figure=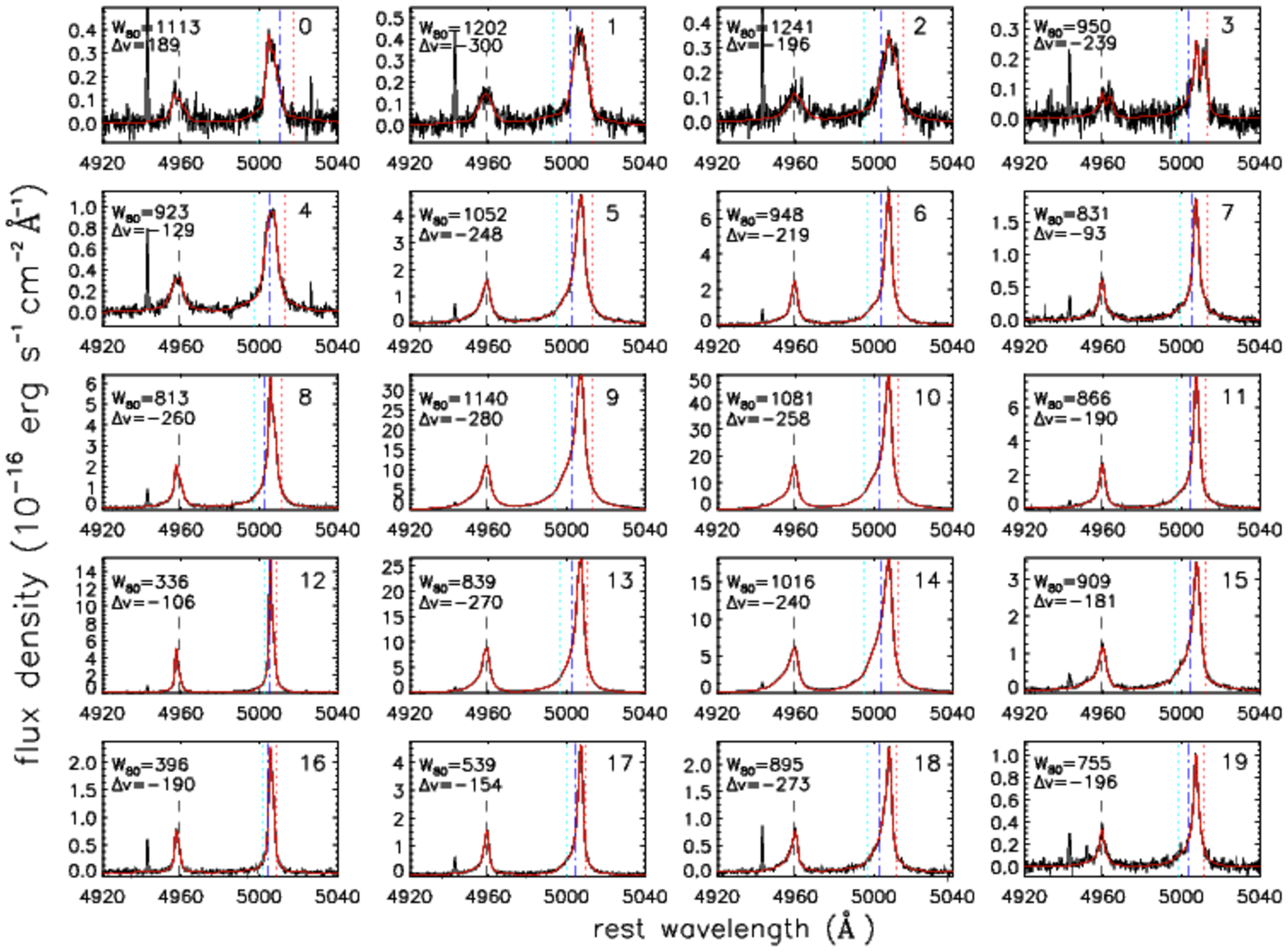,width=6.1in,angle=0}}
\vspace{-0.6cm}
\centerline{\psfig{figure=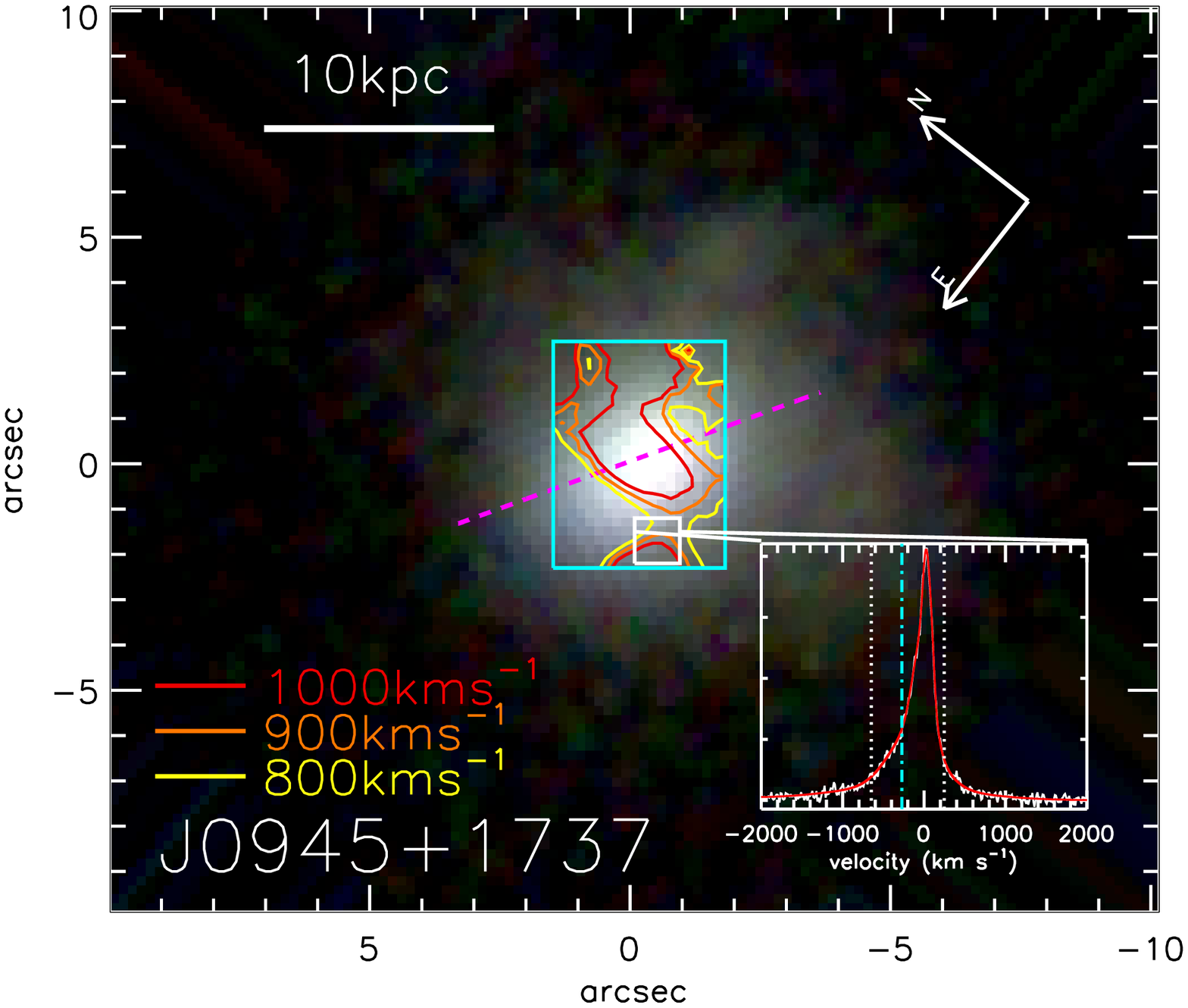,width=2.6in,angle=0}\psfig{figure=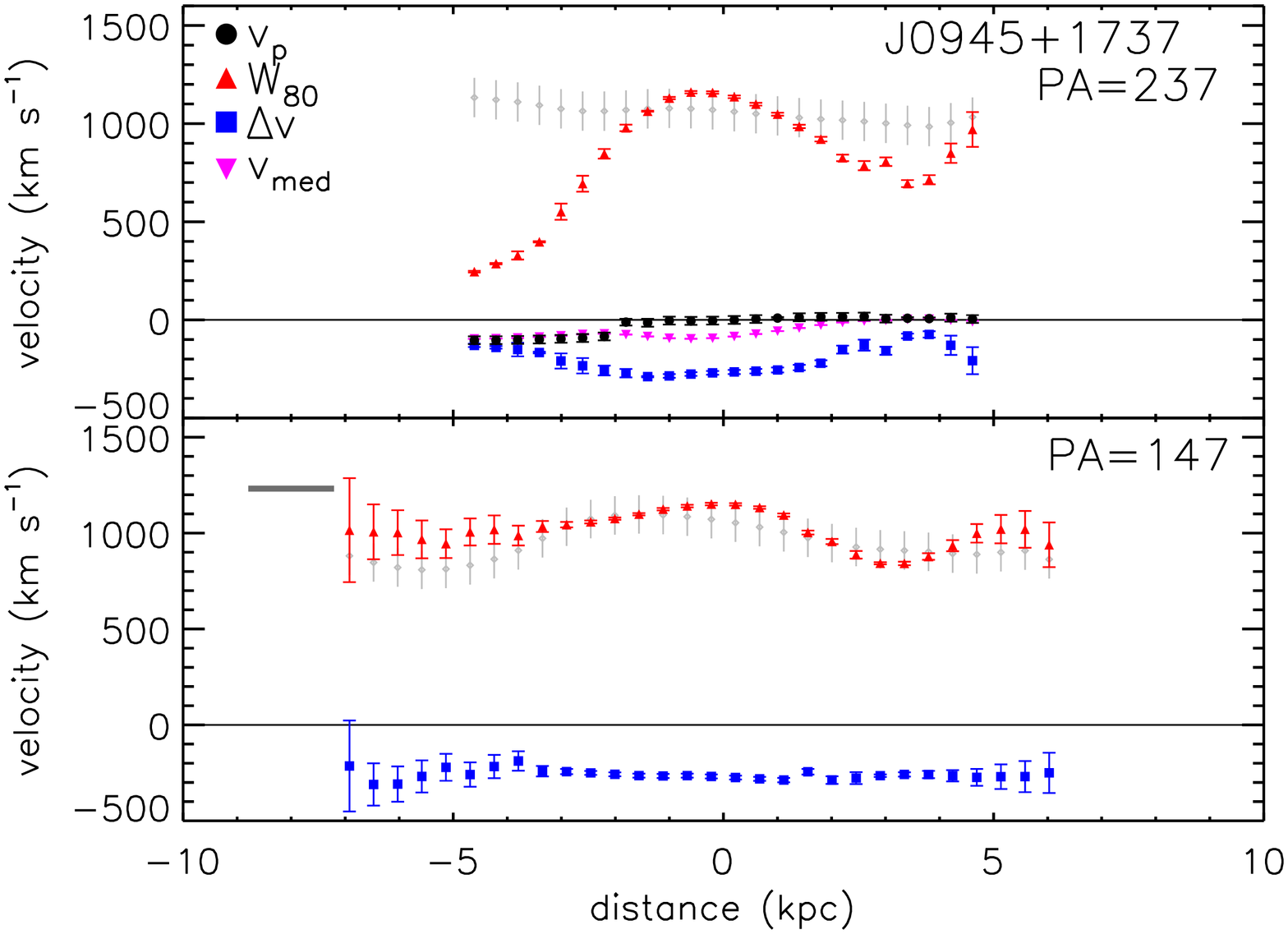,width=3.3in,angle=90}}
\caption{Our IFU data for SDSS\,J0945+1737. In summary (see captions
  in Figure~\ref{Fig:velmaps} and Figure~\ref{Fig:sdss} for a full
  description), top row from left to right: {\em Panel 1:}
  galaxy-integrated [O~{\sc iii}]$\lambda$5007 and H$\beta$
  emission-line profiles; {\em Panel 2:}
  velocity map of the peak in the [O~{\sc iii}] emission-line
  profile ($v_{p}$); {\em Panel 3:} map of the velocity offset ($\Delta v$) of the broad
  [O~{\sc iii}] emission-line wings;
  {\em Panel 4:} [O~{\sc iii}] line-width ($W_{80}$) map; {\em Panel 5:}
  map of the signal-to-noise ratio of the peak in the [O~{\sc iii}] emission-line
  profile. The grid of panels shows [O~{\sc
    iii}]$\lambda\lambda$4959,5007 emission-line profiles extracted
  from the spatial regions illustrated in panel {\em Panel 2}. Bottom left: SDSS image with contours
  overlaid of constant $W_{80}$ from our IFU data. Bottom right:
  velocity-distance profiles along the ``major axis'' shown in {\em
    Panel 2} and an
  axis perpendicular to this major axis. 
}
\label{fig:0945+1737}
\end{figure*}

Our IFU data for this source are shown in Figure~\ref{fig:0945+1737}. The [O~{\sc iii}] peak velocity ($v_{p}$) field is
irregular (Fig.~\ref{fig:0945+1737}) and therefore, based on these data
alone, we do not attribute the $v_{p}$ velocity field to galaxy rotation; however, we do note that the continuum image from our IFU data is
elongated roughly in alignment with our kinematic ``major axis''. The high-velocity broad underlying wings in the [O~{\sc
  iii}] emission-line profile are located over the full
field-of-view ($\approx8\times11$\,kpc) but dominate in the central
regions (with $W_{80}\approx1100$\,km\,s$^{-1}$ and $\Delta
v\approx-260$\,km\,s$^{-1}$), preferentially located along a N--S
axis (i.e., with a PA\,$\approx 0$). This is not parallel to the radio axis based on the FIRST data
(PA\,$\approx$\,110$^{\circ}$) that may disfavour a connection
between a radio jet and the outflow; however, high resolution radio data
are required to accurately measure the morphology and origin of the radio emission.

\subsubsection*{J0958+1439}
\begin{figure*}
\centerline{\psfig{figure=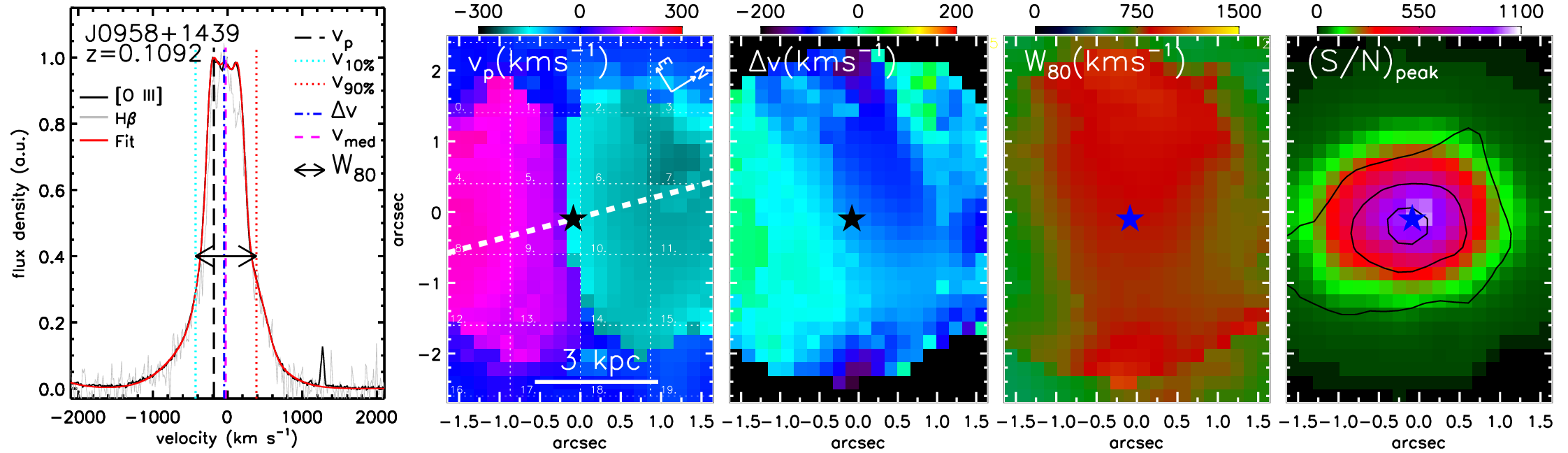,width=6.5in,angle=90}}
\centerline{\psfig{figure=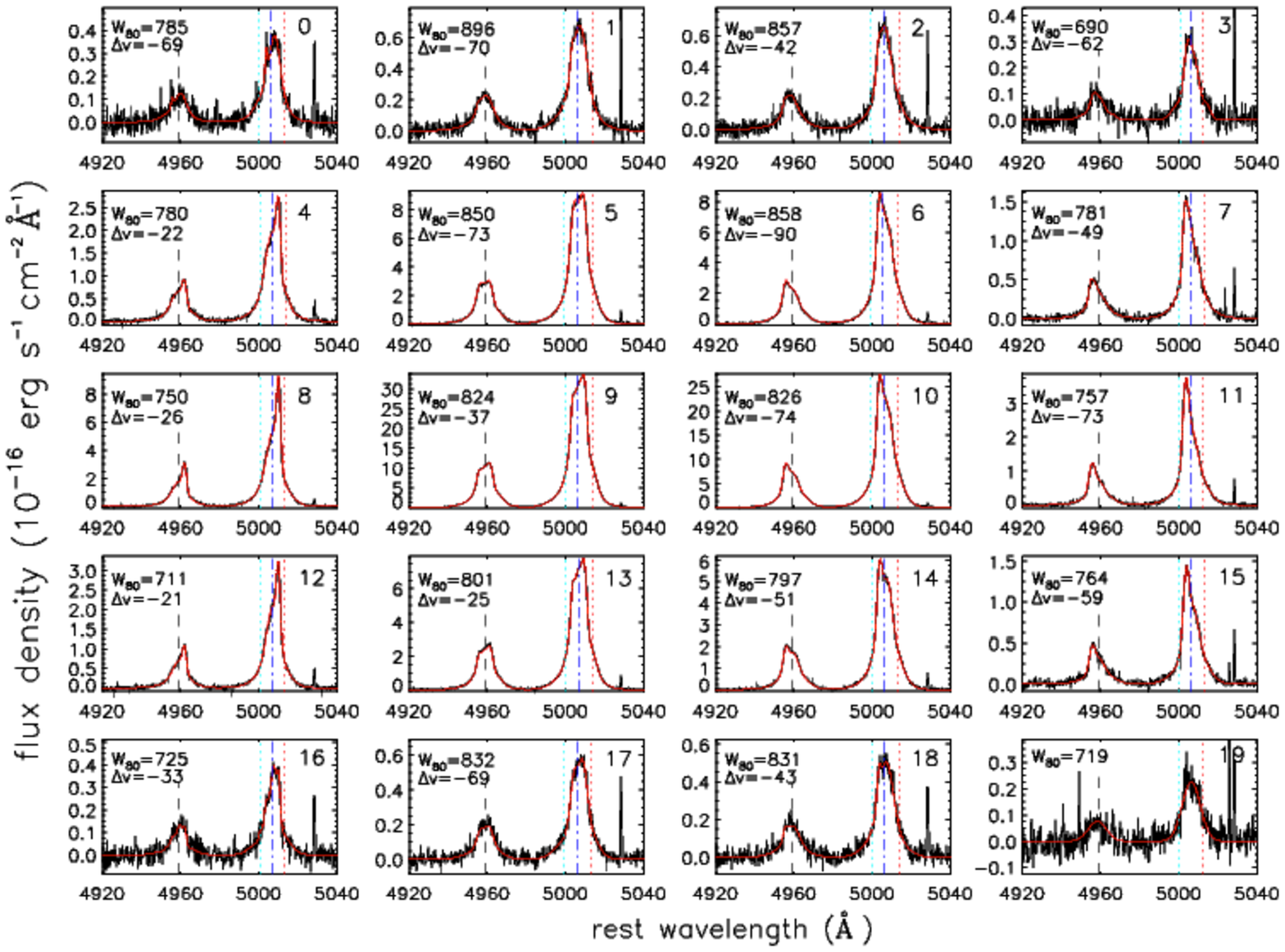,width=6.5in,angle=0}}
\vspace{-0.6cm}
\centerline{\psfig{figure=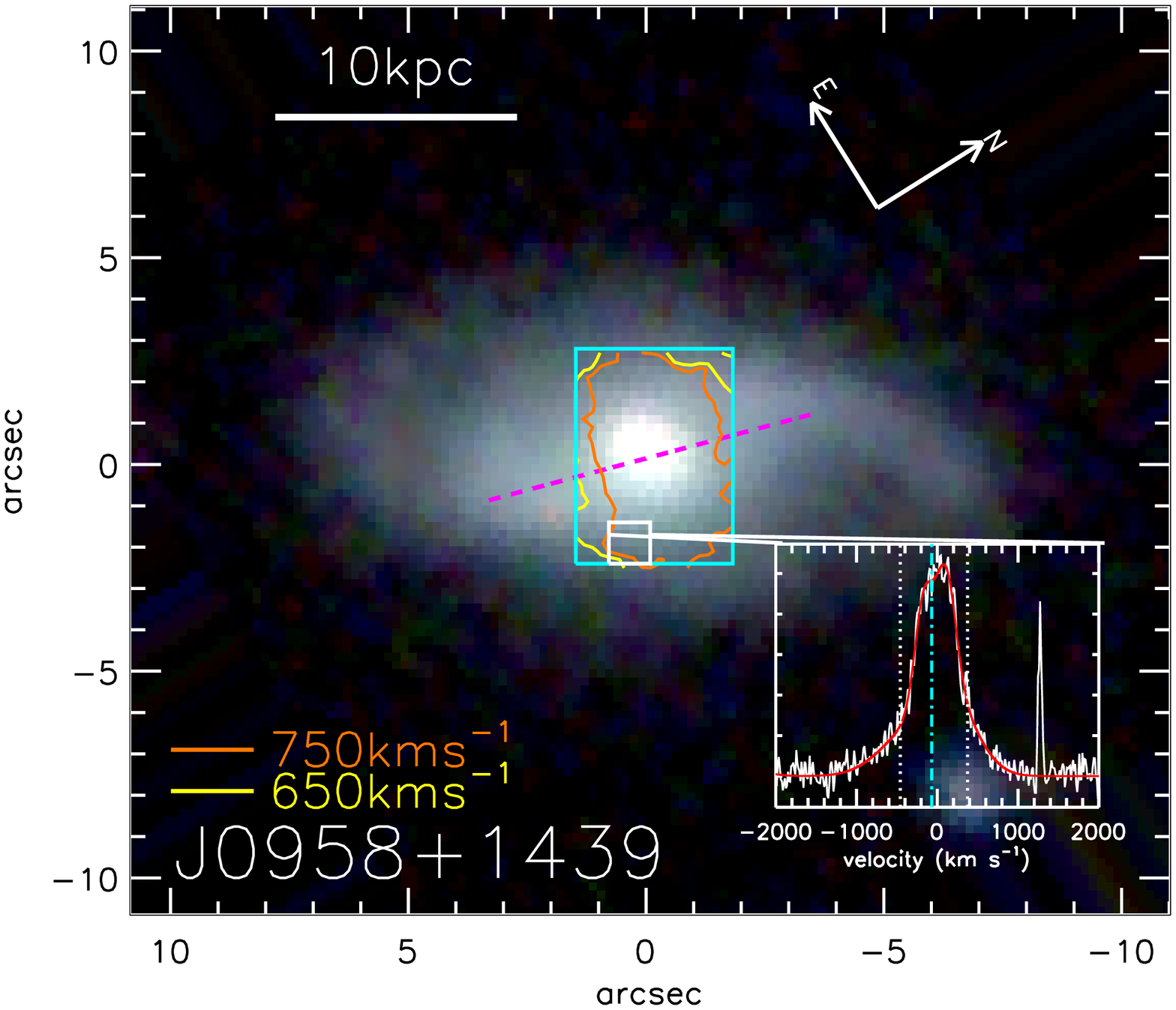,width=2.8in,angle=0}\psfig{figure=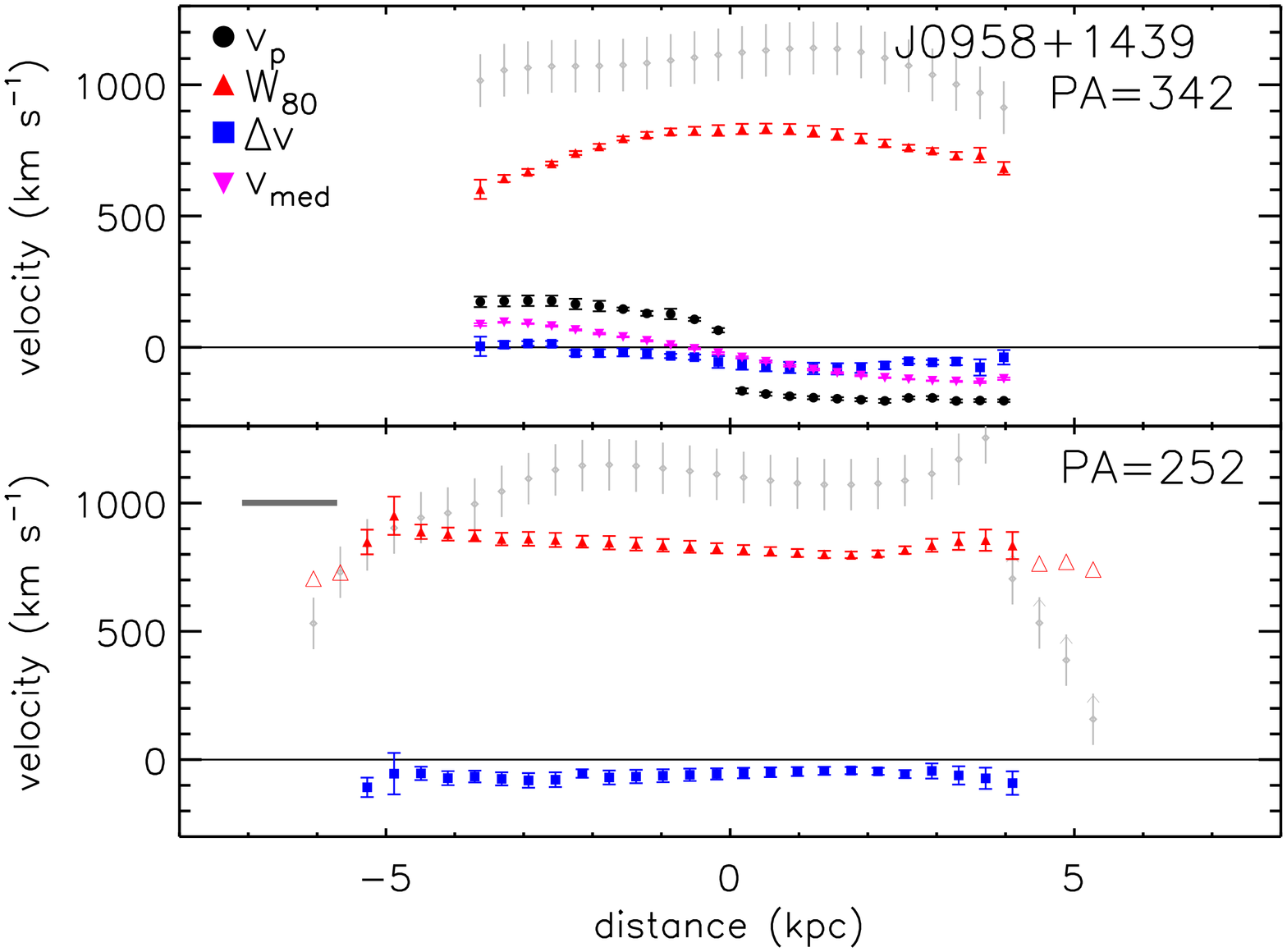,width=3.5in,angle=90}}
\caption{Same as Figure~\ref{Fig:velmaps} and Figure~\ref{Fig:sdss} but
  for SDSS\,J0958+1439}
\label{fig:0958+1439}
\end{figure*}

The IFU data for J0958+1439 are shown in
Figure~\ref{fig:0958+1439}. The data reveal that the ``flat-top''
emission-line profile seen in the galaxy-integrated spectrum is primarily due to the sum of two luminous narrow
kinematic components separated in velocity by
$\approx$\,400\,km\,s$^{-1}$, with the red component dominating in the
south and the blue in the north. At the limit of our resolution
($\approx$1.4\,kpc) we do not see two separate continuum or
emission-line regions. Additionally, the similarity of the H$\beta$ and
[O~{\sc iii}] emission-line profiles indicate similar [O~{\sc
  iii}]/H$\beta$ ratios for each kinematic component (also seen in the
velocity profile; Fig.~\ref{fig:0958+1439}) suggesting that the two
kinematic components are illuminated by a single AGN. The kinematic ``major axis'' we identify
from the peak-velocity ($v_{p}$) map is oriented parallel to the morphological major axis observed in the SDSS
image and our continuum data (Fig.~\ref{fig:0958+1439}). Overall, we favour the interpretation that
the $v_{p}$ velocity map is tracing galaxy rotation, although we cannot rule out a late stage merger without higher spatial resolution
imaging or spectroscopic data. 

In addition to the narrow [O~{\sc iii}] emission-line components there
are underlying broad wings in the emission-line profile over the full
field-of-view ($\approx$\,7$\times$10\,kpc; Fig.~\ref{fig:0958+1439}). These
broad emission-line profiles (with $W_{80}\approx850$\,km\,s$^{-1}$
and $\Delta v\approx-50$\,km\,s$^{-1}$) are more dominant perpendicular to the
kinematic ``major axis''. This potentially indicates that the outflow is
escaping away from a galactic disk. The emission-line flux ratio $\log($[O~{\sc iii}]/H$\beta)\approx1.1$ remains constant over the central
$\approx$\,6\,kpc and then abruptly declines, in a similar
manner to that seen in the type~2 quasars of Liu
et~al. (2013a)\nocite{Liu13a} (see Section~\ref{Sec:properties}). Therefore, to measure the outflow kinetic
energy following Liu et~al. (2013b)\nocite{Liu13b} we adopt a
value of $R_{{\rm br}}=3$\,kpc and measure a H$\beta$ surface
brightness of $\Sigma_{{\rm H}\beta}=1.4\times10^{-15}$\,erg\,s$^{-1}$\,arcsec$^{-2}$ in an
annulus at this radius (corrected for cosmological dimming). Using
this method we obtain an outflow kinetic energy of
${\dot E}\approx2.8\times10^{43}$\,erg\,s$^{-1}$, which is in excellent
agreement with the value that we derive in Section~\ref{Sec:properties} of
${\dot E}_{{\rm out}}\approx2.7\times10^{43}$\,erg\,s$^{-1}$.

\subsubsection*{J1000+1242}
\begin{figure*}
\centerline{\psfig{figure=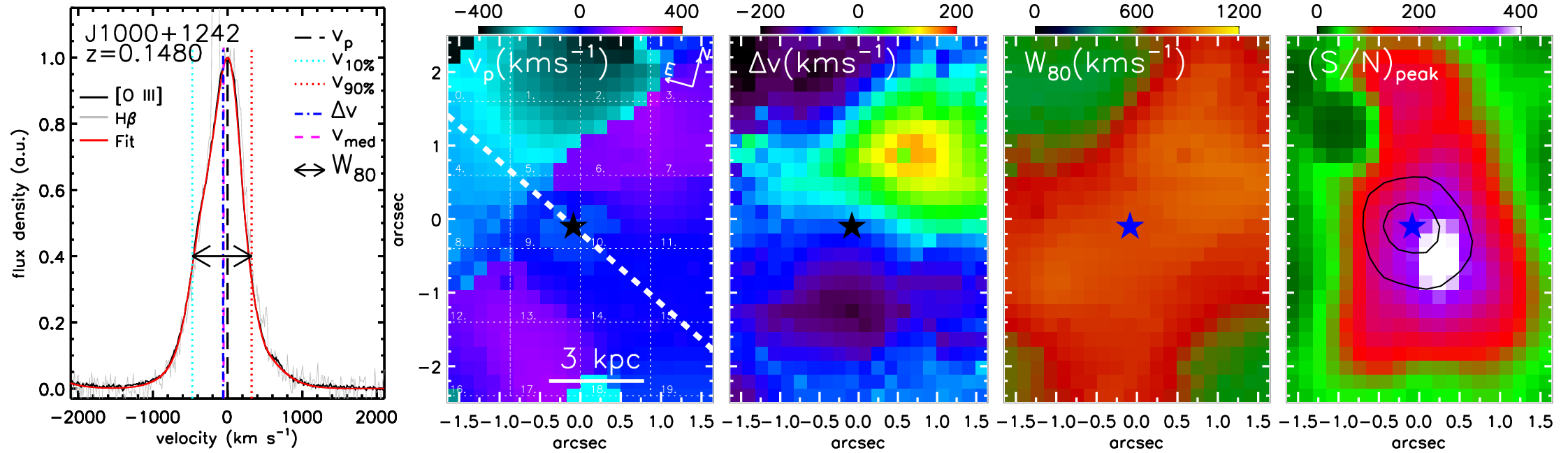,width=6.5in,angle=90}}
\centerline{\psfig{figure=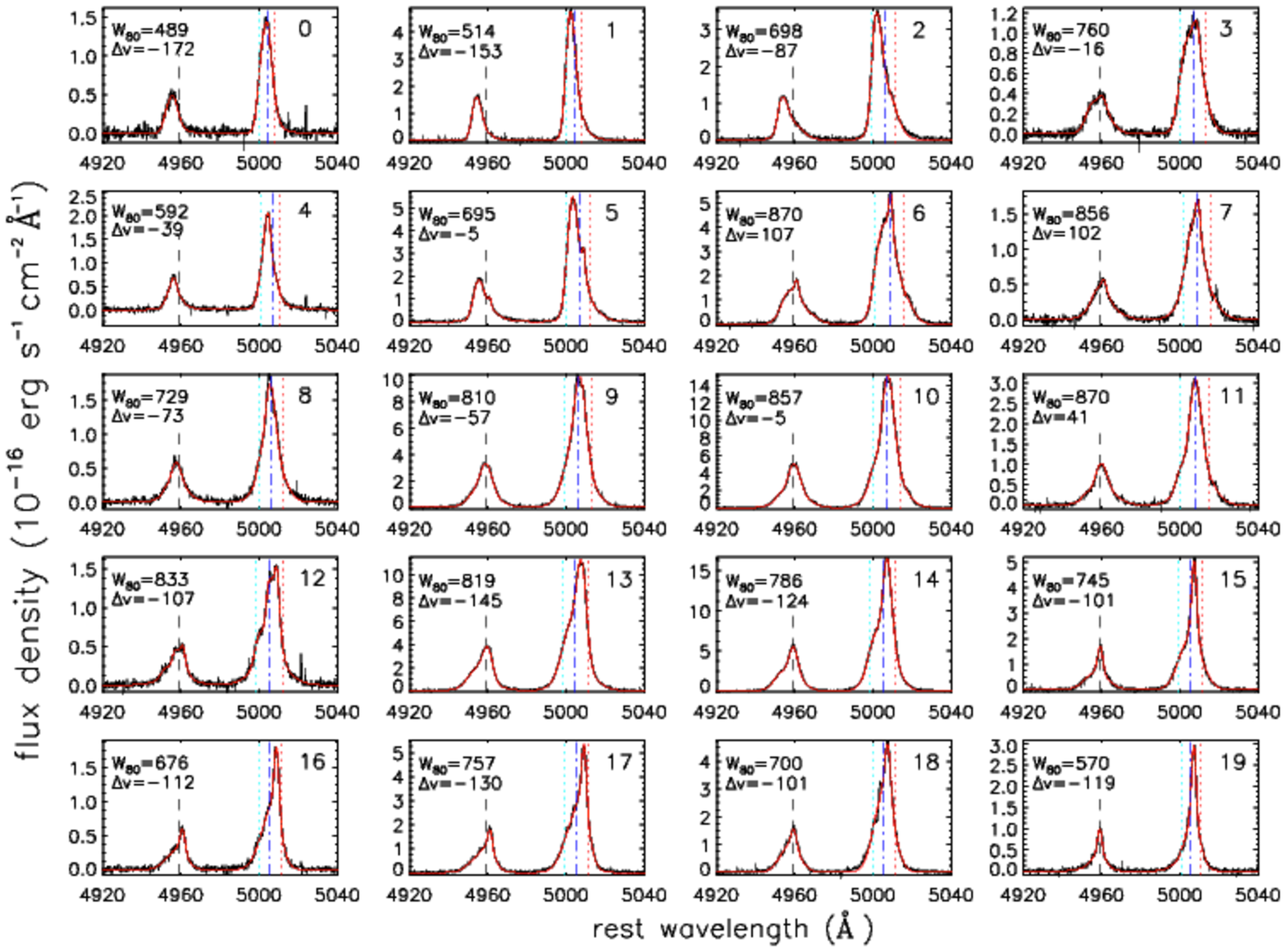,width=6.5in,angle=0}}
\vspace{-0.6cm}
\centerline{\psfig{figure=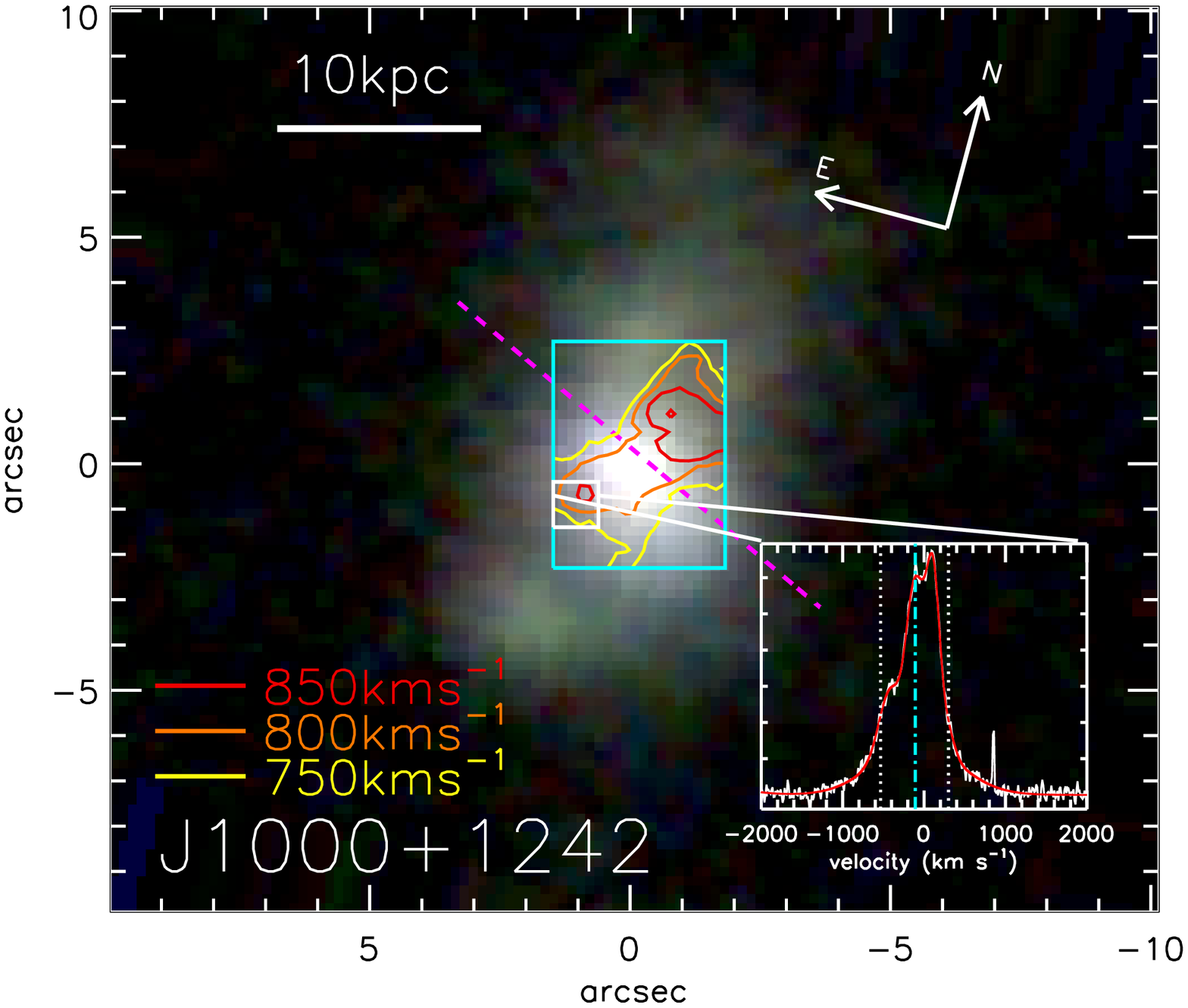,width=2.8in,angle=0}\psfig{figure=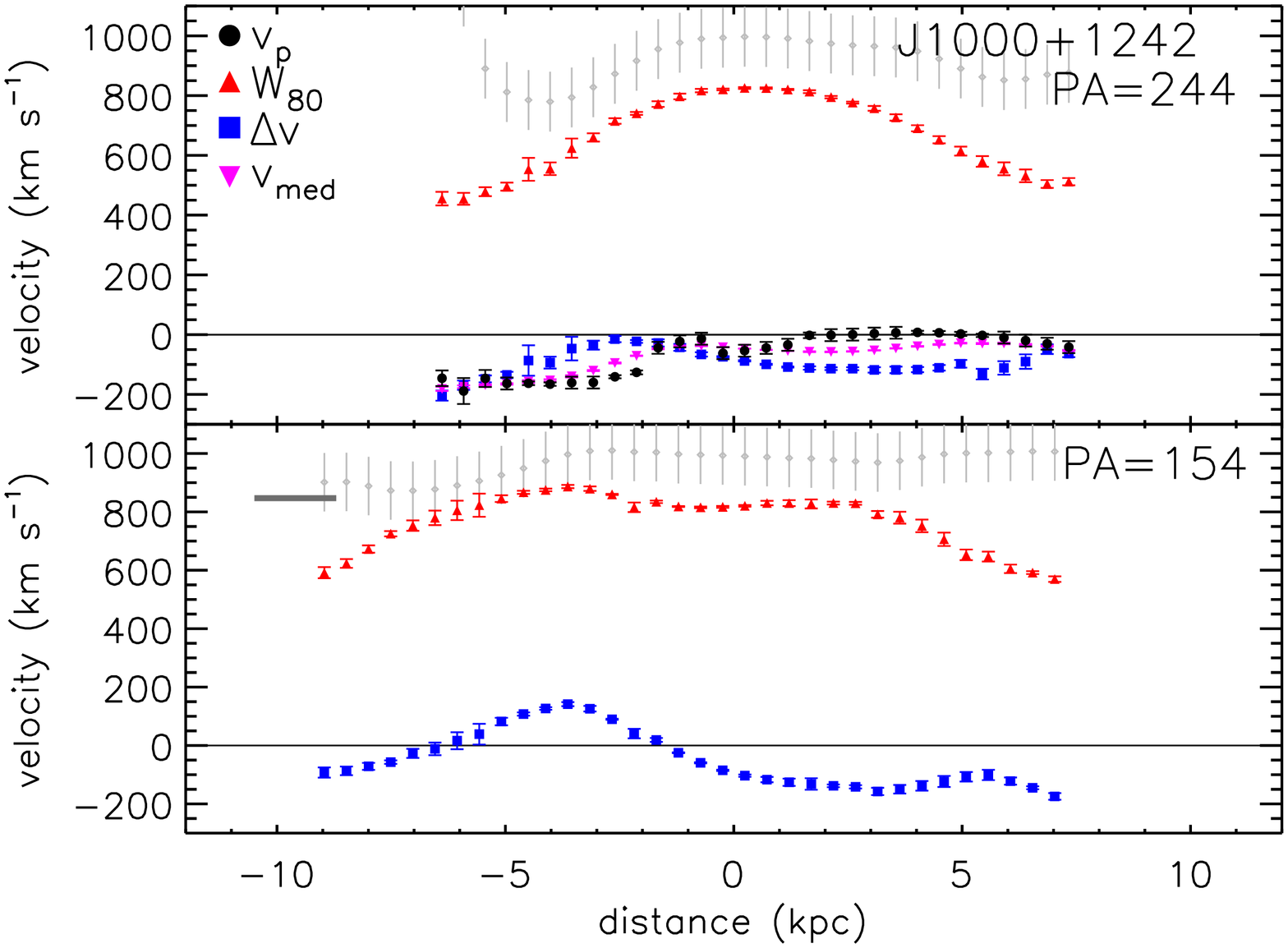,width=3.5in,angle=90}}
\caption{Same as Figure~\ref{Fig:velmaps} and Figure~\ref{Fig:sdss} but
  for SDSS\,J1000+1242}
\label{fig:1000+1242}
\end{figure*}

\cite{Reyes08} previously identified J1000+1242 as a type~2
(``obscured'') quasar. This source has excess radio emission (above that expected from star
 formation; Table~\ref{Tab:observations}) that is marginally resolved based on the FIRST data ($R_{1.4}\approx8.4$\,kpc with a
 PA\,$\approx160^{\circ}$; see Section~\ref{Sec:SEDs} and Table~\ref{Tab:observations}). Although
 fairly irregular, our IFU data reveal a velocity gradient from blue
 to red
 in the peak of the [O~{\sc iii}] emission-line profile ($\Delta
 v_{p}\approx200$\,km\,s$^{-1}$) at a PA of $\approx$\,240$^{\circ}$
 (Fig.~\ref{fig:1000+1242}). The peak signal-to-noise ratio map,
combined with the velocity maps reveal a distinct luminous emission-line
region $\approx$\,3.5\,kpc north-east of the nucleus that has a
reasonably narrow emission-line profile
($W_{80}\lesssim600$\,km\,s$^{-1}$). This feature may be merger
debris, halo gas or an outflow
remnant (see Section~\ref{Sec:GalaxyKinematics}). 

Roughly perpendicular to the $v_{p}$ velocity gradient (with a
PA\,$\approx160$--170$^{\circ}$) we observe regions with
broad underlying wings (with $W_{80}\approx850$\,km\,s$^{-1}$) in the
[O~{\sc iii}] emission-line profile. From the
spectra in boxes 12, 17 and 18 in Figure~\ref{fig:1000+1242} it can be
seen that this broad emission is built up from multiple narrow components. This emission shows a velocity shift from blue
(with $\Delta v\approx-100$\,km\,s$^{-1}$) to red (with $\Delta
v\approx100$\,km\,s$^{-1}$) over $\approx10$--15\,kpc. We also see an
irregular morphology of the emission-line region. All of the above indicates
that we are observing an outflowing bi-polar superbubble, with a strikingly
similar kinematic structure to previously identified superbubbles
(i.e., J1356+1026 from \citealt{Greene12}; J0319-0019 from Liu
et~al. 2013b\nocite{Liu13b}; and Mrk\,273 from \citealt{Rupke13}). The
PA of the superbubbles we observe in J1000+1242 is parallel to the
axis of the extended radio emission and found over a similar spatial extent, potentially
demonstrating a direct connection between a radio jet and the outflow,
although shock-induced radio emission is another possibility. The SDSS image
of this source (Fig.~\ref{fig:1000+1242}) is extended beyond the field-of-view of
our observations and therefore IFU observations over larger regions are required to
trace the full kinematic structure of this target. 

\subsubsection*{J1010+1413}
We show our IFU data for J1010+1413 in Figure~\ref{Fig:velmaps} and
Figure~\ref{Fig:sdss}. This source has the broadest [O~{\sc iii}]
emission-line width of the entire sample ($W_{80}=1450$\,km\,s$^{-1}$)
and is kinematically complex. In the central
regions we observe a smooth gradient from red to blue in the [O~{\sc
  iii}] peak velocity ($v_{p}$; Fig.~\ref{Fig:sdss}), although the overall
velocity field is irregular (Fig.~\ref{Fig:velmaps}). There are luminous narrow [O~{\sc iii}] emission-line
components located to the north and south of the nucleus, separated by
$\approx$16\,kpc in projected distance and $\approx350$\,km\,s$^{-1}$
in projected velocity, that appear to be associated with
emission-line regions that are apparent in the SDSS
image (Fig.~\ref{Fig:sdss}). These features may be merger debris, halo gas or outflow
remnants (see Section~\ref{Sec:GalaxyKinematics}). The extremely broad
emission-line profiles, with velocity offsets reaching up to $\Delta
v\approx-350$\,km\,s$^{-1}$, is located out to the very edge of the field-of-view (i.e., over $\gtrsim16$\,kpc).

\subsubsection*{J1010+0612}
\cite{Reyes08} previously identified J1010+0612 as a type~2
(``obscured'') quasar. This source has a clear radio excess above that expected from star
 formation alone (Table~\ref{Tab:observations}). Based on the FIRST
 data we do not spatially resolve any radio emission on scales $\gtrsim2^{\prime\prime}$ (see
 Section~\ref{Sec:SEDs} and Table~\ref{Tab:observations}). 

\begin{figure*}
\centerline{\psfig{figure=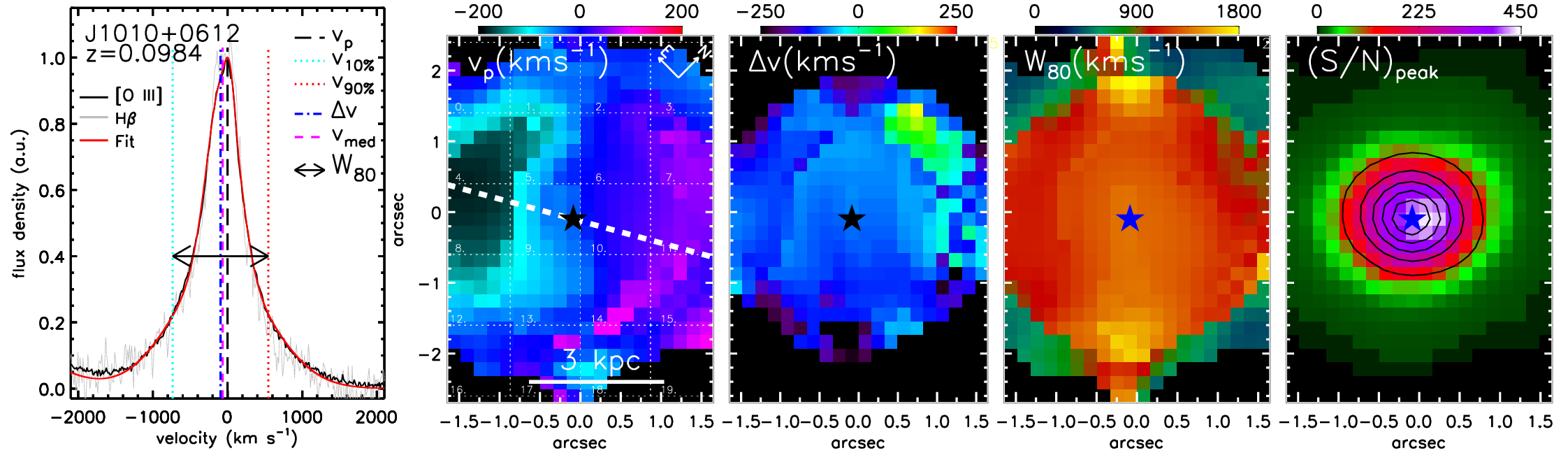,width=6.5in,angle=90}}
\centerline{\psfig{figure=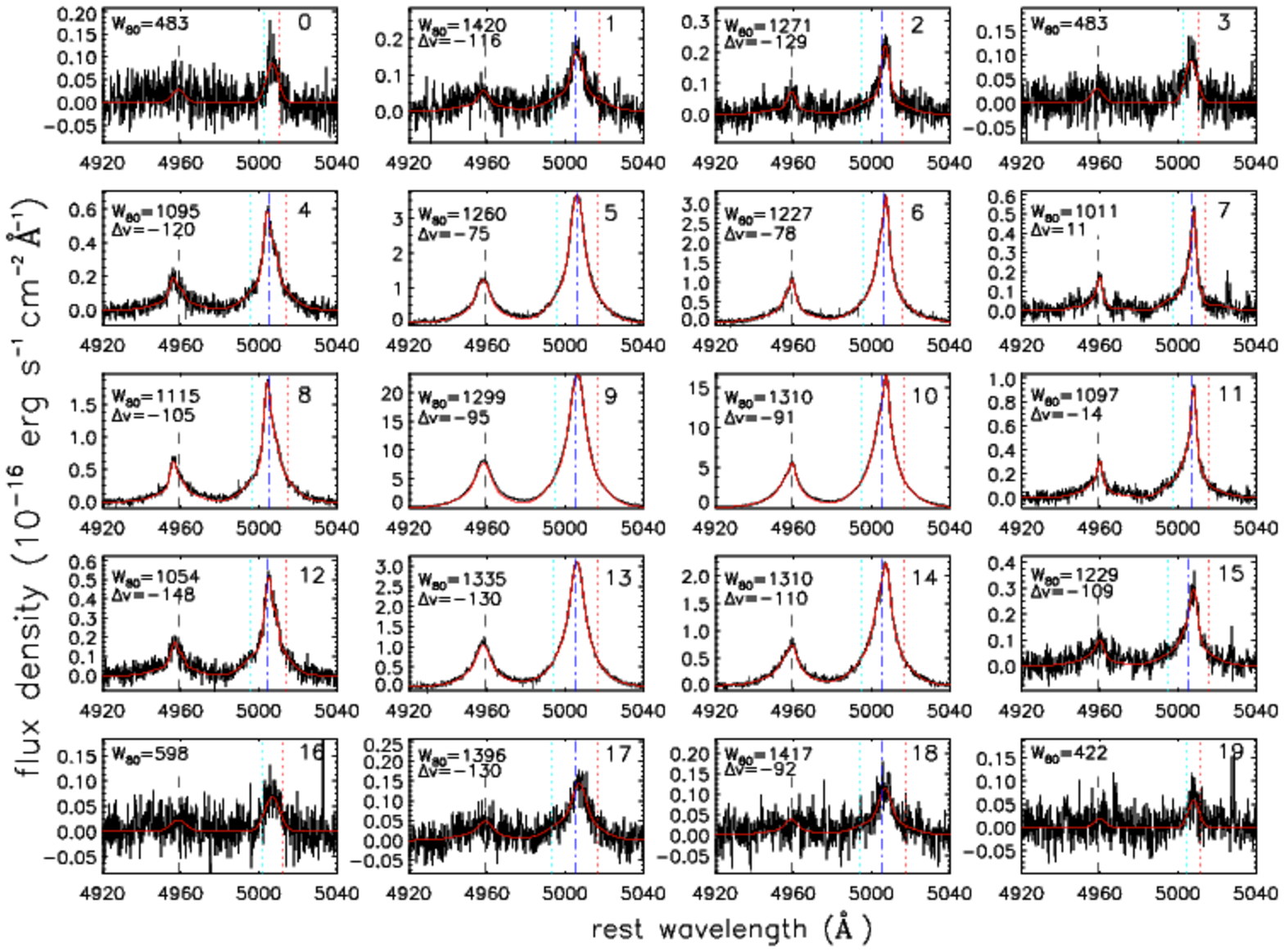,width=6.5in,angle=0}}
\vspace{-0.6cm}
\centerline{\psfig{figure=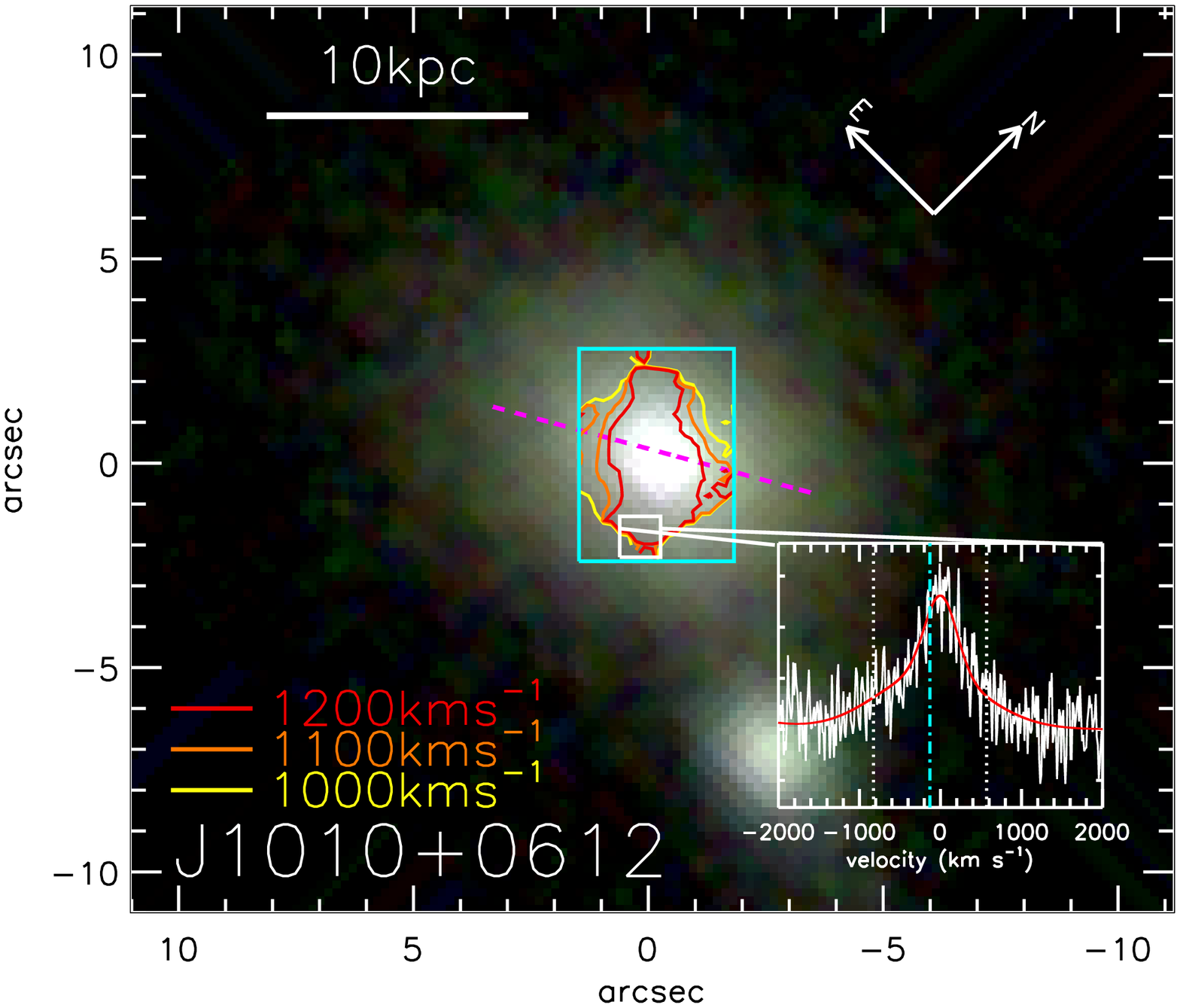,width=2.8in,angle=0}\psfig{figure=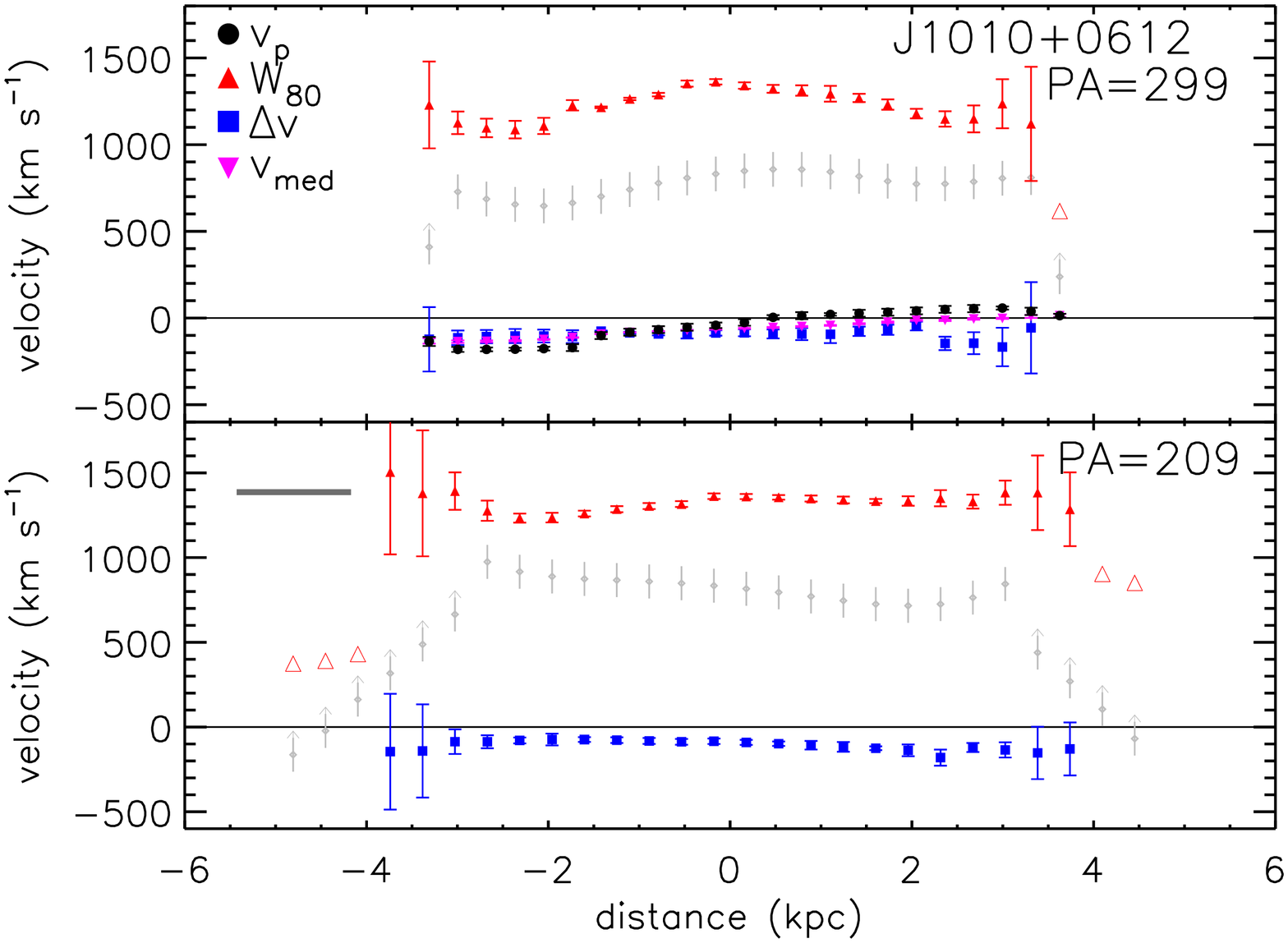,width=3.5in,angle=90}}
\caption{Same as Figure~\ref{Fig:velmaps} and Figure~\ref{Fig:sdss} but
  for SDSS\,J1010+0612}
\label{fig:1010+0612}
\end{figure*}

We show our IFU data for this object in
Figure~\ref{fig:1010+0612}. We identify a regular blue to red velocity field, indicative of galactic rotation, in the
[O~{\sc iii}] peak velocity (with $\Delta v_{p}\approx300$\,km\,s$^{-1}$). Broad
underlying wings lead to emission-line widths of $W_{80}\approx1100-1300$\,km\,s$^{-1}$
and velocity offsets of $\Delta v\approx-100$\,km\,s$^{-1}$ and are
located over the full extent of the field-of-view
($\approx6\times9$\,kpc) with a roughly circular morphology. 

\subsubsection*{J1100+0846}
\begin{figure*}
\centerline{\psfig{figure=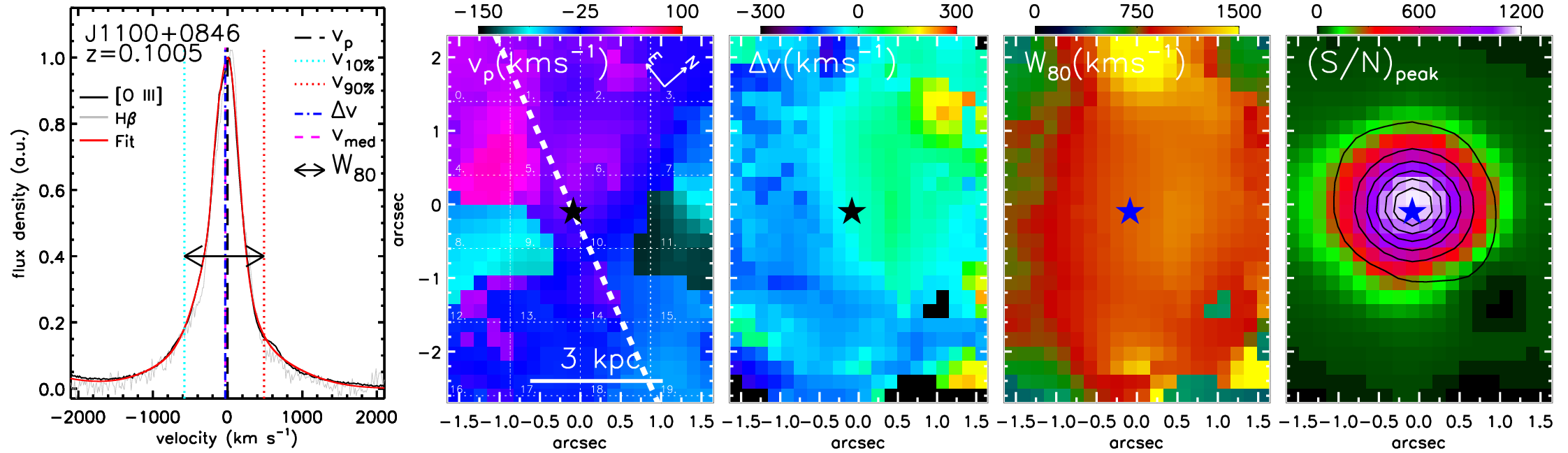,width=6.5in,angle=90}}
\centerline{\psfig{figure=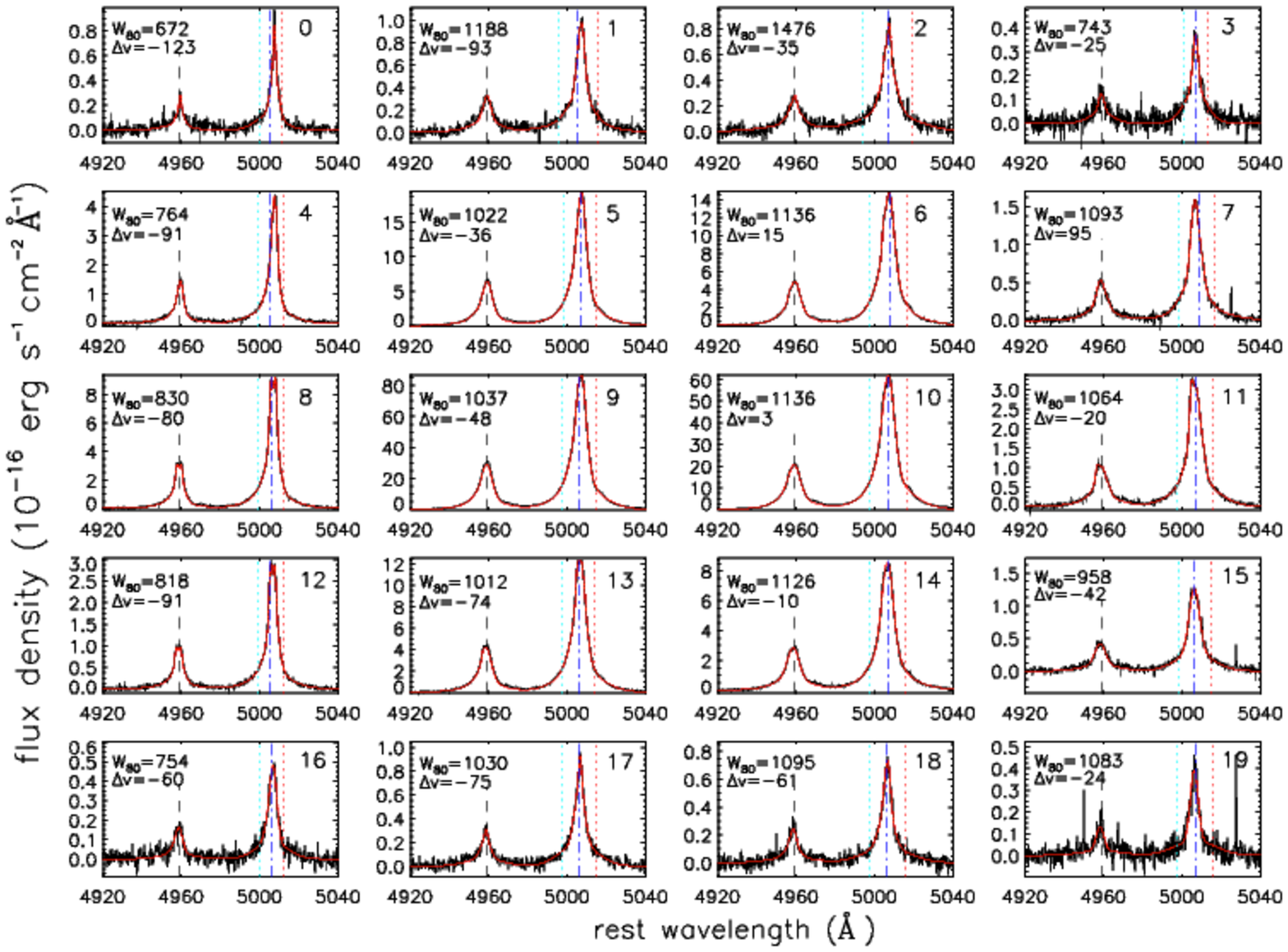,width=6.5in,angle=0}}
\vspace{-0.6cm}
\centerline{\psfig{figure=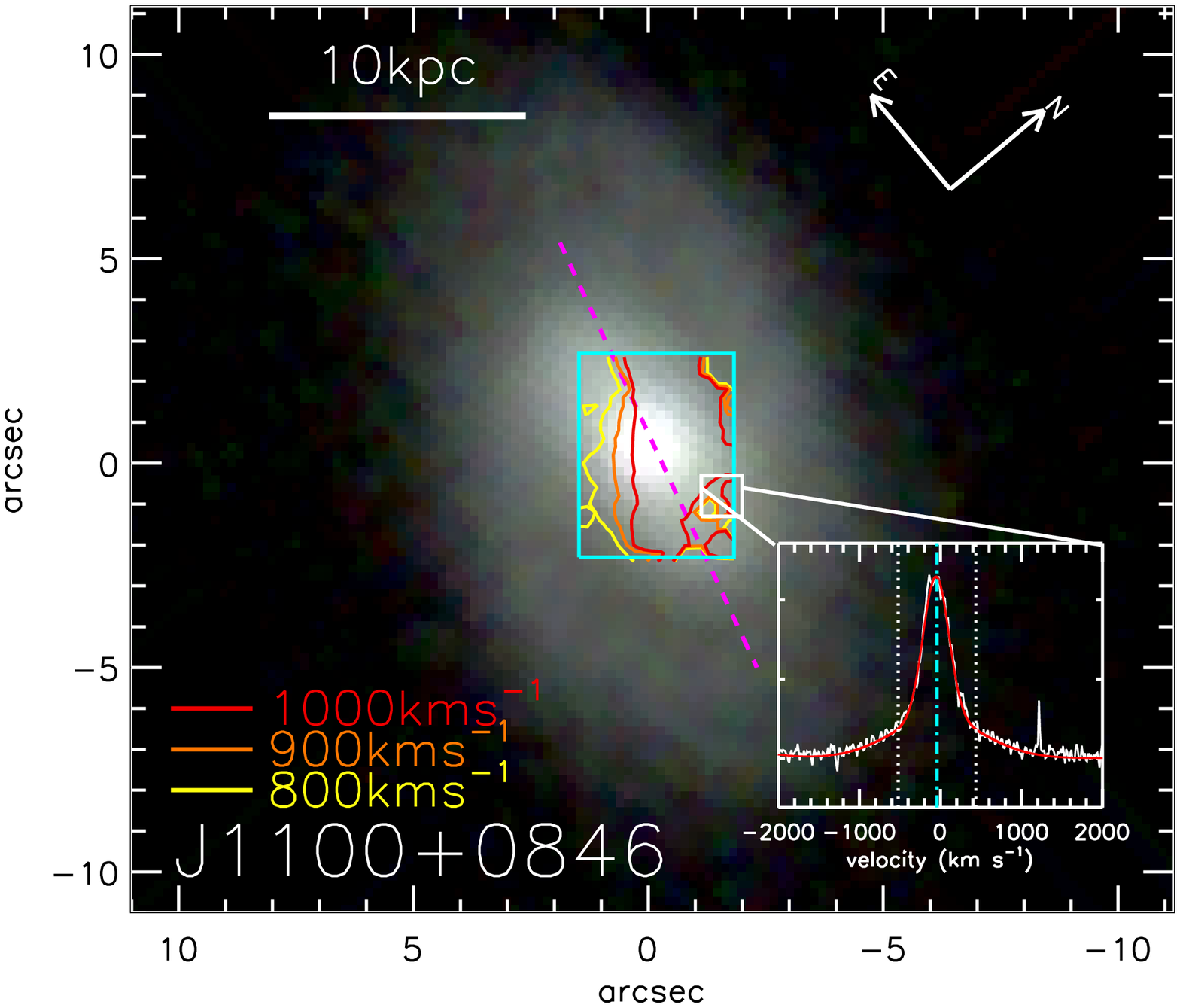,width=2.8in,angle=0}\psfig{figure=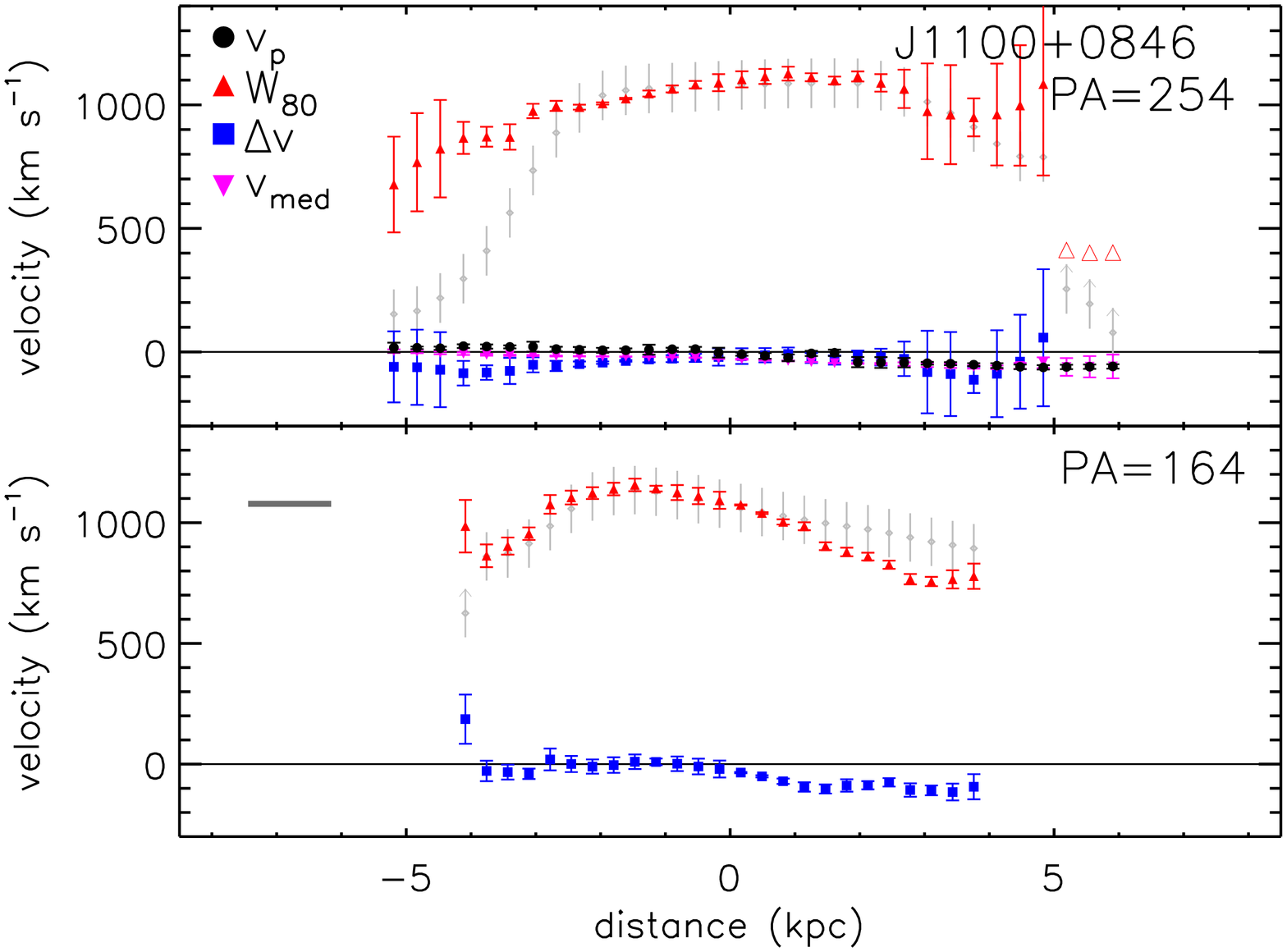,width=3.5in,angle=90}}
\caption{Same as Figure~\ref{Fig:velmaps} and Figure~\ref{Fig:sdss} but
  for SDSS\,J1100+0846}
\label{fig:1100+0846}
\end{figure*}

\cite{Reyes08} previously identified J1100+0846 as a type~2
(``obscured'') quasar and we show our IFU data for this object in
Figure~\ref{fig:1100+0846}. Although the peak velocity ($v_{p}$) map shows some irregular features we see a smooth velocity
gradient ($\Delta v_{p}\approx100$\,km\,s$^{-1}$) along a
PA\,$\approx250^{\circ}$. Additionally, this kinematic ``major axis'' is parallel to the morphological major axis seen in the SDSS
image (with hints of the same axis in our continuum image), providing evidence that kinematics due to galaxy rotation are being
observed. Broad wings (leading to $W_{80}\approx$\,700--1100\,km\,s$^{-1}$) are observed in the emission-line profiles over the
full field-of-view ($\approx9\times6$\,kpc). The
uncertainties on the velocity offset of the broad emission (i.e., $\Delta
v$) are large; however, the velocity seems to be highest in the
southern regions with velocities between $\Delta
v\approx-100$\,km\,s$^{-1}$ and $-80$\,km\,s$^{-1}$. These blue-shifted
velocities could indicate the near-side of an outflow that is moving
away from the almost face on galaxy disc. In the velocity
profile (Fig.~\ref{fig:1100+0846}), we observe that the emission-line
flux ratio of $\log($[O~{\sc iii}]/H$\beta)\approx1.0$--1.1 is constant over the central
regions then declines at a break radius of $R_{{\rm br}}=2.5$\,kpc
(see Section~\ref{Sec:properties}). We therefore  measure a H$\beta$ surface
brightness of $\Sigma_{{\rm H}\beta}=1.7\times10^{-15}$\,erg\,s$^{-1}$\,arcsec$^{-2}$ in an
annulus at this break radius (corrected for cosmological dimming) and
calculate the outflow kinetic
energy following Liu et~al. (2013b)\nocite{Liu13b}. The
kinetic outflow energy we obtain using this method, i.e., ${\dot E}\approx6.3\times10^{43}$\,erg\,s$^{-1}$, is
in excellent agreement with the value that we calculate in
Section~\ref{Sec:properties}, i.e., ${\dot E}_{{\rm out}}\approx6.9\times10^{43}$\,erg\,s$^{-1}$.

\subsubsection*{J1125+1239}
\begin{figure*}
\centerline{\psfig{figure=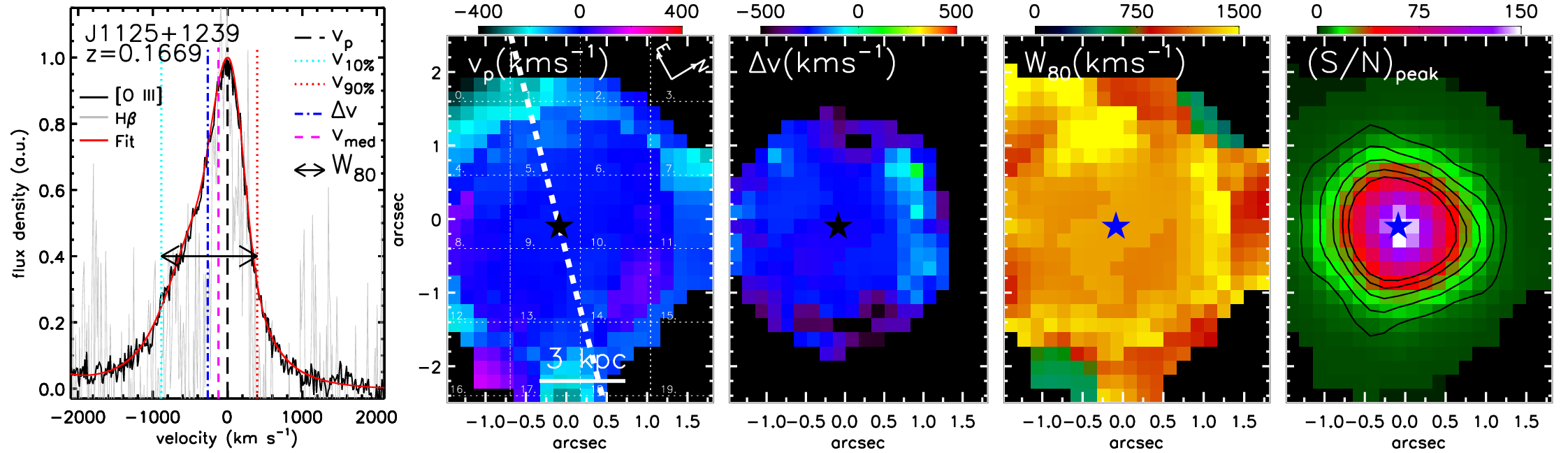,width=6.5in,angle=90}}
\centerline{\psfig{figure=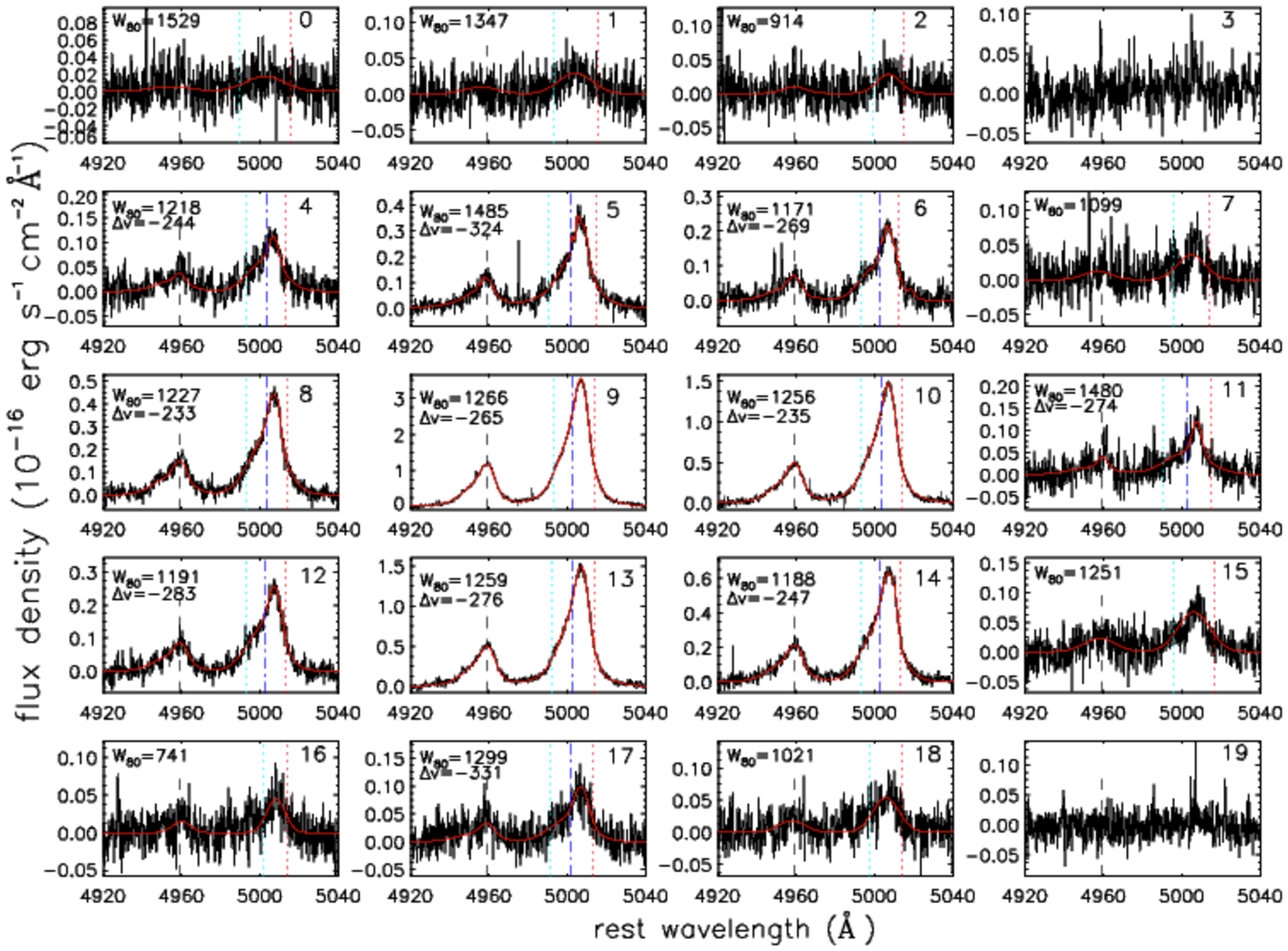,width=6.5in,angle=0}}
\vspace{-0.6cm}
\centerline{\psfig{figure=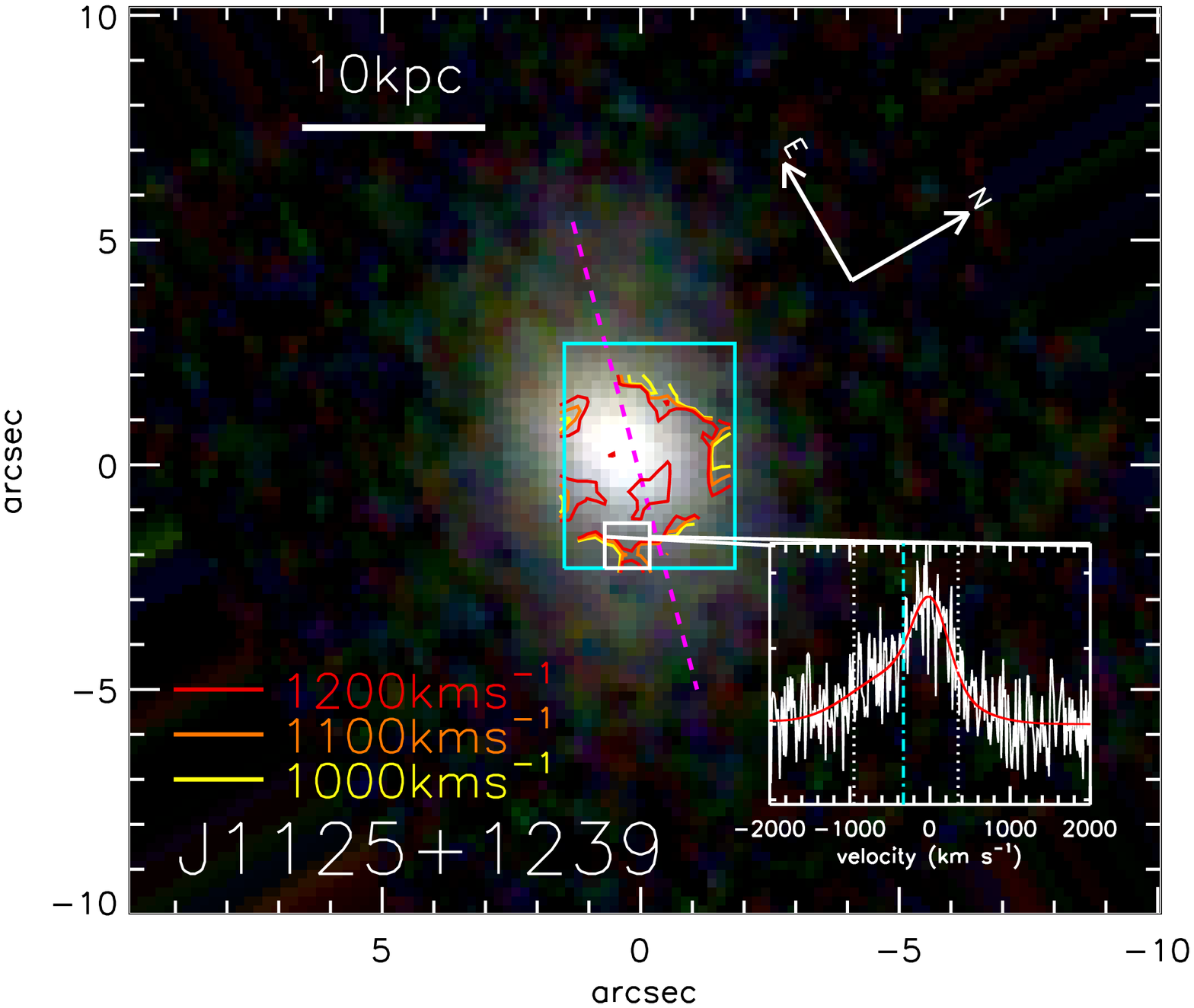,width=2.8in,angle=0}\psfig{figure=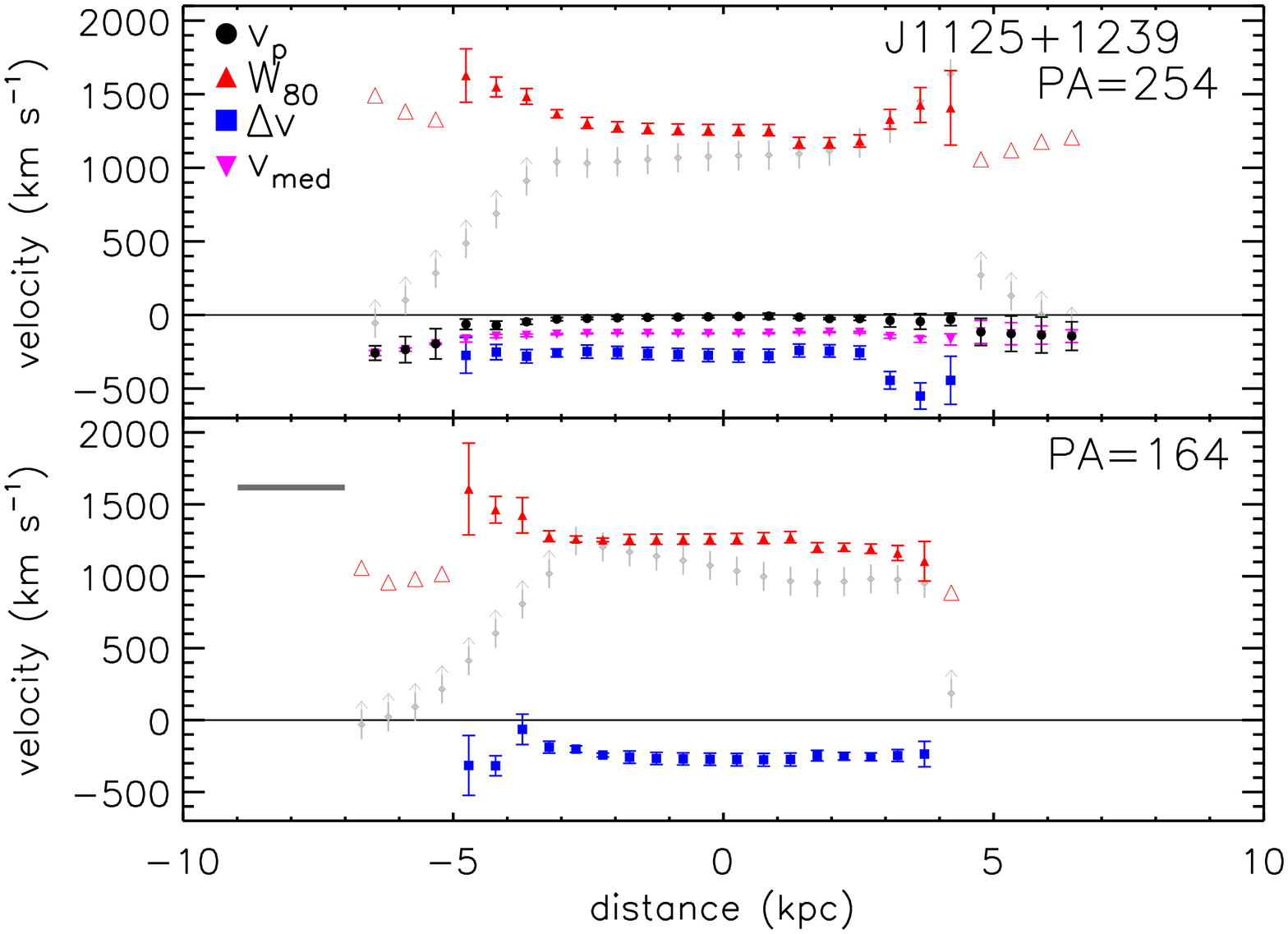,width=3.5in,angle=90}}
\caption{Same as Figure~\ref{Fig:velmaps} and Figure~\ref{Fig:sdss} but
  for SDSS\,J1125+1239}
\label{fig:1125+1239}
\end{figure*}

We show our IFU observations for J1125+1239 in
Figure~\ref{fig:1125+1239}. We observe an emission-line region that is extended right to the edges of
the field-of-view (over $\gtrsim$\,13\,kpc). We do not observe
any velocity gradients in the peak of the [O~{\sc iii}] emission-line profile
(i.e., $v_{p}$; Fig.~\ref{fig:1125+1239}). There is faint, high velocity ($\Delta v\approx-250$ to $-300$\,km\,s$^{-1}$)
and extremely broad [O~{\sc iii}] emission
($W\approx1200$--1500\,km\,s$^{-1}$), out to the very edges of the
field-of-view. Due to the low surface brightness the morphology of
the broad emission is uncertain; however, the SDSS image shows
faint extended emission beyond the IFU field-of-view, 
parallel to the axis where we predominantly observe this broad
emission (i.e., PA\,$\approx250^{\circ}$; Fig.~\ref{fig:1125+1239}).

\subsubsection*{J1130+1301}
\begin{figure*}
\centerline{\psfig{figure=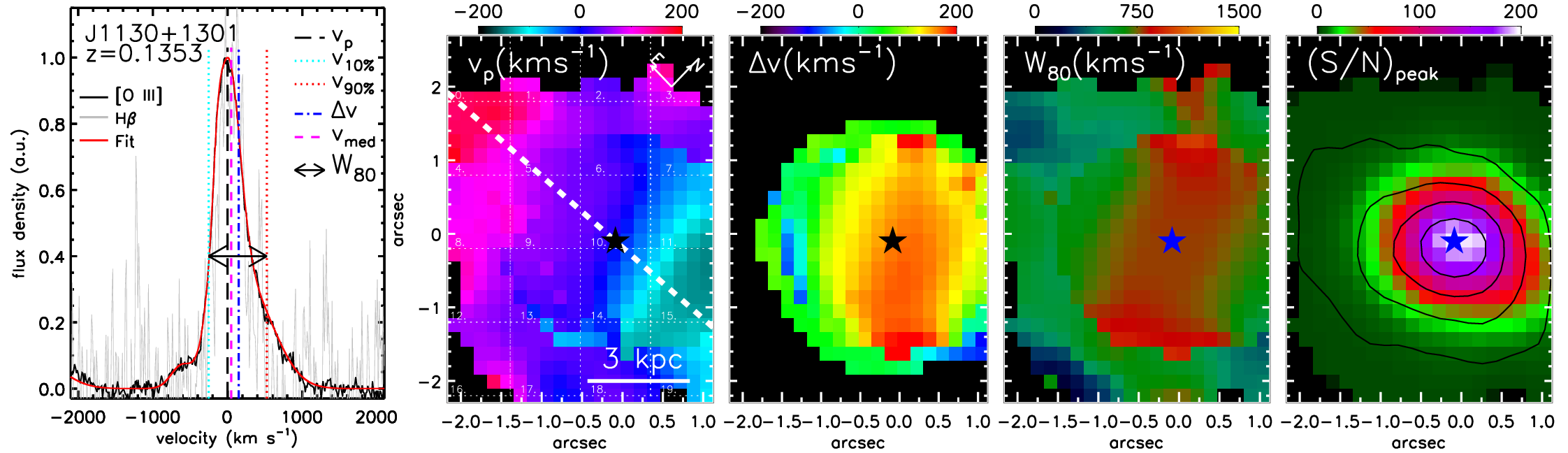,width=6.5in,angle=90}}
\centerline{\psfig{figure=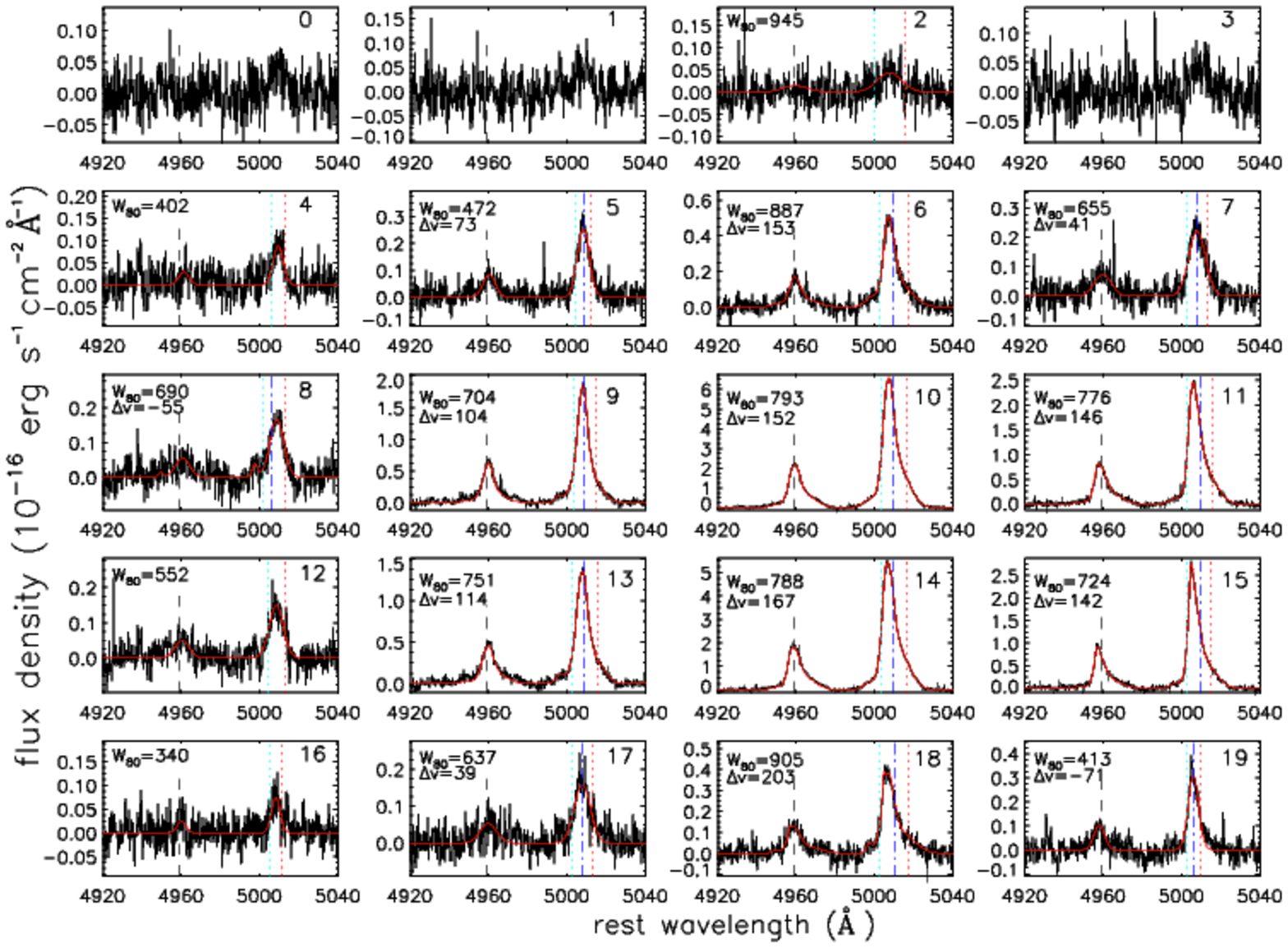,width=6.5in,angle=0}}
\vspace{-0.6cm}
\centerline{\psfig{figure=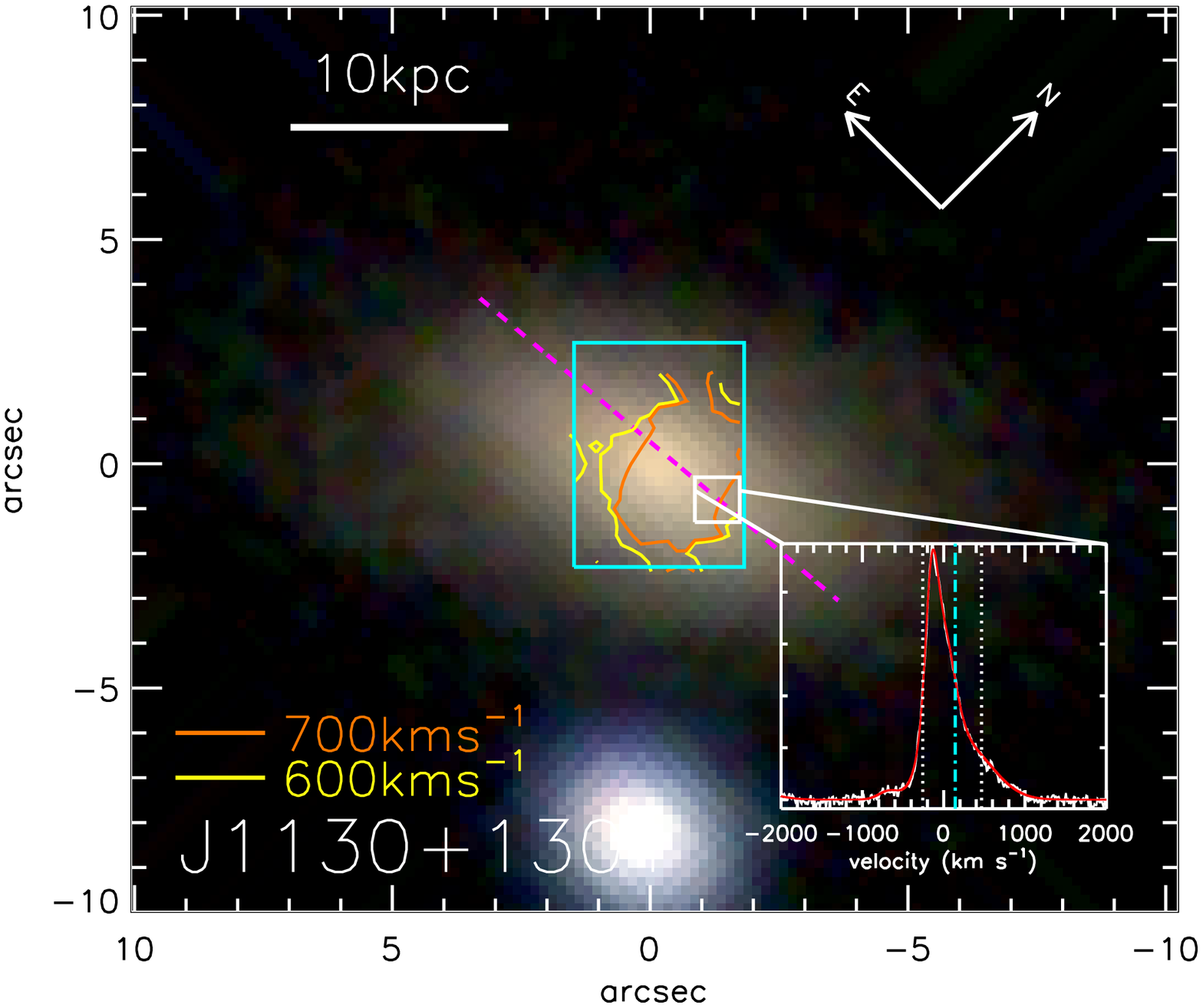,width=2.8in,angle=0}\psfig{figure=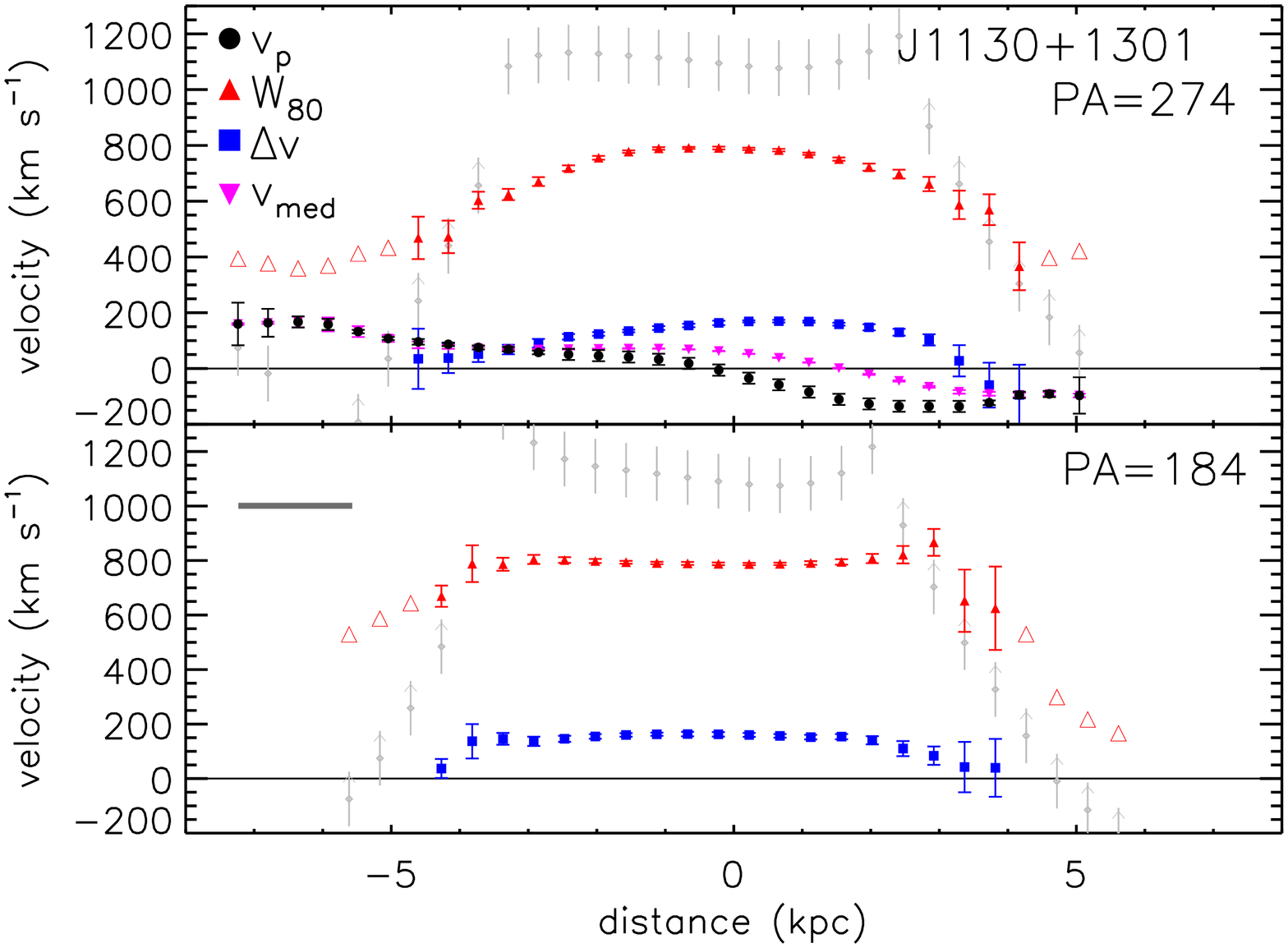,width=3.5in,angle=90}}
\caption{Same as Figure~\ref{Fig:velmaps} and Figure~\ref{Fig:sdss} but
  for SDSS\,J1130+1301}
\label{fig:1130+1301}
\end{figure*}

We show our IFU observations for J1130+1301 in
Figure~\ref{fig:1130+1301}. The peak velocity map ($v_{p}$) is indicative of
galactic rotation showing a smooth velocity gradient
from red to blue with $\Delta v_{p}\approx300$\,km\,s$^{-1}$. This interpretation is
strengthened by the kinematic major axis being aligned with the morphological major axis observed in
the SDSS image and the continuum observed in our data. The broad emission-line wings
(with $W_{80}>600$\,km\,s$^{-1}$) are found perpendicular to the
galaxy major axis over $\approx$\,8\,kpc. Although both blue and red
wings are observed in the emission-line profile the red wing is
more luminous resulting in an overall velocity offset that is positive (i.e., $\Delta
v\approx$\,150\,km\,s$^{-1}$). 

\subsubsection*{J1216+1417}
\begin{figure*}
\centerline{\psfig{figure=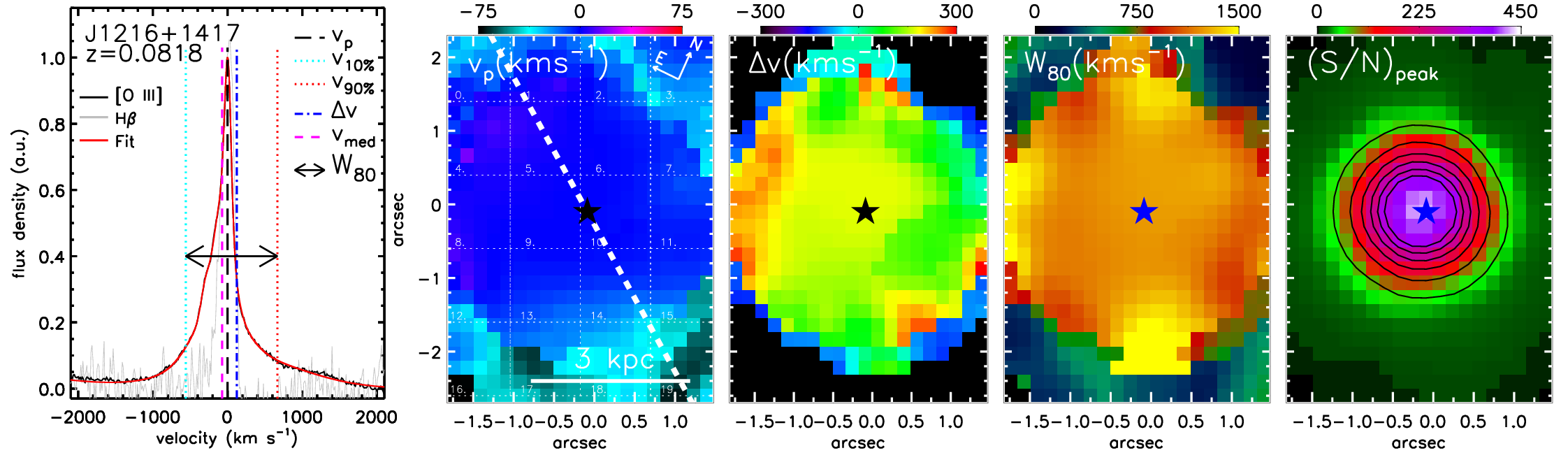,width=6.5in,angle=90}}
\centerline{\psfig{figure=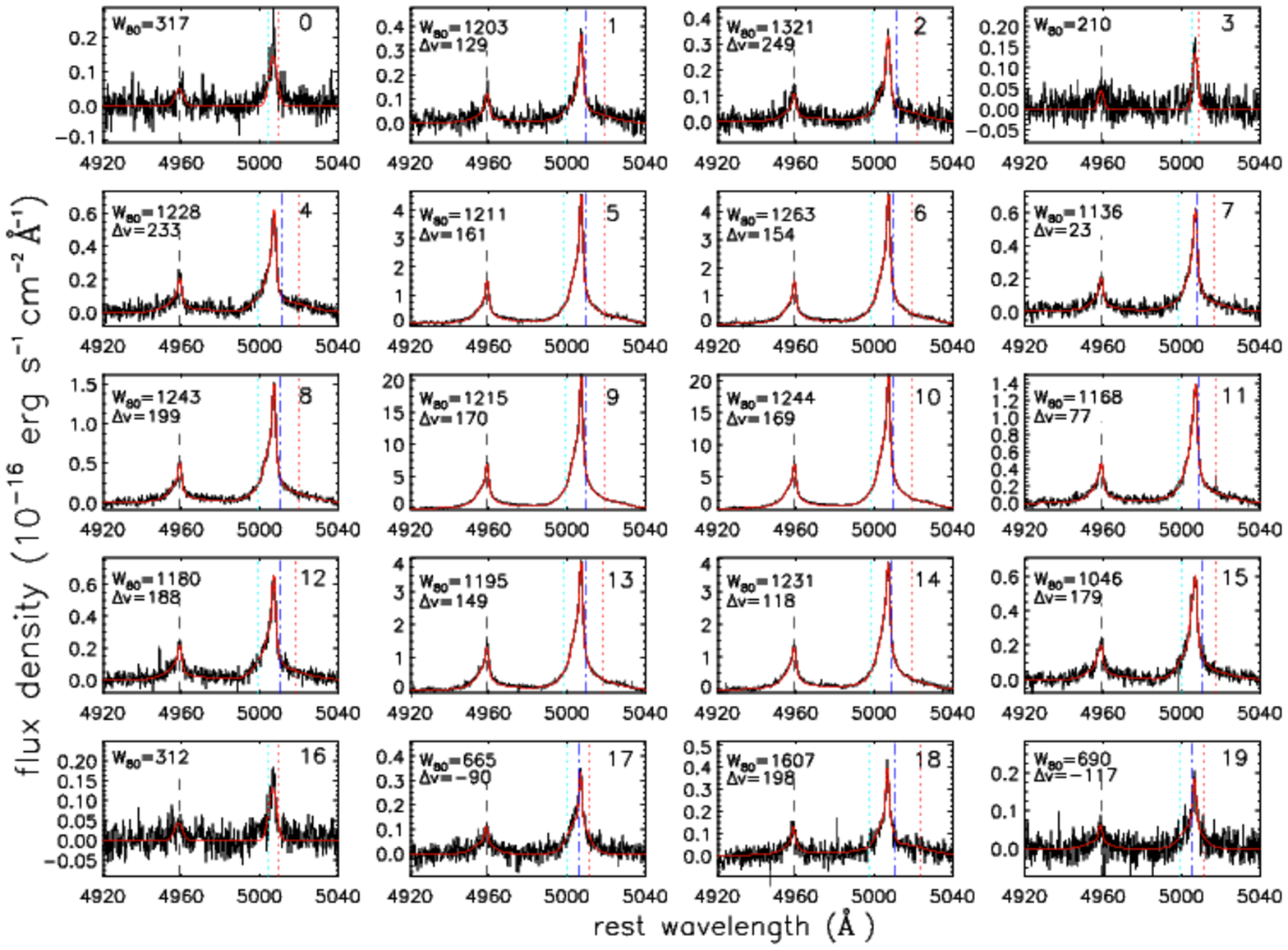,width=6.5in,angle=0}}
\vspace{-0.6cm}
\centerline{\psfig{figure=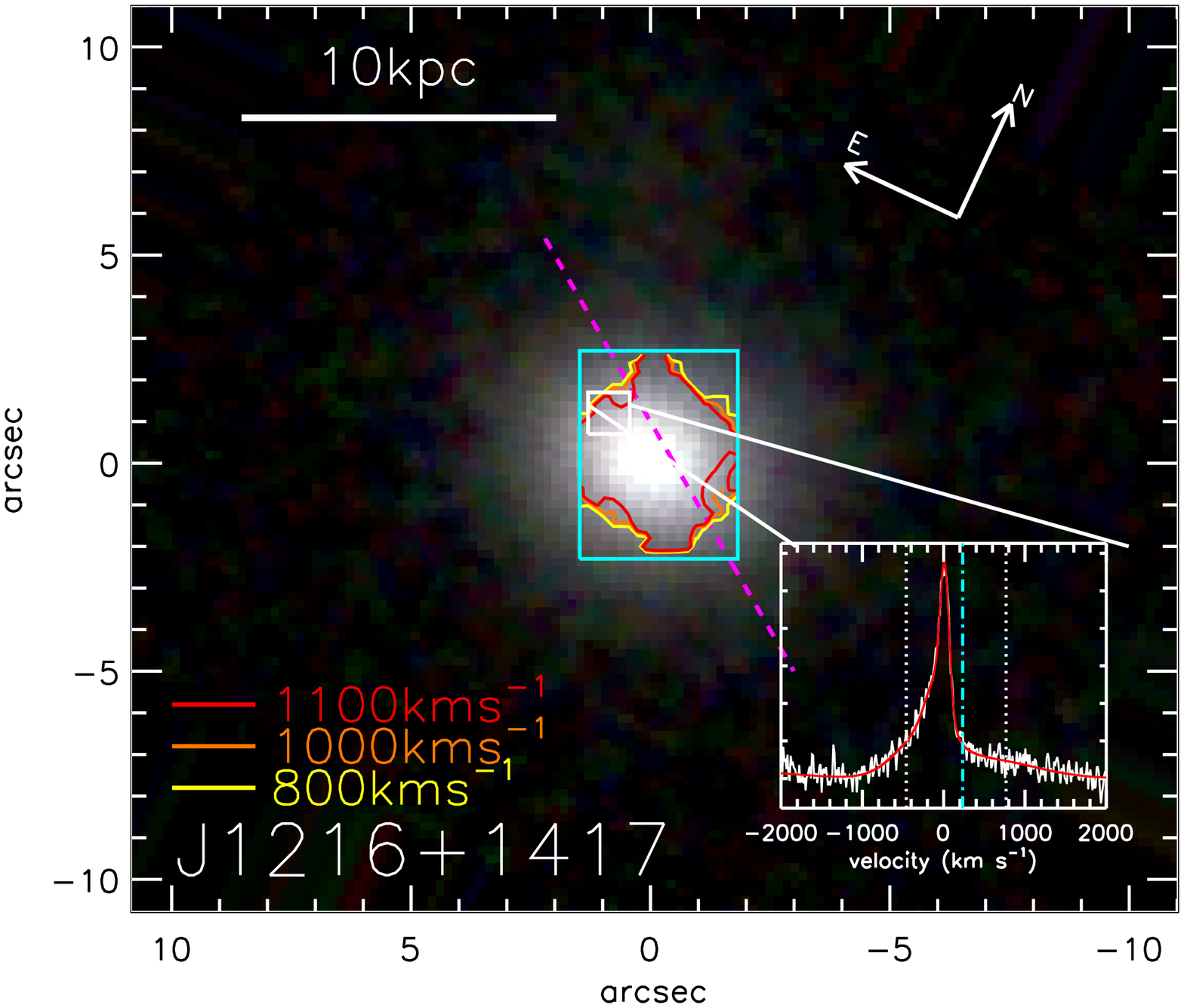,width=2.8in,angle=0}\psfig{figure=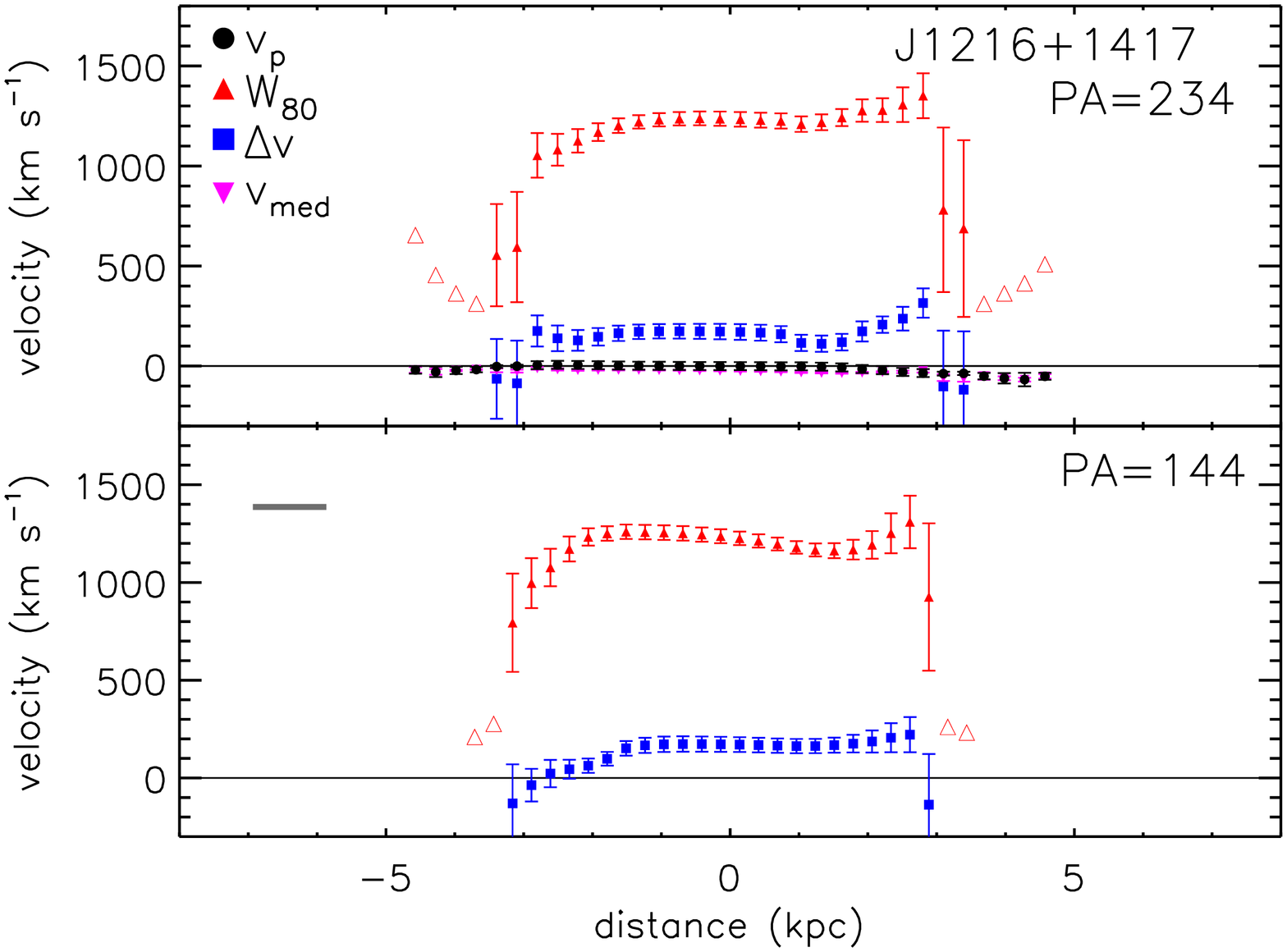,width=3.5in,angle=90}}
\caption{Same as Figure~\ref{Fig:velmaps} and Figure~\ref{Fig:sdss} but
  for SDSS\,J1216+1417}
\label{fig:1216+1417}
\end{figure*}

Spectroscopy of J1216+1417 has previously identified
Wolf-Rayet stars are located in this source and consequently shows
that a recent star formation episode has occurred (i.e.,
$\approx2$--5\,Myrs; \citealt{Brinchmann08}). Furthermore, a subtle
4000\AA\ break and stellar
absorption features are seen in the SDSS spectrum and may explain why the H$\beta$ emission-line profile is different to that seen in [O~{\sc iii}] (Fig.~\ref{fig:1216+1417}). We show our IFU data in
Figure~\ref{fig:1216+1417}. We see little velocity structure in
the $v_{p}$ velocity map. The [O~{\sc iii}] emission line profile is
irregular, with both a blue wing (out to
$v\approx$\,1000\,km\,s$^{-1}$) and a large red wing out to
$v\approx$\,2000\,km\,s$^{-1}$. This is seen in both lines of
the [O~{\sc iii}]$\lambda\lambda$4959,5007 emission-line doublet indicating that
it is a true [O~{\sc iii}] feature associated with this
source. Further kinematic analysis, beyond the scope of this work, is
required to fully identify the origin of these high-velocity
wings. The blue and red wings observed in the
[O~{\sc iii}] emission line are predominantly found in a circular
region across the central $\gtrsim$\,6\,kpc; however, the low surface brightness of the emission in the outer
regions puts some uncertainty on the spatial extent and morphology.  

\subsubsection*{J1316+1753}
\begin{figure*}
\centerline{\psfig{figure=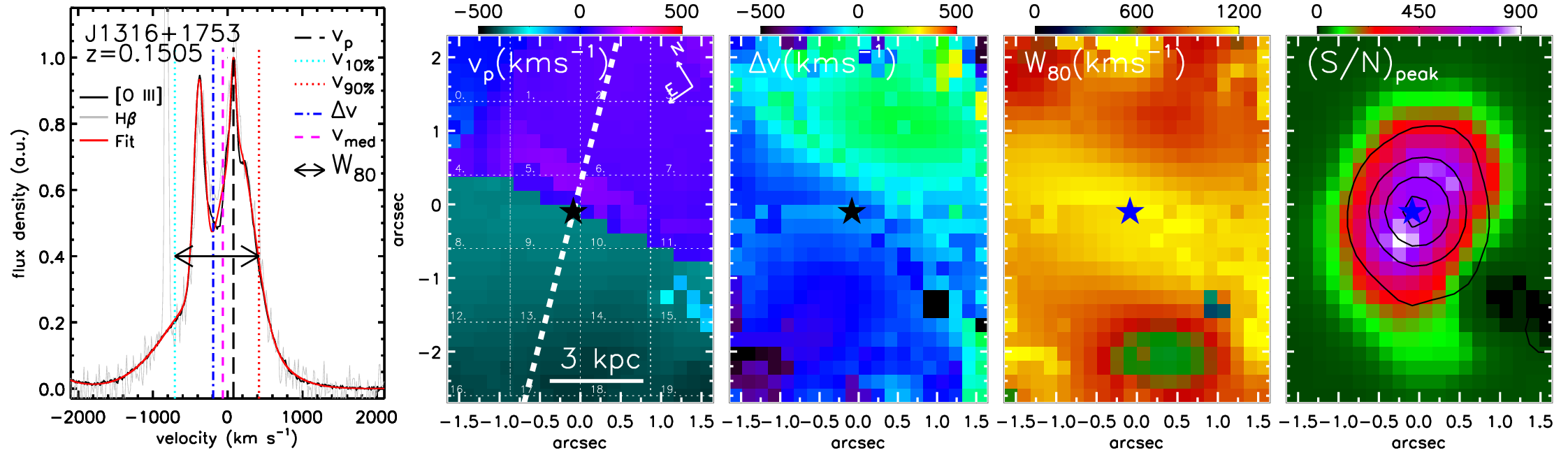,width=6.5in,angle=90}}
\centerline{\psfig{figure=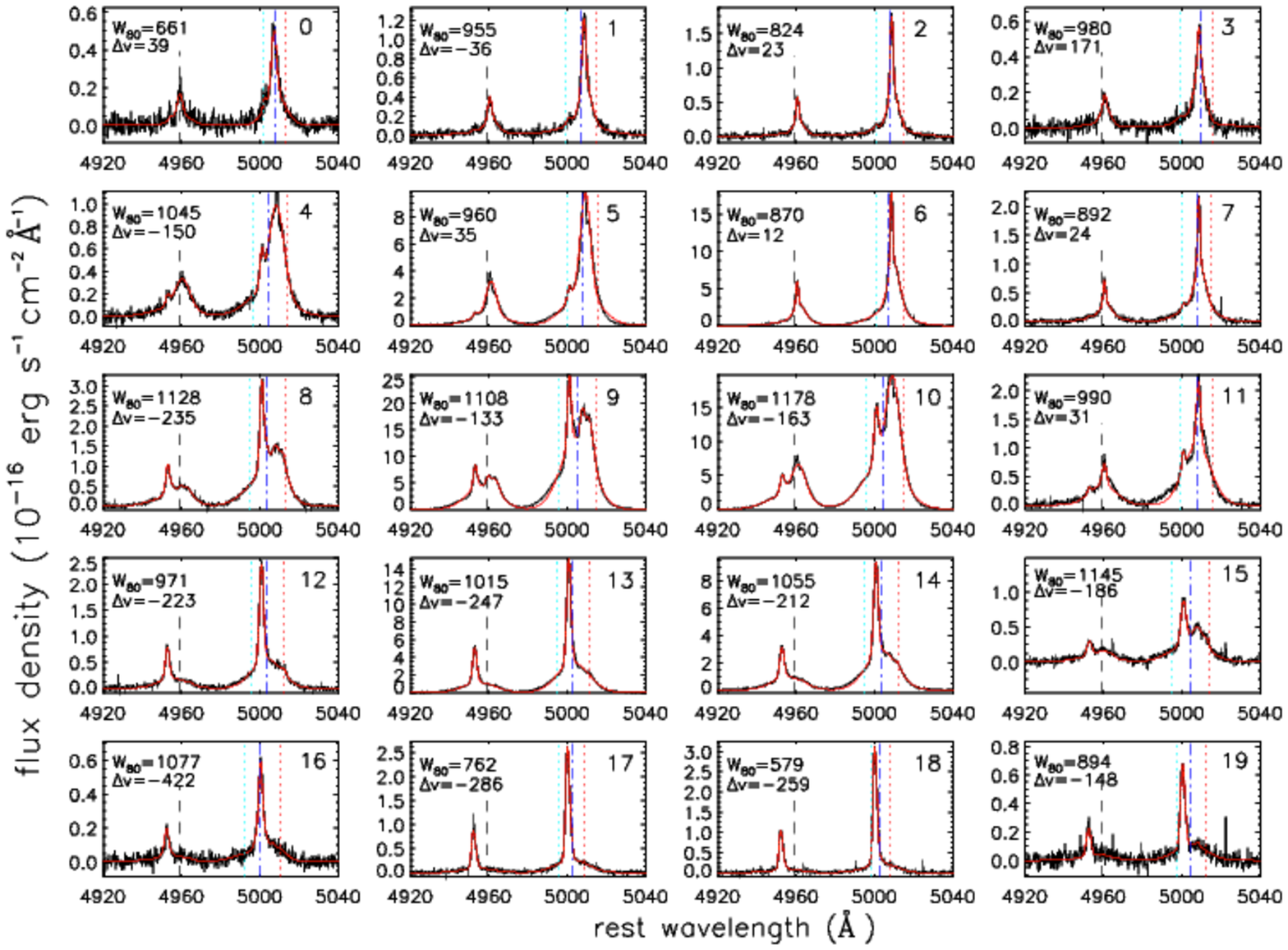,width=6.5in,angle=0}}
\vspace{-0.6cm}
\centerline{\psfig{figure=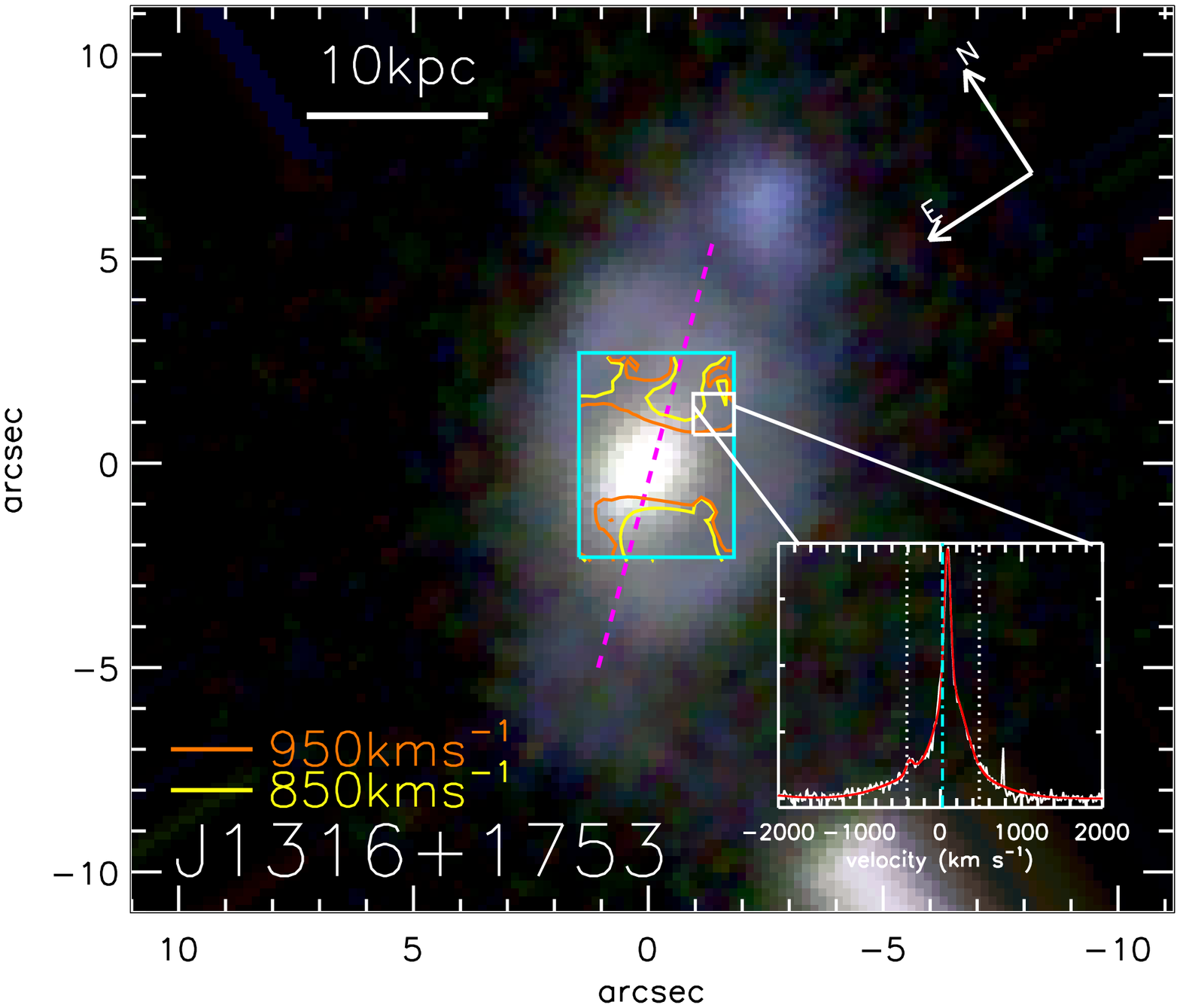,width=2.8in,angle=0}\psfig{figure=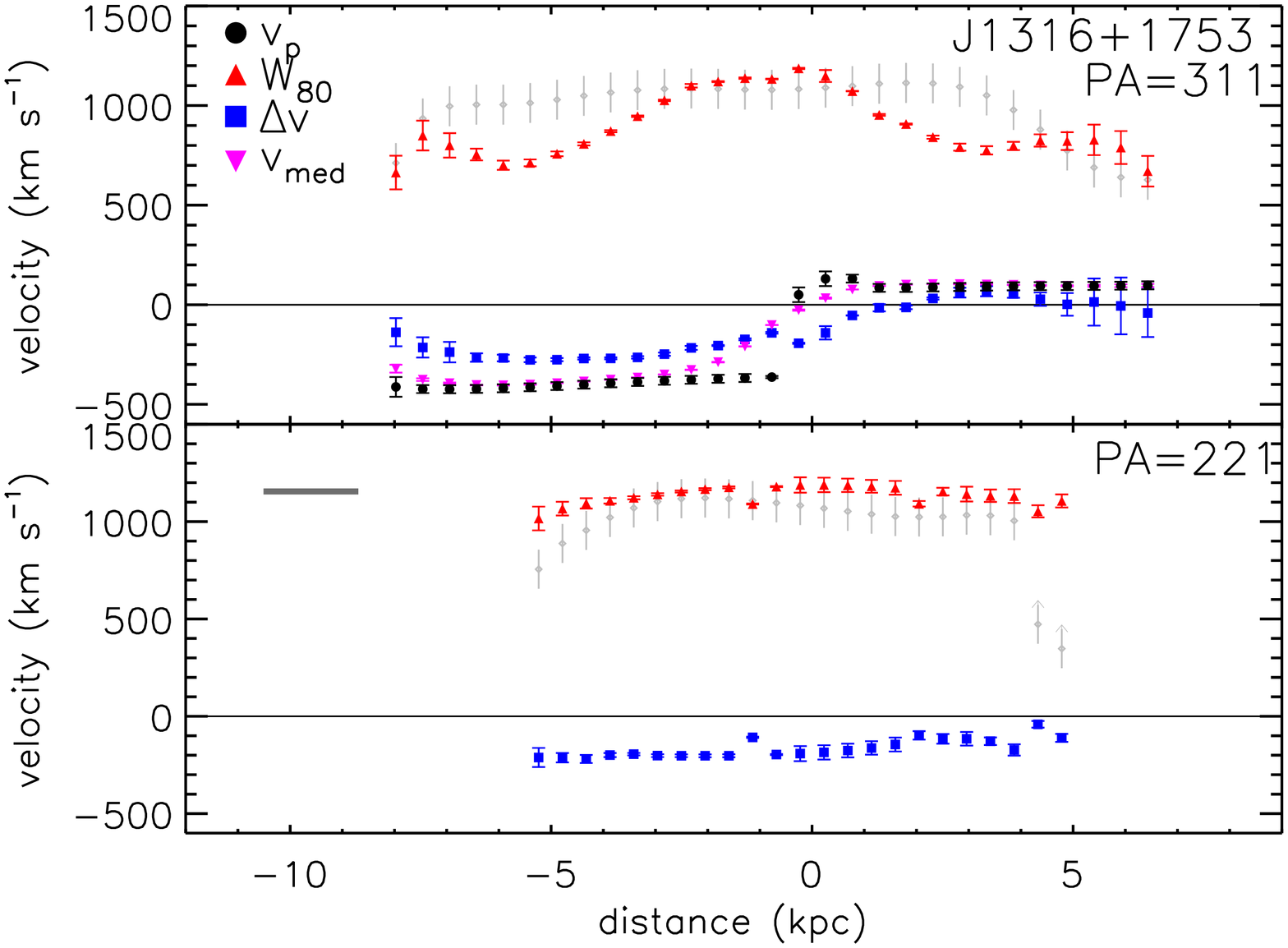,width=3.5in,angle=90}}
\caption{Same as Figure~\ref{Fig:velmaps} and Figure~\ref{Fig:sdss} but
  for SDSS\,J1316+1753}
\label{fig:1316+1753}
\end{figure*}

The target J1316+1753 has a double peaked [O~{\sc iii}] emission-line profile,
with components separated by $\approx$\,400--500\,km\,s$^{-1}$
(Fig.~\ref{fig:1316+1753}). As a result of this, this source has
previously been identified as a dual AGN candidate
(\citealt{Xu09}; also see Section~\ref{Sec:GalaxyKinematics}). In the
peak signal-to-noise ratio map we
observe two spatial regions separated by $\approx2$--3\,kpc and
orientated with a PA\,$\approx130^{\circ}$ associated with these kinematics
components; however, at the resolution of our
observations ($\approx$\,2\,kpc) we do not observe two continuum peaks
(Fig.~\ref{fig:1316+1753}). The similarity of the emission-line ratios
in both systems, and the lack of a double continuum source, suggests that these kinematic components may be
illuminated by a single ionising source. Also, the kinematic components have
similar fluxes and the velocities are independent of projected
distance from the centre; therefore, the double-peaked profile may by the result of gas
kinematics, jet interactions and/or AGN-driven outflows
(Fig.~\ref{fig:1316+1753}; also see \citealt{Xu09}; \citealt{Smith10};
\citealt{Barrows13}). The lack of evidence for
extended luminous radio emission on $\gtrsim2^{\prime\prime}$ scales from the
FIRST data (Table~\ref{Tab:observations}) could argue against a jet
interaction (as seen in other sources e.g., \citealt{Rosario10}); however, deep and high-resolution radio data
would be required to confirm this. Additionally the axis of the two components
 is aligned with the morphological major axis seen in the SDSS image
 and the continuum emission seen in our
data (Fig.~\ref{fig:1316+1753}). Overall, all of these observations
favour the scenarios of a rotating gas disk or two
merging components (separated by $\lesssim2$\,kpc) illuminated by a
single AGN to explain the double peaked [O~{\sc iii}] emission-line
profile in this source (also see \citealt{Smith12} and \citealt{Comerford12}). Combining these IFU
data with deep high-spatial resolution optical, X-ray and radio images would provide the data required to
robustly distinguish between all of the possible scenarios.

In addition to the narrow components our IFU data for J1316+1753
(Fig.~\ref{fig:1316+1753}) reveal a very
broad [O~{\sc iii}] emission-line profile ($W_{80}\approx$\,1100--1200\,km\,s$^{-1}$), preferentially located
perpendicular to the kinematic ``major axis'' defined from the $v_{p}$
map. This [O~{\sc iii}] emission has a velocity offset of $\Delta v\approx-200$\,km\,s$^{-1}$ and is extended right to the edge of the
field-of-view (i.e., over $\gtrsim14$\,kpc). In the velocity profile,
we observe an emission-line
flux ratio $\log($[O~{\sc iii}$]/{\rm H}\beta)\approx1.0$--1.1 over the central
regions and that declines at a break radius of $R_{{\rm br}}=5$\,kpc
(see Section~\ref{Sec:properties}). Therefore we measure a H$\beta$ surface
brightness of $\Sigma_{{\rm H}\beta}=6.2\times10^{-16}$\,erg\,s$^{-1}$\,arcsec$^{-2}$ in an
annulus at this break radius (corrected for cosmological dimming) and
measure the outflow kinetic energy following Liu
et~al. (2013b)\nocite{Liu13b}. Following this method we obtain an
outflow kinetic energy of ${\dot E}\approx5.4\times10^{43}$\,erg\,s$^{-1}$
that is within a factor of 3 of the value we calculated in
Section~\ref{Sec:properties}, i.e., ${\dot E}_{{\rm out}}\approx1.7\times10^{44}$\,erg\,s$^{-1}$.

\subsubsection*{J1338+1503}
\begin{figure*}
\centerline{\psfig{figure=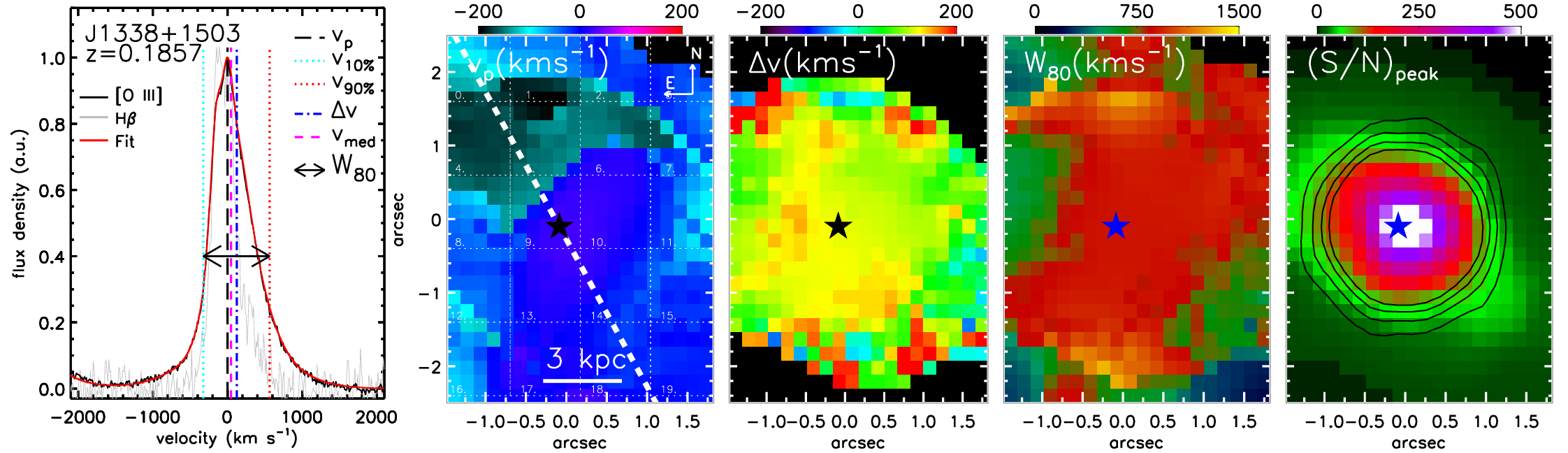,width=6.5in,angle=90}}
\centerline{\psfig{figure=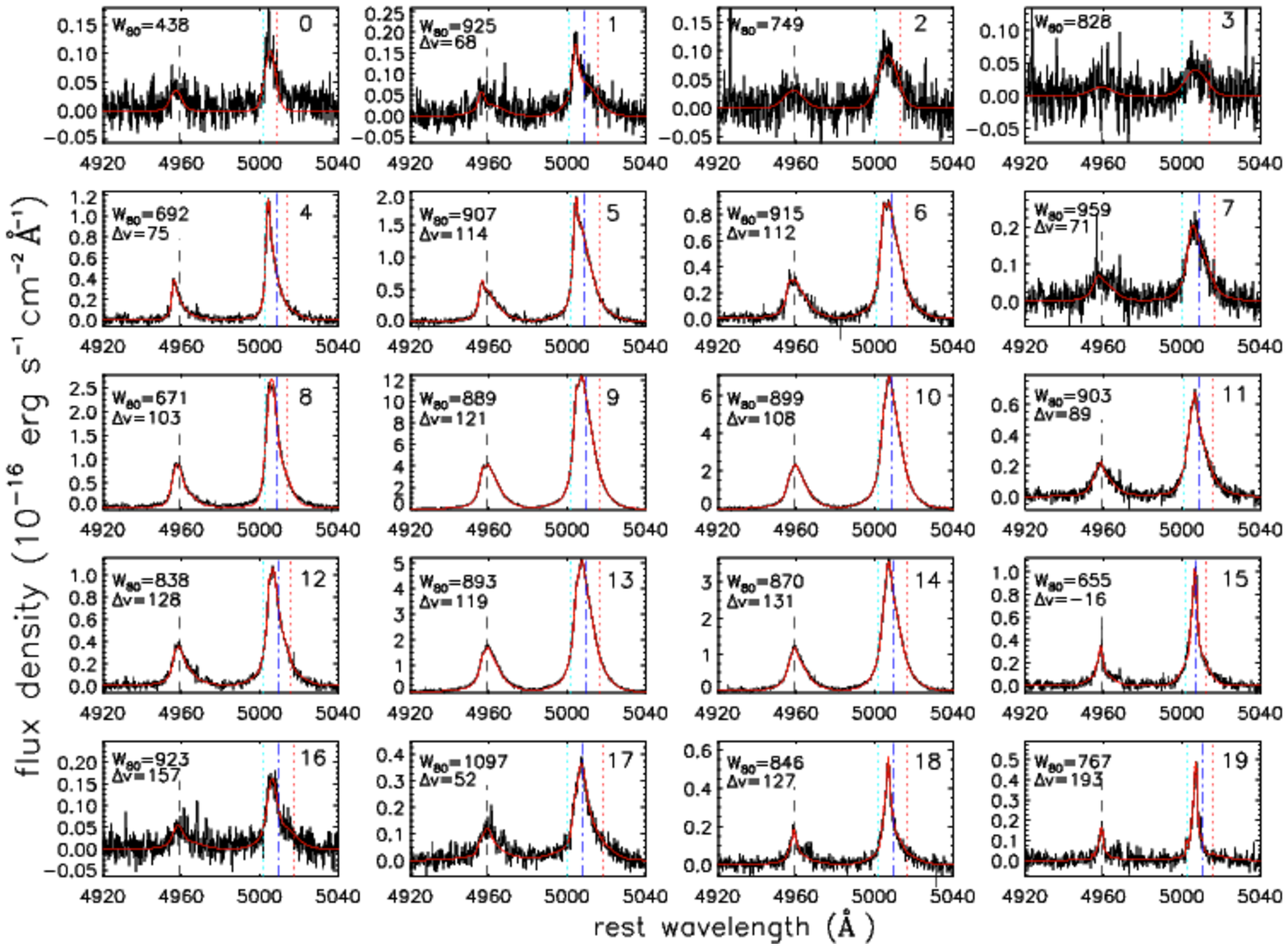,width=6.5in,angle=0}}
\vspace{-0.6cm}
\centerline{\psfig{figure=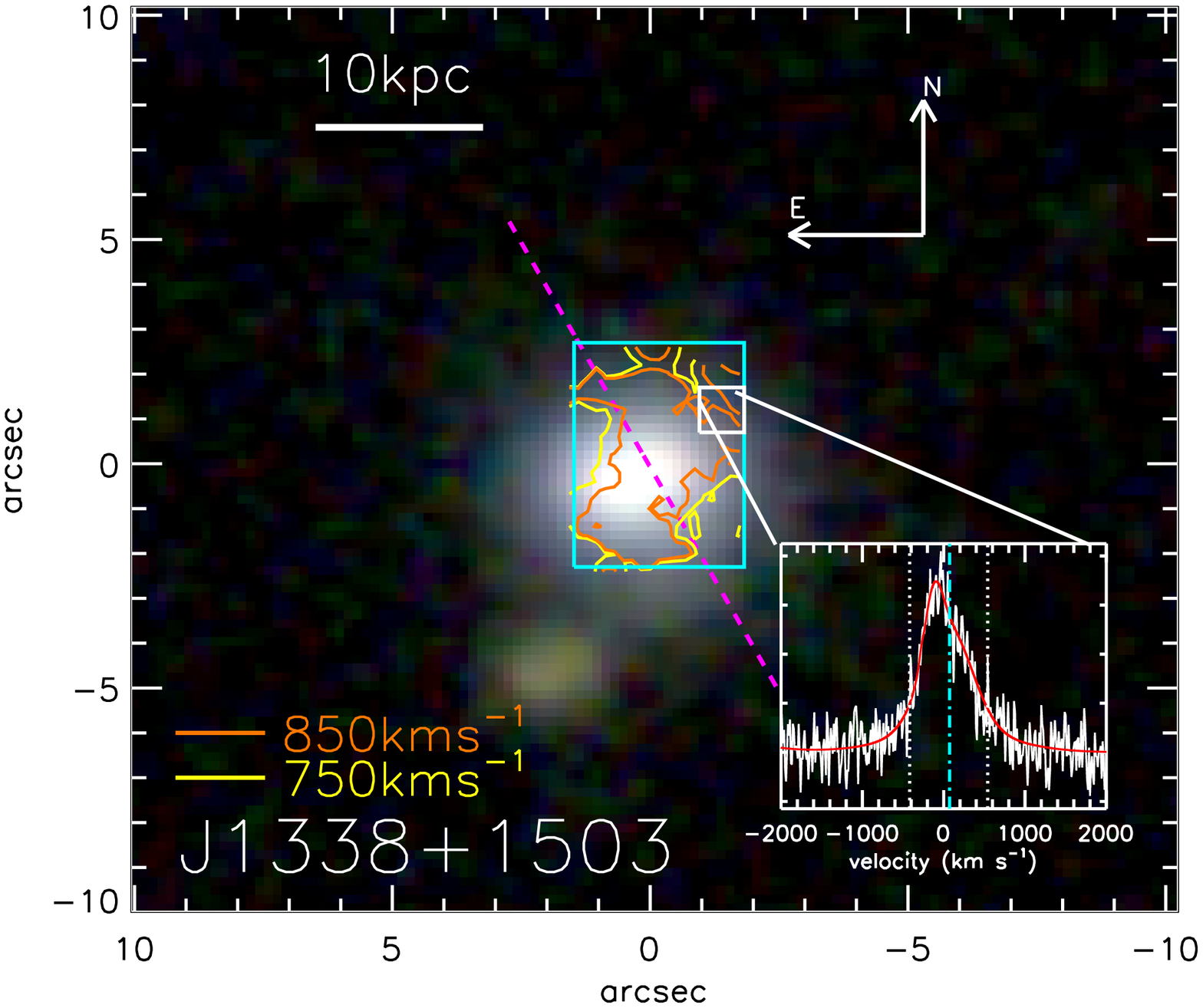,width=2.8in,angle=0}\psfig{figure=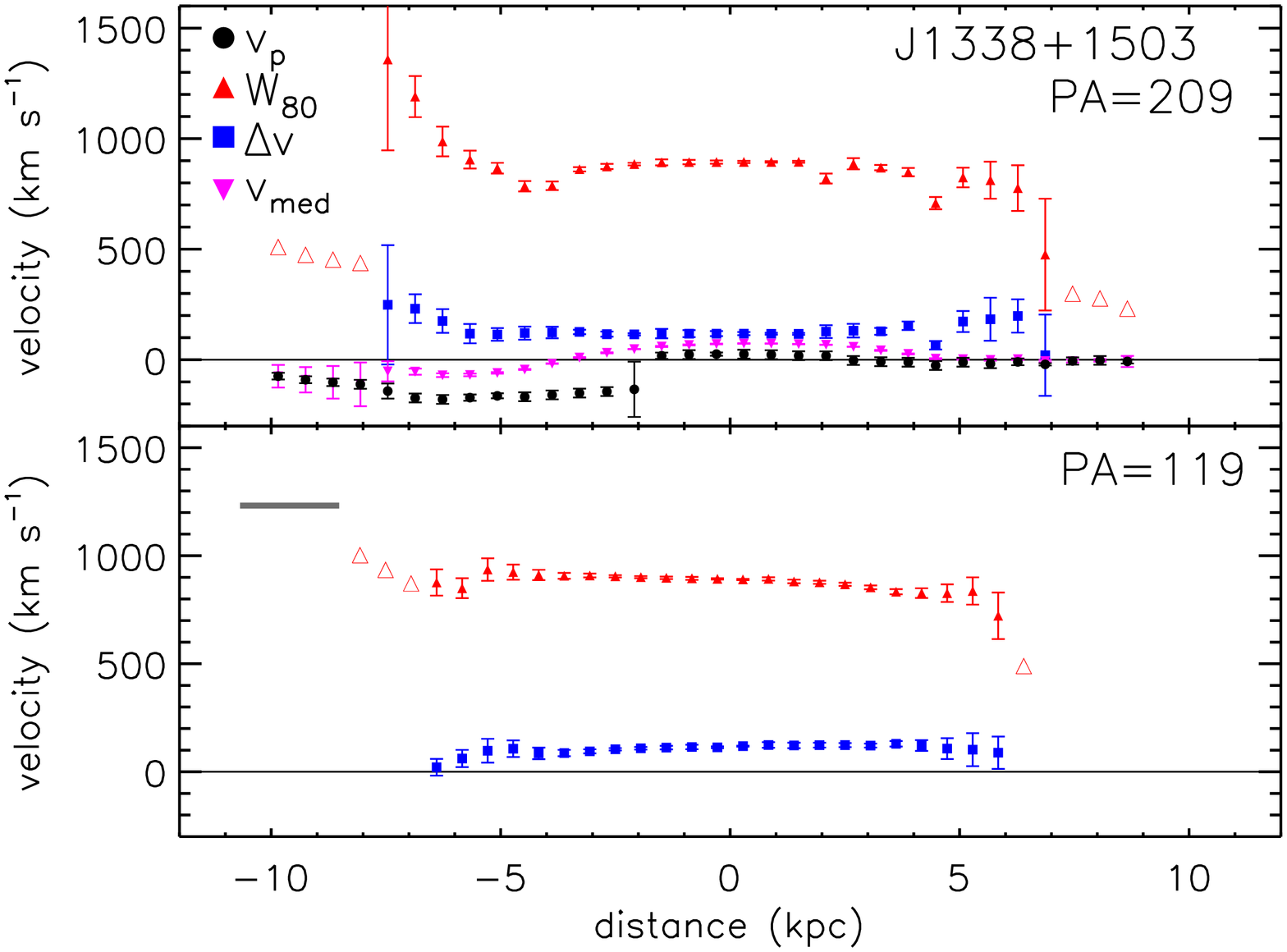,width=3.5in,angle=90}}
\caption{Same as Figure~\ref{Fig:velmaps} and Figure~\ref{Fig:sdss} but
  for SDSS\,J1338+1503}
\label{fig:1338+1503}
\end{figure*}

We show our IFU data for J1338+1503 in
Figure~\ref{fig:1338+1503}. We note that hints of stellar continua are
seen in the SDSS spectrum and contamination from stellar
absorption may cause the difference in the H$\beta$ and [O~{\sc iii}]
emission line profiles (Fig.~\ref{fig:1338+1503}). The peak velocity map ($v_{p}$) for J1338+1503 is irregular; however,
we do see a general shift from blue to red. This source has a
predominantly red wing, in addition to a weaker blue wing leading to an overall
velocity offset of $\Delta v\approx120$\,km\,s$^{-1}$ across the central few
kiloparsecs. The broad emission ($W_{80}\approx900$\,km\,s$^{-1}$) is preferentially found perpendicular to
the velocity gradient (i.e., the kinematic ``major axis'') that we
observe in $v_{p}$ (Fig.~\ref{fig:1338+1503}). 
 
\subsubsection*{J1339+1425}
\begin{figure*}
\centerline{\psfig{figure=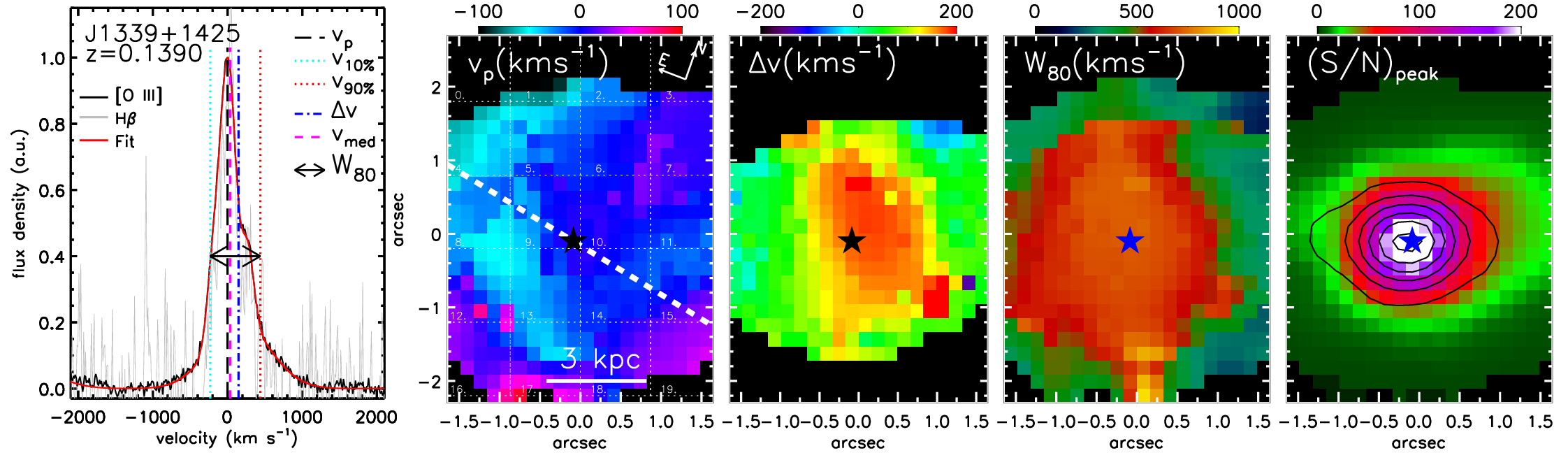,width=6.5in,angle=90}}
\centerline{\psfig{figure=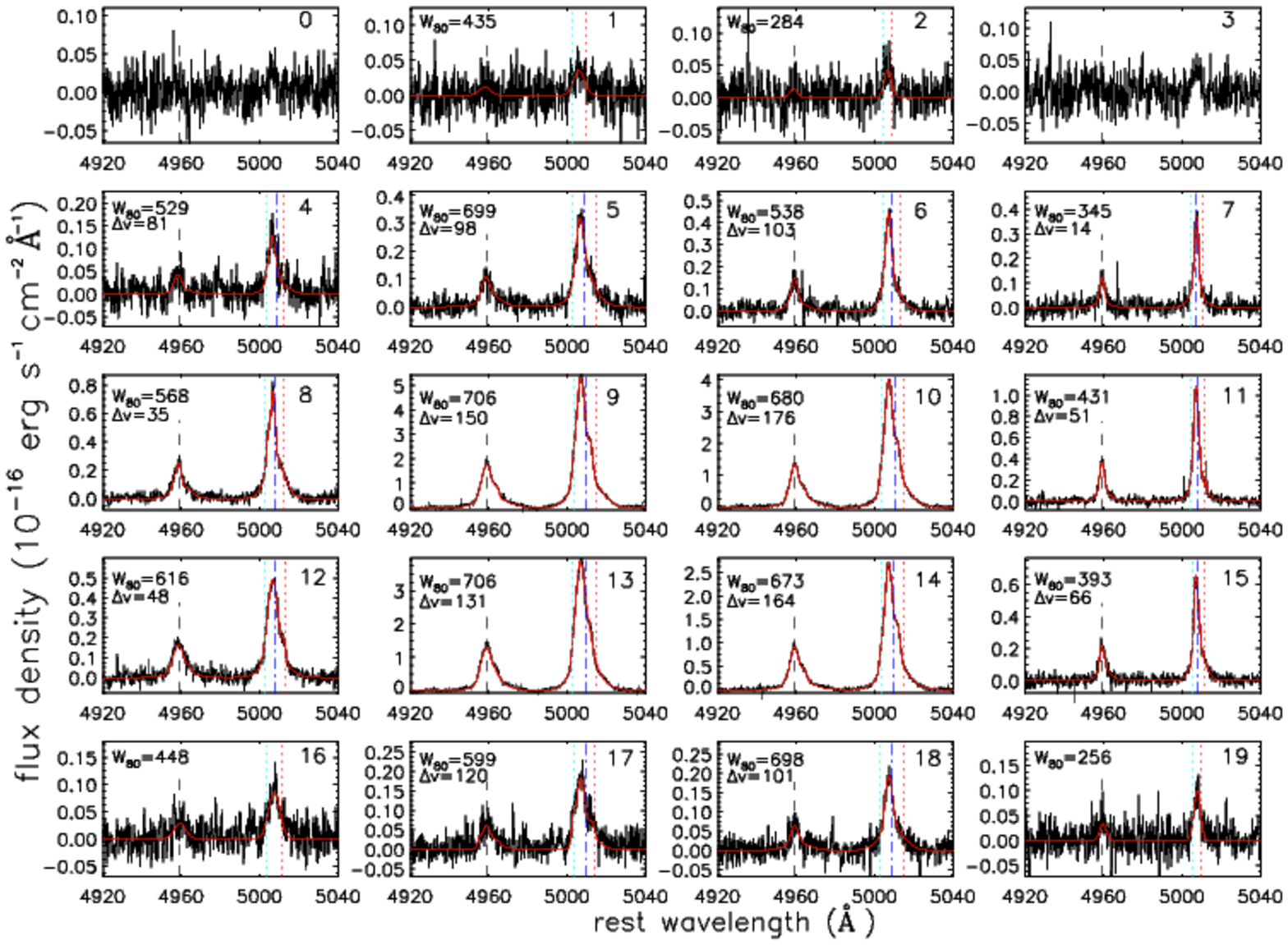,width=6.5in,angle=0}}
\vspace{-0.6cm}
\centerline{\psfig{figure=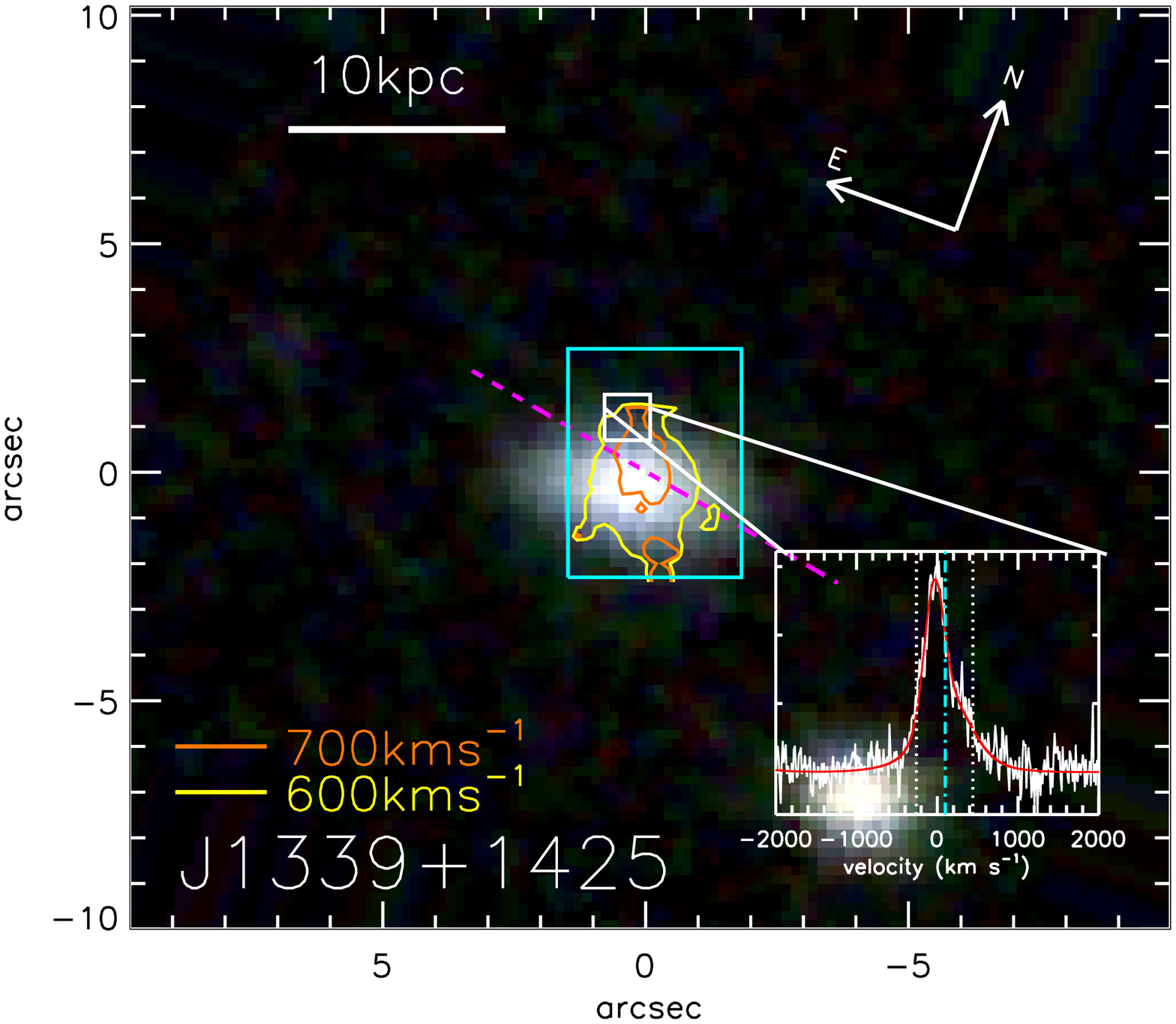,width=2.8in,angle=0}\psfig{figure=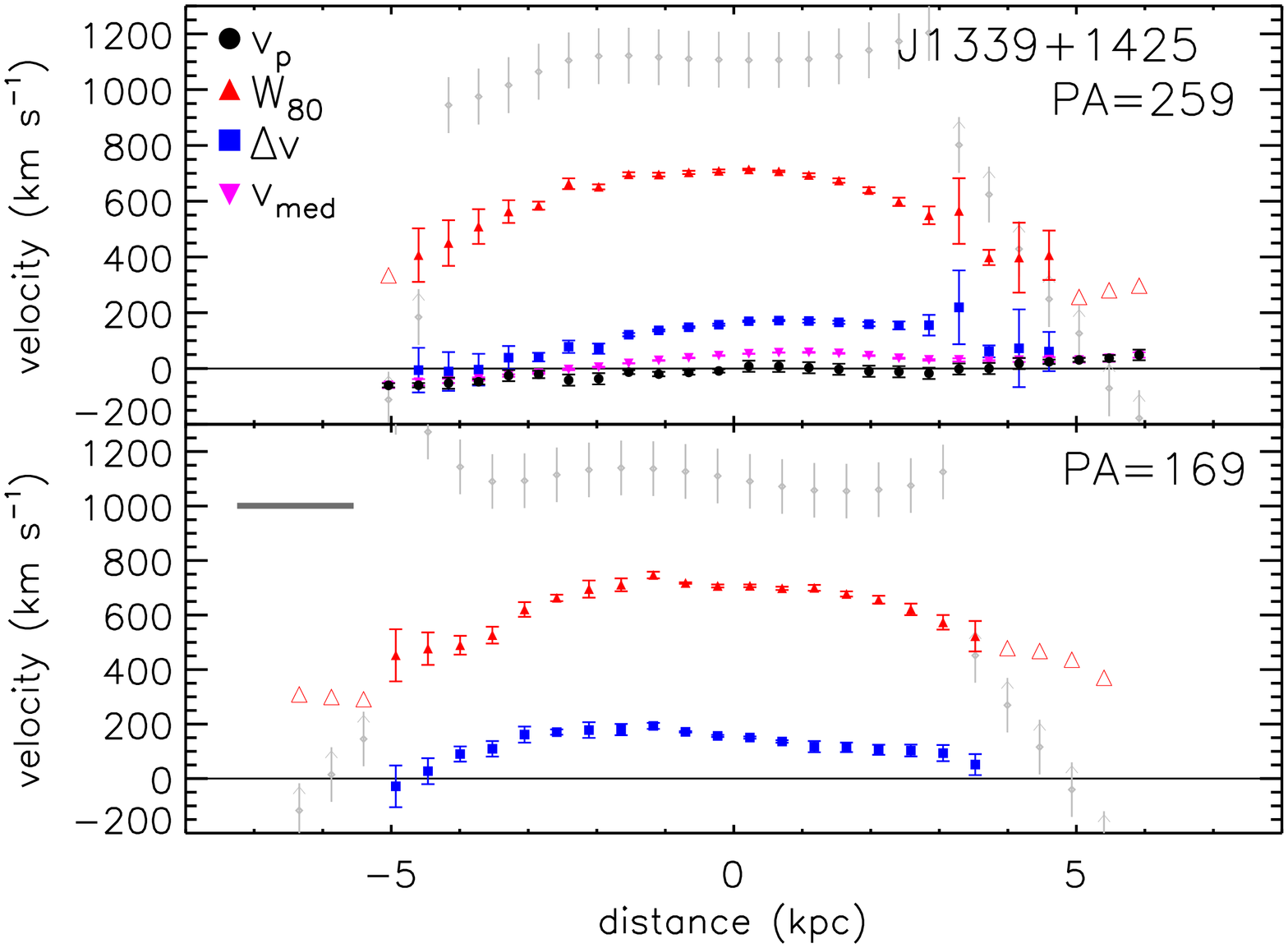,width=3.5in,angle=90}}
\caption{Same as Figure~\ref{Fig:velmaps} and Figure~\ref{Fig:sdss} but
  for SDSS\,J1339+1425}
\label{fig:1339+1425}
\end{figure*}

The source J1339+1425 is the least luminous AGN in our sample (in both
$L_{{\rm AGN}}$ and $L_{{\rm [O~III]}}$) and is the only source
not detected at 1.4\,GHz by FIRST or NVSS
(Table~\ref{Tab:observations}). We show our IFU data for this source
in Figure~\ref{fig:1339+1425} where we observed [O~{\sc iii}] emission right
to the edge of the field-of-view of our observations (i.e., over
$\gtrsim$12\,kpc). This emission-line region is
dominated by narrow kinematic components
(i.e., with $W_{80}<600$\,km\,s$^{-1}$) that trace out a small velocity gradient ($\Delta v_{p}\approx100$\,km\,s$^{-1}$)
from blue to red. We find that
the kinematic major axis is broadly consistent with the continuum
morphological axis observed in our data and the SDSS image
(Fig.~\ref{fig:1339+1425}). We therefore appear to be tracing galactic
rotation in the $v_{p}$ map. 

Across the central $\approx$\,6\,kpc we
observe a blue- and red-wing in the [O~{\sc iii}] emission-line
profile, with an overall velocity offset of
$\Delta v\approx150$\,km\,s$^{-1}$, that is preferentially located at
$\approx50^{\circ}$--70$^{\circ}$ away from this kinematic ``major axis''. In the velocity profile, we observe that the emission-line
flux ratio $\log($[O~{\sc iii}]/H$\beta)\approx1.1$ is constant over the
central regions and then declines at a break radius of $R_{{\rm
    br}}=2$\,kpc (see Section~\ref{Sec:properties}). We therefore measure a H$\beta$ surface
brightness of $\Sigma_{{\rm H}\beta}=7.5\times10^{-16}$\,erg\,s$^{-1}$\,arcsec$^{-2}$ in an
annulus at this break radius (corrected for cosmological dimming) and
calculate the outflow kinetic energy following Liu
et~al. (2013b)\nocite{Liu13b}. Using this method we obtain an outflow
kinetic energy of ${\dot E}\approx5.5\times10^{42}$\,erg\,s$^{-1}$
that is within a factor of $\approx3$ of the value we calculated in
Section~\ref{Sec:properties}, i.e., ${\dot E}_{{\rm out}}\approx1.5\times10^{43}$\,erg\,s$^{-1}$.

\subsubsection*{J1355+1300}
\begin{figure*}
\centerline{\psfig{figure=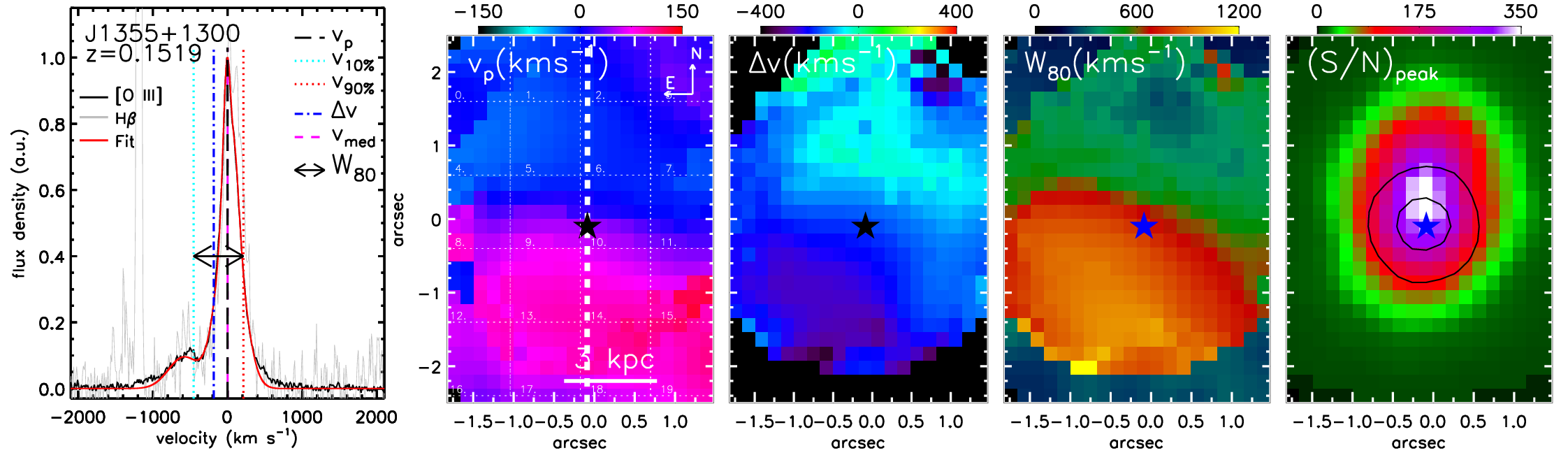,width=6.5in,angle=90}}
\centerline{\psfig{figure=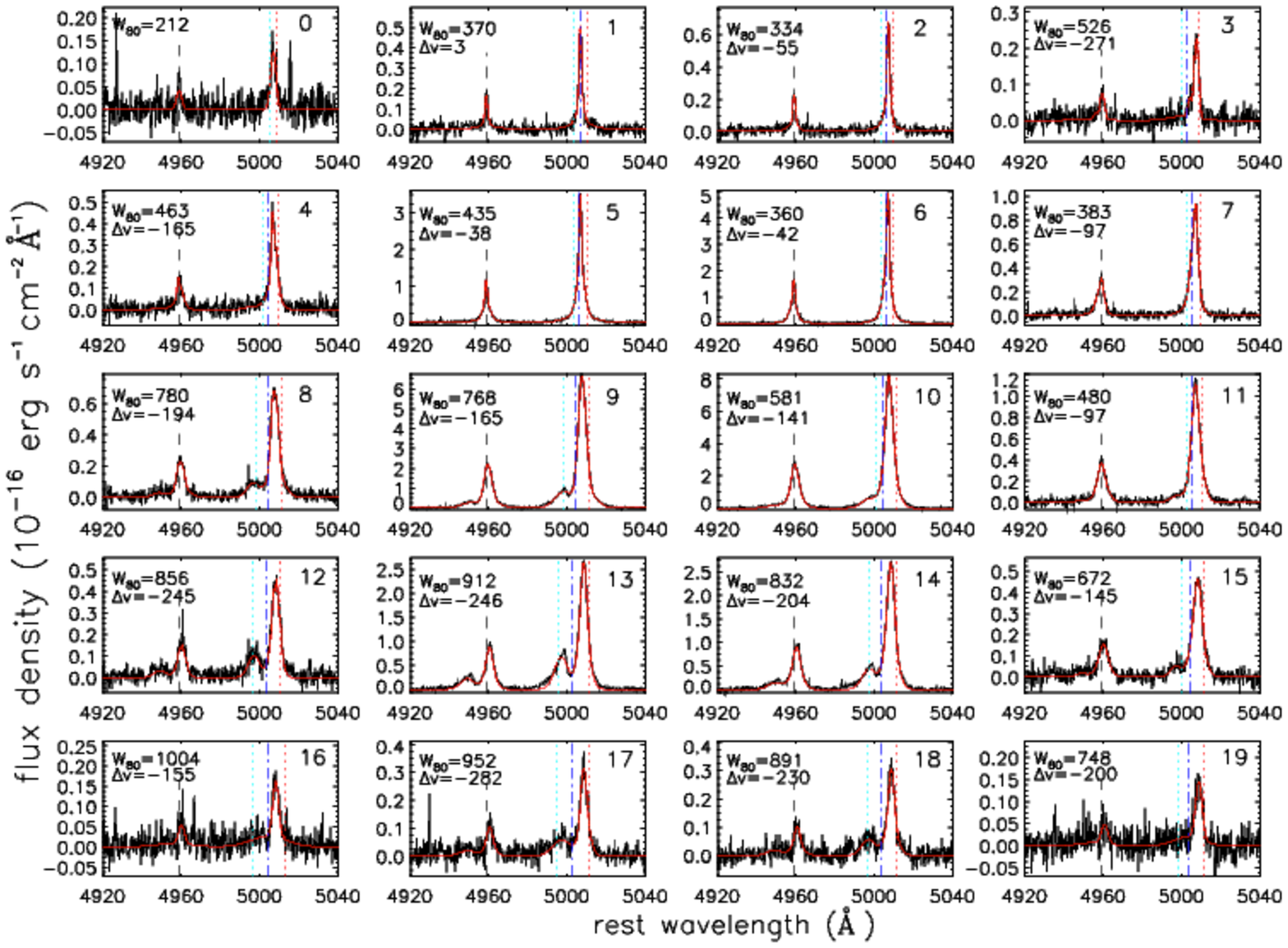,width=6.5in,angle=0}}
\vspace{-0.6cm}
\centerline{\psfig{figure=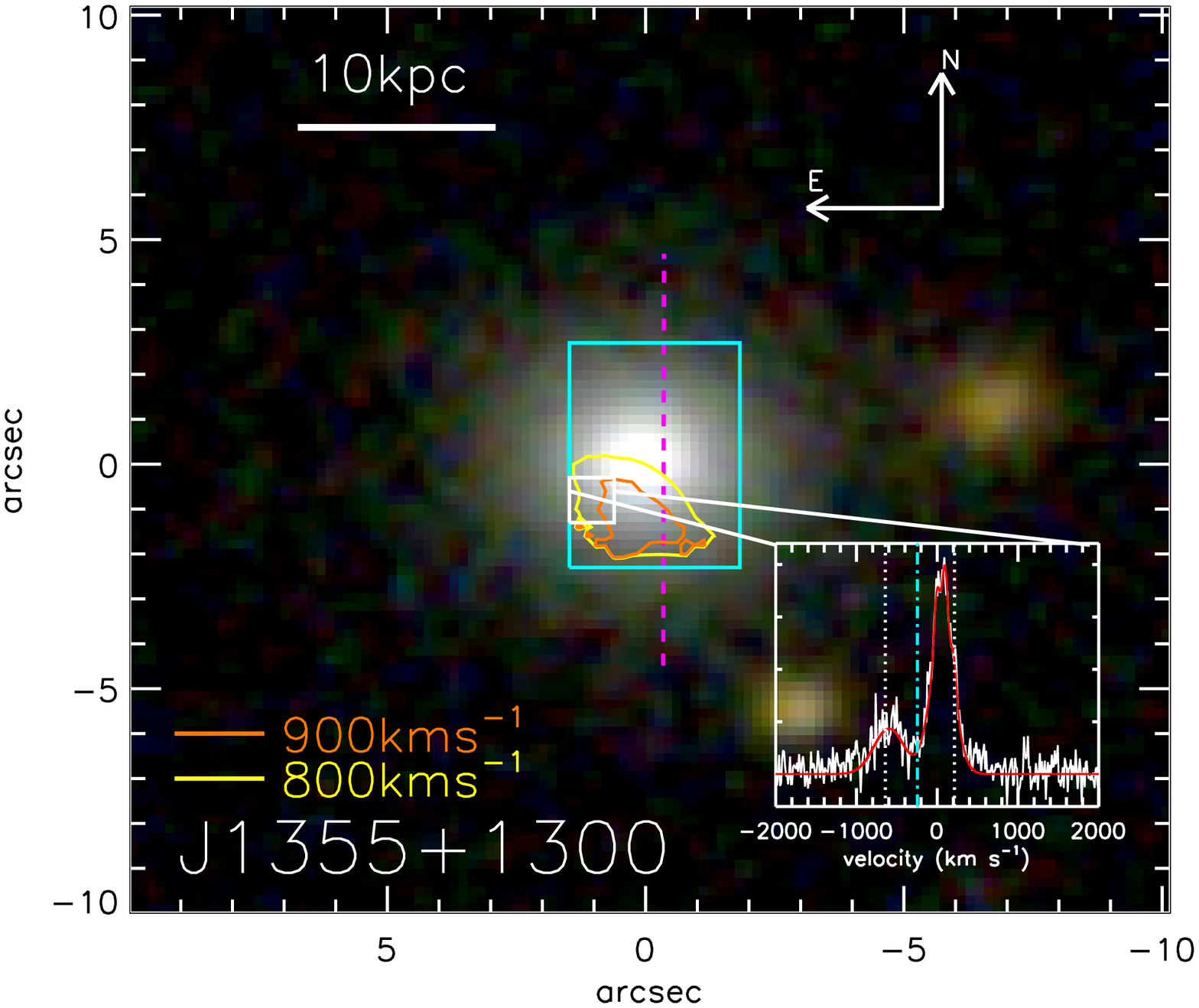,width=2.8in,angle=0}\psfig{figure=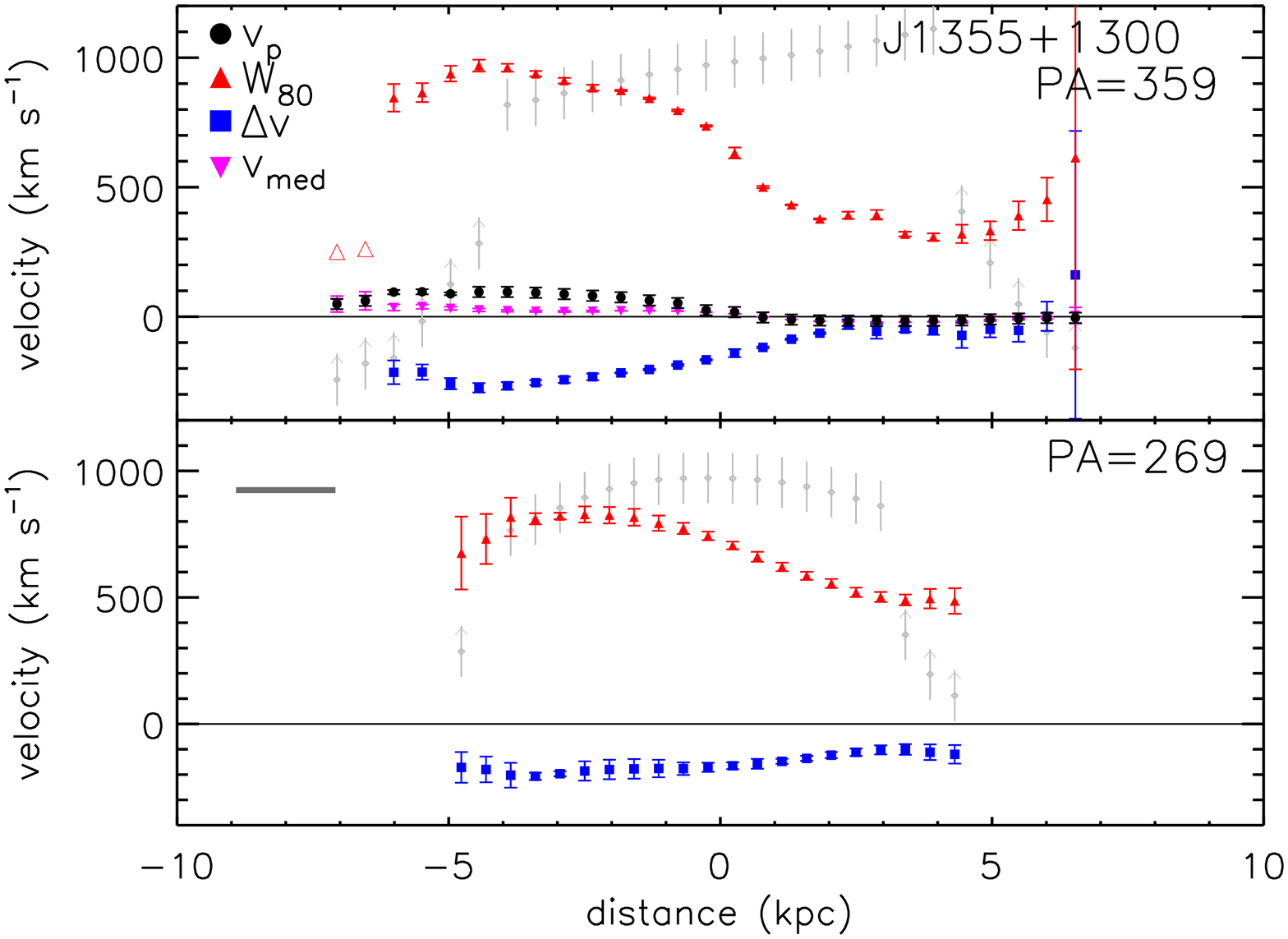,width=3.5in,angle=90}}
\caption{Same as Figure~\ref{Fig:velmaps} and Figure~\ref{Fig:sdss} but
  for SDSS\,J1355+1300}
\label{fig:1355+1300}
\end{figure*}

We show our IFU data for J1355+1300 in Figure~\ref{fig:1355+1300}. The majority of the emission-line
region is dominated by narrow [O~{\sc iii}] emission-line profiles
($W_{80}\approx$\,300--400\,km\,s$^{-1}$) and we observe a small
velocity gradient (with $\Delta v_{p}\approx100$\,km\,s$^{-1}$) from
blue-to-red in a north-south direction in the peak velocity map
($v_{p}$) indicative of galactic rotation. Located $\approx3$\,kpc to the
south-east of the nucleus there is a distinct kinematic component with a velocity between $\approx-500$\,km\,s$^{-1}$ and
$-600$\,km\,s$^{-1}$ which has a high-velocity blue-wing in the emission-line profile
(see grids of spectra in Fig.~\ref{fig:1355+1300}). The similarity of the galaxy-integrated [O~{\sc
  iii}] and H$\beta$ emission-line ratios for this kinematic
component (see Fig.~\ref{fig:1355+1300}) could imply that the ionising
source is the same as that for the rest of the emission-line region
and is therefore likely to be illuminated by the central source. At
the depth of our observations, we observe no spatially distinct continuum source or
emission-line region coincident with this kinematic component (Fig.~\ref{fig:1355+1300}). The
exact origin of this high-velocity feature is difficult to determine
with these observations alone; however, the highly-disturbed nature
and lack of associated continuum strongly suggests an outflow. This
features is similar to that seen in J1430+1339 and therefore could indicate a
smaller-scale outflowing bubble and/or could be due to a jet-ISM interaction at this position
(as is seen in other sources, e.g., \citealt{Emonts05}). Due to the faintness of the radio emission,
and the lack of IRAS detections we are unable to constrain the radio
excess parameter for this source (see Section~\ref{Sec:SEDs} and
Table~\ref{Tab:observations}); therefore, deep and high-resolution radio imaging
is required to determine if radio jets are present and if they are aligned with this high-velocity feature. 

\subsubsection*{J1356+1026}
\begin{figure*}
\centerline{\psfig{figure=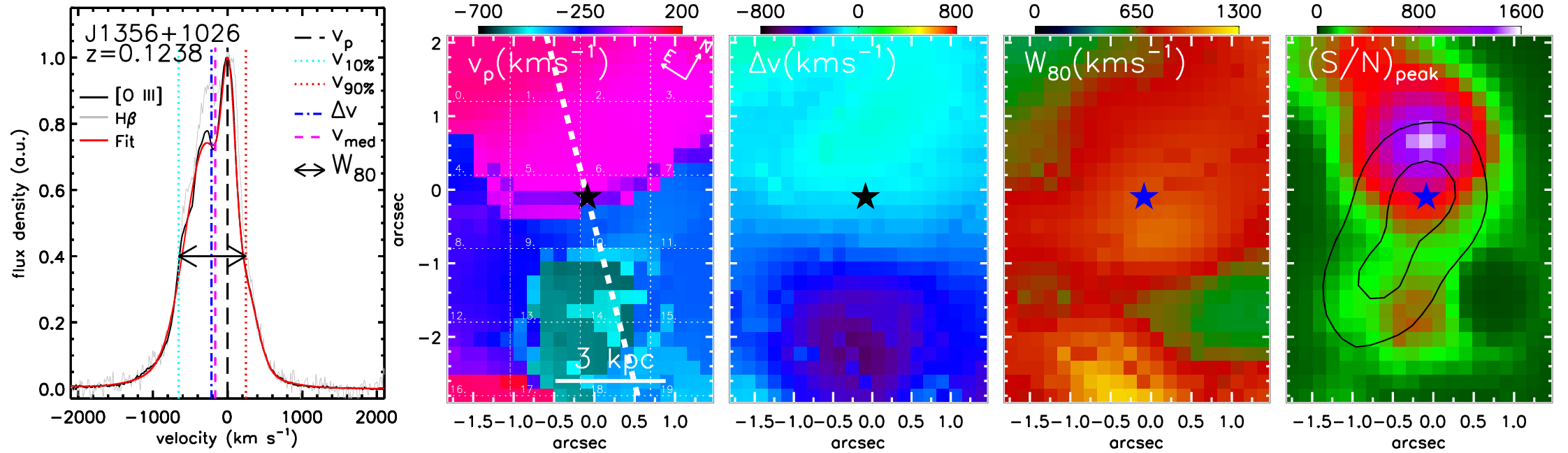,width=6.5in,angle=90}}
\centerline{\psfig{figure=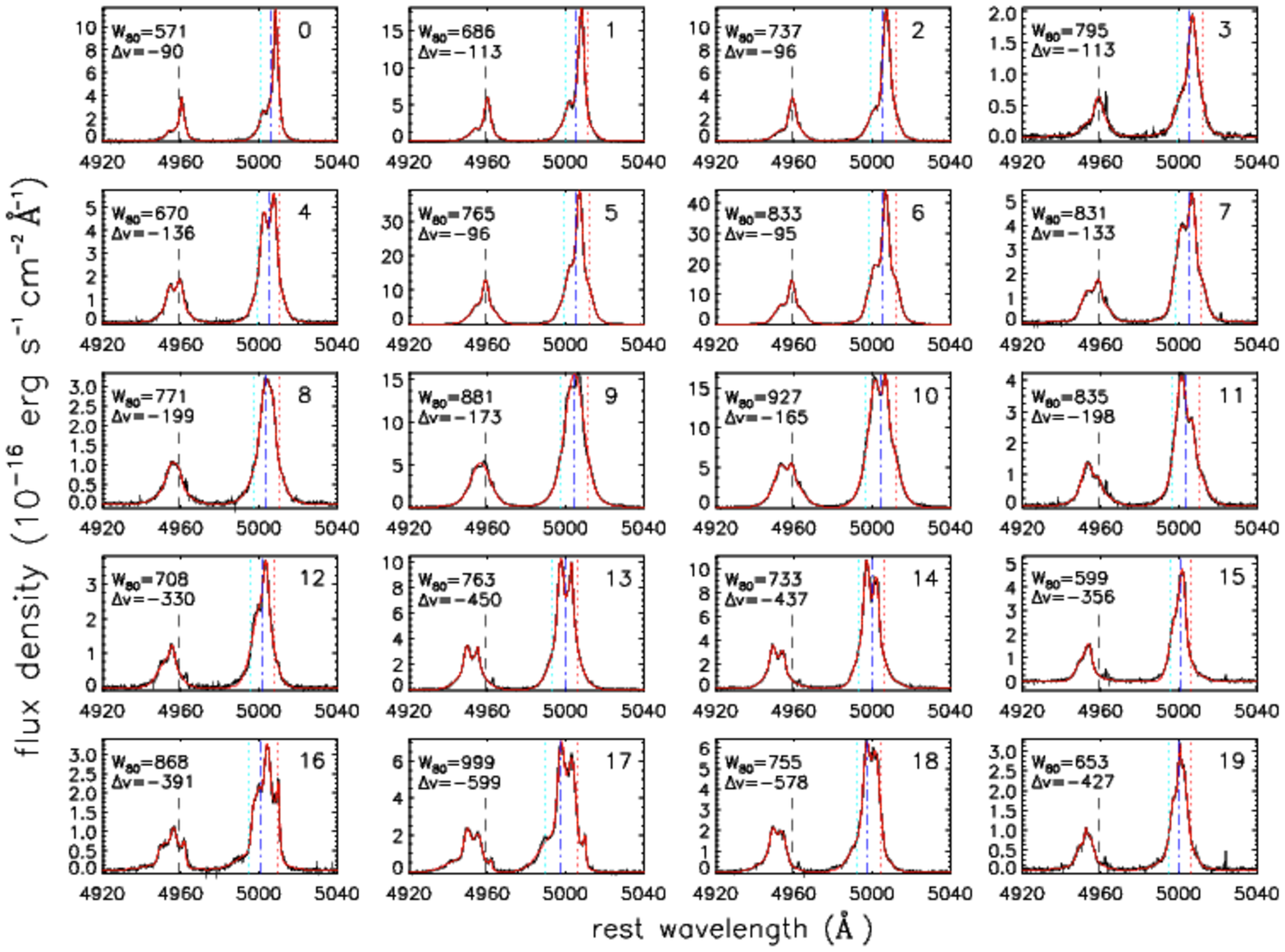,width=6.5in,angle=0}}
\vspace{-0.6cm}
\centerline{\psfig{figure=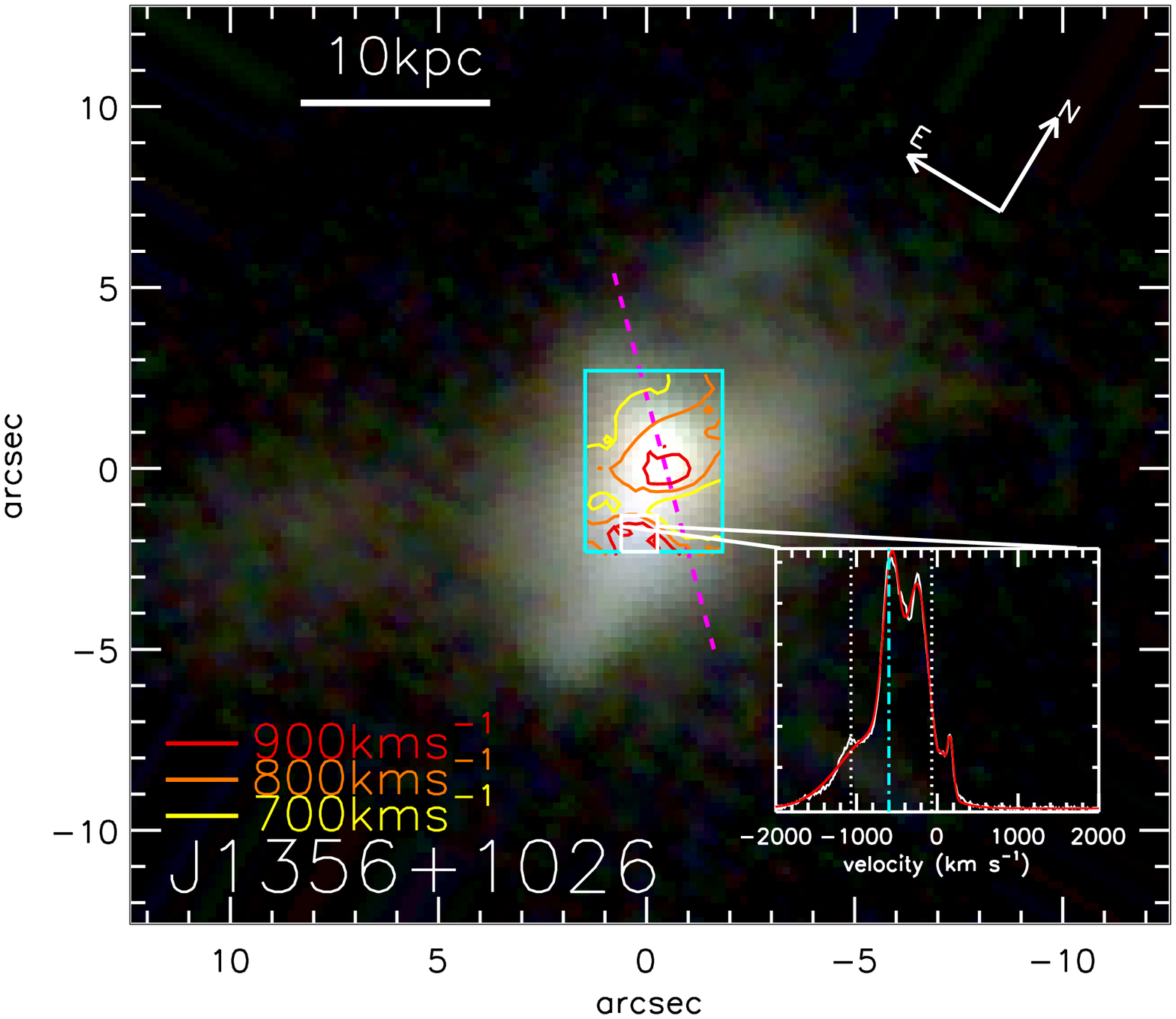,width=2.8in,angle=0}\psfig{figure=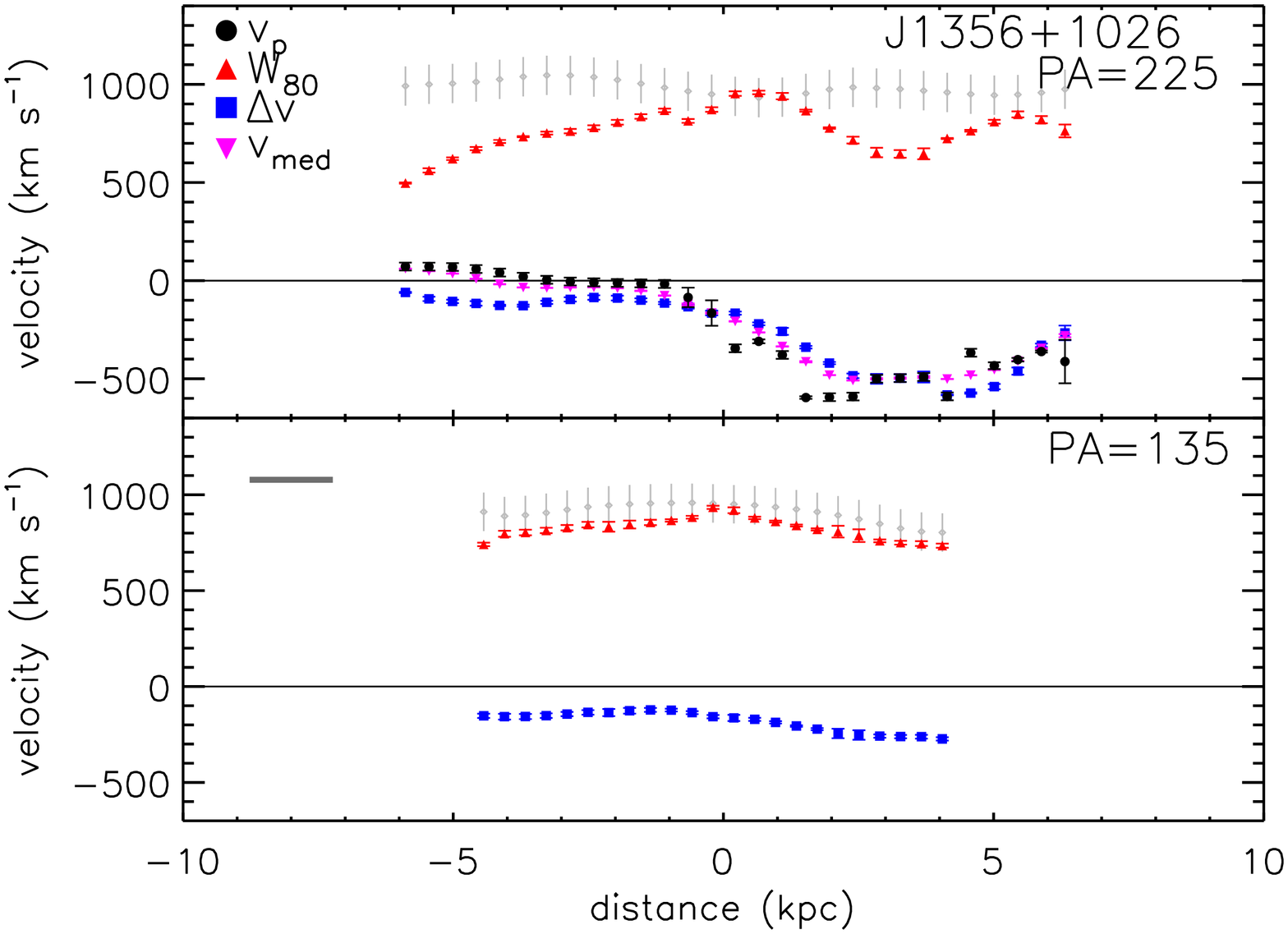,width=3.5in,angle=90}}
\caption{Same as Figure~\ref{Fig:velmaps} and Figure~\ref{Fig:sdss} but
  for SDSS\,J1356+1026}
\label{fig:1356+1026}
\end{figure*}

As a well known type~2 (``obscured'') quasar (\citealt{Reyes08}) this source has
already received a lot of attention in the literature. Two merging
galactic nuclei with a projected separation of $\approx2.5$\,kpc are seen in the optical and near-infrared
(\citealt{Shen11}; \citealt{Fu11}; \citealt{Greene12}). In agreement
with this, our IFU data reveal two continuum sources and, moderately
offset from this, corresponding emission-line regions with distinct
velocities (Fig.~\ref{fig:1356+1026}). We note that \cite{Fu12} did
not confidently associate an emission-line region with the southern
nucleus in their IFU data, although our data are unambiguous (Fig.~\ref{Fig:SNmontage} and Fig.~\ref{fig:1356+1026}). Combined
with the observation that these kinematic
components have slightly different [O~{\sc iii}]/H$\beta$
emission-line flux ratios implies two distinct type~2 quasars
(Fig.~\ref{fig:1356+1026}; also see
\citealt{Greene12}). This source has a clear radio excess above that expected from star
 formation alone (see Section~\ref{Sec:SEDs} and Table~\ref{Tab:observations}). However, there is no
 evidence for {\em extended} radio emission at the few mJy level on scales of
 $\gtrsim2^{\prime\prime}$ based on the FIRST data
 (Section~\ref{Sec:SEDs}; Table~\ref{Tab:observations}). 

Using longslit observations, \cite{Greene12} revealed [O~{\sc iii}] emission extended
over 10s of kiloparsecs, reaching beyond the field-of-view of our
IFU observations. In particular they reveal a ``bubble'' of [O~{\sc
  iii}] emission with a spatial extent of 12\,kpc to the south and
high-velocity ``clumps'' to the north. These ``bubble'' and ``clumps''
reach a projected velocity of $\approx250$\,km\,s$^{-1}$. \cite{Greene12} postulate a quasi-spherical outflow that extends from
the south to the north, where the outflow is forced into this bi-polar
shape due to high density regions of gas and dust in the central
galaxies (e.g., \citealt{FaucherGiguere12b}). Our IFU observations
(Fig.~\ref{fig:1356+1026}) cover the base of these features. It is worth noting that despite
our lack of spatial coverage and different approaches, our outflow energy injection rate estimate
($\dot{E}_{{\rm out}}\approx1.3\times10^{44}$\,erg\,s$^{-1}$; Section~\ref{Sec:properties}) is consistent
with the fiducial range $=10^{44}$--$10^{45}$\,erg\,s$^{-1}$ quoted in
\cite{Greene12}. Based on our SED fitting
(Section~\ref{Sec:SEDs}) the bolometric luminosity of the
AGN in this source is $\approx1\times10^{45}$\,erg\,s$^{-1}$ while the infrared-luminosity from star formation
is $\approx2\times$10$^{45}$\,erg\,s$^{-1}$
(SFR\,$\approx60\Msolyr$; Table~\ref{Tab:observations}). Therefore, in
contrast to the conclusions of \cite{Greene12} we find there
is a comparable amount of energy available from the AGN and star formation
to power the outflow. The observed radio emission in FIRST is on
 much smaller scales than the $\approx10$\,kpc outflowing bubble that
 may argue against a jet-driven outflows; however, deep
 and high-resolution radio imaging is required to determine the origin and morphology of the radio emission in this source. 

\subsubsection*{J1430+1339}
\begin{figure*}
\centerline{\psfig{figure=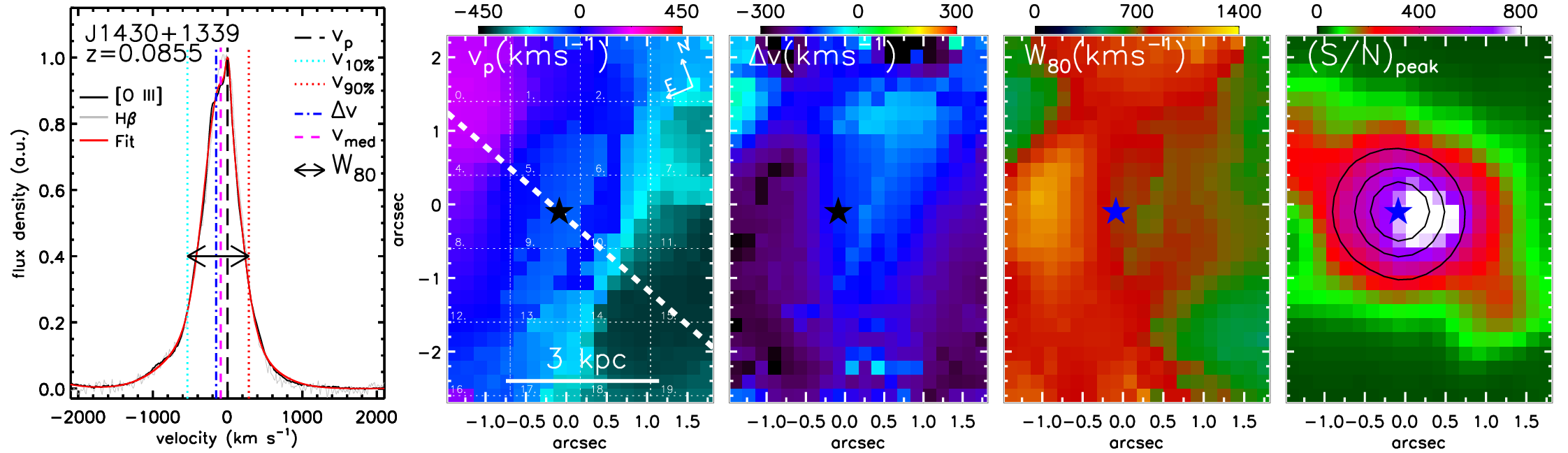,width=6.5in,angle=90}}
\centerline{\psfig{figure=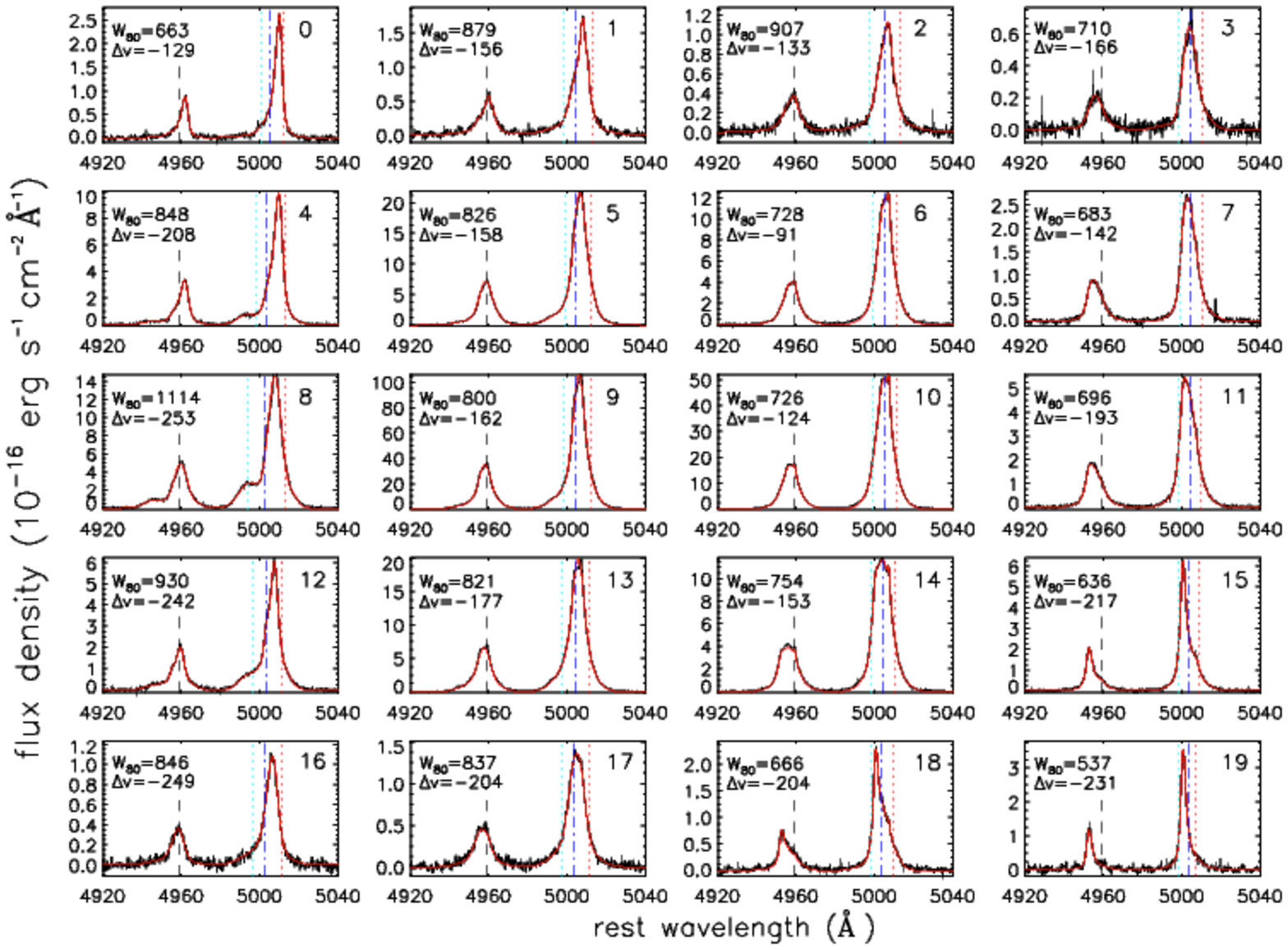,width=6.5in,angle=0}}
\vspace{-0.6cm}
\centerline{\psfig{figure=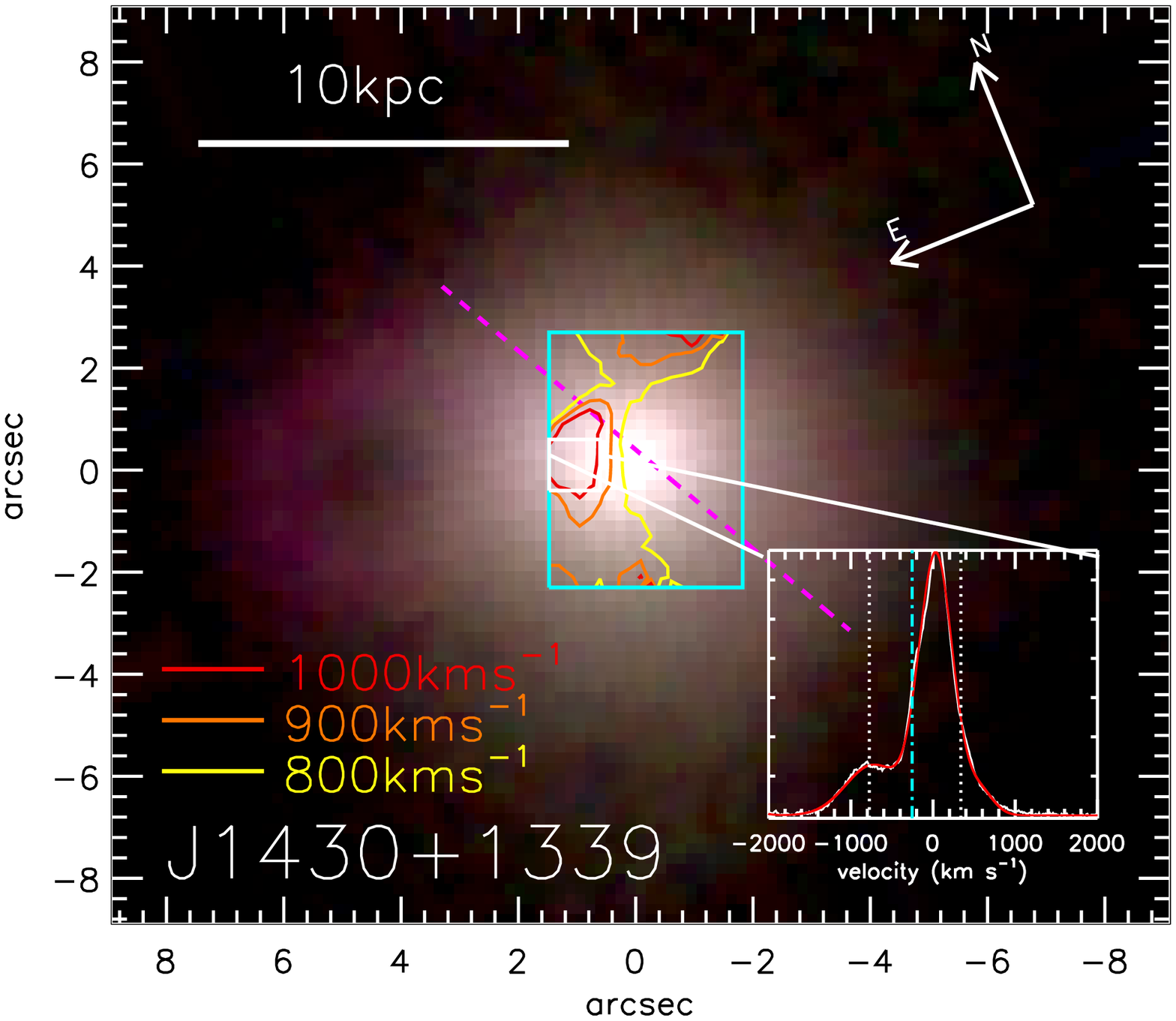,width=2.8in,angle=0}\psfig{figure=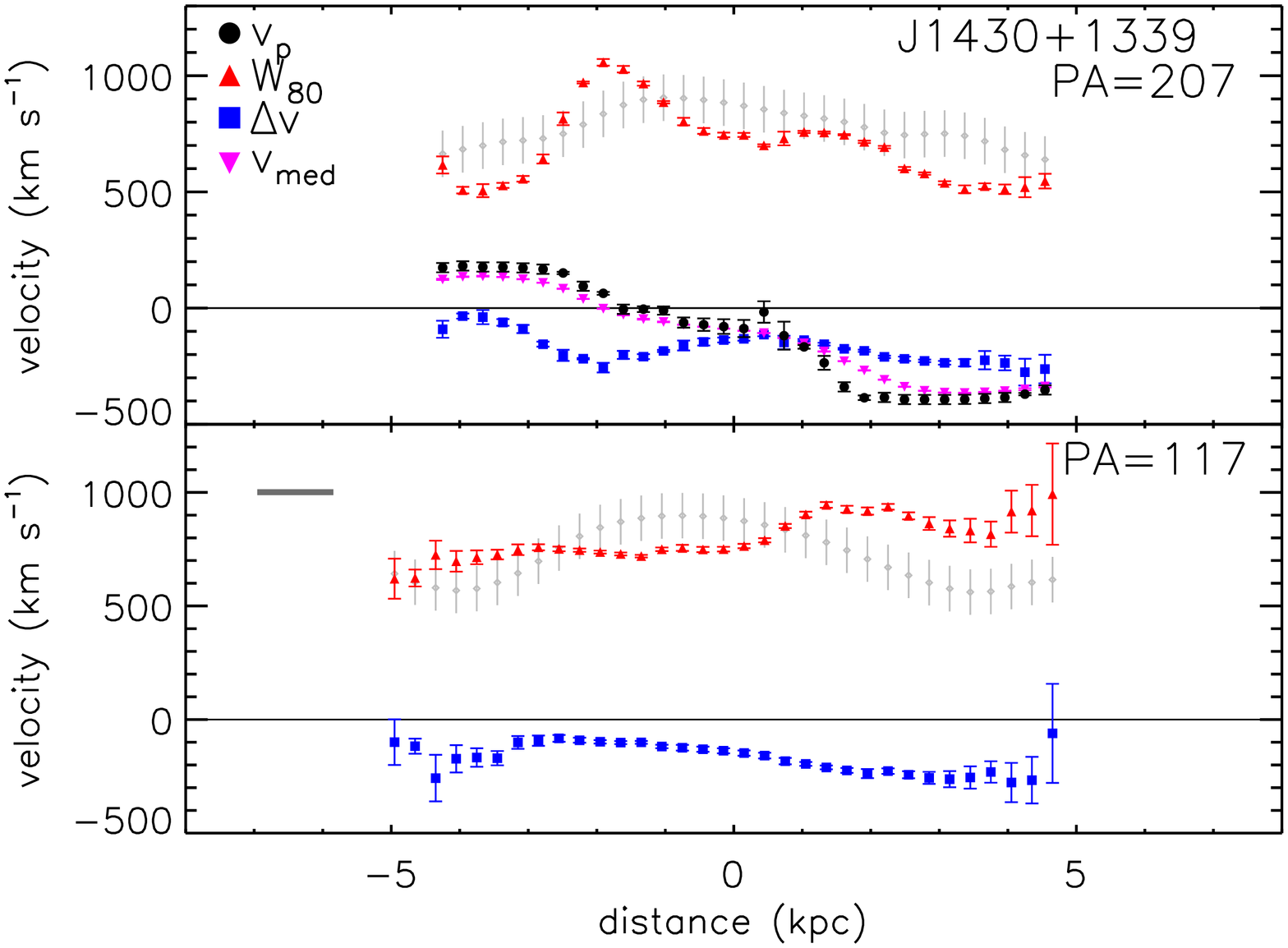,width=3.5in,angle=90}}
\caption{Same as Figure~\ref{Fig:velmaps} and Figure~\ref{Fig:sdss} but
  for SDSS\,J1430+1339}
\label{fig:1430+1339}
\end{figure*}

\cite{Reyes08} previously identified J1430+1339 as a type~2
(``obscured'') quasar. Additionally, this source was identified by Galaxy Zoo
(\citealt{Keel12}) as having an extended emission-line region, due to the arc shaped ``purple
haze'' to the north east of the SDSS image
(Fig.~\ref{fig:1430+1339}). This galaxy has consequently received the
nickname the ``Teacup''. Follow-up {\em
  HST} imaging\footnote{\url{http://blog.galaxyzoo.org/2012/06/14/hubble-spies-the-teacup-and-i-spy-hubble}}
has revealed that this is actually a 5--10\,kpc emission-line ``loop''  accompanied by a
smaller emission-line ``arc'' on the opposite side of the nucleus,
potentially analogous to the bi-polar outflow observed in J1356+1026
(Fig.~\ref{fig:1356+1026}; \citealt{Greene12}). The exact origin of
these emission-line regions are unknown; however, our IFU data
(Fig.~\ref{fig:1430+1339}) cover the base of the north-east loop and reveal it has
a velocity of $\approx-900$\,km\,s$^{-1}$ from the
systemic with a high velocity tail out to $\approx-1400$\,km\,s$^{-1}$. The [O~{\sc iii}] peak signal-to-noise map
show faint extended features similar to those seen in outflowing
superbubbles (e.g., Liu et~al. 2013b\nocite{Liu13b}). Additionally
these features looks remarkably similar to the bi-conal shells of ionised gas
seen in the high-redshift radio galaxy MRC~0406--244 that appear to
be due to gas that has been expelled from the host galaxy by
radio jets and/or star formation (\citealt{Hatch13}; see also \citealt{Nesvadba08}). Indeed, the PA of these
high velocity ionised gas features in J1430+1339 are roughly aligned with,
and found over a similar scale to, the extended radio emission seen in
the FIRST data (see Section~\ref{Fig:SEDs} and Table~\ref{Tab:observations}). High-resolution radio imaging of this
source will play a crucial role in determining the true origin of the
radio emission (e.g., radio jets or shocks). IFU observations covering the full extent of the
emission-line region will reveal the full velocity structure in these
regions. The peak velocity ($v_{p}$) map shows a large velocity gradient from blue to red ($\Delta
 v_{p}\approx600$\,km\,s$^{-1}$) along an axis with
 PA\,$\approx30^{\circ}$ (Fig.~\ref{fig:1430+1339}). 

\subsubsection*{J1504+0151}
\begin{figure*}
\centerline{\psfig{figure=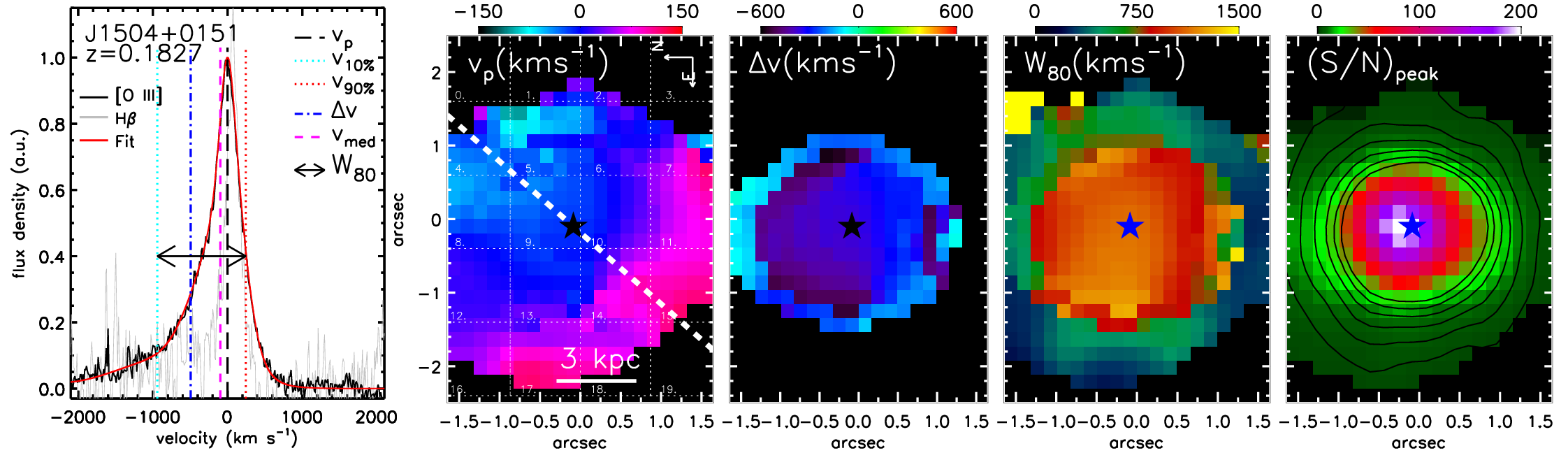,width=6.5in,angle=90}}
\centerline{\psfig{figure=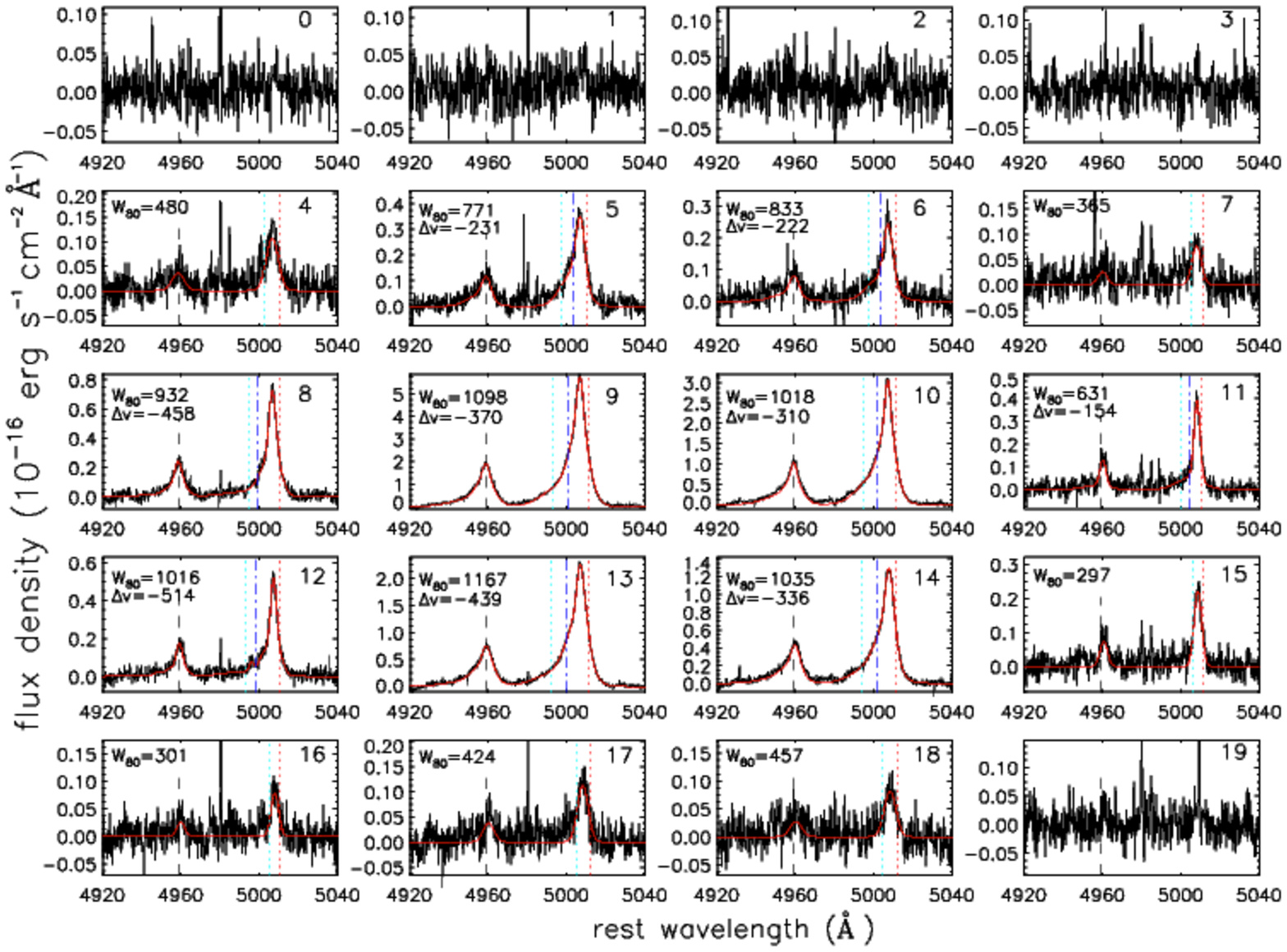,width=6.5in,angle=0}}
\vspace{-0.6cm}
\centerline{\psfig{figure=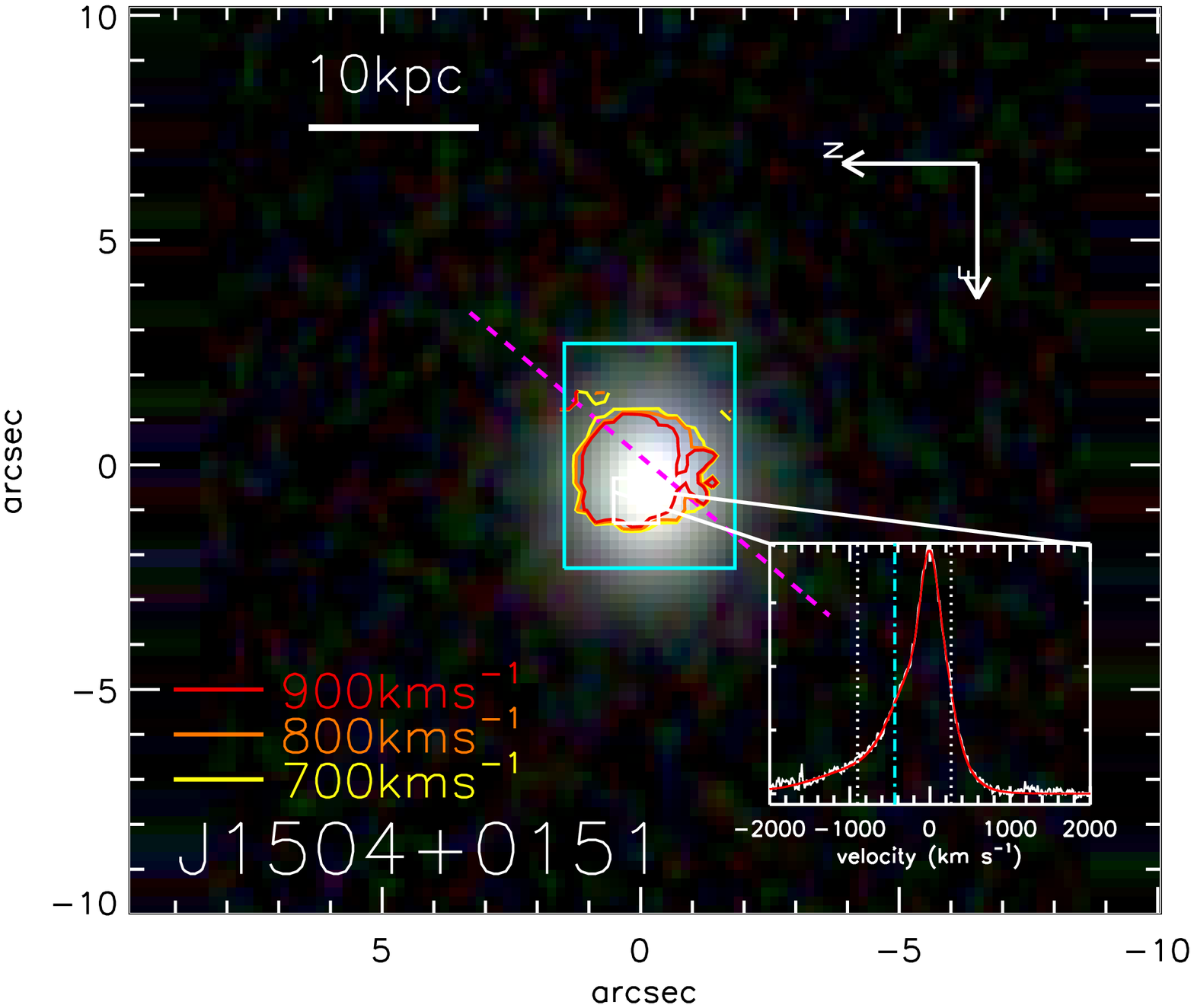,width=2.8in,angle=0}\psfig{figure=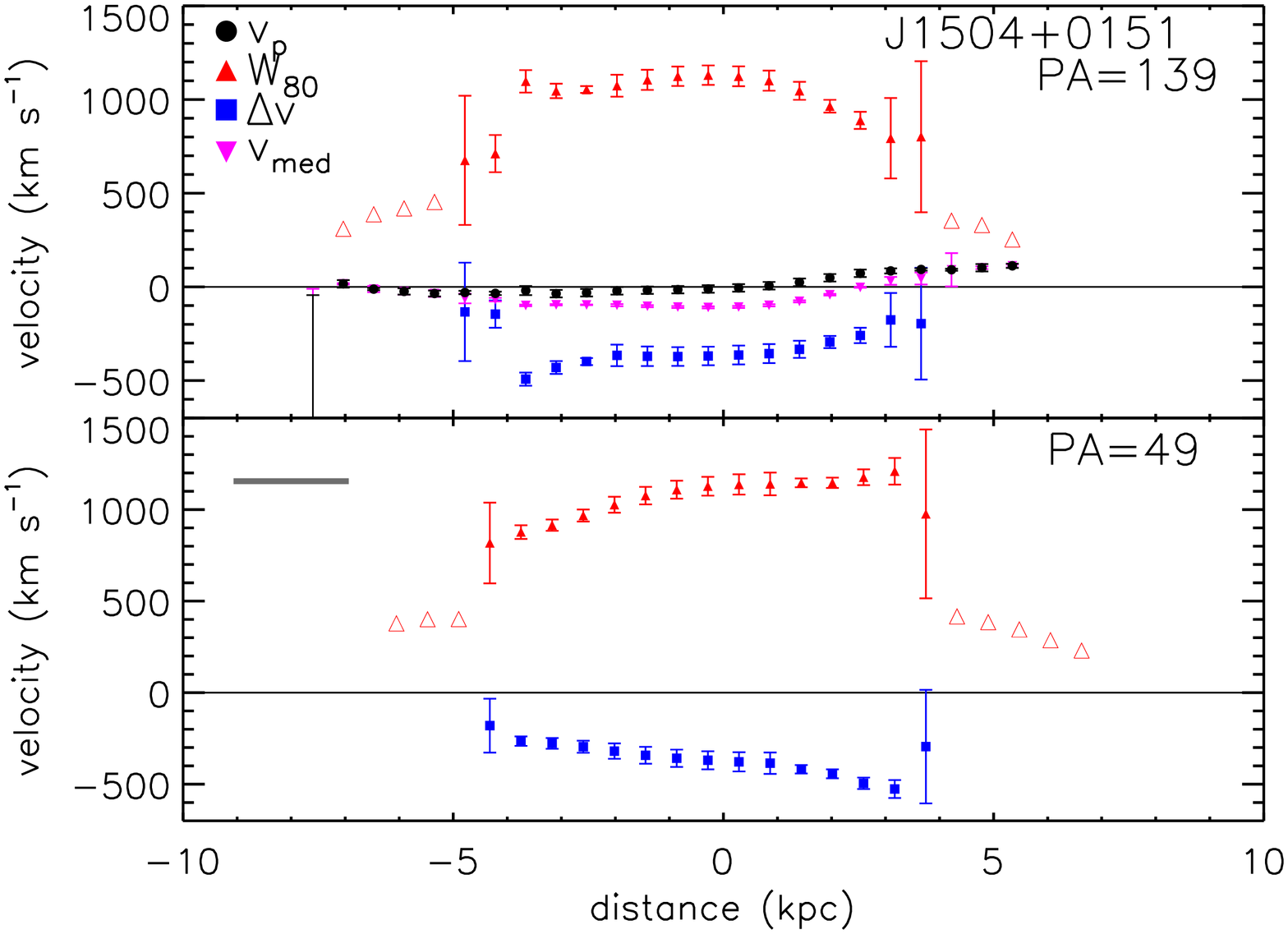,width=3.5in,angle=90}}
\caption{Same as Figure~\ref{Fig:velmaps} and Figure~\ref{Fig:sdss} but for SDSS\,J1504+0151.}
\label{fig:1504+0151}
\end{figure*}

\cite{Reyes08} previously identified this source as being a type~2
(``obscured'') quasar. We note that there are stellar
  continuum and absorption features visible in the SDSS spectrum of this
  source and therefore contamination from stellar
absorption may result in the different H$\beta$ and [O~{\sc iii}]
emission line profiles (Fig.~\ref{fig:1504+0151}). This source is one of the faintest in our sample; however, we
still observe an [O~{\sc iii}] emission-line region out to the edges of the
field-of-view (i.e., over $\gtrsim13$\,kpc; Fig.~\ref{fig:1504+0151}). The peak
velocity map ($v_{p}$) shows a smooth velocity gradient from
blue to red (with $\Delta v_{p}\approx150$\,km\,s$^{-1}$) indicative of
galaxy rotation with a PA\,$\approx140^{\circ}$. Due to the faint
emission lines, it is difficult to establish the true extent and
morphology of the broad, high-velocity emission (which has
$W_{80}\approx1000$\,km\,s$^{-1}$ and a velocity offset between
$\Delta v\approx -300$ and $-500$\,km\,s$^{-1}$); however, it appears
to dominate in the central few kiloparsecs (Fig.~\ref{fig:1504+0151}). 

\nocite{Sutherland07}

\begin{thebibliography}{}

\bibitem[\protect\citeauthoryear{{Abazajian} et~al.,}{{Abazajian}
  et~al.}{2009}]{Abazajian09}
{Abazajian} K.~N.  et~al., 2009, \apjs, 182, 543

\bibitem[\protect\citeauthoryear{{Alatalo} et~al.,}{{Alatalo}
  et~al.}{2011}]{Alatalo11}
{Alatalo} K.  et~al., 2011, \apj, 735, 88

\bibitem[\protect\citeauthoryear{{Alexander} \& {Hickox}}{{Alexander} \&
  {Hickox}}{2012}]{Alexander12}
{Alexander} D.~M.,  {Hickox} R.~C.,  2012, \nar, 56, 93

\bibitem[\protect\citeauthoryear{{Alexander}, {Swinbank}, {Smail}, {McDermid}
  \& {Nesvadba}}{{Alexander} et~al.}{2010}]{Alexander10}
{Alexander} D.~M.,  {Swinbank} A.~M.,  {Smail} I.,  {McDermid} R.,
  {Nesvadba} N.~P.~H.,  2010, \mnras, 402, 2211

\bibitem[\protect\citeauthoryear{{Allington-Smith} et~al.,}{{Allington-Smith}
  et~al.}{2002}]{AllingtonSmith02}
{Allington-Smith} J.  et~al., 2002, \pasp, 114, 892

\bibitem[\protect\citeauthoryear{{Antonucci}}{{Antonucci}}{1993}]{Antonucci93}
{Antonucci} R.,  1993, \araa, 31, 473

\bibitem[\protect\citeauthoryear{{Baldwin}, {Carswell}, {Wampler},
  {Boksenberg}, {Smith} \& {Burbidge}}{{Baldwin} et~al.}{1980}]{Baldwin80}
{Baldwin} J.~A.,  {Carswell} R.~F.,  {Wampler} E.~J.,  {Boksenberg} A.,
  {Smith} H.~E.,    {Burbidge} E.~M.,  1980, \apj, 236, 388

\bibitem[\protect\citeauthoryear{{Baldwin}, {Phillips} \&
  {Terlevich}}{{Baldwin} et~al.}{1981}]{Baldwin81}
{Baldwin} J.~A.,  {Phillips} M.~M.,    {Terlevich} R.,  1981, \pasp, 93, 5

\bibitem[\protect\citeauthoryear{{Barbosa}, {Storchi-Bergmann}, {Cid
  Fernandes}, {Winge} \& {Schmitt}}{{Barbosa} et~al.}{2009}]{Barbosa09}
{Barbosa} F.~K.~B.,  {Storchi-Bergmann} T.,  {Cid Fernandes} R.,  {Winge} C.,
   {Schmitt} H.,  2009, \mnras, 396, 2

\bibitem[\protect\citeauthoryear{{Barrows}, {Sandberg Lacy}, {Kennefick},
  {Comerford}, {Kennefick} \& {Berrier}}{{Barrows} et~al.}{2013}]{Barrows13}
{Barrows} R.~S.,  {Sandberg Lacy} C.~H.,  {Kennefick} J.,  {Comerford} J.~M.,
  {Kennefick} D.,    {Berrier} J.~C.,  2013, \apj, 769, 95

\bibitem[\protect\citeauthoryear{{Barth}, {Greene} \& {Ho}}{{Barth}
  et~al.}{2008}]{Barth08}
{Barth} A.~J.,  {Greene} J.~E.,    {Ho} L.~C.,  2008, \aj, 136, 1179

\bibitem[\protect\citeauthoryear{{Becker}, {White} \& {Helfand}}{{Becker}
  et~al.}{1995}]{Becker95}
{Becker} R.~H.,  {White} R.~L.,    {Helfand} D.~J.,  1995, \apj, 450, 559

\bibitem[\protect\citeauthoryear{{Bennert}, {Falcke}, {Schulz}, {Wilson} \&
  {Wills}}{{Bennert} et~al.}{2002}]{Bennert02}
{Bennert} N.,  {Falcke} H.,  {Schulz} H.,  {Wilson} A.~S.,    {Wills} B.~J.,
  2002, \apjl, 574, L105

\bibitem[\protect\citeauthoryear{{Benson}, {Bower}, {Frenk}, {Lacey}, {Baugh}
  \& {Cole}}{{Benson} et~al.}{2003}]{Benson03}
{Benson} A.~J.,  {Bower} R.~G.,  {Frenk} C.~S.,  {Lacey} C.~G.,  {Baugh} C.~M.,
     {Cole} S.,  2003, \apj, 599, 38

\bibitem[\protect\citeauthoryear{{Blustin et~al.}}{{Blustin
  et~al.}}{2003}]{Blustin03}
{Blustin et~al.} A.~J.,  2003, \aap, 403, 481

\bibitem[\protect\citeauthoryear{{Booth} \& {Schaye}}{{Booth} \&
  {Schaye}}{2010}]{Booth10}
{Booth} C.~M.,  {Schaye} J.,  2010, \mnras, 405, L1

\bibitem[\protect\citeauthoryear{{Boroson}}{{Boroson}}{2005}]{Boroson05}
{Boroson} T.,  2005, \aj, 130, 381

\bibitem[\protect\citeauthoryear{{Boroson} \& {Green}}{{Boroson} \&
  {Green}}{1992}]{Boroson92}
{Boroson} T.~A.,  {Green} R.~F.,  1992, \apjs, 80, 109

\bibitem[\protect\citeauthoryear{{Boroson}, {Persson} \& {Oke}}{{Boroson}
  et~al.}{1985}]{Boroson85}
{Boroson} T.~A.,  {Persson} S.~E.,    {Oke} J.~B.,  1985, \apj, 293, 120

\bibitem[\protect\citeauthoryear{{Bower}, {Benson} \& {Crain}}{{Bower}
  et~al.}{2012}]{Bower12}
{Bower} R.~G.,  {Benson} A.~J.,    {Crain} R.~A.,  2012, \mnras, 422, 2816

\bibitem[\protect\citeauthoryear{{Bower}, {Benson}, {Malbon}, {Helly}, {Frenk},
  {Baugh}, {Cole} \& {Lacey}}{{Bower} et~al.}{2006}]{Bower06}
{Bower} R.~G.,  {Benson} A.~J.,  {Malbon} R.,  {Helly} J.~C.,  {Frenk} C.~S.,
  {Baugh} C.~M.,  {Cole} S.,    {Lacey} C.~G.,  2006, \mnras, 370, 645

\bibitem[\protect\citeauthoryear{{Bradshaw} et~al.,}{{Bradshaw}
  et~al.}{2013}]{Bradshaw13}
{Bradshaw} E.~J.  et~al., 2013, \mnras, 433, 194

\bibitem[\protect\citeauthoryear{{Brinchmann}, {Kunth} \&
  {Durret}}{{Brinchmann} et~al.}{2008}]{Brinchmann08}
{Brinchmann} J.,  {Kunth} D.,    {Durret} F.,  2008, \aap, 485, 657

\bibitem[\protect\citeauthoryear{{Calzetti}, {Armus}, {Bohlin}, {Kinney},
  {Koornneef} \& {Storchi-Bergmann}}{{Calzetti} et~al.}{2000}]{Calzetti00}
{Calzetti} D.,  {Armus} L.,  {Bohlin} R.~C.,  {Kinney} A.~L.,  {Koornneef} J.,
    {Storchi-Bergmann} T.,  2000, \apj, 533, 682

\bibitem[\protect\citeauthoryear{{Cano-D{\'{\i}}az}, {Maiolino}, {Marconi},
  {Netzer}, {Shemmer} \& {Cresci}}{{Cano-D{\'{\i}}az}
  et~al.}{2012}]{CanoDiaz12}
{Cano-D{\'{\i}}az} M.,  {Maiolino} R.,  {Marconi} A.,  {Netzer} H.,  {Shemmer}
  O.,    {Cresci} G.,  2012, \aap, 537, L8

\bibitem[\protect\citeauthoryear{{Capetti}, {Axon}, {Macchetto}, {Marconi} \&
  {Winge}}{{Capetti} et~al.}{1999}]{Capetti99}
{Capetti} A.,  {Axon} D.~J.,  {Macchetto} F.~D.,  {Marconi} A.,    {Winge} C.,
  1999, \apj, 516, 187

\bibitem[\protect\citeauthoryear{{Cattaneo} et~al.,}{{Cattaneo}
  et~al.}{2009}]{Cattaneo09}
{Cattaneo} A.  et~al., 2009, \nat, 460, 213

\bibitem[\protect\citeauthoryear{{Cavagnolo}, {McNamara}, {Nulsen}, {Carilli},
  {Jones} \& {B{\^i}rzan}}{{Cavagnolo} et~al.}{2010}]{Cavagnolo10}
{Cavagnolo} K.~W.,  {McNamara} B.~R.,  {Nulsen} P.~E.~J.,  {Carilli} C.~L.,
  {Jones} C.,    {B{\^i}rzan} L.,  2010, \apj, 720, 1066

\bibitem[\protect\citeauthoryear{{Cecil}, {Bland-Hawthorn}, {Veilleux} \&
  {Filippenko}}{{Cecil} et~al.}{2001}]{Cecil01}
{Cecil} G.,  {Bland-Hawthorn} J.,  {Veilleux} S.,    {Filippenko} A.~V.,  2001,
  \apj, 555, 338

\bibitem[\protect\citeauthoryear{{Chabrier}}{{Chabrier}}{2003}]{Chabrier03}
{Chabrier} G.,  2003, \pasp, 115, 763

\bibitem[\protect\citeauthoryear{{Chen} et~al.,}{{Chen} et~al.}{2013}]{Chen13}
{Chen} C.-T.~J.  et~al., 2013, \apj, 773, 3

\bibitem[\protect\citeauthoryear{{Churazov}, {Sazonov}, {Sunyaev}, {Forman},
  {Jones} \& {B{\"o}hringer}}{{Churazov} et~al.}{2005}]{Churazov05}
{Churazov} E.,  {Sazonov} S.,  {Sunyaev} R.,  {Forman} W.,  {Jones} C.,
  {B{\"o}hringer} H.,  2005, \mnras, 363, L91

\bibitem[\protect\citeauthoryear{{Cicone} et~al.,}{{Cicone}
  et~al.}{2014}]{Cicone14}
{Cicone} C.  et~al., 2014, \aap, 562, A21

\bibitem[\protect\citeauthoryear{{Cimatti} et~al.,}{{Cimatti}
  et~al.}{2013}]{Cimatti13}
{Cimatti} A.  et~al., 2013, \apjl, 779, L13


\bibitem[\protect\citeauthoryear{{Cisternas} et~al.,}{{Cisternas}
  et~al.}{2011}]{Cisternas11}
{Cisternas} M.  et~al., 2011, \apj, 726, 57

\bibitem[\protect\citeauthoryear{{Colina}, {Arribas} \& {Borne}}{{Colina}
  et~al.}{1999}]{Colina99}
{Colina} L.,  {Arribas} S.,    {Borne} K.~D.,  1999, \apjl, 527, L13

\bibitem[\protect\citeauthoryear{{Comerford}, {Gerke}, {Stern}, {Cooper},
  {Weiner}, {Newman}, {Madsen} \& {Barrows}}{{Comerford}
  et~al.}{2012}]{Comerford12}
{Comerford} J.~M.,  {Gerke} B.~F.,  {Stern} D.,  {Cooper} M.~C.,  {Weiner}
  B.~J.,  {Newman} J.~A.,  {Madsen} K.,    {Barrows} R.~S.,  2012, \apj, 753,
  42

\bibitem[\protect\citeauthoryear{{Comerford}, {Schluns}, {Greene} \&
  {Cool}}{{Comerford} et~al.}{2013}]{Comerford13}
{Comerford} J.~M.,  {Schluns} K.,  {Greene} J.~E.,    {Cool} R.~J.,  2013,
  \apj, 777, 64

\bibitem[\protect\citeauthoryear{{Condon}, {Anderson} \& {Broderick}}{{Condon}
  et~al.}{1995}]{Condon95}
{Condon} J.~J.,  {Anderson} E.,    {Broderick} J.~J.,  1995, \aj, 109, 2318

\bibitem[\protect\citeauthoryear{{Condon}, {Cotton}, {Greisen}, {Yin},
  {Perley}, {Taylor} \& {Broderick}}{{Condon} et~al.}{1998}]{Condon98}
{Condon} J.~J.,  {Cotton} W.~D.,  {Greisen} E.~W.,  {Yin} Q.~F.,  {Perley}
  R.~A.,  {Taylor} G.~B.,    {Broderick} J.~J.,  1998, \aj, 115, 1693

\bibitem[\protect\citeauthoryear{{Condon}, {Kellermann}, {Kimball},
  {Ivezi{\'c}} \& {Perley}}{{Condon} et~al.}{2013}]{Condon13}
{Condon} J.~J.,  {Kellermann} K.~I.,  {Kimball} A.~E.,  {Ivezi{\'c}} {\v Z}.,
   {Perley} R.~A.,  2013, \apj, 768, 37

\bibitem[\protect\citeauthoryear{{Courteau}}{{Courteau}}{1997}]{Courteau97}
{Courteau} S.,  1997, \aj, 114, 2402

\bibitem[\protect\citeauthoryear{{Crenshaw} \& {Kraemer}}{{Crenshaw} \&
  {Kraemer}}{2000}]{Crenshaw00a}
{Crenshaw} D.~M.,  {Kraemer} S.~B.,  2000, \apjl, 532, L101

\bibitem[\protect\citeauthoryear{{Crenshaw}, {Schmitt}, {Kraemer}, {Mushotzky}
  \& {Dunn}}{{Crenshaw} et~al.}{2010}]{Crenshaw10}
{Crenshaw} D.~M.,  {Schmitt} H.~R.,  {Kraemer} S.~B.,  {Mushotzky} R.~F.,
  {Dunn} J.~P.,  2010, \apj, 708, 419

\bibitem[\protect\citeauthoryear{{Cui}, {Xia}, {Deng}, {Mao} \& {Zou}}{{Cui}
  et~al.}{2001}]{Cui01}
{Cui} J.,  {Xia} X.-Y.,  {Deng} Z.-G.,  {Mao} S.,    {Zou} Z.-L.,  2001, \aj,
  122, 63

\bibitem[\protect\citeauthoryear{{Dalla Vecchia} \& {Schaye}}{{Dalla Vecchia}
  \& {Schaye}}{2008}]{DallaVecchia08}
{Dalla Vecchia} C.,  {Schaye} J.,  2008, \mnras, 387, 1431

\bibitem[\protect\citeauthoryear{{Debuhr}, {Quataert} \& {Ma}}{{Debuhr}
  et~al.}{2012}]{DeBuhr12}
{Debuhr} J.,  {Quataert} E.,    {Ma} C.-P.,  2012, \mnras, 420, 2221

\bibitem[\protect\citeauthoryear{{Del Moro} et~al.,}{{Del Moro}
  et~al.}{2013}]{DelMoro13}
{Del Moro} A.  et~al., 2013, \aap, 549, A59

\bibitem[\protect\citeauthoryear{{Di Matteo}, {Springel} \& {Hernquist}}{{Di
  Matteo} et~al.}{2005}]{DiMatteo05}
{Di Matteo} T.,  {Springel} V.,    {Hernquist} L.,  2005, \nat, 433, 604

\bibitem[\protect\citeauthoryear{{Diamond-Stanic}, {Moustakas}, {Tremonti},
  {Coil}, {Hickox}, {Robaina}, {Rudnick} \& {Sell}}{{Diamond-Stanic}
  et~al.}{2012}]{DiamondStanic12}
{Diamond-Stanic} A.~M.,  {Moustakas} J.,  {Tremonti} C.~A.,  {Coil} A.~L.,
  {Hickox} R.~C.,  {Robaina} A.~R.,  {Rudnick} G.~H.,    {Sell} P.~H.,  2012,
  \apjl, 755, L26

\bibitem[\protect\citeauthoryear{{Dimitrijevi{\'c}}, {Popovi{\'c}}, {Kova{\v
  c}evi{\'c}}, {Da{\v c}i{\'c}} \& {Ili{\'c}}}{{Dimitrijevi{\'c}}
  et~al.}{2007}]{Dimitrijevic07}
{Dimitrijevi{\'c}} M.~S.,  {Popovi{\'c}} L.~{\v C}.,  {Kova{\v c}evi{\'c}} J.,
  {Da{\v c}i{\'c}} M.,    {Ili{\'c}} D.,  2007, \mnras, 374, 1181

\bibitem[\protect\citeauthoryear{{Emonts}, {Morganti}, {Tadhunter},
  {Oosterloo}, {Holt} \& {van der Hulst}}{{Emonts} et~al.}{2005}]{Emonts05}
{Emonts} B.~H.~C.,  {Morganti} R.,  {Tadhunter} C.~N.,  {Oosterloo} T.~A.,
  {Holt} J.,    {van der Hulst} J.~M.,  2005, \mnras, 362, 931

\bibitem[\protect\citeauthoryear{{Fabian}}{{Fabian}}{1999}]{Fabian99}
{Fabian} A.~C.,  1999, \mnras, 308, L39

\bibitem[\protect\citeauthoryear{{Fabian}}{{Fabian}}{2012}]{Fabian12}
{Fabian} A.~C.,  2012, \araa, 50, 455

\bibitem[\protect\citeauthoryear{{Faucher-Gigu{\`e}re} \&
  {Quataert}}{{Faucher-Gigu{\`e}re} \& {Quataert}}{2012}]{FaucherGiguere12b}
{Faucher-Gigu{\`e}re} C.-A.,  {Quataert} E.,  2012, \mnras, 425, 605

\bibitem[\protect\citeauthoryear{{Feldman}, {Weedman}, {Balzano} \&
  {Ramsey}}{{Feldman} et~al.}{1982}]{Feldman82}
{Feldman} F.~R.,  {Weedman} D.~W.,  {Balzano} V.~A.,    {Ramsey} L.~W.,  1982,
  \apj, 256, 427

\bibitem[\protect\citeauthoryear{{Feruglio}, {Maiolino}, {Piconcelli}, {Menci},
  {Aussel}, {Lamastra} \& {Fiore}}{{Feruglio} et~al.}{2010}]{Feruglio10}
{Feruglio} C.,  {Maiolino} R.,  {Piconcelli} E.,  {Menci} N.,  {Aussel} H.,
  {Lamastra} A.,    {Fiore} F.,  2010, \aap, 518, L155

\bibitem[\protect\citeauthoryear{{Fischer} et~al.,}{{Fischer}
  et~al.}{2010}]{Fischer10}
{Fischer} J.  et~al., 2010, \aap, 518, L41

\bibitem[\protect\citeauthoryear{{Fischer}, {Crenshaw}, {Kraemer} \&
  {Schmitt}}{{Fischer} et~al.}{2013}]{Fischer13}
{Fischer} T.~C.,  {Crenshaw} D.~M.,  {Kraemer} S.~B.,    {Schmitt} H.~R.,
  2013, \apjs, 209, 1

\bibitem[\protect\citeauthoryear{{Fischer}, {Crenshaw}, {Kraemer}, {Schmitt},
  {Mushotsky} \& {Dunn}}{{Fischer} et~al.}{2011}]{Fischer11}
{Fischer} T.~C.,  {Crenshaw} D.~M.,  {Kraemer} S.~B.,  {Schmitt} H.~R.,
  {Mushotsky} R.~F.,    {Dunn} J.~P.,  2011, \apj, 727, 71

\bibitem[\protect\citeauthoryear{{F{\"o}rster Schreiber} et~al.,}{{F{\"o}rster
  Schreiber} et~al.}{2013}]{ForsterSchreiber13}
{F{\"o}rster Schreiber} N.~M.  et~al., 2013, arXiv:1311.2596

\bibitem[\protect\citeauthoryear{{Fu}, {Myers}, {Djorgovski} \& {Yan}}{{Fu}
  et~al.}{2011}]{Fu11}
{Fu} H.,  {Myers} A.~D.,  {Djorgovski} S.~G.,    {Yan} L.,  2011, \apj, 733,
  103

\bibitem[\protect\citeauthoryear{{Fu} \& {Stockton}}{{Fu} \&
  {Stockton}}{2009}]{Fu09}
{Fu} H.,  {Stockton} A.,  2009, \apj, 690, 953

\bibitem[\protect\citeauthoryear{{Fu}, {Yan}, {Myers}, {Stockton},
  {Djorgovski}, {Aldering} \& {Rich}}{{Fu} et~al.}{2012}]{Fu12}
{Fu} H.,  {Yan} L.,  {Myers} A.~D.,  {Stockton} A.,  {Djorgovski} S.~G.,
  {Aldering} G.,    {Rich} J.~A.,  2012, \apj, 745, 67

\bibitem[\protect\citeauthoryear{{Gabor} \& {Bournaud}}{{Gabor} \&
  {Bournaud}}{2014}]{Gabor14}
{Gabor} J.~M.,  {Bournaud} F.,  2014, arXiv:1402.4482

\bibitem[\protect\citeauthoryear{{Gabor}, {Dav{\'e}}, {Oppenheimer} \&
  {Finlator}}{{Gabor} et~al.}{2011}]{Gabor11}
{Gabor} J.~M.,  {Dav{\'e}} R.,  {Oppenheimer} B.~D.,    {Finlator} K.,  2011,
  \mnras, 417, 2676

\bibitem[\protect\citeauthoryear{{Ganguly} \& {Brotherton}}{{Ganguly} \&
  {Brotherton}}{2008}]{Ganguly08}
{Ganguly} R.,  {Brotherton} M.~S.,  2008, \apj, 672, 102

\bibitem[\protect\citeauthoryear{{Gaspari}, {Melioli}, {Brighenti} \&
  {D'Ercole}}{{Gaspari} et~al.}{2011}]{Gaspari11}
{Gaspari} M.,  {Melioli} C.,  {Brighenti} F.,    {D'Ercole} A.,  2011, \mnras,
  411, 349

\bibitem[\protect\citeauthoryear{{Ge}, {Hu}, {Wang}, {Bai} \& {Zhang}}{{Ge}
  et~al.}{2012}]{Ge12}
{Ge} J.-Q.,  {Hu} C.,  {Wang} J.-M.,  {Bai} J.-M.,    {Zhang} S.,  2012, \apjs,
  201, 31

\bibitem[\protect\citeauthoryear{{Gelderman} \& {Whittle}}{{Gelderman} \&
  {Whittle}}{1994}]{Gelderman94}
{Gelderman} R.,  {Whittle} M.,  1994, \apjs, 91, 491

\bibitem[\protect\citeauthoryear{{Genzel} et~al.,}{{Genzel}
  et~al.}{2011}]{Genzel11}
{Genzel} R.  et~al., 2011, \apj, 733, 101

\bibitem[\protect\citeauthoryear{{Gofford et~al.}}{{Gofford
  et~al.}}{2011}]{Gofford11}
{Gofford et~al.} J.,  2011, \mnras, 414, 3307

\bibitem[\protect\citeauthoryear{{Granato}, {De Zotti}, {Silva}, {Bressan} \&
  {Danese}}{{Granato} et~al.}{2004}]{Granato04}
{Granato} G.~L.,  {De Zotti} G.,  {Silva} L.,  {Bressan} A.,    {Danese} L.,
  2004, \apj, 600, 580

\bibitem[\protect\citeauthoryear{{Greene} \& {Ho}}{{Greene} \&
  {Ho}}{2005}]{Greene05a}
{Greene} J.~E.,  {Ho} L.~C.,  2005, \apj, 627, 721

\bibitem[\protect\citeauthoryear{{Greene}, {Zakamska}, {Ho} \&
  {Barth}}{{Greene} et~al.}{2011}]{Greene11}
{Greene} J.~E.,  {Zakamska} N.~L.,  {Ho} L.~C.,    {Barth} A.~J.,  2011, \apj,
  732, 9

\bibitem[\protect\citeauthoryear{{Greene}, {Zakamska} \& {Smith}}{{Greene}
  et~al.}{2012}]{Greene12}
{Greene} J.~E.,  {Zakamska} N.~L.,    {Smith} P.~S.,  2012, \apj, 746, 86

\bibitem[\protect\citeauthoryear{{G{\"u}ltekin} et~al.,}{{G{\"u}ltekin}
  et~al.}{2009}]{Gultekin09}
{G{\"u}ltekin} K.  et~al., 2009, \apj, 698, 198

\bibitem[\protect\citeauthoryear{{Hainline}, {Hickox}, {Greene}, {Myers} \&
  {Zakamska}}{{Hainline} et~al.}{2013}]{Hainline13}
{Hainline} K.~N.,  {Hickox} R.,  {Greene} J.~E.,  {Myers} A.~D.,    {Zakamska}
  N.~L.,  2013, \apj, 774, 145

\bibitem[\protect\citeauthoryear{{Harrison}}{{Harrison}}{2013}]{Harrison14a}
{Harrison} C.~M.,  2014, arXiv:1312.3609

\bibitem[\protect\citeauthoryear{{Harrison} et~al.,}{{Harrison}
  et~al.}{2012}]{Harrison12a}
{Harrison} C.~M.  et~al., 2012, \mnras, 426, 1073

\bibitem[\protect\citeauthoryear{{Hatch} et~al.,}{{Hatch}
  et~al.}{2013}]{Hatch13}
{Hatch} N.~A.  et~al., 2013, \mnras, 436, 2244

\bibitem[\protect\citeauthoryear{{Heckman}, {Armus} \& {Miley}}{{Heckman}
  et~al.}{1987}]{Heckman87}
{Heckman} T.~M.,  {Armus} L.,    {Miley} G.~K.,  1987, \aj, 93, 276

\bibitem[\protect\citeauthoryear{{Heckman}, {Armus} \& {Miley}}{{Heckman}
  et~al.}{1990}]{Heckman90}
{Heckman} T.~M.,  {Armus} L.,    {Miley} G.~K.,  1990, \apjs, 74, 833

\bibitem[\protect\citeauthoryear{{Heckman} et~al.,}{{Heckman}
  et~al.}{2011}]{Heckman11}
{Heckman} T.~M.  et~al., 2011, \apj, 730, 5

\bibitem[\protect\citeauthoryear{{Heckman}, {Kauffmann}, {Brinchmann},
  {Charlot}, {Tremonti} \& {White}}{{Heckman} et~al.}{2004}]{Heckman04}
{Heckman} T.~M.,  {Kauffmann} G.,  {Brinchmann} J.,  {Charlot} S.,  {Tremonti}
  C.,    {White} S.~D.~M.,  2004, \apj, 613, 109

\bibitem[\protect\citeauthoryear{{Heckman}, {Lehnert}, {Strickland} \&
  {Armus}}{{Heckman} et~al.}{2000}]{Heckman00}
{Heckman} T.~M.,  {Lehnert} M.~D.,  {Strickland} D.~K.,    {Armus} L.,  2000,
  \apjs, 129, 493

\bibitem[\protect\citeauthoryear{{Heckman}, {Miley} \& {Green}}{{Heckman}
  et~al.}{1984}]{Heckman84}
{Heckman} T.~M.,  {Miley} G.~K.,    {Green} R.~F.,  1984, \apj, 281, 525

\bibitem[\protect\citeauthoryear{{Heckman}, {Miley}, {van Breugel} \&
  {Butcher}}{{Heckman} et~al.}{1981}]{Heckman81}
{Heckman} T.~M.,  {Miley} G.~K.,  {van Breugel} W.~J.~M.,    {Butcher} H.~R.,
  1981, \apj, 247, 403

\bibitem[\protect\citeauthoryear{{Helou}, {Soifer} \& {Rowan-Robinson}}{{Helou}
  et~al.}{1985}]{Helou85}
{Helou} G.,  {Soifer} B.~T.,    {Rowan-Robinson} M.,  1985, \apjl, 298, L7

\bibitem[\protect\citeauthoryear{{Hill} \& {Zakamska}}{{Hill} \&
  {Zakamska}}{2014}]{Hill14}
{Hill} M.~J.,  {Zakamska} N.~L.,  2014, \mnras, 439, 2701

\bibitem[\protect\citeauthoryear{{Holt}, {Tadhunter}, {Morganti}, {Bellamy},
  {Gonz{\'a}lez Delgado}, {Tzioumis} \& {Inskip}}{{Holt} et~al.}{2006}]{Holt06}
{Holt} J.,  {Tadhunter} C.,  {Morganti} R.,  {Bellamy} M.,  {Gonz{\'a}lez
  Delgado} R.~M.,  {Tzioumis} A.,    {Inskip} K.~J.,  2006, \mnras, 370, 1633

\bibitem[\protect\citeauthoryear{{Holt}, {Tadhunter} \& {Morganti}}{{Holt}
  et~al.}{2008}]{Holt08}
{Holt} J.,  {Tadhunter} C.~N.,    {Morganti} R.,  2008, \mnras, 387, 639

\bibitem[\protect\citeauthoryear{{Holt}, {Tadhunter}, {Morganti} \&
  {Emonts}}{{Holt} et~al.}{2011}]{Holt11}
{Holt} J.,  {Tadhunter} C.~N.,  {Morganti} R.,    {Emonts} B.~H.~C.,  2011,
  \mnras, 410, 1527

\bibitem[\protect\citeauthoryear{{Hopkins}, {Cox}, {Hernquist}, {Narayanan},
  {Hayward} \& {Murray}}{{Hopkins} et~al.}{2013}]{Hopkins13a}
{Hopkins} P.~F.,  {Cox} T.~J.,  {Hernquist} L.,  {Narayanan} D.,  {Hayward}
  C.~C.,    {Murray} N.,  2013, \mnras, 430, 1901

\bibitem[\protect\citeauthoryear{{Hopkins} \& {Elvis}}{{Hopkins} \&
  {Elvis}}{2010}]{Hopkins10}
{Hopkins} P.~F.,  {Elvis} M.,  2010, \mnras, 401, 7

\bibitem[\protect\citeauthoryear{{Hopkins}, {Hernquist}, {Cox}, {Di Matteo},
  {Robertson} \& {Springel}}{{Hopkins} et~al.}{2006}]{Hopkins06}
{Hopkins} P.~F.,  {Hernquist} L.,  {Cox} T.~J.,  {Di Matteo} T.,  {Robertson}
  B.,    {Springel} V.,  2006, \apjs, 163, 1

\bibitem[\protect\citeauthoryear{{Hopkins}, {Hernquist}, {Cox} \& {Kere{\v
  s}}}{{Hopkins} et~al.}{2008}]{Hopkins08a}
{Hopkins} P.~F.,  {Hernquist} L.,  {Cox} T.~J.,    {Kere{\v s}} D.,  2008,
  \apjs, 175, 356

\bibitem[\protect\citeauthoryear{{Hopkins}, {Kere{\v s}}, {Murray},
  {Hernquist}, {Narayanan} \& {Hayward}}{{Hopkins} et~al.}{2013}]{Hopkins13b}
{Hopkins} P.~F.,  {Kere{\v s}} D.,  {Murray} N.,  {Hernquist} L.,  {Narayanan}
  D.,    {Hayward} C.~C.,  2013, \mnras, 433, 78

\bibitem[\protect\citeauthoryear{{Humphrey}, {Villar-Mart{\'{\i}}n},
  {S{\'a}nchez}, {Mart{\'{\i}}nez-Sansigre}, {Delgado}, {P{\'e}rez},
  {Tadhunter} \& {P{\'e}rez-Torres}}{{Humphrey} et~al.}{2010}]{Humphrey10}
{Humphrey} A.,  {Villar-Mart{\'{\i}}n} M.,  {S{\'a}nchez} S.~F.,
  {Mart{\'{\i}}nez-Sansigre} A.,  {Delgado} R.~G.,  {P{\'e}rez} E.,
  {Tadhunter} C.,    {P{\'e}rez-Torres} M.~A.,  2010, \mnras, 408, L1

\bibitem[\protect\citeauthoryear{{Husemann}, {Wisotzki}, {S{\'a}nchez} \&
  {Jahnke}}{{Husemann} et~al.}{2013}]{Husemann13}
{Husemann} B.,  {Wisotzki} L.,  {S{\'a}nchez} S.~F.,    {Jahnke} K.,  2013,
  \aap, 549, A43

\bibitem[\protect\citeauthoryear{{Ivison} et~al.,}{{Ivison}
  et~al.}{2010}]{Ivison10}
{Ivison} R.~J.  et~al., 2010, \aap, 518, L31

\bibitem[\protect\citeauthoryear{{Keel} et~al.,}{{Keel} et~al.}{2012}]{Keel12}
{Keel} W.~C.  et~al., 2012, \mnras, 420, 878

\bibitem[\protect\citeauthoryear{{Kennicutt}
  Jr.}{{Kennicutt}}{1998}]{Kennicutt98}
{Kennicutt} Jr. R.~C.,  1998, \araa, 36, 189

\bibitem[\protect\citeauthoryear{{Kim}, {Sanders}, {Veilleux}, {Mazzarella} \&
  {Soifer}}{{Kim} et~al.}{1995}]{Kim95}
{Kim} D.-C.,  {Sanders} D.~B.,  {Veilleux} S.,  {Mazzarella} J.~M.,    {Soifer}
  B.~T.,  1995, \apjs, 98, 129

\bibitem[\protect\citeauthoryear{{Kim}, {Ho}, {Lonsdale}, {Lacy}, {Blain} \&
  {Kimball}}{{Kim} et~al.}{2013}]{Kim13}
{Kim} M.,  {Ho} L.~C.,  {Lonsdale} C.~J.,  {Lacy} M.,  {Blain} A.~W.,
  {Kimball} A.~E.,  2013, \apjl, 768, L9

\bibitem[\protect\citeauthoryear{{Kimball} \& {Ivezi{\'c}}}{{Kimball} \&
  {Ivezi{\'c}}}{2008}]{Kimball08}
{Kimball} A.~E.,  {Ivezi{\'c}} {\v Z}.,  2008, \aj, 136, 684

\bibitem[\protect\citeauthoryear{{King}}{{King}}{2003}]{King03}
{King} A.,  2003, \apjl, 596, L27

\bibitem[\protect\citeauthoryear{{King}, {Zubovas} \& {Power}}{{King}
  et~al.}{2011}]{King11}
{King} A.~R.,  {Zubovas} K.,    {Power} C.,  2011, \mnras, 415, L6

\bibitem[\protect\citeauthoryear{{Komossa}, {Xu}, {Zhou}, {Storchi-Bergmann} \&
  {Binette}}{{Komossa} et~al.}{2008}]{Komossa08}
{Komossa} S.,  {Xu} D.,  {Zhou} H.,  {Storchi-Bergmann} T.,    {Binette} L.,
  2008, \apj, 680, 926

\bibitem[\protect\citeauthoryear{{Kormendy} \& {Ho}}{{Kormendy} \&
  {Ho}}{2013}]{Kormendy13}
{Kormendy} J.,  {Ho} L.~C.,  2013, \araa, 51, 511

\bibitem[\protect\citeauthoryear{{Kormendy} \& {Richstone}}{{Kormendy} \&
  {Richstone}}{1995}]{Kormendy95}
{Kormendy} J.,  {Richstone} D.,  1995, \araa, 33, 581

\bibitem[\protect\citeauthoryear{{Krajnovi{\'c}}, {Cappellari}, {de Zeeuw} \&
  {Copin}}{{Krajnovi{\'c}} et~al.}{2006}]{Krajnovic06}
{Krajnovi{\'c}} D.,  {Cappellari} M.,  {de Zeeuw} P.~T.,    {Copin} Y.,  2006,
  \mnras, 366, 787

\bibitem[\protect\citeauthoryear{{Lagos}, {Cora} \& {Padilla}}{{Lagos}
  et~al.}{2008}]{Lagos08}
{Lagos} C.~D.~P.,  {Cora} S.~A.,    {Padilla} N.~D.,  2008, \mnras, 388, 587

\bibitem[\protect\citeauthoryear{{Lawrence} \& {Elvis}}{{Lawrence} \&
  {Elvis}}{2010}]{Lawrence10}
{Lawrence} A.,  {Elvis} M.,  2010, \apj, 714, 561

\bibitem[\protect\citeauthoryear{{Lehnert} \& {Heckman}}{{Lehnert} \&
  {Heckman}}{1996}]{Lehnert96}
{Lehnert} M.~D.,  {Heckman} T.~M.,  1996, \apj, 462, 651

\bibitem[\protect\citeauthoryear{{Leipski} \& {Bennert}}{{Leipski} \&
  {Bennert}}{2006}]{Leipski06}
{Leipski} C.,  {Bennert} N.,  2006, \aap, 448, 165

\bibitem[\protect\citeauthoryear{{Leitherer} et~al.,}{{Leitherer}
  et~al.}{1999}]{Leitherer99}
{Leitherer} C.  et~al., 1999, \apjs, 123, 3

\bibitem[\protect\citeauthoryear{{L{\'{\i}}pari} et~al.,}{{L{\'{\i}}pari}
  et~al.}{2009}]{Lipari09b}
{L{\'{\i}}pari} S.  et~al., 2009, \mnras, 398, 658

\bibitem[\protect\citeauthoryear{{Lipari} et~al.,}{{Lipari}
  et~al.}{2009}]{Lipari09a}
{Lipari} S.  et~al., 2009, \mnras, 392, 1295

\bibitem[\protect\citeauthoryear{{Liu}, {Zakamska} \& {Greene}}{{Liu}
  et~al.}{2014}]{Liu14}
{Liu} G.,  {Zakamska} N.~L.,    {Greene} J.~E.,  2014, arXiv:1401.0536

\bibitem[\protect\citeauthoryear{{Liu}, {Zakamska}, {Greene}, {Nesvadba} \&
  {Liu}}{{Liu} et~al.}{2013a}]{Liu13a}
{Liu} G.,  {Zakamska} N.~L.,  {Greene} J.~E.,  {Nesvadba} N.~P.~H.,    {Liu}
  X.,  2013a, \mnras, 430, 2327

\bibitem[\protect\citeauthoryear{{Liu}, {Zakamska}, {Greene}, {Nesvadba} \&
  {Liu}}{{Liu} et~al.}{2013b}]{Liu13b}
{Liu} G.,  {Zakamska} N.~L.,  {Greene} J.~E.,  {Nesvadba} N.~P.~H.,    {Liu}
  X.,  2013b, \mnras, 436, 2576

\bibitem[\protect\citeauthoryear{{Liu}, {Shen}, {Strauss} \& {Greene}}{{Liu}
  et~al.}{2010}]{Liu10}
{Liu} X.,  {Shen} Y.,  {Strauss} M.~A.,    {Greene} J.~E.,  2010, \apj, 708,
  427

\bibitem[\protect\citeauthoryear{{Magorrian} et~al.,}{{Magorrian}
  et~al.}{1998}]{Magorrian98}
{Magorrian} J.  et~al., 1998, \aj, 115, 2285

\bibitem[\protect\citeauthoryear{{Mahony}, {Morganti}, {Emonts}, {Oosterloo} \&
  {Tadhunter}}{{Mahony} et~al.}{2013}]{Mahony13}
{Mahony} E.~K.,  {Morganti} R.,  {Emonts} B.~H.~C.,  {Oosterloo} T.~A.,
  {Tadhunter} C.,  2013, \mnras, 435, L58

\bibitem[\protect\citeauthoryear{{Markwardt}}{{Markwardt}}{2009}]{Markwardt09}
{Markwardt} C.~B.,  2009, in {Bohlender} D.~A.,  {Durand} D.,   {Dowler} P.,
  eds,  Astronomical Society of the Pacific Conference Series Vol. 411,
  Astronomical Data Analysis Software and Systems XVIII. p.~251

\bibitem[\protect\citeauthoryear{{Martin}}{{Martin}}{1999}]{Martin99}
{Martin} C.~L.,  1999, \apj, 513, 156

\bibitem[\protect\citeauthoryear{{Martin}}{{Martin}}{2005}]{Martin05}
{Martin} C.~L.,  2005, \apj, 621, 227

\bibitem[\protect\citeauthoryear{{Mateos} et~al.,}{{Mateos}
  et~al.}{2012}]{Mateos12}
{Mateos} S.  et~al., 2012, \mnras, 426, 3271

\bibitem[\protect\citeauthoryear{{McCarthy}, {Schaye}, {Bower}, {Ponman},
  {Booth}, {Dalla Vecchia} \& {Springel}}{{McCarthy} et~al.}{2011}]{McCarthy11}
{McCarthy} I.~G.,  {Schaye} J.,  {Bower} R.~G.,  {Ponman} T.~J.,  {Booth}
  C.~M.,  {Dalla Vecchia} C.,    {Springel} V.,  2011, \mnras, 412, 1965

\bibitem[\protect\citeauthoryear{{McCarthy} et~al.,}{{McCarthy}
  et~al.}{2010}]{McCarthy10}
{McCarthy} I.~G.  et~al., 2010, \mnras, 406, 822

\bibitem[\protect\citeauthoryear{{McCarthy}, {Baum} \& {Spinrad}}{{McCarthy}
  et~al.}{1996}]{McCarthy96}
{McCarthy} P.~J.,  {Baum} S.~A.,    {Spinrad} H.,  1996, \apjs, 106, 281

\bibitem[\protect\citeauthoryear{{McNamara} \& {Nulsen}}{{McNamara} \&
  {Nulsen}}{2012}]{McNamara12}
{McNamara} B.~R.,  {Nulsen} P.~E.~J.,  2012, New Journal of Physics, 14, 055023

\bibitem[\protect\citeauthoryear{{Morganti}, {Frieswijk}, {Oonk}, {Oosterloo}
  \& {Tadhunter}}{{Morganti} et~al.}{2013}]{Morganti13}
{Morganti} R.,  {Frieswijk} W.,  {Oonk} R.~J.~B.,  {Oosterloo} T.,
  {Tadhunter} C.,  2013, \aap, 552, L4

\bibitem[\protect\citeauthoryear{{Morganti}, {Oosterloo}, {Tadhunter}, {van
  Moorsel} \& {Emonts}}{{Morganti} et~al.}{2005}]{Morganti05b}
{Morganti} R.,  {Oosterloo} T.~A.,  {Tadhunter} C.~N.,  {van Moorsel} G.,
  {Emonts} B.,  2005, \aap, 439, 521

\bibitem[\protect\citeauthoryear{{Moshir}, {Kopman} \& {Conrow}}{{Moshir}
  et~al.}{1992}]{Moshir92}
{Moshir} M.,  {Kopman} G.,    {Conrow} T.~A.~O.,  1992, {IRAS Faint Source
  Survey, Explanatory supplement version 2}

\bibitem[\protect\citeauthoryear{{Mullaney}, {Alexander}, {Fine}, {Goulding},
  {Harrison} \& {Hickox}}{{Mullaney} et~al.}{2013}]{Mullaney13}
{Mullaney} J.~R.,  {Alexander} D.~M.,  {Fine} S.,  {Goulding} A.~D.,
  {Harrison} C.~M.,    {Hickox} R.~C.,  2013, \mnras, 433, 622

\bibitem[\protect\citeauthoryear{{Mullaney}, {Alexander}, {Goulding} \&
  {Hickox}}{{Mullaney} et~al.}{2011}]{Mullaney11}
{Mullaney} J.~R.,  {Alexander} D.~M.,  {Goulding} A.~D.,    {Hickox} R.~C.,
  2011, \mnras, 414, 1082

\bibitem[\protect\citeauthoryear{{Mullaney}, {Alexander}, {Huynh}, {Goulding}
  \& {Frayer}}{{Mullaney} et~al.}{2010}]{Mullaney10}
{Mullaney} J.~R.,  {Alexander} D.~M.,  {Huynh} M.,  {Goulding} A.~D.,
  {Frayer} D.,  2010, \mnras, 401, 995

\bibitem[\protect\citeauthoryear{{Mullaney} et~al.,}{{Mullaney}
  et~al.}{2012}]{Mullaney12b}
{Mullaney} J.~R.  et~al., 2012, \apjl, 753, L30

\bibitem[\protect\citeauthoryear{{Murray}, {M{\'e}nard} \& {Thompson}}{{Murray}
  et~al.}{2011}]{Murray11}
{Murray} N.,  {M{\'e}nard} B.,    {Thompson} T.~A.,  2011, \apj, 735, 66

\bibitem[\protect\citeauthoryear{{Murray}, {Quataert} \& {Thompson}}{{Murray}
  et~al.}{2005}]{Murray05}
{Murray} N.,  {Quataert} E.,    {Thompson} T.~A.,  2005, \apj, 618, 569

\bibitem[\protect\citeauthoryear{{Navarro}, {Frenk} \& {White}}{{Navarro}
  et~al.}{1996}]{Navarro96}
{Navarro} J.~F.,  {Frenk} C.~S.,    {White} S.~D.~M.,  1996, \apj, 462, 563

\bibitem[\protect\citeauthoryear{{Nelson} \& {Whittle}}{{Nelson} \&
  {Whittle}}{1996}]{Nelson96}
{Nelson} C.~H.,  {Whittle} M.,  1996, \apj, 465, 96

\bibitem[\protect\citeauthoryear{{Nesvadba}, {Lehnert}, {De Breuck}, {Gilbert}
  \& {van Breugel}}{{Nesvadba} et~al.}{2007}]{Nesvadba07a}
{Nesvadba} N.~P.~H.,  {Lehnert} M.~D.,  {De Breuck} C.,  {Gilbert} A.,    {van
  Breugel} W.,  2007, \aap, 475, 145

\bibitem[\protect\citeauthoryear{{Nesvadba}, {Lehnert}, {De Breuck}, {Gilbert}
  \& {van Breugel}}{{Nesvadba} et~al.}{2008}]{Nesvadba08}
{Nesvadba} N.~P.~H.,  {Lehnert} M.~D.,  {De Breuck} C.,  {Gilbert} A.~M.,
  {van Breugel} W.,  2008, \aap, 491, 407

\bibitem[\protect\citeauthoryear{{Nesvadba}, {Lehnert}, {Eisenhauer},
  {Gilbert}, {Tecza} \& {Abuter}}{{Nesvadba} et~al.}{2006}]{Nesvadba06}
{Nesvadba} N.~P.~H.,  {Lehnert} M.~D.,  {Eisenhauer} F.,  {Gilbert} A.,
  {Tecza} M.,    {Abuter} R.,  2006, \apj, 650, 693

\bibitem[\protect\citeauthoryear{{Nesvadba}, {Polletta}, {Lehnert}, {Bergeron},
  {De Breuck}, {Lagache} \& {Omont}}{{Nesvadba} et~al.}{2011}]{Nesvadba11}
{Nesvadba} N.~P.~H.,  {Polletta} M.,  {Lehnert} M.~D.,  {Bergeron} J.,  {De
  Breuck} C.,  {Lagache} G.,    {Omont} A.,  2011, \mnras, 415, 2359

\bibitem[\protect\citeauthoryear{{Neugebauer} et~al.,}{{Neugebauer}
  et~al.}{1984}]{Neugebaur84}
{Neugebauer} G.  et~al., 1984, \apjl, 278, L1

\bibitem[\protect\citeauthoryear{{Newman} et~al.,}{{Newman}
  et~al.}{2012}]{Newman12b}
{Newman} S.~F.  et~al., 2012, \apj, 761, 43

\bibitem[\protect\citeauthoryear{{Ohyama} et~al.,}{{Ohyama}
  et~al.}{2002}]{Ohyama02}
{Ohyama} Y.  et~al., 2002, \pasj, 54, 891

\bibitem[\protect\citeauthoryear{{Osterbrock}}{{Osterbrock}}{1989}]{Osterbrock89}
{Osterbrock} D.~E.,  1989, {Astrophysics of gaseous nebulae and active galactic
  nuclei}

\bibitem[\protect\citeauthoryear{{Osterbrock} \& {Ferland}}{{Osterbrock} \&
  {Ferland}}{2006}]{Osterbrock06}
{Osterbrock} D.~E.,  {Ferland} G.~J.,  2006, {Astrophysics of gaseous nebulae
  and active galactic nuclei}

\bibitem[\protect\citeauthoryear{{Ramos Almeida}, {Tadhunter}, {Inskip},
  {Morganti}, {Holt} \& {Dicken}}{{Ramos Almeida}
  et~al.}{2011}]{RamosAlmeida11}
{Ramos Almeida} C.,  {Tadhunter} C.~N.,  {Inskip} K.~J.,  {Morganti} R.,
  {Holt} J.,    {Dicken} D.,  2011, \mnras, 410, 1550

\bibitem[\protect\citeauthoryear{{Reeves}, {O'Brien} \& {Ward}}{{Reeves}
  et~al.}{2003}]{Reeves03}
{Reeves} J.~N.,  {O'Brien} P.~T.,    {Ward} M.~J.,  2003, \apjl, 593, L65

\bibitem[\protect\citeauthoryear{{Reyes} et~al.,}{{Reyes}
  et~al.}{2008}]{Reyes08}
{Reyes} R.  et~al., 2008, \aj, 136, 2373

\bibitem[\protect\citeauthoryear{{Richards} et~al.,}{{Richards}
  et~al.}{2006}]{Richards06}
{Richards} G.~T.  et~al., 2006, \apjs, 166, 470

\bibitem[\protect\citeauthoryear{{Riffel}, {Storchi-Bergmann} \&
  {Riffel}}{{Riffel} et~al.}{2014}]{Riffel14}
{Riffel} R.~A.,  {Storchi-Bergmann} T.,    {Riffel} R.,  2014, \apjl, 780, L24

\bibitem[\protect\citeauthoryear{{Rodr{\'{\i}}guez Zaur{\'{\i}}n}, {Tadhunter},
  {Rose} \& {Holt}}{{Rodr{\'{\i}}guez Zaur{\'{\i}}n}
  et~al.}{2013}]{RodriguezZaurin13}
{Rodr{\'{\i}}guez Zaur{\'{\i}}n} J.,  {Tadhunter} C.~N.,  {Rose} M.,    {Holt}
  J.,  2013, \mnras, 432, 138

\bibitem[\protect\citeauthoryear{{Rosario} et~al.,}{{Rosario}
  et~al.}{2012}]{Rosario12}
{Rosario} D.~J.  et~al., 2012, \aap, 545, A45

\bibitem[\protect\citeauthoryear{{Rosario}, {Shields}, {Taylor}, {Salviander}
  \& {Smith}}{{Rosario} et~al.}{2010}]{Rosario10}
{Rosario} D.~J.,  {Shields} G.~A.,  {Taylor} G.~B.,  {Salviander} S.,
  {Smith} K.~L.,  2010, \apj, 716, 131

\bibitem[\protect\citeauthoryear{{Rosas-Guevara} et~al.,}{{Rosas-Guevara}
  et~al.}{2013}]{RosasGuevara13}
{Rosas-Guevara} Y.~M.  et~al., 2013, arXiv:1312.0598

\bibitem[\protect\citeauthoryear{{Rupke}, {Veilleux} \& {Sanders}}{{Rupke}
  et~al.}{2005}]{Rupke05b}
{Rupke} D.~S.,  {Veilleux} S.,    {Sanders} D.~B.,  2005, \apjs, 160, 115

\bibitem[\protect\citeauthoryear{{Rupke} \& {Veilleux}}{{Rupke} \&
  {Veilleux}}{2011}]{Rupke11}
{Rupke} D.~S.~N.,  {Veilleux} S.,  2011, \apjl, 729, L27+

\bibitem[\protect\citeauthoryear{{Rupke} \& {Veilleux}}{{Rupke} \&
  {Veilleux}}{2013}]{Rupke13}
{Rupke} D.~S.~N.,  {Veilleux} S.,  2013, \apj, 768, 75

\bibitem[\protect\citeauthoryear{{Sanders} \& {Mirabel}}{{Sanders} \&
  {Mirabel}}{1996}]{Sanders96}
{Sanders} D.~B.,  {Mirabel} I.~F.,  1996, \araa, 34, 749

\bibitem[\protect\citeauthoryear{{Sanders}, {Soifer}, {Elias}, {Neugebauer} \&
  {Matthews}}{{Sanders} et~al.}{1988}]{Sanders88}
{Sanders} D.~B.,  {Soifer} B.~T.,  {Elias} J.~H.,  {Neugebauer} G.,
  {Matthews} K.,  1988, \apjl, 328, L35

\bibitem[\protect\citeauthoryear{{Sargsyan}, {Mickaelian}, {Weedman} \&
  {Houck}}{{Sargsyan} et~al.}{2008}]{Sargsyan08}
{Sargsyan} L.,  {Mickaelian} A.,  {Weedman} D.,    {Houck} J.,  2008, \apj,
  683, 114

\bibitem[\protect\citeauthoryear{{Schirmer}, {Diaz}, {Holhjem}, {Levenson} \&
  {Winge}}{{Schirmer} et~al.}{2013}]{Schirmer13}
{Schirmer} M.,  {Diaz} R.,  {Holhjem} K.,  {Levenson} N.~A.,    {Winge} C.,
  2013, \apj, 763, 60

\bibitem[\protect\citeauthoryear{{Schnorr-M{\"u}ller}, {Storchi-Bergmann},
  {Nagar}, {Robinson}, {Lena}, {Riffel} \& {Couto}}{{Schnorr-M{\"u}ller}
  et~al.}{2014}]{SchnorrMuller14}
{Schnorr-M{\"u}ller} A.,  {Storchi-Bergmann} T.,  {Nagar} N.~M.,  {Robinson}
  A.,  {Lena} D.,  {Riffel} R.~A.,    {Couto} G.~S.,  2014, \mnras, 437, 1708

\bibitem[\protect\citeauthoryear{{Shen}, {Liu}, {Greene} \& {Strauss}}{{Shen}
  et~al.}{2011}]{Shen11}
{Shen} Y.,  {Liu} X.,  {Greene} J.~E.,    {Strauss} M.~A.,  2011, \apj, 735, 48

\bibitem[\protect\citeauthoryear{{Shih} \& {Rupke}}{{Shih} \&
  {Rupke}}{2010}]{Shih10}
{Shih} H.-Y.,  {Rupke} D.~S.~N.,  2010, \apj, 724, 1430

\bibitem[\protect\citeauthoryear{{Shih}, {Stockton} \& {Kewley}}{{Shih}
  et~al.}{2013}]{Shih13}
{Shih} H.-Y.,  {Stockton} A.,    {Kewley} L.,  2013, \apj, 772, 138

\bibitem[\protect\citeauthoryear{{Silk} \& {Rees}}{{Silk} \&
  {Rees}}{1998}]{Silk98}
{Silk} J.,  {Rees} M.~J.,  1998, \aap, 331, L1

\bibitem[\protect\citeauthoryear{{Simpson} et~al.,}{{Simpson}
  et~al.}{2012}]{Simpson12}
{Simpson} C.  et~al., 2012, \mnras, 421, 3060

\bibitem[\protect\citeauthoryear{{Smith}, {Shields}, {Bonning}, {McMullen},
  {Rosario} \& {Salviander}}{{Smith} et~al.}{2010}]{Smith10}
{Smith} K.~L.,  {Shields} G.~A.,  {Bonning} E.~W.,  {McMullen} C.~C.,
  {Rosario} D.~J.,    {Salviander} S.,  2010, \apj, 716, 866

\bibitem[\protect\citeauthoryear{{Smith}, {Shields}, {Salviander}, {Stevens} \&
  {Rosario}}{{Smith} et~al.}{2012}]{Smith12}
{Smith} K.~L.,  {Shields} G.~A.,  {Salviander} S.,  {Stevens} A.~C.,
  {Rosario} D.~J.,  2012, \apj, 752, 63

\bibitem[\protect\citeauthoryear{{Soto} \& {Martin}}{{Soto} \&
  {Martin}}{2010}]{Soto10}
{Soto} K.~T.,  {Martin} C.~L.,  2010, \apj, 716, 332

\bibitem[\protect\citeauthoryear{{Springel}, {Di Matteo} \&
  {Hernquist}}{{Springel} et~al.}{2005}]{Springel05}
{Springel} V.,  {Di Matteo} T.,    {Hernquist} L.,  2005, \mnras, 361, 776

\bibitem[\protect\citeauthoryear{{Stockton}}{{Stockton}}{1976}]{Stockton76}
{Stockton} A.,  1976, \apjl, 205, L113

\bibitem[\protect\citeauthoryear{{Stockton} \& {MacKenty}}{{Stockton} \&
  {MacKenty}}{1987}]{Stockton87}
{Stockton} A.,  {MacKenty} J.~W.,  1987, \apj, 316, 584

\bibitem[\protect\citeauthoryear{{Storchi-Bergmann}, {Lopes}, {McGregor},
  {Riffel}, {Beck} \& {Martini}}{{Storchi-Bergmann}
  et~al.}{2010}]{StorchiBergmann10}
{Storchi-Bergmann} T.,  {Lopes} R.~D.~S.,  {McGregor} P.~J.,  {Riffel} R.~A.,
  {Beck} T.,    {Martini} P.,  2010, \mnras, 402, 819

\bibitem[\protect\citeauthoryear{{Sutherland} \& {Bicknell}}{{Sutherland} \&
  {Bicknell}}{2007}]{Sutherland07}
{Sutherland} R.~S.,  {Bicknell} G.~V.,  2007, \apjs, 173, 37

\bibitem[\protect\citeauthoryear{{Swinbank} et~al.,}{{Swinbank}
  et~al.}{2009}]{Swinbank09}
{Swinbank} A.~M.  et~al., 2009, \mnras, 400, 1121

\bibitem[\protect\citeauthoryear{{Tadhunter}, {Morganti}, {Robinson},
  {Dickson}, {Villar-Martin} \& {Fosbury}}{{Tadhunter}
  et~al.}{1998}]{Tadhunter98}
{Tadhunter} C.~N.,  {Morganti} R.,  {Robinson} A.,  {Dickson} R.,
  {Villar-Martin} M.,    {Fosbury} R.~A.~E.,  1998, \mnras, 298, 1035

\bibitem[\protect\citeauthoryear{{Tombesi}, {Cappi}, {Reeves}, {Palumbo},
  {Yaqoob}, {Braito} \& {Dadina}}{{Tombesi} et~al.}{2010}]{Tombesi10}
{Tombesi} F.,  {Cappi} M.,  {Reeves} J.~N.,  {Palumbo} G.~G.~C.,  {Yaqoob} T.,
  {Braito} V.,    {Dadina} M.,  2010, \aap, 521, A57

\bibitem[\protect\citeauthoryear{{Tremaine} et~al.,}{{Tremaine}
  et~al.}{2002}]{Tremaine02}
{Tremaine} S.  et~al., 2002, \apj, 574, 740

\bibitem[\protect\citeauthoryear{{Veilleux}}{{Veilleux}}{1991}]{Veilleux91}
{Veilleux} S.,  1991, \apjs, 75, 383

\bibitem[\protect\citeauthoryear{{Veilleux}, {Cecil} \&
  {Bland-Hawthorn}}{{Veilleux} et~al.}{2005}]{Veilleux05}
{Veilleux} S.,  {Cecil} G.,    {Bland-Hawthorn} J.,  2005, \araa, 43, 769

\bibitem[\protect\citeauthoryear{{Veilleux}, {Cecil}, {Bland-Hawthorn},
  {Tully}, {Filippenko} \& {Sargent}}{{Veilleux} et~al.}{1994}]{Veilleux94}
{Veilleux} S.,  {Cecil} G.,  {Bland-Hawthorn} J.,  {Tully} R.~B.,  {Filippenko}
  A.~V.,    {Sargent} W.~L.~W.,  1994, \apj, 433, 48

\bibitem[\protect\citeauthoryear{{Veilleux}, {Kim}, {Sanders}, {Mazzarella} \&
  {Soifer}}{{Veilleux} et~al.}{1995}]{Veilleux95}
{Veilleux} S.,  {Kim} D.-C.,  {Sanders} D.~B.,  {Mazzarella} J.~M.,    {Soifer}
  B.~T.,  1995, \apjs, 98, 171

\bibitem[\protect\citeauthoryear{{Veilleux} et~al.,}{{Veilleux}
  et~al.}{2013}]{Veilleux13}
{Veilleux} S.  et~al., 2013, \apj, 776, 27

\bibitem[\protect\citeauthoryear{{Veilleux} et~al.,}{{Veilleux}
  et~al.}{2009}]{Veilleux09}
{Veilleux} S.  et~al., 2009, \apjs, 182, 628

\bibitem[\protect\citeauthoryear{{Veron}}{{Veron}}{1981}]{Veron81}
{Veron} M.~P.,  1981, \aap, 100, 12

\bibitem[\protect\citeauthoryear{{Villar-Mart{\'{\i}}n}, {Binette} \&
  {Fosbury}}{{Villar-Mart{\'{\i}}n} et~al.}{1999}]{VillarMartin99}
{Villar-Mart{\'{\i}}n} M.,  {Binette} L.,    {Fosbury} R.~A.~E.,  1999, \aap,
  346, 7

\bibitem[\protect\citeauthoryear{{Villar-Mart{\'{\i}}n}, {Cabrera Lavers},
  {Bessiere}, {Tadhunter}, {Rose} \& {de Breuck}}{{Villar-Mart{\'{\i}}n}
  et~al.}{2012}]{VillarMartin12}
{Villar-Mart{\'{\i}}n} M.,  {Cabrera Lavers} A.,  {Bessiere} P.,  {Tadhunter}
  C.,  {Rose} M.,    {de Breuck} C.,  2012, \mnras, 423, 80

\bibitem[\protect\citeauthoryear{{Villar-Mart{\'{\i}}n}, {Humphrey}, {Delgado},
  {Colina} \& {Arribas}}{{Villar-Mart{\'{\i}}n} et~al.}{2011}]{VillarMartin11b}
{Villar-Mart{\'{\i}}n} M.,  {Humphrey} A.,  {Delgado} R.~G.,  {Colina} L.,
  {Arribas} S.,  2011, \mnras, 418, 2032

\bibitem[\protect\citeauthoryear{{Villar-Mart{\'{\i}}n}, {Tadhunter},
  {Humphrey}, {Encina}, {Delgado}, {Torres} \&
  {Mart{\'{\i}}nez-Sansigre}}{{Villar-Mart{\'{\i}}n}
  et~al.}{2011}]{VillarMartin11a}
{Villar-Mart{\'{\i}}n} M.,  {Tadhunter} C.,  {Humphrey} A.,  {Encina} R.~F.,
  {Delgado} R.~G.,  {Torres} M.~P.,    {Mart{\'{\i}}nez-Sansigre} A.,  2011,
  \mnras, 416, 262

\bibitem[\protect\citeauthoryear{{Vrtilek}}{{Vrtilek}}{1985}]{Vrtilek85b}
{Vrtilek} J.~M.,  1985, \apj, 294, 121

\bibitem[\protect\citeauthoryear{{Wagner}, {Bicknell} \& {Umemura}}{{Wagner}
  et~al.}{2012}]{Wagner12}
{Wagner} A.~Y.,  {Bicknell} G.~V.,    {Umemura} M.,  2012, \apj, 757, 136

\bibitem[\protect\citeauthoryear{{Wampler}, {Burbidge}, {Baldwin} \&
  {Robinson}}{{Wampler} et~al.}{1975}]{Wampler75}
{Wampler} E.~J.,  {Burbidge} E.~M.,  {Baldwin} J.~A.,    {Robinson} L.~B.,
  1975, \apjl, 198, L49

\bibitem[\protect\citeauthoryear{{Wang}, {Mao} \& {Wei}}{{Wang}
  et~al.}{2011}]{Wang11}
{Wang} J.,  {Mao} Y.~F.,    {Wei} J.~Y.,  2011, \apj, 741, 50

\bibitem[\protect\citeauthoryear{{Wang}, {Chen}, {Hu}, {Mao}, {Zhang} \&
  {Bian}}{{Wang} et~al.}{2009}]{Wang09}
{Wang} J.-M.,  {Chen} Y.-M.,  {Hu} C.,  {Mao} W.-M.,  {Zhang} S.,    {Bian}
  W.-H.,  2009, \apjl, 705, L76

\bibitem[\protect\citeauthoryear{{Wang}, {Rowan-Robinson}, {Norberg}, {Heinis}
  \& {Han}}{{Wang} et~al.}{2014}]{Wang14}
{Wang} L.,  {Rowan-Robinson} M.,  {Norberg} P.,  {Heinis} S.,    {Han} J.,
  2014, ArXiv::1402.4991

\bibitem[\protect\citeauthoryear{{Weedman}}{{Weedman}}{1970}]{Weedman70}
{Weedman} D.~W.,  1970, \apj, 159, 405

\bibitem[\protect\citeauthoryear{{Westmoquette}, {Clements}, {Bendo} \&
  {Khan}}{{Westmoquette} et~al.}{2012}]{Westmoquette12}
{Westmoquette} M.~S.,  {Clements} D.~L.,  {Bendo} G.~J.,    {Khan} S.~A.,
  2012, \mnras, 424, 416

\bibitem[\protect\citeauthoryear{{Whittle}}{{Whittle}}{1985}]{Whittle85}
{Whittle} M.,  1985, \mnras, 213, 1

\bibitem[\protect\citeauthoryear{{Whittle}}{{Whittle}}{1992}]{Whittle92}
{Whittle} M.,  1992, \apj, 387, 109

\bibitem[\protect\citeauthoryear{{Whittle}, {Pedlar}, {Meurs}, {Unger}, {Axon}
  \& {Ward}}{{Whittle} et~al.}{1988}]{Whittle88}
{Whittle} M.,  {Pedlar} A.,  {Meurs} E.~J.~A.,  {Unger} S.~W.,  {Axon} D.~J.,
   {Ward} M.~J.,  1988, \apj, 326, 125

\bibitem[\protect\citeauthoryear{{Whittle} \& {Wilson}}{{Whittle} \&
  {Wilson}}{2004}]{Whittle04}
{Whittle} M.,  {Wilson} A.~S.,  2004, \aj, 127, 606

\bibitem[\protect\citeauthoryear{{Wilson} \& {Heckman}}{{Wilson} \&
  {Heckman}}{1985}]{Wilson85}
{Wilson} A.~S.,  {Heckman} T.~M.,  1985, in {Miller} J.~S.,  ed., Astrophysics
  of Active Galaxies and Quasi-Stellar Objects. pp 39--109

\bibitem[\protect\citeauthoryear{{Wright} et~al.,}{{Wright}
  et~al.}{2010}]{Wright10}
{Wright} E.~L.  et~al., 2010, \aj, 140, 1868

\bibitem[\protect\citeauthoryear{{Xu}, {Livio} \& {Baum}}{{Xu}
  et~al.}{1999}]{Xu99}
{Xu} C.,  {Livio} M.,    {Baum} S.,  1999, \aj, 118, 1169

\bibitem[\protect\citeauthoryear{{Xu} \& {Komossa}}{{Xu} \&
  {Komossa}}{2009}]{Xu09}
{Xu} D.,  {Komossa} S.,  2009, \apjl, 705, L20

\bibitem[\protect\citeauthoryear{{York} et~al.,}{{York} et~al.}{2000}]{York00}
{York} D.~G.  et~al., 2000, \aj, 120, 1579

\bibitem[\protect\citeauthoryear{{Zakamska}, {G{\'o}mez}, {Strauss} \&
  {Krolik}}{{Zakamska} et~al.}{2008}]{Zakamska08}
{Zakamska} N.~L.,  {G{\'o}mez} L.,  {Strauss} M.~A.,    {Krolik} J.~H.,  2008,
  \aj, 136, 1607

\bibitem[\protect\citeauthoryear{{Zakamska} \& {Greene}}{{Zakamska} \&
  {Greene}}{2014}]{Zakamska14}
{Zakamska} N.~L.,  {Greene} J.~E.,  2014, arXiv:1402.6736

\bibitem[\protect\citeauthoryear{{Zakamska}, {Strauss}, {Heckman}, {Ivezi{\'c}}
  \& {Krolik}}{{Zakamska} et~al.}{2004}]{Zakamska04}
{Zakamska} N.~L.,  {Strauss} M.~A.,  {Heckman} T.~M.,  {Ivezi{\'c}} {\v Z}.,
  {Krolik} J.~H.,  2004, \aj, 128, 1002

\bibitem[\protect\citeauthoryear{{Zhang} et~al.}{{Zhang} et~al.}{2011}]{Zhang11}
{Zhang} K. et~al.,  2011, \apj, 737, 71

\bibitem[\protect\citeauthoryear{{Zubovas} \& {King}}{{Zubovas} \&
  {King}}{2012}]{Zubovas12}
{Zubovas} K.,  {King} A.,  2012, \apjl, 745, L34

\bibitem[\protect\citeauthoryear{{Zubovas} \& {King}}{{Zubovas} \&
  {King}}{2014}]{Zubovas14}
{Zubovas} K.,  {King} A.,  2014, arXiv:1401.0392

\end{thebibliography}
\end{document}